\documentclass[twocolumn,preprintnumbers,amsmath,amssymb]{revtex4}
\usepackage{graphicx}
\usepackage{dcolumn}
\usepackage{bm}
\usepackage{lscape} 

\def\be{\begin{equation}}
\def\ee{\end{equation}}
\def\br{\begin{eqnarray}}
\def\er{\end{eqnarray}}
\def\brn{\begin{eqnarray*}}
\def\ern{\end{eqnarray*}}

\begin{document}

\title{Theory and Applications of Coulomb Excitation}
\author{Carlos Bertulani}\email{carlos_bertulani@tamu-commerce.edu}
\affiliation{Texas A\& M University-Commerce, Commerce, TX 75429, USA}


\begin{abstract}
Because the interaction is well-known, Coulomb excitation is one of the best tools for the investigation of nuclear properties. In the last 3 decades new reaction theories for Coulomb excitation have been developed such as: (a) relativistic Coulomb excitation, (b) Coulomb excitation at intermediate energies, and (c) multistep coupling in the continuum. These developments are timely with the advent of rare isotope facilities. Of special interest is the Coulomb excitation and dissociation of weakly-bound systems.
I review the Coulomb excitation theory, from low to relativistic collision energies. Several applications of the theory to situations of interest in nuclear physics and nuclear astrophysics are discussed. 
\end{abstract}
\maketitle

{\sl \small Lecture notes presented at the 8th CNS-EFES Summer School,  held at Center for Nuclear Study (CNS), the University of Tokyo, and at the RIKEN Wako Campus,
August 26 - September 1, 2009. Supported by the Japan-US Theory Institute for Physics with Exotic Nuclei ({\bf JUSTIPEN}).

\vspace{2cm}
\noindent ``I don't know what they have to say, \\
It makes no difference anyway -- \\
Whatever it is, I'm against it! "  - Groucho Marx}

\vspace{2cm}

\tableofcontents

\section{What is Coulomb excitation?}

Coulomb excitation is a process of inelastic scattering in which a charged
particle transmits energy to the nucleus through the electromagnetic field.
This process can happen at a much lower energy than the necessary for the
particle to overcome the Coulomb barrier; the nuclear force is, in this way,
excluded in the process.

Let $v$ be the relative velocity of two nuclei at infinity which
determines the energy of relative motion $E=mv^{2}/2$ where $m$ is the
reduced mass. The strength of Coulomb
interaction can be measured by the {\it Sommerfeld parameter}
\begin{equation}
\eta=\frac{Z_{1}Z_{2}e^{2}}{\hbar v}                 \label{127}
\end{equation}
where $Z_{1,2}$ are charges of the nuclei. For $Z_{1}Z_{2}>137$, or $v\ll c$, the
parameter (\ref{127})  can easily reach values $\eta\gg 1$. 

\begin{figure}
[ptb]
\begin{center}
\includegraphics[
width=3.5in
]%
{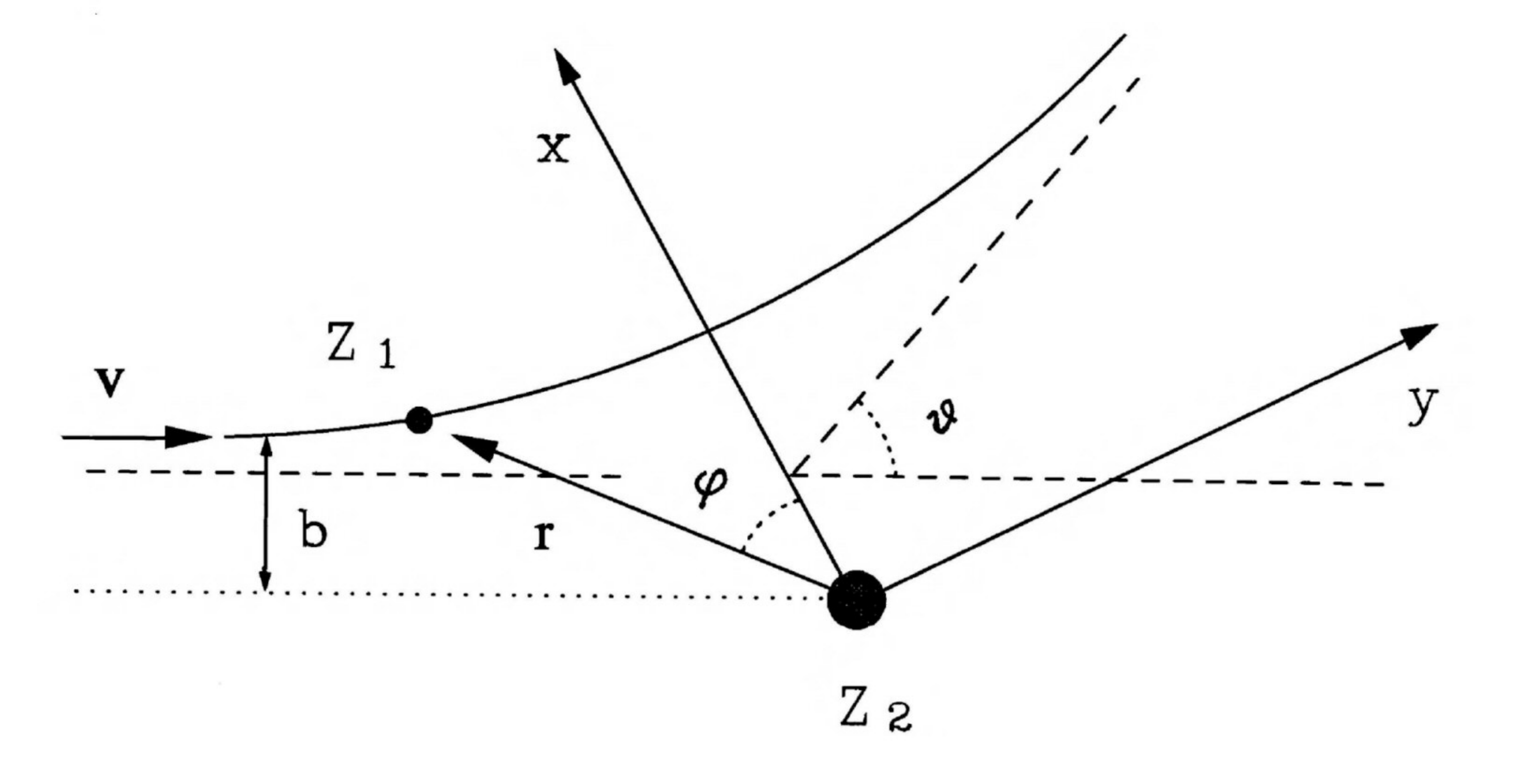}%
\caption{\sl Coulomb excitation occurs when a  charged projectile
passes by a target nucleus along a
Rutherford trajectory. The coordinates used in text are shown. }%
\label{coulex}%
\end{center}
\end{figure}

This situation allows for the use of the semiclassical approximation:
the Coulomb interaction is taken into account exactly in
determining the classical Rutherford trajectory of relative motion ${\bf R}(t)$
where ${\bf R}$ is the distance between the centers of the colliding
nuclei, Fig. \ref{coulex}. The relative energy $E$ is assumed to be
large enough so that we
can neglect the feedback from the intrinsic excitations to relative
motion. Then the trajectory is fixed by energy and impact parameter
or deflection angle.

The {\it  classical distance of closest approach},
\begin{equation}
a=2a_0={2Z_{1}Z_{2}e^{2}\over mv^{2}}                    \label{128}
\end{equation}
is larger than $R_{1}+R_{2}$ at
relative energy lower than the Coulomb barrier
\begin{equation}
E_{B}=\frac{Z_{1}Z_{2}e^{2}}{R_{1}+R_{2}}.        \label{129}
\end{equation}
The excitation is generated by the time-dependent field and the
probability of the process is determined by the presence in this field
of Fourier harmonics with the excitation frequencies
\begin{equation}\omega= {E_{f}-E_{i}\over \hbar}.\end{equation}
If the motion is too slow, the field acts adiabatically,
the intrinsic wave function is changing reversibly and the probability
of excitation is low. The corresponding {\it adiabaticity parameter}
is the ratio $\zeta$ of the time scales for the Coulomb collision, $\sim a_0/v$,
and for the nuclear excitation, $\sim 1/\omega$:
\begin{equation}\zeta={\omega a_0\over v}.\end{equation}
When $\zeta>1$,
the situation is adiabatic and transition probabilities are small.

The simplest treatment that one can give to the problem is a
semi-classical calculation, where the incident particle describes
a well defined trajectory, which is a classic hyperbolic
trajectory of Rutherford scattering (see figure \ref{coulex}).  It has been proven that this treatment is
valid in almost all situations studied in Coulomb excitation at low energies [AW75]. For high energy collisions, 
because the nuclear interaction distorts the scattering waves appreciably, a quantum treatment might be necessary
for some observables, e.g. angular distributions [BN93]. Fortunately, at high energies, one can use the eikonal approximation
for the scattering waves, which simplifies enormously the calculations.

The fundamental
review paper [Ald56] contains a great deal of information on the subject
and  even now
does not look obsolete, 40 years after its publication. However, in the last decades collisions between relativistic nuclei, with energies 
$E_{lab} > 50$ MeV/nucleon, have become a main tool of investigation in nuclear physics, in particular for the study of nuclei far from the
stability. Many new aspects of Coulomb excitation theory, such as the inclusion of transitions in/into the continuum, have been developed in the last two
decades. It is thus timely to discuss the theory of Coulomb excitation from low to relativistic energies. These notes, far from being complete, reviews the main aspects of the theory of Coulomb excitation. As the reader will notice, very little is said about the nuclear interaction as I want to emphasize the role of the Coulomb interaction in the excitation process. Also, nuclear  structure and nuclear excitation models are discussed only schematically and focused mainly on collective properties, such as the giant and pigmy resonances. 

The readers are encouraged to study the Appendix section, where many basic quantum mechanics tools are discussed. These tools will be used throughout the text.

Beforehand, I would like to thank the JUSTIPEN (Japan-USA Theory Institute for Physics with Exotic Nuclei) for the financial support. Special thanks to  Takaharu Otsuka, Takashi Nakatsukasa and Naoyuki Itagaki for the invitation to participate in this school and  for hosting my visit to Japan.

\section{Interaction of photons with matter}

\subsection{Electrostatic multipoles}

Electromagnetic multipoles  appear in classical field theory as a result of the
multipole expansion of the fields created by a finite system of
charges and currents. We start with the system of point-like classical
particles with electric charges $e_{i}$ located at the points $\mathbf{r}_{i}%
$.

The electrostatic potential of this system measured at the point
$\mathbf{r}$ is given by the Coulomb law,
\begin{equation}
V(\mathbf{r})=\sum_{i}\frac{e_{i}}{|\mathbf{r}-\mathbf{r}_{i}%
|}.\label{1.2.25}%
\end{equation}
The function
\begin{equation}
\frac{1}{|\mathbf{r}-\mathbf{r}^{\prime}|}=\frac{1}{\sqrt{r^{2}+r^{\prime
2}-2rr^{\prime}\cos\gamma}}\label{1.2.26}%
\end{equation}
depends on the lengths $r,r^{\prime}$ of two vectors and the angle $\gamma$
between them rather than on the angles of the vectors $\mathbf{r}$ and
$\mathbf{r}^{\prime}$ separately. If $\mathbf{r}\neq\mathbf{r}^{\prime}$, this
function has no singularities and can be expressed with the aid of the
expansion over the infinite set of Legendre polynomials with the coefficients
depending on $r$ and $r^{\prime}$,
\begin{equation}
\frac{1}{|\mathbf{r}-\mathbf{r}^{\prime}|}=\sum_{L=0}^{\infty}P_{L}(\cos
\gamma)f_{L}(r,r^{\prime}).\label{1.2.27}%
\end{equation}
Using the notations $r_{<}$ and $r_{>}$ for the smaller and the
greater $r$ and $r'$, one can show that the expansion (\ref{1.2.27})
takes the form
\begin{equation}
\frac{1}{|\mathbf{r}-\mathbf{r}^{\prime}|}=\sum_{L}\frac{r_{<}^{L}}%
{r_{>}^{L+1}}P_{L} (\cos\gamma).\label{1.2.32}%
\end{equation}

The applications of the multipole expansion usually consider the potential
(\ref{1.2.25}) outside the system, i.e. at the point $\mathbf{r}$
with $r>r_{i}$. Then we can use the expansion (\ref{1.2.32}) and the  addition
theorem 
\begin{equation}
P_{L}(\cos\gamma)= \frac{4\pi}{2L+1}\sum_{M}%
Y_{LM}(\mathbf{n}) Y^{\ast}_{LM}(\mathbf{n^{\prime}}),\label{C33}%
\end{equation}
where $\mathbf{n}$ ($\mathbf{n^{\prime}}$) is the direction of ${\bf r}$ (${\bf r^\prime}$). With that we
 get
\begin{equation}
V(\mathbf{r})=\sum_{LM}\frac{4\pi}{2L+1}\frac{1}{r^{L+1}}Y_{LM}^{\ast
}(\mathbf{n})\mathcal{M}(EL,M).\label{1.2.33}%
\end{equation}
Here the {\it electric multipole moment of rank} $L,\; L
=0,1,...$, is defined for a system of point-like charges
$i=1,2,...,A$ as a set of $(2L+1)$ quantities
\begin{equation}
\mathcal{M}(EL,M)=\sum_{i}e_{i}r_{i}^{L}Y_{LM}(\mathbf{n}_{i}),\quad
M=-L,-L+1,...,+L,\label{1.2.34}%
\end{equation}
where the sum runs over all charges $e_{i}$ located at $\mathbf{r}_{i}%
=(r_{i},\theta_{i},\varphi_{i})\equiv(r_{i},\mathbf{n}_{i})$. Exactly in the
same way one can define, instead of the charge distribution, multipole moments
for any other property of the particles, for example for the mass
distribution, $e_{i}\rightarrow \rho_{i}$.

In quantum theory, multipole moments are to be considered as operators acting
on the variables of the particles. Containing explicitly the spherical
functions, the operator $\mathcal{M}(EL,M)$ has the necessary features of the
tensor operator of rank $L$. Introducing the charge density operator
\begin{equation}
\rho(\mathbf{r})=\sum_{i}e_{i}\delta(\mathbf{r}-\mathbf{r}_{i}),\label{1.2.35}%
\end{equation}
we come to a more general form of the multipole moment,
\begin{equation}
\mathcal{M}(EL,M)=\int d^{3}r\,\rho(\mathbf{r})r^{L}Y_{LM}(\mathbf{n}%
),\quad\mathbf{n}=\frac{\mathbf{r}}{r}.\label{1.2.36}%
\end{equation}
In this form we even do not need to make an assumption of existence of
point-like constituents in the system; for example, in the nucleus charged
pions and other mediators of nuclear forces are included here along with the
nucleons if $\rho(\mathbf{r})$ is the total operator of electric charge
density. As expected, we can separate the geometry of multipole operators from
their dynamical origin. 

The lowest multipole moment $L=0$ is the \textit{monopole} one. It determines
the scalar part, the total electric charge $Ze$,
\begin{equation}
\mathcal{M}(E0,0)=\frac{1}{\sqrt{4\pi}}\sum_{i}e_{i}=\frac{1}{\sqrt{4\pi}}\int
d^{3}r\,\rho(\mathbf{r})=\frac{1}{\sqrt{4\pi}}Ze.\label{1.2.37}%
\end{equation}

The next term, $L=1$, defines the vector of the dipole moment
\begin{equation}
\mathbf{d}=\sum_{i}e_{i}\mathbf{r}_{i}=\int d^{3}r\,\rho(\mathbf{r}%
)\mathbf{r}.\label{1.2.38}%
\end{equation}
Taking into account the relation between the vectors and the
spherical functions of rank $L=1$, we obtain
\begin{equation}
\mathcal{M}(E1,M)=\sqrt{\frac{3}{4\pi}}\sum_{i}e_{i}r_{i}(n_{i})_{M}%
=\sqrt{\frac{3}{4\pi}}d_{M}.\label{1.2.39}%
\end{equation}

Subsequent terms of the multipole expansion determine the quadrupole ($L=2$),
octupole ($L=3$), hexadecapole ($L=4$), and higher moments.

In a similar way one can define magnetic multipoles $\mathcal{M}(ML,M)$
related to the distribution of currents. The convection current due to orbital
motion and the magnetization current generated by the spin magnetic moments
determine corresponding contributions to the {\it magnetic multipole moment of rank}
$L$,
\begin{equation}
\mathcal{M}(ML,M)=\sum_{i}\Bigl(g_{i}^{s}\mathbf{s}_{i}+\frac{2}{L+1}g_{i}%
^{L}\mathbf{l}_{i}\Bigr)\cdot\mbox{\boldmath$\nabla$}\Bigl(r_{i}^{L}Y_{LM}(\mathbf{n}%
_{i})\Bigr).\label{1.2.40}%
\end{equation}
Here $\mathbf{s}_{i}$ and $\mathbf{l}_{i}$ stand for the spin and orbital
angular momentum of a particle $i$, respectively; $g_{i}^{s}$ and $g_{i}^{L}$
are corresponding gyromagnetic ratios. (In this section, we
measure all angular momenta in units of $\hbar$ and the gyromagnetic ratios in
the magnetons $e\hbar/(2m_{i}c)$). 

The expression (\ref{1.2.40}) vanishes for
$L=0$ demonstrating the absence of magnetic monopoles. At $L=1$, we come to
the spherical components $\mu_{M}$ of the \textit{magnetic
moment }$\mbox{\boldmath$\mu$}$,
\begin{equation}
\mathcal{M}(M1,M)=\sqrt{\frac{3}{4\pi}}\mu_{M},\label{1.2.41}%
\end{equation}%
\begin{equation}
\mbox{\boldmath$\mu$}=\sum_{i}(g_{i}^{s}\mathbf{s}_{i}+g_{i}^{L}\mathbf{l}_{i}%
).\label{1.2.42}%
\end{equation}
Higher terms determine magnetic quadrupole, $L=2$, magnetic octupole, $L=3$,
and so on.

\subsection{Real photons}
\subsubsection{Radiative decay}
The probability for the radiative (by emission of a photon) transition for a nuclear transition
$i\rightarrow f$ integrated over angles of the emitted photon and summed
over its polarizations involves the same electromagnetic matrix elements as discussed above.
We will not derive the equations but only quote the results for the the probability for radiative decay [Ald56]:
\begin{eqnarray}
&&w_{fi}=\frac{8\pi(L+1)}{L[(2L+1)!!]^{2}\hbar}\kappa^{2L+1}\nonumber \\
&\times&\sum_{LM}\Bigl\{
\Bigl|\Bigl({\cal M}({\rm EL},M)\Bigr)_{fi}\Bigr|^{2}+\Bigl|\Bigl({\cal M}
({\rm ML},M)\Bigr)_{fi}\Bigr|^{2}\Bigr\}          \nonumber \\      \label{81}
\end{eqnarray}
where $\kappa=\omega/c$, and the multipole matrix elements for electric and magnetic transitions
are expressed via the current operator ${\bf j}({\bf r})$,
\begin{equation}
{\cal M}({\rm ML},M)=\int d^{3}r ({\bf j}\cdot
\hat{{\bf l}})\Phi_{LM},                              \label{82}
\end{equation}
\begin{equation}
{\cal M}({\rm EL},M)=\int d^{3}r \Bigl[(\mbox{\boldmath$\nabla$} \cdot {\bf j})\frac{1}{\kappa}\frac{
\partial}{\partial r}r - \kappa({\bf r}\cdot {\bf j})\Bigr]\Phi_{LM},
                                                        \label{83}
\end{equation}
${\bf l}=-i[{\bf r}\times\mbox{\boldmath$\nabla$}]$ is the orbital momentum
operator and the function $\Phi_{LM}$ arises from the partial wave
expansion of the plane wave,
\begin{equation}
\Phi_{LM}=-i\frac{(2L+1)!!}{L+1}\frac{1}{c\kappa^{L}}j_{L}(\kappa r)Y_{LM}(
{\bf n}),                                             \label{84}
\end{equation}
where $j_{L}(\kappa r)$ are spherical Bessel functions.
Of course,
for a given pair of states $f,i$ and a given multipolarity $L$, only
one term, either electric or magnetic, in (\ref{81}) works if parity is
conserved.

An important practical case is connected to the long wavelength
radiation: $\lambda\sim1/\kappa \gg R$ where $R$ is a size of the system.
In nuclei $R\approx 1.2\,A^{1/3}$ fm so that the condition
\begin{equation}
\kappa R\ll 1                                   \label{41}
\end{equation}
is equivalent to $\hbar \omega\ll 165 A^{-1/3}$ MeV which is usually
fulfilled. 
In the long wavelength approximation,  we can use the limiting values of
spherical Bessel functions $j_{L}(x)\approx x^{L}/(2L+1)!!$
at small arguments $x$. Then the
{\it magnetic multipoles} (\ref{82}) become
\begin{equation}
{\cal M}({\rm ML},M)=-\frac{i}{c(L+1)}\int d^{3}r \Bigl({\bf j}({\bf r})
\cdot \hat{{\bf l}}\Bigr)r^{L}Y_{LM}({\bf n})             \label{85}
\end{equation}
One can show that for the orbital current in Eq. \eqref{1.2.40}, this expression coincides with the
orbital part of the static magnetic moment [EG88]. The presence of spin implies magnetization
currents. In macroscopic electrodynamics such a current is $c [\mbox{\boldmath$\nabla$}
\times{\bf m}]$ where ${\bf m}$ stands for the magnetization density. The
analogous quantity for a quantum particle is
\begin{equation}
{\bf m}({\bf r})=\sum_{a}g^{s}_{a}{\bf s}_{a}\delta({\bf r}-
{\bf r}_{a}).                                      \label{86}
\end{equation}
The induced spin current is
\begin{equation}
{\bf j}_{{\rm spin}}=c\sum_{a}g_{a}^{s}[\mbox{\boldmath$\nabla$}\times{\bf s}_{a}]\delta({\bf r}
-{\bf r}_{a}).                        \label{87}
\end{equation}
Being substituted into (\ref{85}), this current reproduces the spin part
of static magnetic multipoles as in  \eqref{1.2.40}.

In the long wavelength limit, the second term of the electric multipole
(\ref{83}) is  smaller by a factor $(kR)^{2}$ than the first one. In the first
term $\boldsymbol\nabla \cdot {\bf j}$ reduces to the time derivative of the charge
density $\rho^{ch}$ with the aid of the continuity equation, \begin{equation}\boldsymbol\nabla \cdot {\bf j}+\partial \rho^{ch}/\partial t=0.\end{equation}
This
time derivative can be expressed through the  difference of energies between the initial and final
states gives $\hbar\omega_{{\bf k}}$. Performing the expansion of the
spherical Bessel functions we come to
\begin{equation}
{\cal M}({\rm EL},M)=\int d^{3}r \rho^{ch}({\bf r})r^{L}Y_{LM}({\bf n})
                                                            \label{88}
\end{equation}
which is a standard definition of the {\it electric multipoles}.

Usually the angular momentum projection $M_{f}$ of the final state is
not measured. Then we have to sum the transition rate over all $M_{f}$
defining the {\it reduced transition probability}
\begin{equation}
B({\rm EL};i\rightarrow f)=\sum_{MM_{f}}\Bigl|\Bigl({\cal M}({\rm EL},M)
\Bigr)_{fi}\Bigr|^{2}.                                \label{90}
\end{equation}
According to the {\it Wigner-Eckart theorem} [BM75],
the entire dependence of
the matrix element on the magnetic quantum numbers is concentrated in
the Clebsch-Gordan coefficients of vector addition. Using instead the Wigner
$3j$-symbols, we have for any tensor operator $T_{LM}$
\begin{eqnarray}
&&\langle J_{f}M_{f}|T_{LM}|J_{i}M_{i}\rangle = \nonumber \\
&&(-)^{J_{f}-M_{f}}
\left(\begin{array}{ccc}
J_{f}&L&J_{i}\\
-M_{f}&M&M_{i}\end{array}\right)\left<f\|T_{L}\|i\right>,             \label{90a}
\end{eqnarray}
where the reduced matrix element $\left<f\|T_{L}\|i\right>$ does not depend on projections.
We can use the orthogonality of the vector coupling coefficients
and perform the summation
over $M$ and $M_{f}$. 

The reduced transition probability is then related
to the reduced matrix element of the multipole operator,
\begin{equation}
B({\rm EL};i\rightarrow f)=\frac{1}{2J_{i}+1}\Bigl|(f\|{\cal M}
\left<{\rm EL})\|i\right>\Bigr|^{2}.                              \label{91}
\end{equation}
This quantity is convenient because it does not depend on the initial
population of various projections $M_{i}$. Note that for the inverse
transition induced by the same operator, the detailed balance
relation is valid,
\begin{equation}
B({\rm EL};f\rightarrow i)=\frac{2J_{i}+1}{2J_{f}+1}B({\rm EL};i\rightarrow f).
                                                       \label{92}
\end{equation}

The reduced transition probability determines the {\it partial lifetime} of a
given initial state with respect to a specific radiative decay (all
$M_{f}$ summed up),
\begin{equation}
\tau^{-1}_{i\rightarrow f}=\sum_{M_{f}}w_{fi}
=\frac{8\pi}{\hbar}\frac{L+1}{L[(2L+1)!!]^{2}}k^{2L+1}B({\rm EL};
i\rightarrow f).                                   \label{93}
\end{equation}
Here the kinematic factors are singled out. They are associated with the
geometry and phase space volume of the emitted photon. Information
concerning structure of the radiating system is accumulated in the
reduced transition probability. With the substitution EL$\rightarrow$ML,
the same expressions are valid for magnetic multipoles.

Classically, the electromagnetic radiation emitted by a system is the result
of the variation in time of the charge density or of the distribution of
charge currents in the system. The energy is emitted in two types of multipole
radiation: the electric and the magnetic. Each one of them is expressed as
function of the corresponding multipole moments, being the quantities which
contain the variables (charge and current) of the system. If the wavelength of
the emitted radiation is long in comparison to the dimensions of the system
(which is valid for a $\gamma$-ray of $\lesssim\,10$ MeV energy) the power
emitted by each multipole is given by ([Ja75], chap.16):
\begin{equation}
P_{E}(LM)={\frac{8\pi(L+1)c}{L[(2L+1)!!]^{2}}}\left(  {\frac{\omega}{c}%
}\right)  ^{2L+2}|{\cal M}(EL,M)|^{2}\label{8.1}%
\end{equation}
for electric multipole radiation and
\begin{equation}
P_{M}(LM)={\frac{8\pi(L+1)c}{L[(2L+1)!!]^{2}}}\left(  {\frac{\omega}{c}%
}\right)  ^{2L+2}|{\cal M}(ML,M)|^{2}.\label{8.2}
\end{equation}

In quantum mechanics, the energy is not emitted continually but in packets of
energy $\hbar\omega$. In a quantum calculation the disintegration constant is
the same as the number of quanta emitted per unit of time when the
power is given by the classical expressions \eqref{8.1} and \eqref{8.2}. Thus,
\begin{eqnarray}
\lambda_{E}(LM)&=&{\frac{P_{E}(LM)}{\hbar\omega}}\nonumber\\
&=&{\frac{8\pi(L+1)}{\hbar
L[(2L+1)!!]^{2}}}\left(  {\frac{\omega}{c}}\right)  ^{2L+1}|{\cal M}(EL;M)%
|^{2}\nonumber \\
\label{8.7}%
\end{eqnarray}
and
\begin{eqnarray}
\lambda_{M}(LM)&=&{\frac{P_{M}(LM)}{\hbar\omega}}\nonumber \\
&=&{\frac{8\pi(L+1)}{\hbar
L[(2L+1)!!]^{2}}}\left(  {\frac{\omega}{c}}\right)  ^{2L+1}|{\cal M}(M,LM)%
|^{2}.\nonumber \\\label{8.8}%
\end{eqnarray}
The decaying nucleus should also be treated as a quantum system. In this
sense, the expressions for the multipole moments
${\cal M}(EL,M)$ and ${\cal M}(ML,M)$ continue to
have validity if we use for the charge and current densities the quantum
expressions
\begin{equation}
\rho_{fi}(\mathbf{r})=\Psi_{f}^{\ast}(\mathbf{r})\Psi_{i}(\mathbf{r}%
),\label{8.9}%
\end{equation}%
\begin{equation}
\mathbf{j}_{fi}(\mathbf{r})={\frac{e\hbar}{2im}}\left[  \Psi_{f}^{\ast
}(\mathbf{\mbox{\boldmath$\nabla$}}\Psi_{i})-(\mathbf{\mbox{\boldmath$\nabla$}}\Psi_{f}^{\ast})\Psi_{i}\right]
,\label{8.10}%
\end{equation}
where the argument $i$ ($f$) denotes the initial (final) state described by the
wavefunction $\Psi_{i}\ (\Psi_{f}).$ Equations \eqref{8.9} and \eqref{8.10} refer
to a single nucleon with mass $m$ that emits radiation in its passage from the
state $i$ to the state $f$. Thus, a sum over all the protons should be
incorporated to the result, when we do the substitution of eqs. \eqref{8.9} and
\eqref{8.10} in eqs. \eqref{8.7} and \eqref{8.8} ([Ja75], chap. 16).

Not only are the values of the magnetic multipole moments are small in
comparison to the electric moments of same order, but also the
transition probability decreases quickly with increasing  $L$,
restricting the multipole orders that give significant contribution. To show
this, it is sufficient to observe that the product $(\omega/c)^{L}{\cal M}(EL,M)$
in \eqref{8.7} is at most equal to $Ze(\omega R/c)^{L}$, where $R$ is the radius
of the nucleus. For the energy that we consider, $\omega R/c$ is very small
implying that, for the larger powers of $L$, the disintegration constant is
also very small. These facts imply, in principle, that the electric dipole is
always the dominant radiation. But the selection rules that we will see next
can modify this situation.

\subsubsection{Selection Rules}
The conservation of
angular momentum and parity can prohibit certain $\gamma$ transitions between
two states. The selection rules for the $\gamma$-radiation are easy to
establish if we accept the fact that a quantum of radiation carries
an angular momentum \textbf{L} of module $\sqrt{L(L+1)}\,\hbar$ and component
$z$ equal to $M\hbar$, where $L$ is the multipolar order. Thus, in transition
between an initial spin $\mathbf{I}_{i}$ and a final spin $\mathbf{I}_{f}$ the
conservation of angular momentum imposes $\mathbf{I}_{i}=\mathbf{I}%
_{f}+\mathbf{L}$ and, in this way, the possible values for the multipole order
$L$ should obey
\begin{equation}
|I_{i}-I_{f}|\leq l\leq I_{i}+I_{f},\label{8.15}%
\end{equation}
where $|\mathbf{I}_{i}|=\sqrt{I_{i}(I_{i}+1)}\,\hbar$, etc. A special case is
the transition $0^{+}\rightarrow0^{+}$: as multipole radiation of order zero,
these transitions do not exist and they are effectively impossible through the
emission of a $\gamma$-ray. But in this case a process of internal
conversion can happen, where the energy is released by the ejection of an
atomic electron. 

Transitions between states of same parity can only
be accomplished by electric multipole radiation of even number or by magnetic
radiation of odd number. The inverse is valid for transitions where there is
parity change. Why this happens can be understood examining the definitions
of the multipole moments. The functions that compose the integrand have
definite parity and it is necessary that the integrand has even parity,
otherwise the contribution in \textbf{r} cancels with the contribution in
$-$\textbf{r} and the integral over the whole space vanishes. Let us look at
the case of  Eq.  \eqref{88}: $r^{L}$ is always positive and the spherical
harmonic $Y_{L}^{M}(\theta,\phi)$ is even if $L$ is even. For Eq.  \eqref{88}
to be non-zero, $\Psi_{i}$ should have the same parity of $\Psi_{f}$ for $L$ even
and opposite parity for $L$ odd. This justifies the transition rule for the
electric multipole radiation. A similar procedure applied to Eq.  \eqref{85}
justifies the selection rules for the magnetic multipole radiation.

Let us take an example: if the initial state is a $3^{+}$ and the
final a $2^{-}$, the possible values of $L$ will be 1, 2, 3, 4 and
5. But the change of parity restricts the transitions to E1, M2, E3,
M4 and E5, where E1 symbolizes an electric dipole transition
($L=1$), etc.

\subsubsection{Estimate of the Disintegration Constants}

\begin{figure}[t]
\begin{center}
\includegraphics[
height=0.9254in,
width=3.0372in
]%
{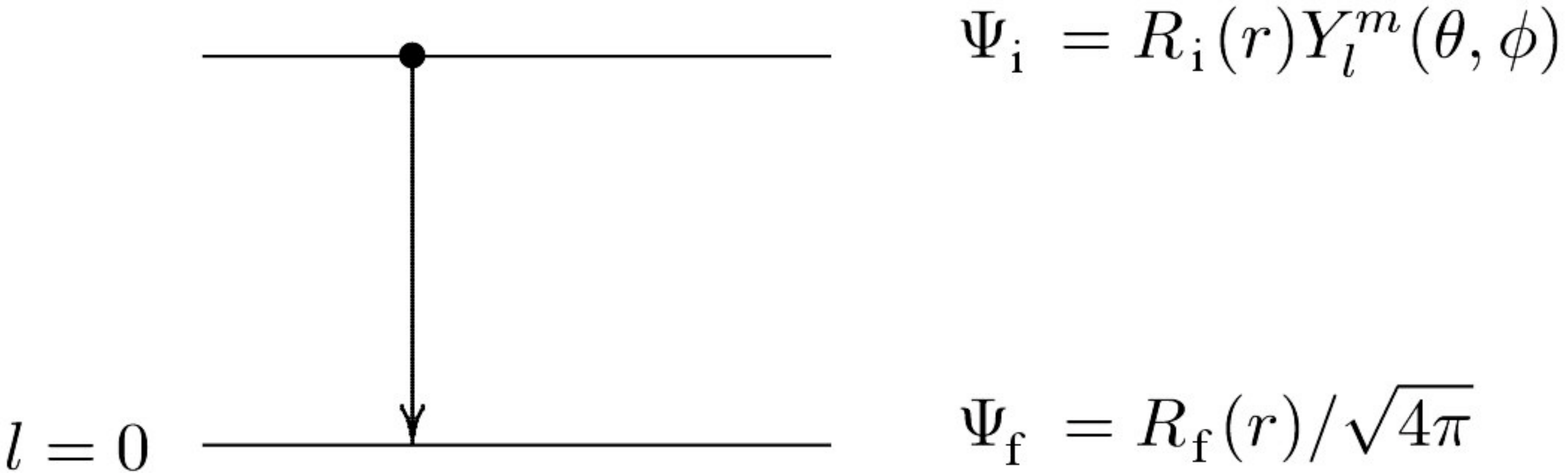}
\caption{\textsl{De-excitation of a proton to a level with }$L=0$\textsl{.}}%
\label{fig:8.1}%
\end{center}
\end{figure}

The use of Eq.  \eqref{8.7} and \eqref{8.8} for the calculation of the transition
probabilities in a real nucleus has the difficulty that wavefunctions that
appear in eqs. \eqref{8.9} and \eqref{8.10} are not known. But, a prediction of
their order of magnitude for the several modes can be done by assuming a single 
proton decaying from an
excited state described by the wavefunction $\Psi_{i}$ to a final state $f$
with $L=0$.  as shown in figure
\ref{fig:8.1}.%

For an approximate calculation it is enough to do: %
\begin{align}
R_{i}(r)  & =\text{const.}=R_{i}\qquad(r<R),\nonumber\\
R_{i}(r)  & =0\qquad(r>R)\label{8.16}%
\end{align}
and to use the same approach for $R_{f}(r)$. The normalization
yields immediately the values for the constants $R_{i}$ and $R_{f}$:
\begin{equation}
R_{i}=R_{f}=\sqrt{\frac{3}{R^{3}}},\label{8.17}%
\end{equation}
where $R$ is the nuclear radius. In this way, it is not difficult to calculate
the electric multipole moment ${\cal M}(EL,M)$:
\begin{equation}
{\cal M}(EL,M)=e\int r^{L}Y_{L}^{M\ast}(\theta,\phi){\frac{3}{R^{3}}}{\frac{Y_{L}%
^{M}(\theta,\phi)}{\sqrt{4\pi}}}r^{2}\,dr\,d\Omega,\label{8.18}%
\end{equation}
which yields
\begin{equation}
{\cal M}(EL,M)={\frac{3eR^{L}}{\sqrt{4\pi}(L+3)}}.\label{8.19}%
\end{equation}
Thus, in this approximation, the disintegration constant in Eq.  \eqref{8.7} is
\begin{equation}
\lambda_{E}={\frac{2(L+1)}{L[(2L+1)!!]^{2}}}\left(  {\frac{3}{L+3}}\right)
^{2}{\frac{e^{2}}{\hbar}}\left(  {\frac{E_{\gamma}}{\hbar c}}\right)
^{2L+1}R^{2L},\label{8.20}%
\end{equation}
where we wrote explicitly the disintegration energy
$E_{\gamma}=\hbar\omega$. A similar calculation for the magnetic
disintegration constant in Eq.  \eqref{8.8} yields
\begin{eqnarray}
\lambda_{M}&=&{\frac{20(L+1)}{L[(2L+1)!!]^{2}}}\left(  {\frac{3}{L+3}}\right)
^{2}{\frac{e^{2}}{\hbar}}\left(  {\frac{E_{\gamma}}{\hbar c}}\right)
^{2L+1}\nonumber \\
&\times&R^{2L}\left(  {\frac{\hbar}{mcR}}\right)  ^{2},\label{8.21}%
\end{eqnarray}
\noindent where $m$ is the mass of a nucleon. For a typical nucleus of
intermediate mass of $A=120$, it is easy to see that $\lambda_{E}/\lambda
_{M}\cong100$, independently of the multipolar order $L$. Evidently, both
constants decrease rapidly when the value of $L$ increases.

\begin{figure}[t]
\begin{center}
\includegraphics[
width=3.5in
]%
{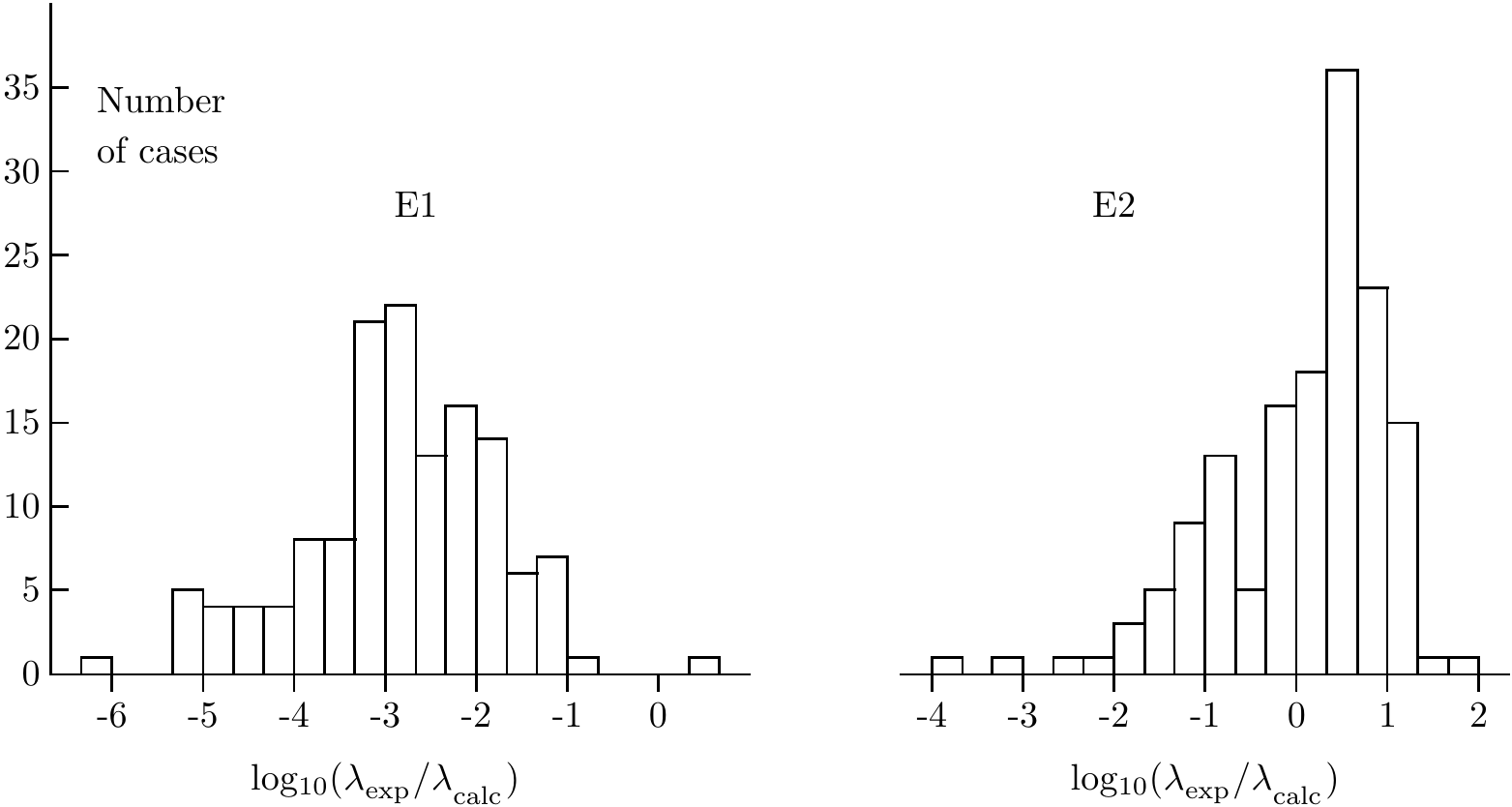}

\caption{\textsl{Distribution of the ratio between the experimental and
theoretical disintegration constants for transitions of the E1 and E2 type}
[Sk66]. }%
\label{fig:8.2}%
\end{center}
\end{figure}

Disagreements of several orders of magnitude
between the result of the calculation above and the corresponding experimental
values can happen. In particular, if the experimental disintegration rates are
smaller than the ones predicted by eqs. \eqref{8.20} and \eqref{8.21}, that can
mean that Eq.  \eqref{8.16} is not very reasonable and that the small
interception of the wavefunctions $\Psi_{i}$ and $\Psi_{f}$ decrease the
values of $\lambda$. Experimental values higher than predicted by eqs.
\eqref{8.20} and \eqref{8.21} can mean, on the other hand, that the transition
involves the participation of more than one nucleon or even a collective
participation of the whole nucleus.

\begin{figure}[t]
\begin{center}
\includegraphics[
width=3.5in
]%
{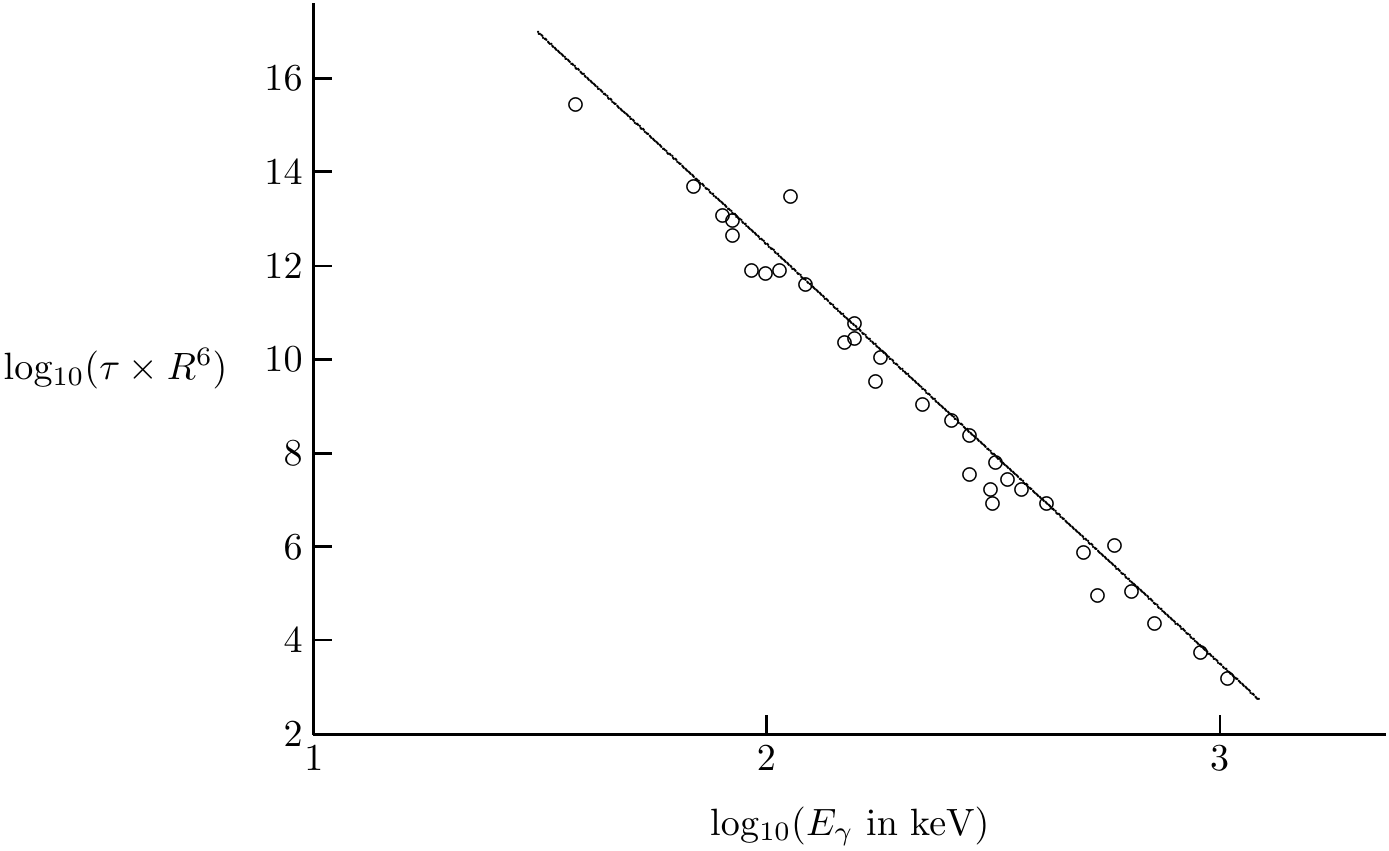}
\caption{$\tau_{(half-life)} >R^{6}$%
\textsl{\ (R=\ nuclear\ radius, $c=1$)\ as\ a\ function\ of\ the\ energy\ of\ the\ }%
$\gamma$%
\textsl{-ray\ for\ a\ series\ of\ transitions\ of\ the\ M4\ type.\ One\ sees\  good\ agreement\ between\ the\ theoretical\ estimate\ (straight\ line)\ and\ the\ experimental\ data [GS51].}%
}%
\label{fig:8.3}%
\end{center}
\end{figure}

Figures \ref{fig:8.2} and \ref{fig:8.3} illustrate the two situations: in
figure \ref{fig:8.2} the experimental values of $\lambda$ for transitions of
E1 multipolarity are orders of magnitude smaller than that calculated from Eq. 
\eqref{8.20}. The opposite happens with the E2 multipolarity where in most cases
the experimental rate is larger than calculated; this is due to the fact that
E2 transitions are common among levels of collective bands, especially
rotational bands in deformed nuclei. In figure \ref{fig:8.3}, on the other
hand, what one notices is  very good agreement between theoretical values and
experimental ones for M4. This behavior is typical of transitions of high
multipolarity.%

\section{Low energy collisions}
Coulomb excitation involves the same nuclear matrix elements as in radiative decay, but with different phase space factors.
The reason is that Coulomb excitation is a process in which the excitation occurs when nuclei
are outside their mutual charge distributions. Therefore, the Maxwell equation $\boldsymbol\nabla \cdot {\bf E} =0$ applies for the electric field inducing the transition. This means that only transverse photons fields, the same as for real photons, appear in Coulomb excitation [EG88]. 
In this section, we will prove this statement, and we will describe the semiclassical theory of Coulomb excitation for low energy collisions.

\subsection{Central collisions}

According to Eq.  (\ref{5.46}) of the Appendix C,
the probability of exciting the nucleus to a state $f$
above the ground state $i$ is%
\begin{equation}
a_{if}=-{\frac{i}{\hbar}}\int V_{if}\,e^{i\omega t}\,dt,\label{5.44a}%
\end{equation}
with $\omega=(E_{f}-E_{i})/\hbar$, is the probability amplitude that there
will be a transition $i\rightarrow f$. The matrix element
\begin{equation}
V_{if}=\int\Psi_{f}^{\ast}V\Psi_{i}\,d\tau\label{5.44b}%
\end{equation}
contains a potential $V$ of interaction of the incident particle with the
nucleus. The square of $a_{if}$ measures the transition probability from $i$
to $f$ and this probability should be integrated along the trajectory.

A simple calculation can be done in the case of the excitation of the ground
state $J=0$ of a deformed nucleus to an excited state with $J=2$ as a result
of a frontal collision with scattering angle of $\theta=180^{\circ}$. The
perturbation $V$ comes, in this case, from the interaction of the charge
$Z_{p}e$ of the projectile with the quadrupole moment of the target nucleus.
This quadrupole moment should work as an operator that acts between the
initial and final states. The way of adapting (\ref{1.2.33}) is
writing
\begin{equation}
V={\frac{1}{2}}{\frac{Z_{p}e^{2}Q_{if}}{r^{3}}},\label{10.94}%
\end{equation}
with
\begin{equation}
Q_{if}=\sum_{i}\int\Psi_{f}^{\ast}(3z_{i}^{2}-r_{i}^{2})\Psi_{i}%
\,d\tau,\label{10.95}%
\end{equation}
where the sum extends to all protons. The excitation amplitude is then written
as
\begin{equation}
a_{if}={\frac{Z_{p}e^{2}Q_{if}}{2i\hbar}}\int{\frac{e^{i\omega t}}{r^{3}}%
}\,dt.\label{10.96}%
\end{equation}

At an scattering of $\theta=180^{\circ}$ a relationship exists between the
separation $r$, the velocity $v$, the initial velocity $v_{0}$ and the
distance of closest approach $a=2a_0$:
\begin{equation}
v={\frac{dr}{dt}}=\pm v_{0}\left(  1-{\frac{a}{r}}\right)^{1/2}  ,\label{10.97}%
\end{equation}
which is obtained easily from the conservation of energy. Besides, if the
excitation energy is small, we can assume that the factor $e^{i\omega t}$ in
(\ref{10.96}) does not vary much during the time that the projectile is close
to the nucleus. Thus, this factor can be placed outside the integral and it
does not contribute to the cross section. One gets
\begin{equation}
a_{if}={\frac{Z_{p}e^{2}Q_{if}}{2i\hbar v_{0}}}\times2\int{\frac{dr}%
{r^{3}(1-a/r)^{1/2}}}.\label{10.98}%
\end{equation}
The integral is solved easily by the substitution $u=1-a/r$, in what results%
\begin{equation}
a_{if}={\frac{4Z_{p}e^{2}Q_{if}}{3i\hbar v_{0}a^{2}}}={\frac{4Q_{if}E^{2}%
}{3Z_{p}e^{2}\hbar v_{0}Z_{T}^{2}}},\label{10.99}%
\end{equation}
where the conservation of energy, $E={\frac{1}{2}}m_{0}v_{0}^{2}=Z_{p}%
Z_{T}e^{2}/a$ was used, with $m_{0}$ being the reduced mass of the
projectile+target system and $Z_{T}$ the atomic number of the target. The
differential cross section is given by the product of the Rutherford
differential cross section at 180$^{\circ}$ and the excitation probability
along the trajectory, measured by the square of $a_{if}$:
\begin{equation}
\left.  {\frac{d\sigma}{d\Omega}}\right\vert _{\theta=180^{\circ}}%
={\frac{d\sigma_{R}}{d\Omega}}\left\vert _{\theta=180^{\circ}}\right.
\times|a_{if}|^{2}.\label{10.100}%
\end{equation}
The Rutherford differential cross section is the classic expression
\begin{equation}
{\frac{d\sigma_{R}}{d\Omega}}=\left(  {\frac{Z_{p}Z_{T}e^{2}}{4E}}\right)
^{2}\sin^{-4}\left(  {\frac{\theta}{2}}\right) \label{10.101}%
\end{equation}
and, at $\theta=180^{\circ}$, we obtain
\begin{equation}
\left.  {\frac{d\sigma}{d\Omega}}\right\vert _{\theta=180^{\circ}}%
={\frac{m_{0}E|Q_{if}|^{2}}{18\hbar^{2}Z_{T}^{2}}},\label{10.102}%
\end{equation}
an expression that is independent of the charge of the projectile. It is, on
the other hand, proportional to the mass of the projectile, indicating that
heavy ions are more effective for Coulomb excitation.

The quadrupole moment operator $Q_{if}$ uses, as we saw, the wavefunctions
$\Psi_{i}$ and $\Psi_{f}$ of the initial and final states. If those two
wavefunctions are similar, as is the case of an excitation to the first level
of a rotational band, the operator $Q_{if}$ can be replaced by the intrinsic
quadrupole moment $Q$. The expression (\ref{10.102}) translates, in this way,
the possibility to evaluate the quadrupole moment from a measurement of a
value of the cross section.

\subsection{Electric excitations}

The interaction hamiltonian responsible for the excitation processes
in the nonrelativistic case can be written as
\begin{equation}
H'=\int d^{3}r_{1}d^{3}r_{2}\frac{\rho^{ch}_{1}({\bf r}_{1}-{\bf R}_{1})
\rho^{ch}_{2}({\bf r}_{2}-{\bf R}_{2})}{|{\bf r}_{1}-{\bf r}_{2}|}
- \frac{Z_{1}Z_{2}e^{2}}{R(t)}        \label{130}
\end{equation}
where charge densities of unperturbed nuclei depend on the distances from
the corresponding centers. We subtracted the interaction between nuclei
as a whole which determines the trajectory (time dependence of
${\bf R}={\bf R}_{1}-{\bf R}_{2}$) but does not
contribute to intrinsic excitations. We introduce the intrinsic coordinates
for each nucleus ${\bf x}_{1,2}={\bf r}_{1,2}-{\bf R}_{1,2}$,
\begin{equation}
H'=\int d^{3}x_{1}d^{3}x_{2}\frac{\rho_{1}^{ch}({\bf x}_{1})
\rho_{2}^{ch}({\bf x}_{2})}{|{\bf R}(t)+{\bf x}_{1}-{\bf x}_{2}|}
-\frac{Z_{1}Z_{2}e^{2}}{R(t)},                    \label{131}
\end{equation}
and carry out the multipole expansion for a large
distance between the centers, $R\gg x_{1,2}$,
\begin{eqnarray}
H'&=&\int d^{3}x_{1}d^{3}x_{2}\rho_{1}^{ch}({\bf x}_{1}) \rho_{2}^{ch}
({\bf x}_{2})\nonumber \\
&\times&\sum_{L >0,M }\frac{4\pi}{2L +1}\frac{x_{12}^{L}}
{R^{L +1}(t)}Y_{L M }({\bf x}_{12})Y_{L M }^{\ast}({\bf R}).  \label{132}
\end{eqnarray}
Here ${\bf x}_{12}\equiv {\bf x}_{1}-{\bf x}_{2}$.

This hamiltonian is rather complicated due to the correlations
associated with the mutual excitation of the nuclei. It becomes much
simpler if we are interested in the excitation of one of the partners
only. Let the ``projectile" {\sl 2} be not excited and we can neglect effects
related to its structure as an extended object. Then $\rho^{ch}_{2}
({\bf x}_{2})\approx Z_{2}e\delta({\bf x}_{2})$, and the hamiltonian is
expressed in terms of the electric
multipole moments of the ``target" {\sl 1},
\begin{equation}
H'=Z_{2}e\sum_{L >0,M }\frac{4\pi}{2L +1}\frac{1}{R^{L +1}(t)}Y^{\ast}_{L M }
({\bf R}(t)){\cal M}({\rm EL },M ).                   \label{133}
\end{equation}
The hamiltonian (\ref{133}) is time-dependent since the trajectory is
considered as a given function of time. According to Eq.  \eqref{5.46}
the transition amplitude
$i\rightarrow f$ with excitation by $\hbar \omega=E_{f}-E_{i}$ is
the Fourier component for the transition frequency
of the interaction hamiltonian taken along the unperturbed trajectory,
\begin{equation}
a_{fi}=-\frac{i}{\hbar}\int_{-\infty}^{\infty}dt H_{fi}'(t)e^{i\omega t}.
                                                  \label{134}
\end{equation}
For the unpolarized initial nuclei and with no final polarization
registered, the transition rate is to be averaged over initial
projections and summed over final projections of the target,
\begin{equation}
w_{fi}=\frac{1}{2J_{i}+1}\sum_{M_{f}M_{i}}|a_{fi}|^{2};   \label{135}
\end{equation}
the polarization state of the projectile is assumed to be unchanged.

The trajectory enters the result via the time integral
\begin{equation}
I_{LM}(\omega)=\int_{-\infty}^{\infty}dt \frac{1}{R^{L+1}(t)}Y_{LM}
({\bf R}(t))e^{i\omega t}.                          \label{136}
\end{equation}
This Fourier component
becomes small, $\propto \exp(-{\rm const}\cdot\zeta)$, if the
trajectory changes at too slowly a rate compared to the needed transition
frequency and the parameter of adiabaticity $\zeta>1$. We will discuss more about this
integral later.

The intrinsic matrix elements of multipole moments appear in the
transition rate (\ref{135}) in sums over magnetic quantum numbers
\begin{equation}
\Sigma_{LM,L'M'}=
\frac{1}{2J_{i}+1}\sum_{M_{f}M_{i}}{\cal M}_{fi}({\rm EL},M)
{\cal M}_{fi}^{\ast}({\rm EL'},M').       \label{137}
\end{equation}
The
summation in Eq.  \eqref{137} selects $L'=L, M'=M$ and the result does not depend on
$M$. 
\begin{eqnarray}
\Sigma_{LM,L'M'}&=&\frac{1}{2J_{i}+1}\sum_{M_{f}M_{i}}\Bigl|
{\cal M}_{fi}({\rm EL},M)\Bigr|^{2}\delta_{LL'}\delta_{MM'}\nonumber \\
&=&\frac{B({\rm EL};i\rightarrow f)}{2L+1}\delta_{LL'}\delta_{MM'},
                                                        \label{138}
\end{eqnarray}
where $B({\rm EL};i\rightarrow f)$ are given by Eq. \eqref{91}.

The total excitation probability is therefore
\begin{equation}
w_{fi}=\Bigl(\frac{4\pi Z_{2}e}{\hbar}\Bigr)^{2}\sum_{L>0}
\frac{B({\rm EL};i\rightarrow f)}{(2L+1)^{3}}\sum_{M}|I_{LM}
(\omega_{fi})|^{2}.                                  \label{139}
\end{equation}

From the viewpoint of the projectile the process is inelastic
scattering. The Coulomb trajectory defines the deflection angle $\theta$
and the  Rutherford cross section, Eq.  \eqref{10.101}.
In our approximation, the trajectory is not influenced by the target
excitation so that the inelastic cross section is factorized into the
product of the Rutherford cross section (\ref{10.101}) and the excitation
probability (\ref{139}),
\begin{equation}
d\sigma_{fi}=d\sigma_{R}w_{fi}=\sum_{L>0}d\sigma_{fi}({\rm EL})  \label{141}
\end{equation}
where the cross section for the excitation of multipolarity $L$ is equal
to
\begin{equation}
\frac{d\sigma_{fi}({\rm L})}{d\Omega}=\Bigl(\frac{\pi Z_{2}ea}
{\hbar\sin^{2}(\theta/2)}\Bigr)^{2}\frac{B({\rm EL};i\rightarrow f)}
{(2L+1)^{3}}\sum_{M}|I_{LM}(\omega_{fi})|^{2},            \label{142}
\end{equation}
where $a$ is the distance of closest approach, Eq. \eqref{128}.

This theory can be extended, considering quantum scattering instead of
classical trajectories, using relativistic kinematics, taking into account
magnetic multipoles which become equally important for relativistic
velocities, and including higher order processes of sequential excitation of
nuclear states. The last generalization is necessary for excitation of
rotational bands and overtones of giant resonances (quantum states
with several vibrational quanta). The mutual excitation of the
projectile and of the target can also be studied.

\subsection{Estimates}

We can make a crude estimate of the cross section of
Coulomb excitation. The trajectory integral (\ref{136}), after
changing the variable to $dR=v dt$, gives the dimensional factor
$a^{-L}/v$. It has to be taken near the closest approach point
(\ref{128}) which is the most effective for excitation. The constants
from the Rutherford cross section can be combined into the Coulomb
parameter (\ref{127}). As a result,
\begin{equation}
\sigma({\rm EL})\simeq \eta^{2}\frac{B({\rm EL})}{a^{2L-2}}
f_{L}(\zeta)                                                    \label{143}
\end{equation}
where the function $f_{L}(\zeta)$ depends smoothly on $L$ but contains
the exponential cut-off at large values of the adiabaticity parameter
$\zeta$. We remember, Eq.  (\ref{8.20}),
that in photoabsorption each consecutive multipole
was suppressed by a factor $(kR)^{2}$. The situation for exciting higher
multipoles is easier in the Coulomb excitation because here
\begin{equation}
\frac{\sigma({\rm E,L+1})}{\sigma({\rm EL})}\sim \frac{B({\rm E,L+1})}
{B({\rm EL})a^{2}}\sim \Bigl(\frac{R}{a}\Bigl)^{2}.
                                                            \label{144}
\end{equation}
This ratio is significantly larger than $(kR)^{2}$.

\subsection{Inclusion of magnetic interactions}

A more accurate description of Coulomb excitation for all scattering angles
requires the correct treatment of magnetic interactions and the Coulomb recoil
of the classical trajectories.  Here we will discuss the role of magnetic interactions.
Note that magnetic interactions induce electric transitions, too. And electric interactions
also induce magnetic transitions. Thus, in electromagnetic excitation, both interactions mix
and can only be isolated under special circumstances. We now show how magnetic interactions
modify the results obtained in the last section.

Following the derivation of the electromagnetic
interaction presented above, the excitation a target nucleus from
an initial rate $\mid i>$ to a final state $\mid f>$ is, to first order, given
by
\begin{equation}
a_{fi}=\frac{1}{i\hbar}\int_{-\infty}^{\infty}dt\;e^{i(E_{f}-E_{i})t/\hbar
}<\;f\mid\mathcal{H}_{int}\mid i>\label{(B.4)}%
\end{equation}
where (see Appendix D)
\begin{equation}
<f\mid\mathcal{H}_{int}\mid i>\;=\int\;\left[  \rho_{fi}\phi(\mathbf{r}%
,t)-\frac{1}{c}\;\mathbf{j}_{fi}(\mathbf{r})\cdot\mathbf{A}(\mathbf{r}%
,t)\right]  d^{3}r\label{(B.5)}%
\end{equation}
with $\rho_{fi}({\bf r})$ and ${\bf j}_{fi}({\bf r})$ given by Eqs. \eqref{8.9} and
\eqref{8.10}, respectively.

Thus,%
\begin{equation}
a_{fi}=\frac{1}{i\hbar}\int_{-\infty}^{\infty}dt\;e^{i\omega t}\left[
\rho_{fi}(\mathbf{r})-\frac{\mathbf{v}\left(  t\right)  }{c^{2}}%
\cdot\mathbf{j}_{fi}(\mathbf{r})\right]  \phi\left(  \mathbf{r},t\right)
d^{3}r.\label{(B.7)}%
\end{equation}
where \textbf{v} is the velocity of the projectile, and we used \begin{equation}\mathbf{A}%
(\mathbf{r},t)=\dfrac{\mathbf{v}\left(  t\right)  }{c}\phi\left(
\mathbf{r},t\right),  \end{equation} valid for a spinless projectile following a classical trajectory.\

The scalar and vector potentials at the target nucleus interior,
\begin{align}
\phi(\omega,\,\mathbf{r})  & =Z_{p}e\int_{-\infty}^{\infty}e^{i\omega
t}\,{\frac{1}{\mid
\mathbf{r-r^{\prime}}(t)\mid}}\,dt\label{3.14a}\\
\mathbf{A}(\omega,\,\mathbf{r})  & ={\frac{Z_{p}e}{c}}\int_{-\infty}^{\infty
}\mathbf{v^{\prime}}(t)\,e^{i\omega t}\,{\frac{1}{\mid\mathbf{r-r^{\prime}}(t)\mid}%
}\,dt\label{3.14b}%
\end{align}
are  generated by a
projectile with charge Z$_{p}$ following a Coulomb trajectory, described by
its time-dependent position $\mathbf{r}(t)$.

We now use the expansions of Eqs. \eqref{1.2.32} and \eqref{C33} and, 
assuming that the projectile does not penetrate the target, we use
$\mathbf{r}_{>}$ ($\mathbf{r}_{<}$) for the projectile (target) coordinates.

To perform the time-integrals, we need the time dependence, ${\bf r}'(t)$,
for a particle moving along the Rutherford
trajectory, which can be directly obtained by solving the equation of
angular momentum conservation (see [Go02]) for a given
scattering angle $\vartheta$\ in the center of mass system (see
figure \ref{coulex}). Introducing the parametrization
\begin{equation}
r(\chi)={a_{0}}\,\left[  \epsilon\,\cosh\chi+1\right] \label{3.10c}%
\end{equation}
where%
\begin{equation}
a_{0}={\frac{Z_{p}Z_{T}e^{2}}{m_{0}v^{2}}=}\frac{a}{2}{,\ \ \ \ \ \ \ }%
\text{{and}}{\ \ \ \ \ }\epsilon={1\over \sin\left(  \vartheta/2\right)} \label{3.5}%
\end{equation}
one obtains [Go02]%
\begin{equation}
t={\frac{a_{0}}{v}}\,\left[  \chi+\epsilon\,\sinh\chi\right]  \,.\label{3.11}%
\end{equation}
Using the scattering plane perpendicular to the z-axis, one finds that the
corresponding components of $\mathbf{r}$ may be written as
\begin{align}
x  & =a_{0}\,\left[  \cosh\chi+\epsilon\right],  \nonumber \\
y  & =a_{0}\,\sqrt{\epsilon^{2}-1}\,\sinh\chi ,\nonumber \\
z  & =0\ .\label{3.12c}%
\end{align}
The impact parameter $b$ in figure \ref{coulex} is related to the
scattering
angle ${\vartheta}$\ by%
\begin{equation}
b=a_{0}\cot\left(  {\vartheta/2}\right) \label{bimpct},
\end{equation}
and the eccentricity parameter $\epsilon$ is related to it by means of
\begin{equation}
\epsilon=\sqrt{1+{b^2\over a_0^2}} \label{bimpct2},
\end{equation}

In the limit of straight-line motion $\epsilon\simeq b/a_{0}\gg1$, and the
equations above reduce to the simple straight-line parametrization,
\begin{equation}
y=vt\ ,\ \ \ \ \ \ \ x=b\ ,\ \ \ \mathrm{and}\ \ \ z=0\ .
\end{equation}

Using the continuity equation, $\mathbf{\nabla}\cdot\mathbf{j}_{fi}%
=-i\omega\rho_{fi}$, for the nuclear transition current,  the terms in the expansion of the
scalar and vector potentials mix up. After a long but straightforward calculation, one can show that the
result can be expressed in terms of spherical tensors (see,
e.g., Ref. [EG88], Vol. II) and Eq.  (\ref{(B.7)}) becomes%
\begin{eqnarray}
a_{fi}&=&{\frac{Z_{p}e}{i\hbar}}\sum_{L M }\,{\frac{4\pi}{2L +1}%
}\,(-1)^{M }\nonumber \\
&\times& \Big[  S(EL ,\,M )\mathbf{\mathcal{M}}_{fi}
(EL ,\,-M )\nonumber\\
&+&S(ML ,\,M )\mathbf{\mathcal{M}}_{fi}(ML 
,\,-M )\Big] \label{a1st}%
\end{eqnarray}
where 
\begin{eqnarray}
\mathcal{M}_{fi}(EL ,M )  & =&{\frac{(2L +1)!!}{\kappa^{L 
+1}c(L +1)}}\nonumber \\
&\times&\int\mathbf{j}_{fi}(\mathbf{r})\cdot\boldsymbol\nabla \times
\mathbf{L}\left[  j_{L }(\kappa r)Y_{L M }(\hat{\mathbf{r}%
})\right]  d^{3}r\ ,\label{ME1lm} \nonumber \\ \\ 
\mathcal{M}_{fi}(ML ,M )  & =&-i{\frac{(2L +1)!!}{\kappa^{L 
}c(L +1)}}\nonumber \\
&\times&\int\mathbf{j}_{fi}(\mathbf{r})\cdot\mathbf{L}\left[
j_{L }(\kappa r)Y_{L M }(\hat{\mathbf{r}})\right]  d^{3}%
r\ .\label{MM1lm}%
\end{eqnarray}

The orbital integrals $S( \pi L  ,M $) are given by
\begin{align}
S(EL ,M )  & =-{\frac{i\kappa^{L +1}}{L (2L -1)!!}}%
\,\int_{-\infty}^{\infty}{\frac{\partial}{\partial r^{\prime}}}\Big\{
r^{\prime}(t)h_{L }\left[  \kappa r^{\prime}(t)\right]  \Big\}\nonumber \\
&\times Y_{L M }\left[  \theta^{\prime}(t),\phi^{\prime}(t)\right]
\,e^{i\omega t}\,dt\nonumber\label{3.19a}\\
& -{\frac{\kappa^{L +2}}{cL (2L -1)!!}}\,\int_{-\infty}%
^{\infty}\mathbf{v^{\prime}}(t)\cdot\mathbf{r}^{\prime}(t)\,h_{L }[\kappa
r^{\prime}(t)]\nonumber \\
&\times Y_{L M }[\theta^{\prime}(t),\phi^{\prime}%
(t)]\,e^{i\omega t}\,dt
\end{align}
and
\begin{eqnarray}
S(ML ,M )&=&-{\frac{i}{m_{0}c}}\,{\frac{\kappa^{L +1}}{L 
(2L -1)!!}} \mathbf{L}_{0}\cdot\int_{-\infty}^{\infty}\nabla^{\prime
}\Big\{ h_{L }[\kappa
r^{\prime}(t)]\nonumber\\
&\times& Y_{L M }[\theta^{\prime
}(t),\phi^{\prime}(t)]\Big\}  \,e^{i\omega t}\,dt\label{3.19b}%
\end{eqnarray}
where $\mathbf{L}_{0}$ is the angular momentum of relative motion, which is
constant:
\begin{equation}
L_{0}=a_{0}m_{0}v\cot{\frac{\vartheta}{2}\ .}\label{3.20}%
\end{equation}

In non-relativistic collisions
\begin{equation}
\kappa r^{\prime}={\frac{\omega r^{\prime}}{c}}={\frac{v}{c}}{\frac{\omega
r^{\prime}}{v}}<{\frac{v}{c}}\ll1\label{3.20b}%
\end{equation}
because when the relative distance $r^{\prime}$ obeys the
relations $\omega r^{\prime}/v\geq1$ the interaction becomes
adiabatic. Eq. \eqref{3.20b} is long wavelength
approximation, as discussed in connection to Eq. \eqref{41}. Then one uses the 
limiting form of $h_{L }$
for small values of its argument [AS64] to show that
\begin{equation}
S(EL ,M )=\int_{-\infty}^{\infty}{r^{\prime}}^{-L 
-1}(t)\,Y_{L M }\left\{  \theta^{\prime}(t),\phi^{\prime}(t)\right\}
\,e^{i\omega t}\,dt\label{3.21a}%
\end{equation}
and
\begin{eqnarray}
S(ML ,M )&=&-{\frac{1}{L  m_{o}c}}\,\mathbf{L}_{0}\cdot
\int_{-\infty}^{\infty}\nabla^{\prime}\Big\{  {r^{\prime}}^{-L 
-1}(t)\nonumber \\
&\times&Y_{L M }\left[  \theta^{\prime}(t),\phi^{\prime}(t)\right]
\Big\}  \,e^{i\omega t}\,dt\label{3.21b}%
\end{eqnarray}
which are the usual orbital integrals in the non-relativistic
Coulomb excitation theory with hyperbolic trajectories (see eqs.
(II.A.43) of Ref. {AW75]).

It is convenient to perform a translation of the integrand 
by $\chi\rightarrow\chi+i\left(
\pi/2\right)  $ [AW75]. This renders many simplifications in the calculations of the {\it orbital integrals}
$S( \pi L  ,M $), which become
\begin{align}
S(EL ,M ) &  =\dfrac{\mathcal{C}_{L M }}{va_{0}^{L }%
}\;I(EL ,M ),\nonumber\\
S(ML ,M ) &  =-\dfrac{{\mathcal{C}_{L +1,M }\,}}{L 
ca_{0}^{L }} [(2L +1)/(2L +3)]^{1/2}\nonumber \\
& \times [(L +1)^{2}%
-M ^{2}]^{1/2}\, \cot \left(  {\vartheta/2}\right)  {\ \
}I(ML 
,M )\ ,\label{S}%
\end{align}
with
\begin{equation}
\mathcal{C}_{L M }=\left\{
\begin{array}
[c]{ccl}%
\sqrt{\dfrac{2L +1}{4\pi}}\,\dfrac{\sqrt{(L -M )!(L +M )!}%
}{(L -M )!!(L +M )!!}\;(-1)^{(L +M )/2} &  &
\\
0 &, & 
\end{array}
\right. \label{Clm}%
\end{equation}
where the upper (lower) form is valid for $ L  +M =\mathrm{even}$  $({\rm odd})$
The (reduced) orbital integrals are given by
\begin{align}
I(EL ,M ) &
=\,e^{-\pi{\eta}/2}\,\int_{-\infty}^{\infty}d\chi \;\exp\left[
-{\zeta}\epsilon\cosh\chi+i{\zeta}\chi\right]  \nonumber \\
& \times \dfrac
{(\epsilon+i\sinh\chi-\sqrt{\epsilon^{2}-1}\cosh\chi)^{M }}{(i\epsilon
\sinh\chi+1)^{L +M }},\nonumber\\
I(ML ,M ) &  =I(E,L +1,M )\ ,\label{Obtnr}%
\end{align}
where%
\[
\zeta=\omega a_{0}/v.
\]

The square modulus of Eq. ~(\ref{a1st}) gives the probability of
exciting the target from the initial state $\mid
I_{i}M_{i}\rangle$ to the final state $\mid I_{f}M_{f}\rangle$ in
a collision with the center of mass scattering angle $\vartheta$.
If the orientation of the initial state is not specified, the
cross section for exciting the nuclear state of spin $I_{f}$ is
\begin{equation}
d\sigma_{i\rightarrow f}=\frac{a_{0}^{2}\epsilon^{4}}{4}\;\frac{1}{2I_{i}%
+1}\;\sum_{M_{i},M_{f}}\mid a_{fi}\mid^{2}\,d\Omega\,,\label{cross_1}%
\end{equation}
where $a_{0}^{2}\epsilon^{4}d\Omega/4$ is the elastic (Rutherford) cross
section. Using the Wigner-Eckart theorem, Eq. \eqref{90a}
and the orthogonality properties of the Clebsch-Gordan coefficients, one gets
(for more details, see [BP99])
\begin{equation}
{\frac{d\sigma_{i\rightarrow f}}{d\Omega}}=\frac{4\pi^{2}Z_{P}^{2}e^{2}}%
{\hbar^{2}}\;a_{0}^{2}\epsilon^{4}\;\sum_{ \pi L  M }{\dfrac{B(\pi
L ,I_{i}\rightarrow I_{f})}{(2L +1)^{3}}}\;\mid S( \pi L  
,M )\mid^{2}\ ,\label{cross_2.6}%
\end{equation}
where $\pi=E$ or $M$ stands for the electric or magnetic multipolarity, and
$B( \pi L  ,I_{i})$ are the reduced transition probability of Eq. \eqref{91}.

\subsection{Virtual photon numbers}

\subsubsection{Angular dependence}

Integration of (\ref{cross_2.6}) over all energy transfers
$E_\gamma =\hbar\omega$, and summation over all possible final
states of the projectile nucleus leads to
\begin{equation}
{\frac{d\sigma_{C}}{d\Omega}}=\sum_{f}\int{\frac{d\sigma_{i\rightarrow f}%
}{d\Omega}}\;\rho_{f}(E_\gamma)\;dE_\gamma\,,\label{4.4.6}%
\end{equation}
where $\rho_{f}(E_\gamma)$ is the density of final states of
the target with energy $E_{f}=E_{i}+E_\gamma$. Inserting Eq. 
(\ref{cross_2.6}) into Eq.  (\ref{4.4.6}) one finds
\begin{equation}
{\frac{d\sigma_{C}}{d\Omega}}=\sum_{ \pi L  }\;{\frac{d\sigma_{ \pi L  }%
}{d\Omega}}=\sum_{ \pi L  }\int{\frac{dE_\gamma}{E_\gamma}}%
\;{\frac{dn_{ \pi L  }}{d\Omega}}(E_\gamma)\;\sigma_{\gamma}^{ \pi L  
}(E_\gamma)\,,\label{4.5a}%
\end{equation}
where $\sigma_{\gamma}^{ \pi L  }$ are the \textit{photonuclear cross
sections} for a given multipolarity $ \pi L  $, given by%
\begin{equation}
\sigma_{\gamma}^{ \pi L  }(E_\gamma)={\frac{(2\pi)^{3}(L 
+1)}{L \left[  (2L +1)!!\right]  ^{2}}}\;\sum_{f}\rho_{f}%
(E_\gamma)\,\kappa^{2L -1}\;B( \pi L  ,I_{i}\rightarrow I_{f})\ .\label{photonn}
\end{equation}
The \textit{virtual photon numbers}, $n_{ \pi L  }(E_\gamma)$, are given
by
\begin{equation}
{\frac{dn_{ \pi L  }}{d\Omega}}={\frac{Z_{p}^{2}\alpha}{2\pi}}%
\;{\frac{L \left[  (2L +1)!!\right]  ^{2}}{(L +1)(2L 
+1)^{3}}}\;{\frac{c^{2}a_{0}^{2}\epsilon^{4}}{\kappa^{2(L -1)}}}\sum
_{M }\mid S( \pi L  ,M )\mid^{2}\,,\label{4.6.6}%
\end{equation}
where $\kappa=E_\gamma/\hbar c$, and $\alpha=1/137$.

In terms of the orbital integrals $I(EL ,M )$, given by Eq. 
(\ref{Obtnr}), and using the Eq.  (\ref{4.6.6}), we find for the
electric multipolarities
\begin{align}
{\frac{dn_{EL }}{d\Omega}}  & ={\frac{Z_{p}^{2}\alpha}{8\pi^{2}}%
}\;({\frac{c}{v}})^{2L }\;{\frac{L \left[  (2L +1)!!\right]
^{2}}{(L +1)(2L +1)^{2}}}\;\epsilon^{4}\,\zeta^{-2L 
+2}\nonumber\\
& \times\sum_{%
\genfrac{}{}{0pt}{}{M }{{L +M =even}}%
}{\frac{(L -M )!(L +M )!}{\left[
(L -M )!!(L 
+M )!!\right]  ^{2}}}\;\mid I(EL ,M )\mid^{2}\,.\label{4.7.6}%
\end{align}
In the case of magnetic excitations one obtains
\begin{align}
&{\frac{dn_{ML }}{d\Omega}}=   {\frac{Z_{p}^{2}\alpha}{8\pi^{2}}%
}\;({\frac{c}{v}})^{2(L -1)}\;{\frac{\left[  (2L +1)!!\right]  ^{2}%
}{L (L +1)(2L +1)^{2}}}\nonumber
\\
&\times \; \zeta^{-2L +2} \epsilon
^{4}\;(\epsilon^{2}-1)\nonumber\\
&\times  \sum_{%
\genfrac{}{}{0pt}{}{M }{{L +M =odd}}%
}{\frac{\left[  (L +1)^{2}-M ^{2}\right]  \,(L +1-M 
)!(L +1+M )!}{\left[  (L +1-M )!!(L +1+M )!!\right]  ^{2}}%
}\nonumber\\
&\times \mid I(ML ,M )\mid^{2}\,.\label{4.7.6b}
\end{align}

\subsubsection{Impact parameter dependence}

Since the impact parameter is related to the scattering angle by Eqs. \eqref{bimpct}
and \eqref{bimpct2}, we can also write
\begin{equation}
n_{ \pi L  }(E_\gamma,b)\equiv{\frac{dn_{ \pi L  }}{2\pi bdb}}%
={\frac{4}{a_{0}^{2}\epsilon^{4}}}\;{\frac{dn_{ \pi L  }}{d\Omega}%
}\label{4.9.6}%
\end{equation}
which are interpreted as the number of equivalent photons of energy
$E_\gamma=\hbar\omega$, incident on the target per unit area, in a
collision with impact parameter $b$. The impact parameter dependence
of the Coulomb excitation cross section is
\begin{equation}
{\frac{d\sigma_{C}}{2\pi b db}}=\sum_{ \pi L  }\int{\frac{dE_\gamma}{E_\gamma}}
\;{n_{ \pi L  }(E_\gamma,b)}\;\sigma_{\gamma}^{ \pi L  
}(E_\gamma)\,.\label{4.5d}%
\end{equation}

The total cross section for Coulomb excitation is obtained by
integrating Eq.  (\ref{4.5d}) over $b$ from a minimum impact
parameter $b_{\min}$, or equivalently, integrating Eq. \eqref{4.5a} over
$\vartheta$ up to a maximum scattering angle $\vartheta_{\max}$, i.e.
\begin{eqnarray}
\sigma_C=\sum_{\pi L}\int {dE_\gamma\over E_\gamma} N_{\pi L} (E_\gamma)
\sigma^{\pi L}_\gamma (E_\gamma),\label{virtt1}%
\end{eqnarray}
where the total number of virtual photons is
\begin{eqnarray}
 N_{\pi L} (E_\gamma) &=&2\pi\int_{b_{\min}}^\infty db\ b\ n_{\pi L} (E_\gamma , b)\nonumber \\&=&
2\pi \int_0^{\vartheta_{\max}} d\theta \sin\theta {dn_{ \pi L  }\over d\Omega}.
\label{virtt2}%
\end{eqnarray}

This condition is necessary for collisions at high energies to
avoid the situation in which the nuclear interaction with the
target becomes important. At very low energies, below the Coulomb
barrier, $b_{\min}=0$ and $\vartheta_{\max}=180^{0}$.

\begin{figure}
[ptb]
\begin{center}
\includegraphics[
width=2.9132in
]%
{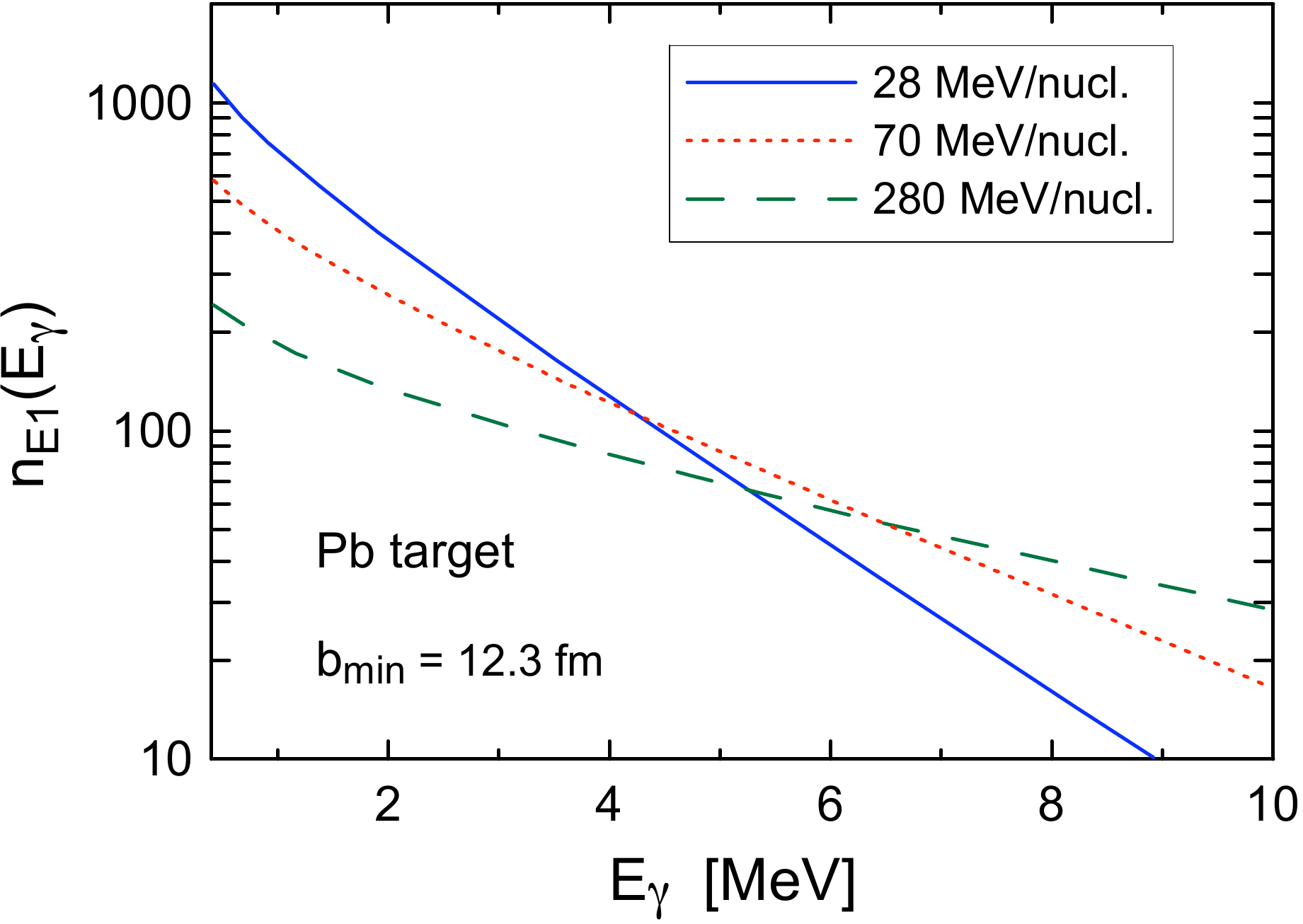}%
\caption{\sl
Total number of virtual photons for the $E1$ multipolarity, ``as seen'' by a
projectile passing by a lead target  at impact parameters $b_{min}=12.3$ fm and
larger (i.e., integrated over impact parameters), for three typical bombarding energies.} 
\label{vpn}
\end{center}
\end{figure}

The concept of virtual photon numbers is very useful, specially
in high energy collisions. In such collisions\ the momentum and the energy
transfer due to the Coulomb interaction are related by $\Delta p=\Delta
E/v\simeq\Delta E/c$. This means that the virtual photons are almost real. One
usually explores this fact to extract information about real photon processes
from the reactions induced by relativistic charges, and vice-versa. This is
the basis of the \textit{Weizs\"{a}cker-Williams method } [Fe24,WW34] (it should be called {\it Fermi's method} - see historical note later),
used to calculate cross sections for Coulomb excitation, particle production,
Bremsstrahlung, etc., (see, e.g., Ref. [Ja75,BB88]). 

\subsubsection{Virtual and real photons}
We have shown that even at low energies 
the cross sections for Coulomb excitation can 
be written as a product of  equivalent photon numbers and the cross sections induced by real
photons. The reason for this is the assumption that
Coulomb excitation is a process which involves only collisions for which the
nuclear matter distributions do not overlap at any point of the classical
trajectory. The excitation of the target nucleus thus occurs in a region where
the divergence of the electric field is zero, i.e. $\mathbf{\nabla\cdot E}%
_{p}\left(  t\right)  =0$, where $\mathbf{E}_{p}\left(  t\right)  $ is the
electric field generated by the projectile at the target's position. This
condition implies that the electromagnetic fields involved in Coulomb
excitation are exactly the same as those involved in the absorption of a real
photon [EG88].

\subsubsection{Analytical expression for E1 excitations}

For the $E1$ multipolarity the orbital integrals, Eq.
(\ref{Obtnr}), assumes a particularly simple form. It can be performed analytically 
for  $M=0,\pm 1$ [AW66]. Using these results, one gets the
compact expression for the E1 virtual photon numbers
\begin{eqnarray}
{\frac{dn_{E1}}{d\Omega}}&=&{\frac{Z_{p}^{2}\alpha}{4\pi^{2}}}\;\left(
{\frac{c}{v}}\right)  ^{2}\;{\epsilon^{4}\;\zeta^{2}}\;\;e^{-\pi{\zeta}%
}\;\Big\{  {\frac{\epsilon^{2}-1}{\epsilon^{2}}}\;\left[  K_{i{\zeta}%
}({\epsilon\zeta})\right]  ^{2}\nonumber \\
&+&\left[  K_{i{\zeta}}^{\prime}({\epsilon\zeta
})\right]  ^{2}\Big\}  \ ,\label{nE1ann1}%
\end{eqnarray}
where $K_{i\zeta}$ is the modified Bessel function with imaginary index, 
\begin{equation}
K_{i\zeta}(z)={\Gamma(i\zeta +{1\over 2}) (2z)^{i\zeta}\over \sqrt{\pi}}
\int_0^\infty dt {\cos t \over (t^2+z^2)^{i\zeta+{1\over 2}}},
\end{equation}
$K_{i\zeta
}^{\prime}$ is the derivative with respect to its argument.%

This result is not particular useful, as one still has to perform a time integration in the equation above. However, as we will see later, the above formula will help us to understand the connection with relativistic 
Coulomb exctitation.

In Figure \ref{vpn} we show a calculation (with $E_\gamma \equiv E_x$ =  excitation energy) of the
virtual photons for the $E1$ multipolarity, ``as seen'' by a  
projectile passing by a lead target at impact parameters equal to and exceeding
$b=12.3$ fm, for three  typical bombarding energies. As the projectile energy  
increases, more virtual photons of large energy are available for the
reaction. This increases the number of states accessed in the excitation
process. 

\subsection{Higher order corrections and quantum scattering}

The results presented in this section are only valid if the excitation is of
first-order. For higher-order excitations one has to use the coupled-channels
equations (\ref{5.43}) of Appendix B, with the excitation amplitudes in each time interval
given by Eq.  (\ref{a1st}). Important cases of applications of Coulomb excitation require a non-perturbative treatment of the collision process. We will discuss some of these cases in later sections.

A full quantum calculation for Coulomb
excitation uses scattering waves for the projectile (and target).
The cross section in first order perturbation theory is given by Eq. \eqref{10.76} of Appendix B,
with the transition matrix element of Eq. \eqref{10.77} given by
\begin{eqnarray}
&&V_{fi}=<f\mid\mathcal{H}_{int}\mid i>\nonumber \\
&=&\int\;\left[  \rho_{fi}({\bf r})\phi_{\beta\alpha}(\mathbf{r}%
,{\bf r}')-\frac{1}{c}\mathbf{j}_{fi}(\mathbf{r})\cdot\mathbf{A}_{\beta\alpha}(\mathbf{r}%
,{\bf r}')\right]  d^{3}r d^3r' ,\nonumber \\\label{(higho1)}%
\end{eqnarray}
where $\phi_{\beta\alpha}({\bf r},{\bf r}')$ and ${\bf A}_{\beta\alpha}({\bf r},{\bf r}')$ being the electromagnetic potential generated by scattering waves $\chi_{\alpha , \beta}({\bf r}')$ in the initial and final
scattering states, $\alpha$  and $\beta$, respectively.

Instead of the Coulomb gauge, i.e. with the Coulomb potential proportional to $1/|{\bf r}-{\bf r}'|$,
it is better to adopt the Lorentz gauge, for which the scalar potential is given by [AW75]
\begin{align}
\phi({\bf r},\mathbf{r}')   =Z_{p}e\chi_\beta^{(-)*}({\bf r}'){\frac{e^{i\kappa\mid\mathbf{r-r^{\prime}}(t)\mid}}{\mid
\mathbf{r-r^{\prime}}\mid}}\chi_\alpha^{(+)}({\bf r}'),\label{higho2}
\end{align}
and a similar expression for ${\bf A}({\bf r},{\bf r}')$, with the expression for the transition current replacing the product of the outgoing and incoming waves, i.e. $\chi_\beta^{(-)*}\chi_\alpha^{(+)} \rightarrow 
(\hbar/2i\mu)[\chi^{(-)*}_\beta\boldsymbol\nabla \chi_\alpha^{(+)}-\chi^{(+)}_\alpha\boldsymbol\nabla \chi_\beta^{(-)*}]$.

In Eq. \eqref{higho2} one uses well-known expressions for Coulomb waves [BD04]. 
One also uses the expansion
\begin{equation}
{\frac{e^{i\kappa\mid\mathbf{r-r^{\prime}}\mid}}{\mid\mathbf{r-r^{\prime}}%
\mid}}=4\pi\ i\kappa\sum_{LM}j_{L}(\kappa r_{<})Y_{LM
}^{\ast}(\hat{\mathbf{r}}_{<})h_{L}(\kappa r_{>})\,Y_{LM}%
(\hat{\mathbf{r}}_{>})\ ,\label{3.15b}%
\end{equation}
where $j_{L}$ ($h_{L}$) denotes the spherical Bessel (Hankel)
functions (of first kind), $\mathbf{r}_{>}$ ($\mathbf{r}_{<}$) refers to
whichever of $\mathbf{r}$ and $\mathbf{r}^{\prime}$ has the larger (smaller)
magnitude. Assuming that the projectile does not penetrate the target, one uses
$\mathbf{r}_{>}$ ($\mathbf{r}_{<}$) for the projectile (target) coordinates.

From here on, the calculation is tedious, but  straightforward. A detailed description is found in Ref. [AW75]. More insight into this calculation are not useful because on can show that the quantum treatment of the relative motion between 
the nuclei does not alter the results of the semiclassical calculations for $\eta \gg 1$. Thus, we can safely
use the machinery of the previous sections to have an accurate and reliable description of Coulomb excitation at low energies. For collisions at high energies, the distortion of the scattering waves due to the nuclear interaction yield important modifications of the angular distribution of the relative motion. We will discuss this in more details later.

\begin{figure}
[t]
\begin{center}
\includegraphics[
width=3.1695in
]%
{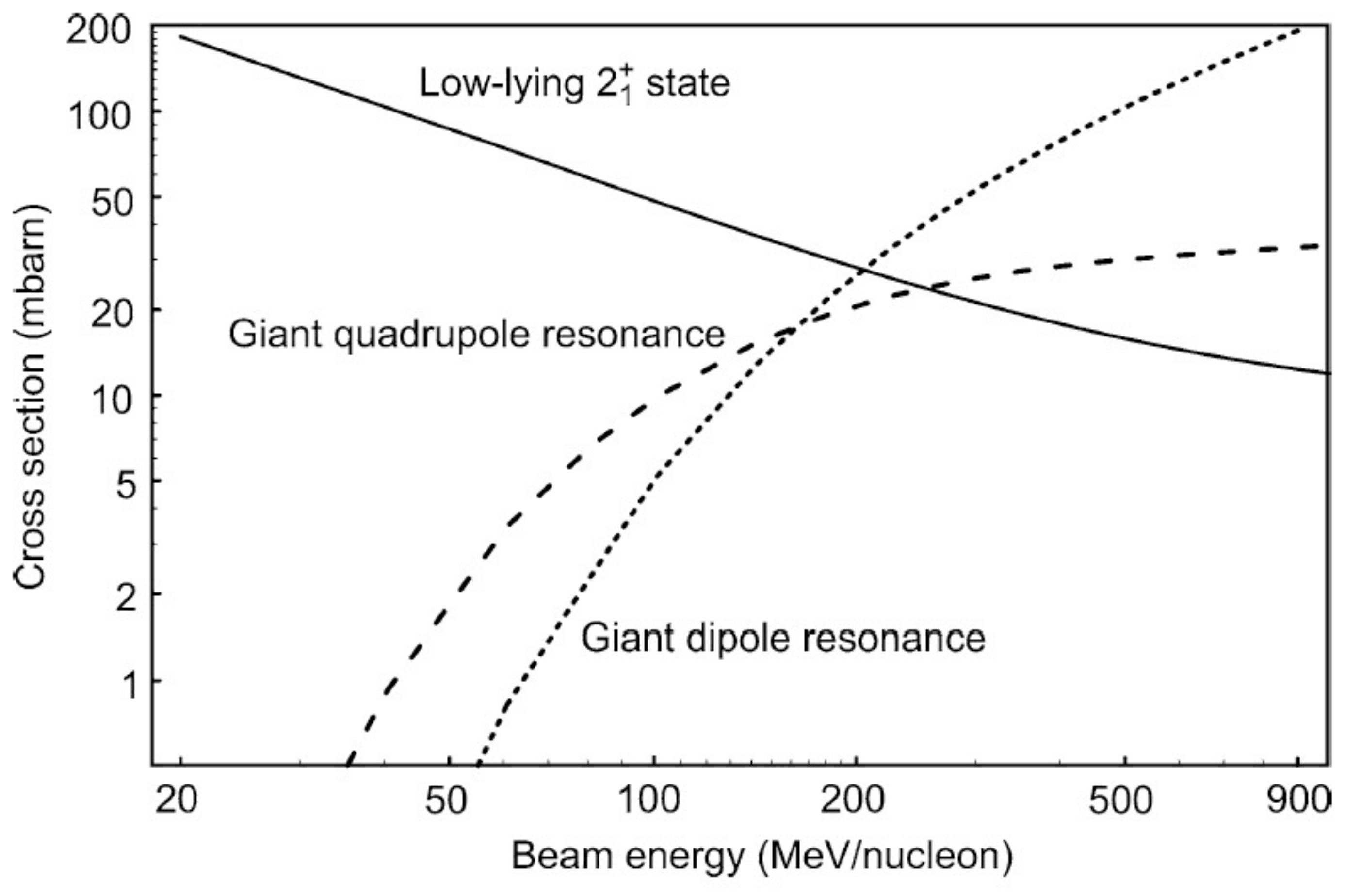}%
\caption{\sl Cross sections for the Coulomb excitation versus incident
beam energy for different collective states in the $^{40}$S$\
+^{197}$Au reaction,
assuming a minimum impact parameter $b_{\min}=16$ fm (from Ref. [Gla01]%
).}%
\label{glasmacher}%
\end{center}
\end{figure}

As the bombarding energy increases Coulomb excitation
predominantly favors the excitation of high lying states, e.g.
giant resonances. This is shown in figure \ref{glasmacher} for the
cross sections of Coulomb excitation versus
incident beam energy for different collective states in the $^{40}$%
S$\ +^{197}$Au reaction, assuming a minimum impact parameter $b_{\min}=16$ fm.
Later we will discuss more
about the excitation of giant resonances.

\section{Angular distribution of $\gamma$-rays}

Coulomb excitation is a useful method to obtain static quadrupole
moments as well as the reduced probabilities for several nuclear
transitions. In order to identify the multipolarity of the
excitation it is often necessary to study the de-excitation of the
excited state by measuring a $\gamma$-ray from its decay (see
figure \ref{cascade1}).  

A detailed description of $\gamma$-ray emission following
excitation is give in Appendix E. The angular distribution of the gamma rays
emitted into solid angle $\Omega_{\gamma}$, as a function of the
scattering angle of the projectile $\vartheta$, is given by Eq. \eqref{watheta}, i.e.,
\begin{equation}
W(\theta_{\gamma})=1+\sum_{k=2,4,...}b_{k}^{ \pi L  }\left(  \vartheta
\right)  P_{k}\left(  \cos\theta_{\gamma}\right)  \ ,\label{stat0}%
\end{equation}
where $P_{k}\left(  \cos\theta_{\gamma}\right)  $ are the Legendre
polynomials. The coefficients $b_{k}^{ \pi L  }\left(  \vartheta\right)
$\ are related to the Coulomb excitation amplitudes (\ref{a1st}) and to the
B-values (\ref{91})\ for the transition from the excited state $f$ to a
final state $g$.

\begin{figure}[tb]
\begin{center}
\includegraphics[width=3.5in]{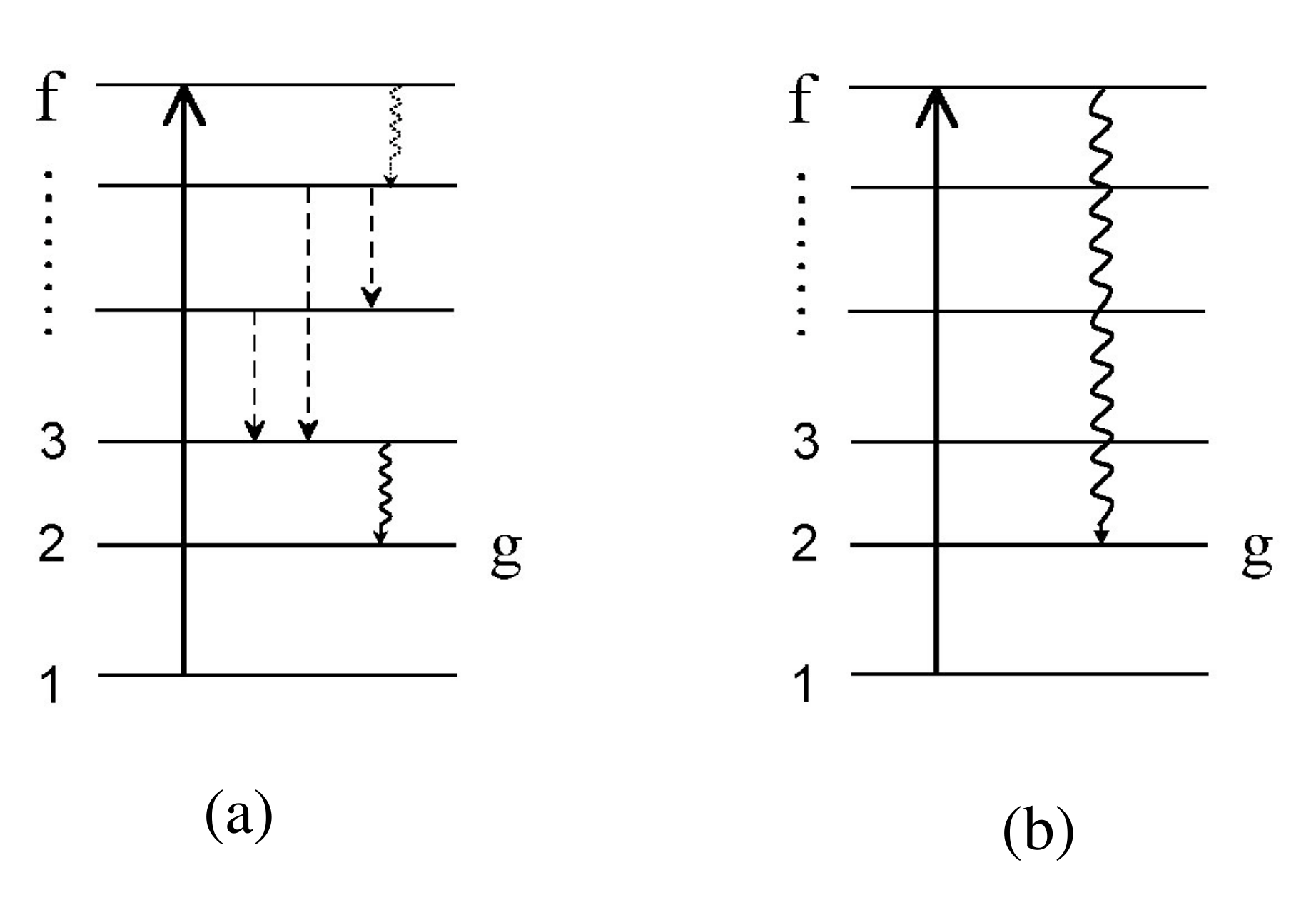}
\end{center}
\par
\vspace*{0pt}\caption{\sl  Schematic description of a nuclear excitation (solid
line) followed by $\gamma$-decay (solid wavy line). (a) The dashed lines are
transitions due to internal conversion (unobserved). The dashed wavy line is
an unobserved $\gamma$-decay. (b) Direct emission of an observed gamma ray. }%
\label{cascade1}%
\end{figure}

Figure \ref{wa2} shows the Coulomb excitation of sodium by protons. The yield
of the 446 keV $\gamma$-rays is shown [Tem55]. Between the resonances due
to compound nucleus formation one observes a smoothly rising background yield
which may be ascribed to Coulomb excitation. It is possible to determine the
multipole order of the Coulomb excitation by comparing with the yield observed
in the Coulomb excitation with $\alpha$-particles [Tem55]. The dashed
curves correspond to the cross sections expected for $L =1$ and 2 on the
basis of the observed cross section for the excitation with $\alpha
$-particles. The close agreement of the measured cross section with the
theoretical curve for $E2$ excitation also confirms that the yield away from
resonances is primarily due to Coulomb excitation.%

\begin{figure}
[ptb]
\begin{center}
\includegraphics[
width=2.9132in
]%
{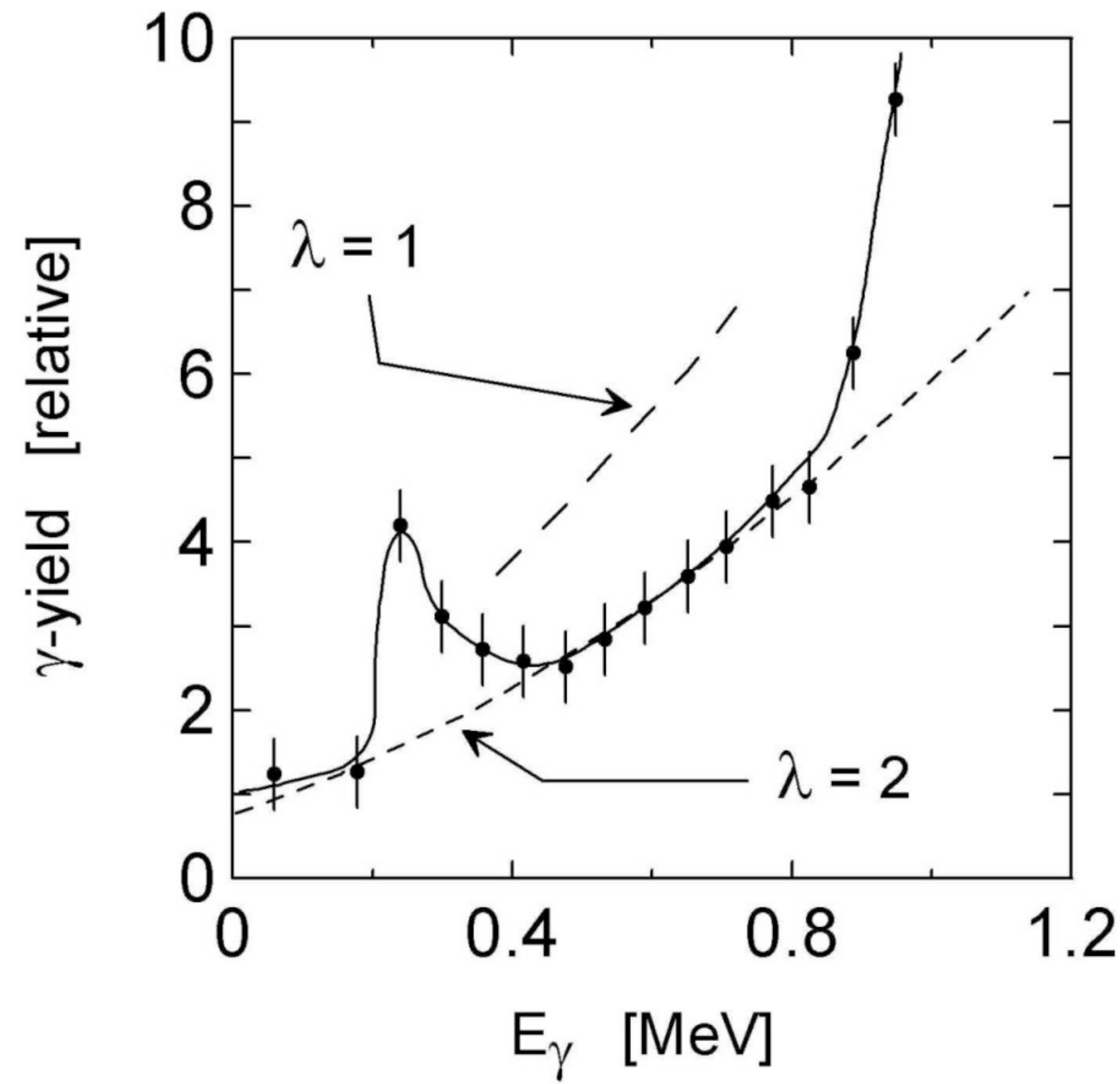}%
\caption{\sl  Coulomb excitation of sodium by protons. The yield of the
446 keV $\gamma$-rays is shown [Tem55]. The dashed curves
correspond to the cross sections expected for $L =1$ and 2 on
the basis of the observed cross
section for the excitation with $\alpha$-particles.}%
\label{wa2}%
\end{center}
\end{figure}

Figure \ref{wa1} shows the $\gamma$-ray yield from Coulomb excitation and
compound nucleus formation in $^{19}$F bombarded with $\alpha$-particles. The
dashed curve shows the yields of the 114 keV\ $\gamma$-ray from the first
excited state in $^{19}$F and the solid curve shows the 1.28 MeV\ $\gamma$-ray
from the first excited state of $^{22}$Ne formed by an $\left(  \alpha
,\text{p'}\right)  $ process on $^{19}$F [She54]. For bombarding energies
\ below 1.2 MeV, the penetration of the $\alpha$-particle through the Coulomb
barrier is very small and the cross section for compound nucleus formation is
small compared to that for Coulomb excitation. With increasing energy, the cross section for compound nucleus formation,
$\sigma_{\mathrm{CN}}$, increases rapidly and soon becomes larger than the Coulomb excitation cross section,
$\sigma_{C}$. However, even for $E_{\alpha}\sim2$ MeV, at which energy the
average value of $\sigma_{\mathrm{CN}}$ is an order of magnitude larger than
$\sigma_{C}$, the yield of the 114 keV $\gamma$-ray is only very little
affected by the compound nucleus formation, since the probability that the
compound nucleus decays by inelastic $\alpha$-emission is small. Finally, for
$E_{\alpha}\geq2$ MeV, the Coulomb excitation yield of the 114 keV $\gamma
$-ray is overshadowed by the resonance yield from compound nucleus formation.%

\begin{figure}
[ptb]
\begin{center}
\includegraphics[
width=3.4in
]%
{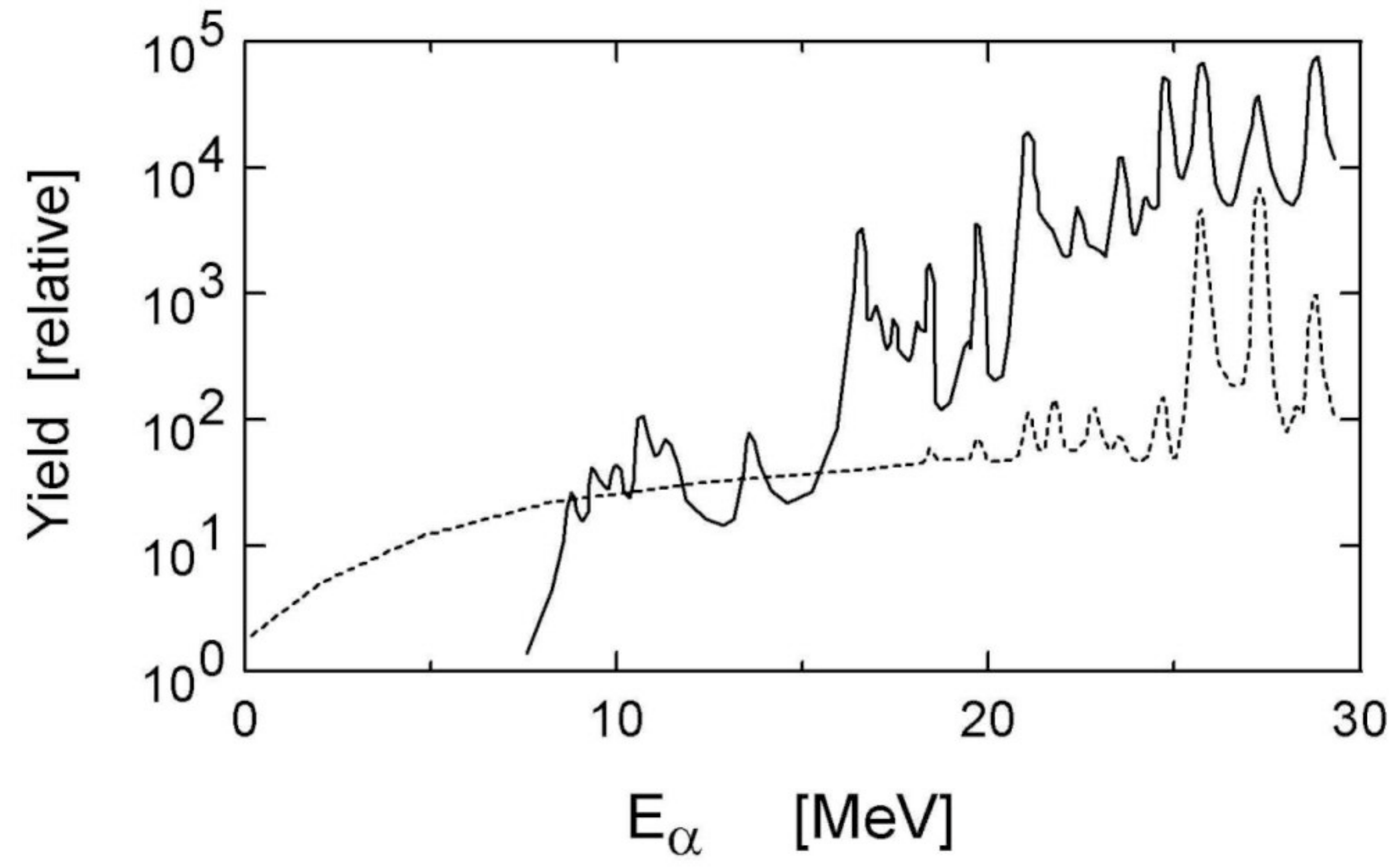}%
\caption{\sl  $\gamma$-rays from Coulomb excitation and compound
nucleus formation in $^{19}$F bombarded with $\alpha$-particles.
The dashed curve shows the yields of the 114 keV\ $\gamma$-ray
from the first excited state in $^{19}$F and the solid curve shows
the 1.28 MeV\ $\gamma$-ray from the first excited state of
$^{22}$Ne formed by an $\left(  \alpha,\text{p'}\right)  $
process on $^{19}$F [She54].}%
\label{wa1}%
\end{center}
\end{figure}

\section{Relativistic collisions}

\subsection{Multipole expansion}

When the projectile has very high energies, e.g. $E_{lab}>100 $ MeV/nucleon, there is very little
deflection of the ion trajectory. The recoil by the target is also minimal. We  thus assume that the projectile moves on a
straight-line trajectory of impact parameter $b$, which is also the distance of
closest approach between the centers of mass of the two nuclei at time $t = 0$. We
will consider the situation where $b$ is larger than the sum of the two nuclear
radii, $R$, such that the charge distributions of the two nuclei do not overlap at any
time. We will use a coordinate system with origin in the center of mass of
the excited nucleus and with z-axis along the projectile velocity $v$, which is assumed to be
constant. In this coordinate system the electromagnetic field from this
other nucleus is given by the Lienard-Wiechert expression
\begin{equation}
\phi \left( \mathbf{r},t\right) =\frac{\gamma Ze}{\sqrt{(x-b)^{2}+y^{2}+\gamma^2(z-vt)^{2}}}.  \label{(B.1)}
\end{equation}

We have chosen the x-axis in the plane of the trajectory such that the x-component
of  the trajectory is $b$. This expression reduces to the non-reletivistic Coulomb field of an low energy
charge given by $={Ze}/{\mid \mathbf{r}-\mathbf{R}(t)\mid 
}$. The appearance of the factors $\gamma=(1-v^2/c^2)^{-1/2}$ are due to retardation (see [Ja75]).
The vector potential is
\begin{equation}
\mathbf{A}(\mathbf{r},t)=\frac{\mathbf{v}}{c}\phi (\mathbf{r},t),
\end{equation}
with ${\bf v}=v\hat{\bf z}$ being a constant velocity vector.

The procedure for obtaining the excitation amplitude in \eqref{(B.7)} should be the same as adopted
before: first we make a multipole expansion of \eqref{(B.1)}, then we separate the intrinsic from the
relative motion coordinates, and perform the time integrals for the trajectories (we named them orbital integrals). 
However, as shown in Ref. [AW79], it is better to first calculate the time integrals and then perform the multipole
expansion. In fact, this seems to be the only way to get  analytical expressions, except when one uses a more 
complicated approach as we will discuss later.

The integral in Eq.  \eqref{3.14a} yields
\begin{equation}
\phi (\mathbf{r},\omega)={2Ze\over v} e^{i\omega v/z} K_0 \left( {\omega \over v} q\right), \label{relc3}
\end{equation}
where $K_0$ is a modified Bessel function and 
\begin{equation}
q={1\over \gamma^2}\left[(b-x)^2+y^2\right].
\end{equation}

In Ref. [AW79] it was shown that the multipole expansion of \eqref{relc3} is given by
\begin{equation}
\phi (\mathbf{r},\omega)=\sum_{LM} W_{LM}({\bf r},\omega) Y^\ast_{LM}(\hat{\bf r}),
\end{equation}
with
\begin{eqnarray}
&&W_{LM}({\bf r},\omega)=\sqrt{16\pi(2L+1)}\left({(L-M)!\over
(L+M)!}\right)^{1/2}(2M-1)!!\nonumber\\
&&\times i^{L+M}{Ze\over v} \left({c\over \gamma v}\right)^M K_M\left({\omega b\over \gamma v}\right)
C_{L-M}^{M+1/2}\left({c\over v}\right)
j_L({\kappa r}).\nonumber \\
\end{eqnarray}

The quantity $C_{M}^{N}(x)$ is the Gegenbauer polynomial [AS64], while $j_m(\kappa r)$ is a spherical Bessel
function, with $\kappa=\omega/c$. For $M<0$,
\begin{equation}
W_{L,-M}({\bf r},\omega)=(-1)^M W_{LM}({\bf r},\omega)
\end{equation}

After separation of the internal degrees of freedom, one gets [AW79]
\begin{eqnarray}
a_{fi}&=&-i{Ze\over \hbar v\gamma} \sum_{\pi LM} G_{\pi LM}\left({c\over v}\right)
(-1)^MK_M\left({\omega b\over \gamma v}\right) \nonumber \\
&\times& \sqrt{2L+1} \kappa^L {\cal M}_{fi}(\pi L,-M) \label{relcoul5}
\end{eqnarray}
where the transition matrix elements are given by ${\cal M}_{fi}(\pi L,-M)$ are given by Eqs. \eqref{ME1lm} and \eqref{MM1lm}.

The functions $G_{\pi LM}$ have analytical expressions [WA79].
For the $E1$, $E2$, $E3$, $M1$ and $M2$ multipolarities, they are given by  
\begin{eqnarray}
G_{E10}(x)&=&-i{4\over 3} \sqrt{\pi (x^{2}-1)}; \notag \\
G_{E11}(x) &=&-G_{E1-1}(x)={1\over 3} x\sqrt{8\pi };\notag \\
G_{M11}(x) &=&G_{M1-1}(x)=-i{1\over 3} \sqrt{8\pi };\ \ \
G_{M10}(x)=0;  \notag \\
G_{E22}(x) &=&G_{E2-2}(x)=-{ 2\over 5} x\sqrt{{\pi\over 6} (x^{2}-1)};  \notag
\\
G_{E21}(x) &=&-G_{E2-1}(x)=i{ 2\over 5} \sqrt{\pi \over 6}(2x^{2}-1);\ \ \ 
\notag \\
G_{E20}(x) &=&{ 2\over 5} x\sqrt{\pi (x^{2}-1)};\notag \\
G_{M22}(x)&=&-G_{M2-2}(x)=i{2\over 5}\sqrt{\pi\over 6}\sqrt{x^2-1};\notag\\
G_{M21}(x)&=&G_{M2-1}(x)={2\over 5}x\sqrt{\pi\over 6};\ \ \ G_{M20}=0; \notag\\
G_{E33}(x)&=&-G_{E3-3}(x)={1\over 21}\sqrt{\pi\over 5}x(x^2-1);\notag\\
G_{E32}(x)&=&G_{E3-2}(x)=-i{1\over 21}\sqrt{2\pi \over 15}(3x^2-1)\sqrt{x^2-1};\notag\\
G_{E31}(x)&=&-G_{E3-1}(x)=-{1\over 105}\sqrt{\pi\over 3}x(15x^2-11)\nonumber\\
G_{E30}(x)&=&i{2\over 105}\sqrt{\pi} (5x^2-1)\sqrt{x^2-1}.  \label{Glm}\nonumber \\
\end{eqnarray}

From the excitation amplitude \eqref{relcoul5} one finds the total cross section
for exciting the nuclear state of spin $I_f$ in collisions with impact parameters larger
than $R$ given by,
\begin{eqnarray}
\sigma_{fi}&=&\left({Ze\over \hbar c}\right)^2 \sum_{\pi LM} \kappa^{2(L-1)} B(\pi L, I_i \rightarrow I_f)\nonumber \\
&\times&  \left| G_{\pi LM}\left({c\over v}\right) \right|^2
g_M\left(\xi (R) \right) ,
\label{relcoul6}
\end{eqnarray}
where 
\begin{eqnarray}
\xi(R)={\omega R\over \gamma v}, 
\label{relcoul7}
\end{eqnarray}
and, for $M\ge 0$,
\begin{eqnarray}
g_M(\xi)&=&g_{-M}(\xi)=2\pi\left( {\omega \over \gamma v}\right)^2 \int_R^\infty db \ b\ K_M^2 \left(\xi(b)\right)\nonumber \\
&=&\pi \xi^2\left[ K_{M+1}^2-K_M^2-{2M\over \xi}K_{M+1}K_M\right],
\label{relcoul7b}
\end{eqnarray}
where all $K_{M }$'s are functions of $\xi (b)={\omega b/}\gamma v$.

\subsection{Excitation probabilities and virtual photon numbers}

The theoretical results for relativistic Coulomb excitation can be rewritten in terms of the
virtual photon numbers, as shown in Ref. [BB85]. The probability for exciting the nuclear state of energy $I_{f}$ is obtained directly
from Eq. \eqref{relcoul5} by using 
\begin{equation}
P_{fi}={\frac{1}{2I_{i}+1}}\;\sum_{M_{i},M_{f}}\mid a_{fi}\mid
^{2}.  \label{(4.1)}
\end{equation}
As with the low-energy case, we can rewrite the final result in terms of the reduced transition probabilities for photo-excitation. It can be cast in the form 
\begin{eqnarray}
P_{fi}(b,E_\gamma )&=&\sum_{\pi L}\;P_{\pi L }(b,E_\gamma)\nonumber\\
&=&\sum_{\pi L }\int {\frac{dE_\gamma }{E_\gamma }}\;n_{\pi
L }(E_\gamma,b )\;\sigma _{\gamma }^{\pi L }(E_\gamma ),\nonumber \\
\label{(4.5a)}
\end{eqnarray}
where  $ \sigma _{\gamma }^{\pi L }(E_\gamma )$
is the photonuclear absorption cross sections for a given multipolarity $%
\pi L $ given by Eq. \eqref{photonn}. 
The total photonuclear cross section is a sum of all these multipolarities, 
\begin{equation}
\sigma _{\gamma }=\sum_{\pi L }\sigma _{\gamma }^{\pi L
}(E_\gamma )\,.  \label{(4.5c)}
\end{equation}

\begin{figure}
[t]
\begin{center}
\includegraphics[
width=3.2in
]%
{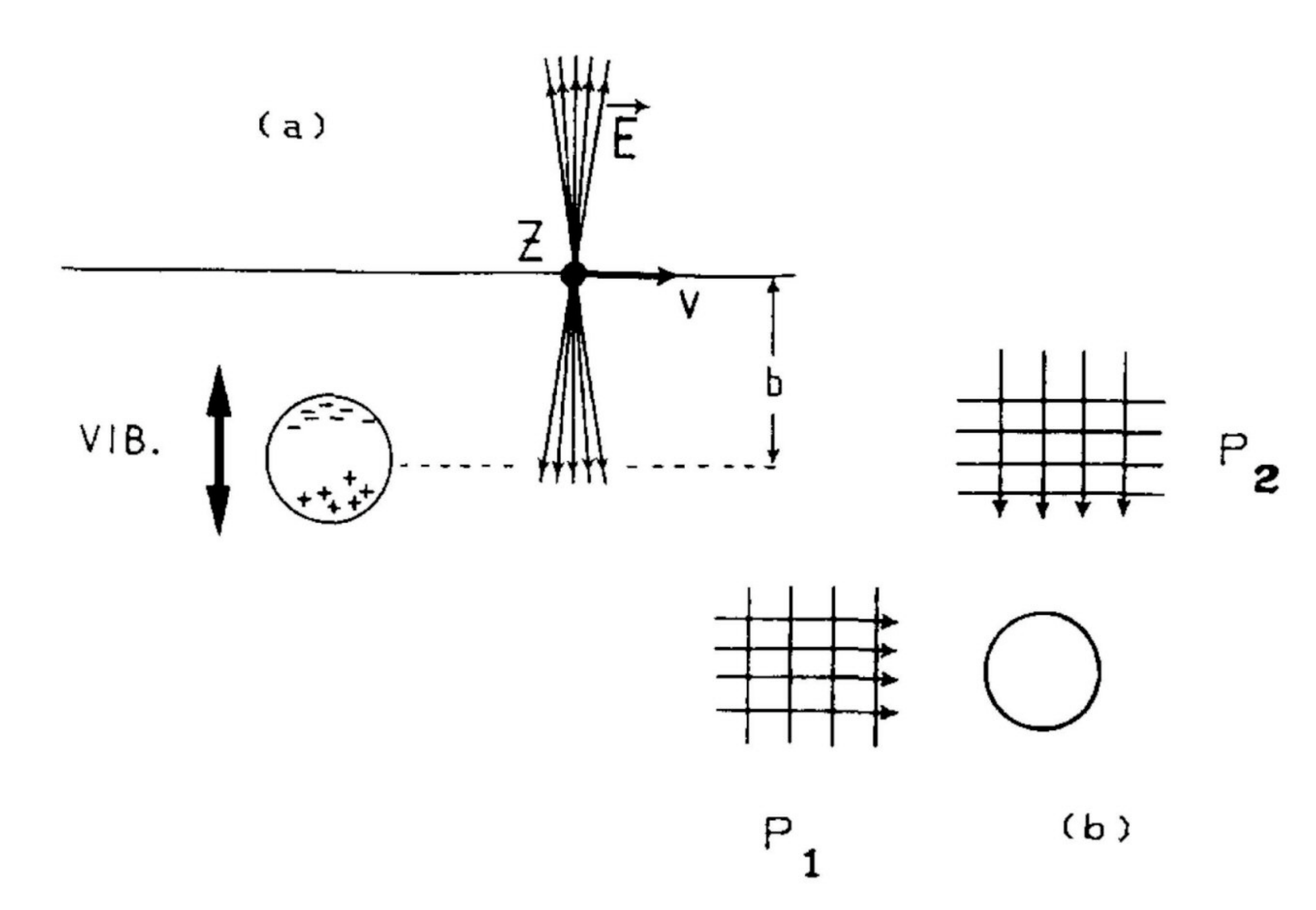}%
\caption{\sl  (a) A relativistic charged projectile incident on a target with impact parameter larger than the strong interaction radius. A sketch of the
electric field generated by it is also shown. One of the effects of this field is to induce collective vibrations of the nuclear charges. (b) Two pulses of
plane wave of light which produce the same effect on the target as the electromagnetic field created by the projectile's motion. (from [BB88]).}%
\label{WW}%
\end{center}
\end{figure}

The functions $n_{\pi L }(E_\gamma )$ are called the \textit{%
virtual photon numbers}, and are given by [BB85] 
\begin{eqnarray}
n_{E1}(b,E_\gamma) &=&{\frac{Z^{2}\alpha }{\pi ^{2}}}\,{\frac{\xi
^{2}}{b^{2}}}\,({\frac{c}{v}})^{2}\,\left\{ K_{1}^{2}+{\frac{1}{\gamma ^{2}}}%
\,K_{0}^{2}\right\} , \label{(A.3a)} \\
n_{E2}(b,E_\gamma ) &=&{\frac{Z^{2}\alpha }{\pi ^{2}b^{2}}}\,({\frac{c%
}{v}})^{4}\,\Bigg\{ {\frac{4}{\gamma ^{2}}}\,\left[ K_{1}^{2}+\xi
K_{0}K_{1}+\xi ^{2}K_{0}^{2}\right] \nonumber \\
&+&\xi
^{2}(2-v^{2}/c^{2})^{2}K_{1}^{2}\Bigg\}  ,\label{(A.3b)} \\
n_{M1}(b,E_\gamma ) &=&{\frac{Z^{2}\alpha }{\pi ^{2}}}\,{\frac{\xi
^{2}}{b^{2}}}\,K_{1}^{2}  ,\label{(A.3c)}
\end{eqnarray}
where all $K_{M }$'s are functions of $\xi (b)={\omega b/}\gamma v$.

Since all nuclear excitation dynamics is contained in the photoabsorption
cross section, the virtual photon numbers, Eqs. \eqref{(A.3a)}, \eqref{(A.3b)}
and \eqref{(A.3c)}, do not depend on the nuclear structure. They are
kinematical factors, depending on the orbital motion. They may be
interpreted as the number of equivalent (virtual) photons that hit the
target per unit area.

The cross section is obtained by the impact parameter integral of the
excitation probabilities. Eq. {\eqref{(4.5a)} shows that we only need to
integrate the number of virtual photons over impact parameter. One has to
introduce a minimum impact parameter $b_{0}$ in the integration. Impact
parameters smaller than $b_{0}$ are dominated by nuclear fragmentation
processes. One finds 
\begin{equation}
\sigma _{C}=\sum_{\pi L }\;\sigma _{\pi L }=\sum_{\pi L
}\int {\frac{dE_\gamma }{E_\gamma }}\;N_{\pi L }(E_\gamma
)\;\sigma _{\gamma }^{\pi L }(E_\gamma )\,,  \label{(4.5e)}
\end{equation}
where the \textit{total virtual photon numbers} $N_{\pi L
}(E_\gamma )=2\pi \int db\ b\ n_{\pi L}(b,E_\gamma )$ are given analytically
by } 
\begin{eqnarray}
N_{E1}(E_\gamma) &=&{\frac{2Z^{2}\alpha }{\pi }}\,({\frac{c}{v}}%
)^{2}\,\left[ \xi K_{0}K_{1}-{\frac{v^{2}\xi ^{2}}{2c^{2}}}%
\,(K_{1}^{2}-K_{0}^{2})\right] , \nonumber \\ \label{(A.6a)} \\
N_{E2}(E_\gamma ) &=&{\frac{2Z^{2}\alpha }{\pi }}\,({\frac{c}{v}}%
)^{4}\,\Big[ {2\over \gamma^2}K_{1}^{2}+{\xi \over \gamma^4} K_{0}K_{1}\notag \\
&&+{\frac{\xi ^{2}v^{4}}{2c^{4}}}(K_{1}^{2}-K_{0}^{2})
+ \xi ^{2}(2-v^{2}/c^{2})^{2}K_{1}^{2}\Big] , \nonumber \\
\label{(A.6b)} \\
N_{M1}(E_\gamma ) &=&{\frac{2Z^{2}\alpha }{\pi }}\,{\frac{\xi ^{2}}{%
R^{2}}}\,\left[ \xi K_{0}K_{1}-{\frac{\xi ^{2}}{2}}\,(K_{1}^{2}-K_{0}^{2})%
\right] ,\nonumber \\
 \label{(A.6c)}
\end{eqnarray}
where all $K_{M }$'s are now functions of $\xi (R)={\omega R/}\gamma v$%

The usefulness of Coulomb excitation, even in first order processes, is
displayed in Eq. \eqref{(4.5a)}. The field of a real photon contains all
multipolarities with the same weight and the photonuclear cross section, Eq. \eqref
{(4.5c)} is a mixture of the contributions from all multipolarities,
although only a few contribute in most processes. In the case of Coulomb
excitation the total cross section is weighted by kinematical factors which
are different for each projectile or bombarding energy. This allows one to
disentangle the multipolarities when several ones are involved in the
excitation process, except for the very high bombarding energies $\gamma \gg
1$ for which all virtual photon numbers can be shown to be all the same 
[BB85] to
\begin{eqnarray}
n_{\pi L}={2\over \pi} Z^2\alpha \ln \left({\delta \over \xi}\right) .
\label{WWvirtlim}
\end{eqnarray}
Since $\xi=\omega R/\gamma c \ll 1$, we have a
logarithmic rise of the cross section for all multipolarities with $\gamma$. The impinging projectile acts like a
spectrum of plane wave photons with helicity $m = \mp 1$. Such a photon spectrum contains equally all
multipolarities $\pi L$.

\subsection{Historical note: Fermi's forgotten papers}
In 1924, Enrico Fermi, then 23 years old, submitted a paper ``On the Theory of
Collisions Between Atoms and Elastically Charged ParticlesÓ to Zeitschrift f\"ur
Physik [Fe24]. This paper does not appear in his ``Collected Works.Ó Nevertheless, it
is said that this was one of FermiÕs favorite ideas and that he often used it later in
life. In this publication, Fermi devised a method known as the equivalent (or
virtual) photon method, where he treated the electromagnetic fields of a charged
particle as a flux of virtual photons. It is also interesting that Fermi published the same paper, but in the Italian language, in Nuovo Cimento [Fe25] (it is rather uncommon that the same paper is published twice!). Ten years later, Weiszs\"acker and Williams [WW34]
extended this approach to include ultra-relativistic particles, basically restoring the Lorentz $\gamma$ factors in the right places. However, it is indisputable that Fermi's papers [Fe24,Fe25] introduced the ingenious virtual photon method.

\begin{figure}
[t]
\begin{center}
\includegraphics[
width=3.2in
]%
{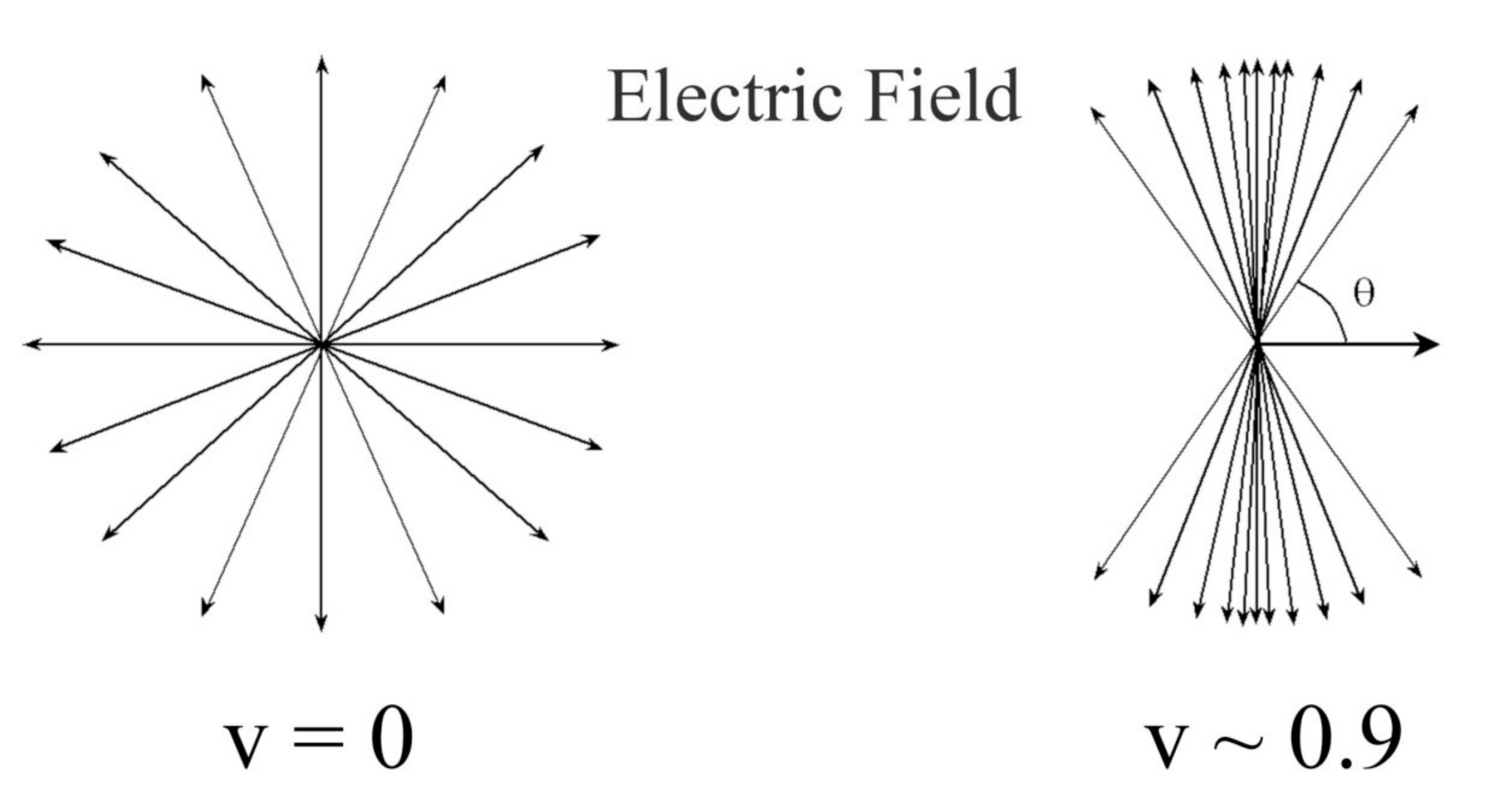}%
\caption{\sl  Left: Electric field of a slowly moving charge. Right: Electric field of a charge moving with $v=0.9$ (in units of $c$). 
}%
\label{WW2}%
\end{center}
\end{figure}

A fast-moving charged particle has electric field vectors pointing radially outward
and magnetic fields circling it. The field at a point some distance away from
the trajectory of the particle resembles that of a real photon. Thus, Fermi replaced
the electromagnetic fields from a fast particle with an equivalent flux of photons, as shown 
schematically in figure \ref{WW}.

Fermi's virtual photon method is based on the idea that, when $v\sim c$, where $c$ is the velocity of light, the electromagnetic field generated by the projectile looks contracted in the direction perpendicular to its motion (see figures \ref{WW2} and \ref{ETEZ}) and is given by
\begin{eqnarray}
E_z&=&-{Ze\gamma vt\over (b^2+\gamma^2v^2t^2)^{3/2}},
\notag \\
{\bf E}_T&=&{Ze\gamma {\bf b}\over (b^2+\gamma^2v^2t^2)^{3/2}},
 \notag \\
{\bf B}_T&=&{{\bf v}\over c} \otimes {\bf E}_T,    \ \ \ \ {\rm and} \ \  B_z=0. \label{WWfields}
\end{eqnarray}
where the $z$ ($T$) indices denote the direction parallel (transverse) to the velocity of the projectile.

\begin{figure}
[t]
\begin{center}
\includegraphics[
width=3.2in
]%
{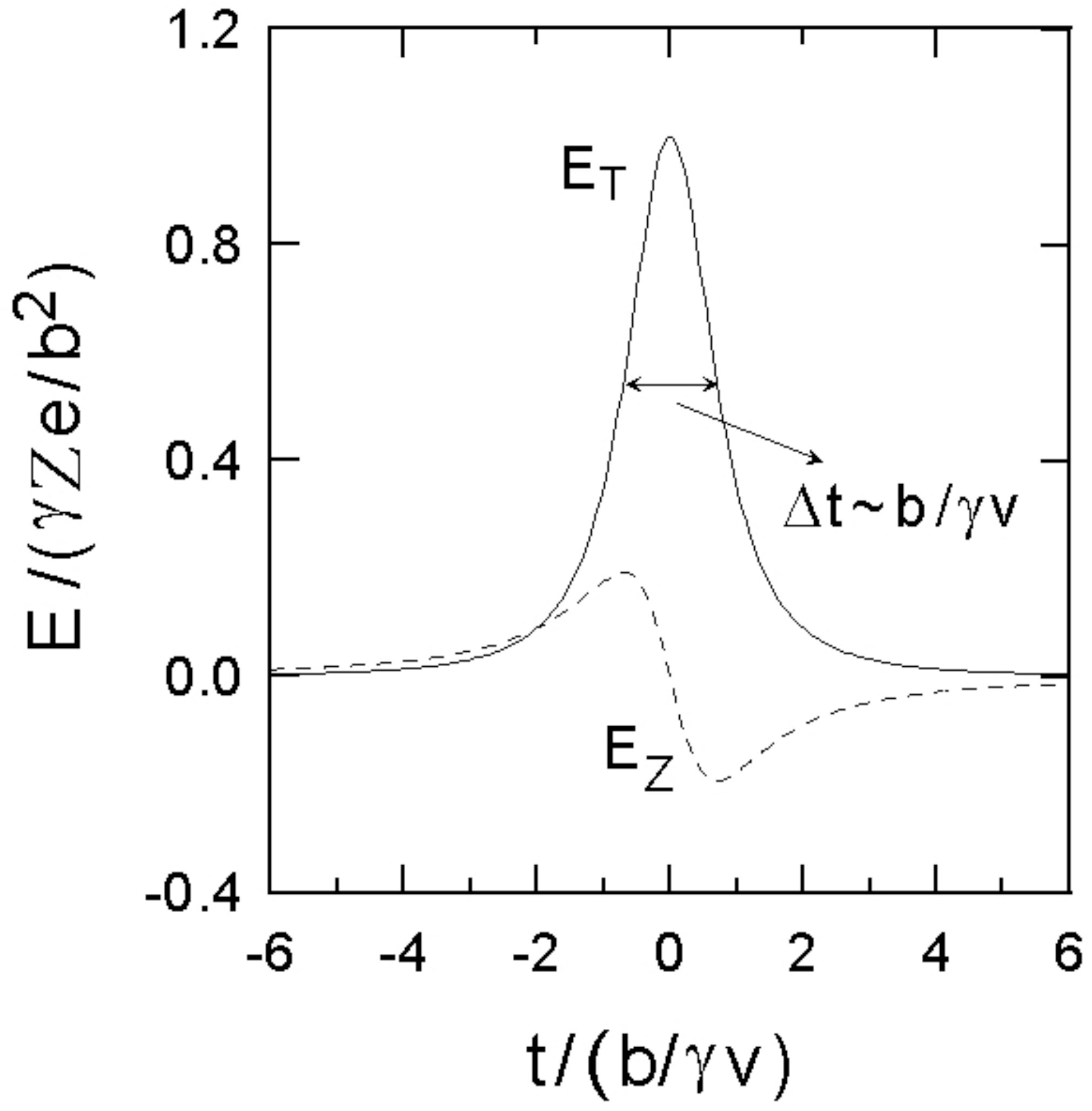}%
\caption{\sl  Longitudinal and transverse electric field of a straight-line moving charge at high energies as a function of time. Both variables are gauged in appropriate units, as sown in the axis labels.}%
\label{ETEZ}%
\end{center}
\end{figure}

When  $\gamma \gg 1$, these fields will act during a very short time, of order
\begin{equation}
t={b\over \gamma v} \simeq {b\over \gamma c}, \label{WWt} 
\end{equation}
and they are equivalent to two pulses of plane-polarized radiation incident on the target (see fig. \ref{WW}):
one in the beam direction (P1), and the other perpendicular to it (P2). In the case of the pulse P1 the
equivalency is exact. Since the electric field in the $z$-direction is not accompanied by a corresponding
magnetic field, the equivalency is not complete for pulse P2, but this will not appreciably affect the
dynamics of the problem since the effects of the field $E_z$ are of minor relevance when v = c. Therefore, we add a field $B= vE_z/c$ to Eq.  \eqref{WWfields} in order to treat also P2 as a plane-wave pulse of radiation. This
analogy permits one to calculate the amount of energy incident on the target per unit area and per
by Fourier transforming the Poynting  vector ${\bf S}= {\bf E} \otimes {\bf B}$ and calculating the intensity of the virtual radiation, $I(\omega,b$), with $E_\gamma =\hbar \omega$. This procedure is nicely explained in Ref. [Ja75].

Fermi associated the spectrum of the virtual radiation as described above to the one
of a real pulse of light incident on the target. Then he obtained the probability for nuclear (in fact, Fermi was interested in atomic transitions) by a fast charge, in terms of the cross sections for the
same process generated by an equivalent pulse of light, i.e.
\[
P(b)=\int I(\omega, b)\sigma_\gamma (E_\gamma) dE_\gamma = \int n(\omega, b) \sigma_\gamma (E_\gamma)
{d\omega \over \omega},
\]
where $\sigma_\gamma (E_\gamma) $ is the photo cross-section for the photon energy $E_\gamma$, and the integral runs over all the
frequency spectrum of the virtual radiation. The quantities $n(\omega, b)$ can be interpreted as the number of equivalent photons incident on the target per unit area.

Following this procedure, Fermi obtained the Eq. \eqref{(A.3a)} for $ n(\omega, b)\equiv  n_{E1}(\omega, b)$, without the $\gamma$ factors (Fermi was not interested in relativistic collisions in 1924!). The $\gamma$ factors were found in the proper places by Weizs\"acker and Williams [WW34]. It is somewhat surprising that Fermi,  Weizs\"acker, and Williams obtained the ``exact" result of the virtual photons for the $E1$ multipolarity. Their method was completely classical and approximate (in adding the $B_z$ field). To reach Eq. \eqref{(A.3a)} some quantum mechanics was used (e.g., the continuity equation for the nuclear transitions). Eq.  \eqref{(A.3a)} is also the result of a multipole expansion, which was not used in the classical prescription. In fact, the Eqs. (\ref{(A.3a)}-\ref{(A.3c)}) are an improvement of Fermi's method for higher multipolarities, and were obtained in Ref. [BB85]. As we show later, a quantum mechanical description of high-energy scattering leads to the same expressions  in Eqs. (\ref{(A.3a)}-\ref{(A.3c)}).

\subsection{Spectrum of virtual photons}

In Eq. \eqref{(A.3a)} the first term inside parentheses comes from the contribution of the
pulse P1, whereas the second term comes from the contribution of the pulse P2. One immediately sees
that the contribution of pulse P2 becomes negligible for $\gamma \gg 1$.
The shape of the equivalent photon
spectrum for a given impact parameter can be expressed in terms of the dimensionless function
$\phi(x) = x^2 K^2_1(x)$, if we neglect the pulse P2. In a crude approximation, $\phi\sim 0$  for $x > 1$, and $\phi = 1$ for
$x< 1$. This reflects in a sudden cutoff of the virtual photon spectra, as can be seen from figure \ref{wwphi}. This implies that, in a collision with impact parameter $b$, the spectrum will contain equivalent photons with energies up to a maximum value of order
\begin{equation}
E^{\max}_\gamma\simeq{\gamma \hbar v\over b}, \label{WWemax} 
\end{equation}
which we call by {\it adiabatic cutoff energy}. This means that in an electromagnetic collision of two nuclei
the excitation of states with energies up to the above value can be reached. We can explain this result by observing that in a collision with interaction time given by Eq. \eqref{WWt} only states satisfying the condition $T/\Delta t \gg 1$, where $T$ is the period of
the quantum states, will have an appreciable chance to be excited. Otherwise, the quantum system will
respond adiabatically to the interaction.  In a collision with a typical impact parameter of $b  10$ fm one can reach
states with energy around $E_{\max}\sim 20\gamma$ MeV. Among the many possibilities, we cite the following: for
$E_\gamma = 10-20$ MeV (already small values of $\gamma$), excitation of giant resonances, with subsequent nucleon
emission; for $E = 20-100$ MeV, the quasideuteron effect which corresponds to a photon absorption of a
correlated N-N pair in the nucleus; and for $E_\gamma > 100$ MeV, pion production through A-isobar excitation
which has a maximum at $E \simeq 200$ MeV. Also the production of lepton pairs ($e^+e^-$, $q\hat{q}$ (mesons), etc.) are
accessible with increasing value of $\gamma$.

\begin{figure}
[t]
\begin{center}
\includegraphics[
width=3.5in
]%
{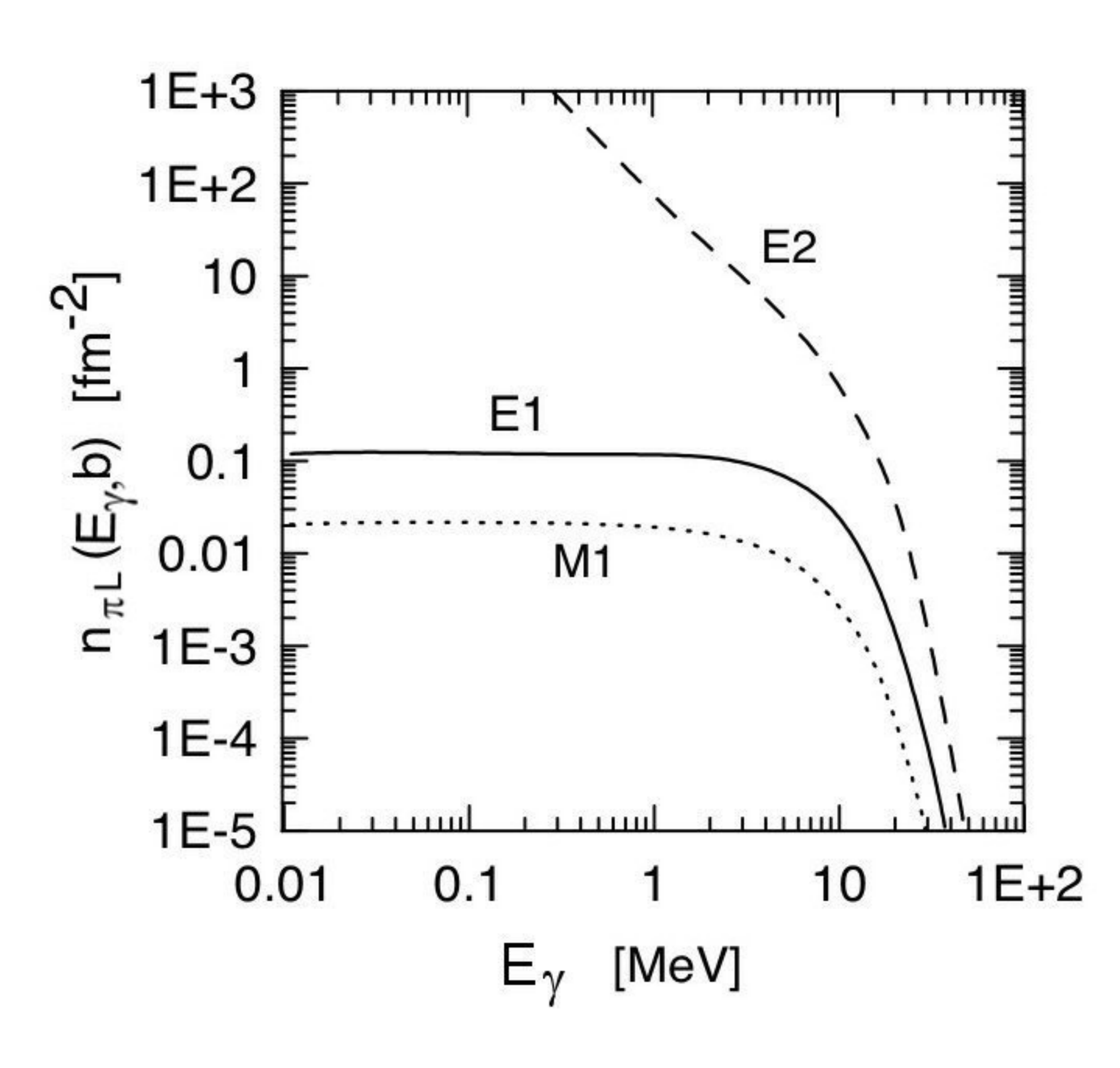}%
\caption{\sl  Equivalent photon numbers per unit area incident on $^{208}$Pb, in a collision with $^{16}$O at 100 MeV/nucleon and with
impact parameter $b=15$ fm, as a function of the photon energy $E=\hbar \omega$. The curves for the E1, E2 and M1 multipolarities
are shown.}%
\label{wwphi}%
\end{center}
\end{figure}

In figure \ref{EPAf} we show $n_{\pi L}$ (with $Z= {\rm unity}$) as given by Eqs. (\ref{(A.3a)}-\ref{(A.3c)}), as a function of $\omega R/c$. We see that
$n_{E2}\simeq  n_{E1} \simeq n_{M1}$ for small values of $\gamma$, in contrast to the limit $\gamma \sim 1$. The physical reason for these two
different behaviours of the equivalent photon spectrum is the following. The electric field of a charged
particle moving at low energies is approximately radial and the lines of force of the field are
isotropically distributed, with their relative spacing increasing with the radial distance (see figure \ref{WW2}). When interacting
with a target of finite dimension, the non-uniformity of the field inside the target is responsible for the
large electric quadrupole interaction between them. The same lines of force of an ultrarelativistic
($\gamma \gg 1$) charged particle appear more parallel and compressed in the direction transverse to the particle's motion, due to the Lorentz contraction (see figure \ref{WW2}). As seen from the target, this field looks like a pulse of a plane wave. But plane waves contain all electric and magnetic multipolarities with the same weight. This is the cause for the equality between the equivalent photon numbers as $\gamma \rightarrow \infty$.

\begin{figure}
[t]
\begin{center}
\includegraphics[
width=3.2in
]%
{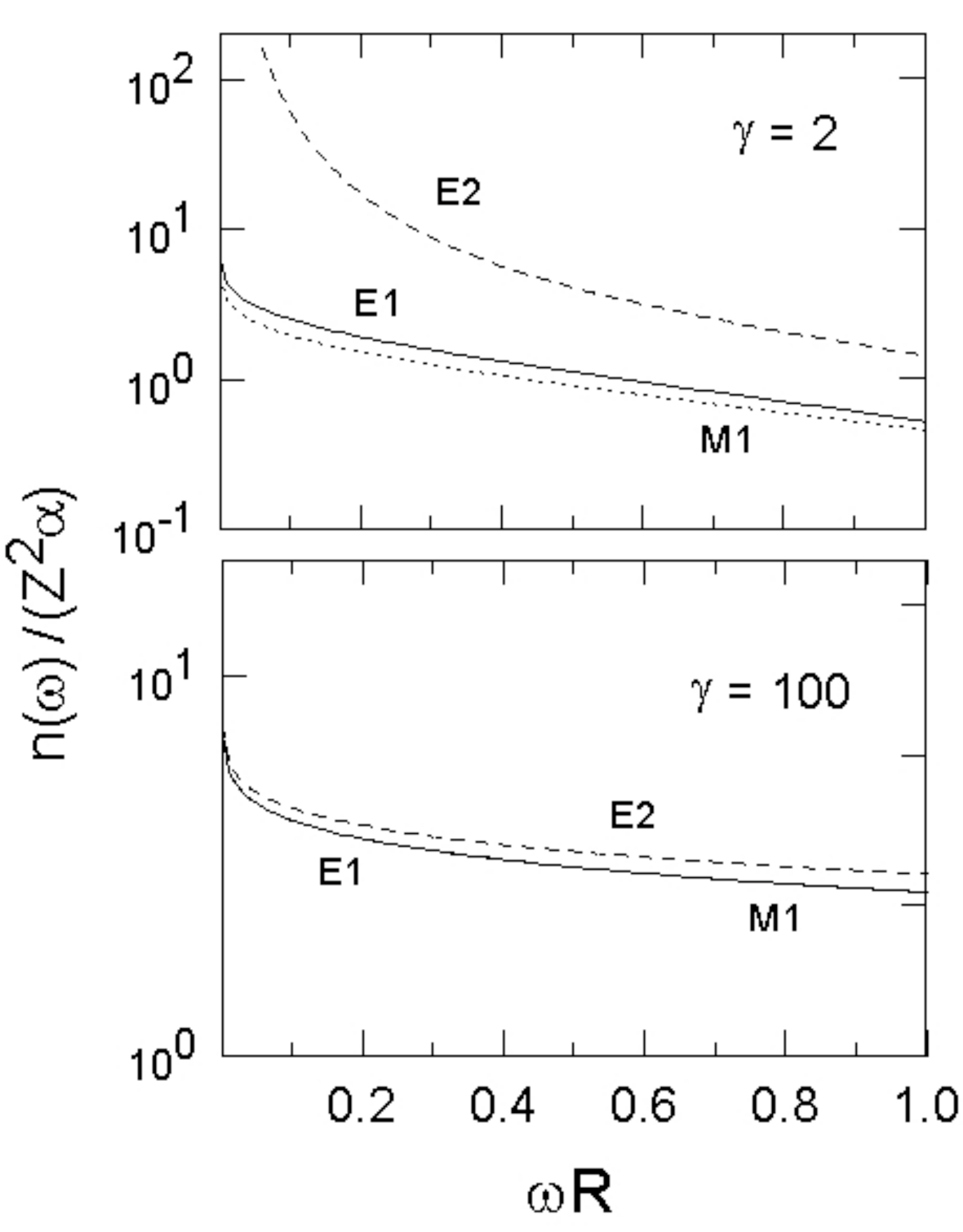}%
\caption{\sl  Equivalent photon number per unit projectile charge, for El, M1 and E2 radiation, and as a function of the ratio between $R$ and the
photon wavelength. $\gamma$ is the ratio of the projectile energy to its rest energy. (Here, $c=1$.)}%
\label{EPAf}%
\end{center}
\end{figure}

\section{Quantum treatment of relativistic Coulomb excitation}

In contrast to sub-barrier Coulomb excitation, at relativistic energies the effects of the nuclear interaction are
visible due to the distortion it causes on the scattered waves. Fortunately, at high energies, this can be treated in a simple manner by using the eikonal approximation. Our discussion will be general, with very little emphasis on the details of the nuclear interaction.

\subsection{The eikonal wavefunction}

The free-particle wavefunction
\begin{equation}
\psi\sim e^{i\mathbf{k}\cdot\mathbf{r}}\label{planew}%
\end{equation}
becomes \textquotedblleft distorted\textquotedblright\ in the
presence of a potential $\,V(\mathbf{r}\,)\,$. The distorted wave can be calculated 
numerically by performing a partial wave-expansion [Ber07]
solving the Schr\"{o}dinger equation for each partial wave, i.e.
\begin{equation}
\left[  \frac{d^{2}}{dr^{2}}+k_{l}^{2}(r)\right]  \chi_{l}%
(r)=0\;,\label{schrol}%
\end{equation}
where
\begin{equation}
k_{l}(r)=\left\{  \frac{2\mu}{\hbar^{2}}\left[  E-V(r)-\frac{l(l+1)\hbar^{2}%
}{2\mu r^{2}}\right]  \right\}  ^{1/2}.\label{kl}%
\end{equation}
with the condition that asymptotically $\,\psi(\mathbf{r}\,)$ behaves as
(\ref{planew}).

The solution of (\ref{schrol}) involves a great numerical effort at large
bombarding energies $\,E\,$. Fortunately, at large energies $\,E\,$ a very
useful approximation is valid when the excitation energies $\,\Delta E\,$ are
much smaller than $\,E\,$ and the nuclei (or nucleons) move in forward
directions, i.e., $\,\theta\ll1\,$.

Calling $\,\mathbf{r}=(z,\mathbf{b}\,)$, where $\,z\,$ is the coordinate along
the beam direction, we can assume that
\begin{equation}
\psi(\mathbf{r}\,)=e^{ikz}\,\phi(z,\mathbf{b}\,),\label{psireik}%
\end{equation}
where $\,\phi\,$ is a slowly varying function of $\,z\,$ and $\,b\,$, so that
\begin{equation}
\left\vert \nabla^{2}\phi\,\right\vert \ll\,k\left\vert \nabla\phi\right\vert
.\label{cond}%
\end{equation}
In cylindrical coordinates the Schr\"{o}dinger equation%
\[
-\frac{\hbar^{2}}{2\mu}\nabla^{2}\psi(\mathbf{r}\,)+V(\mathbf{r}%
\,)\psi(\mathbf{r}\,)=E\,\psi(\mathbf{r}\,)
\]
becomes
\[
2ik\,e^{ikz}\frac{\partial\phi}{\partial z}+e^{ikz}\frac{\partial^{2}\phi
}{\partial z^{2}}+e^{ikz}\,\nabla_{b}^{2}\phi-\frac{2m}{\hbar^{2}}%
\,V\,e^{ikz}\,\phi=0
\]
or, neglecting the 2nd and 3rd terms because of (\ref{cond}),
\begin{equation}
\frac{\partial\phi}{\partial z}=-\frac{i}{\hbar\mathrm{v}}V(\mathbf{r}%
\,)\phi\label{phizeq}%
\end{equation}
whose solution is
\begin{equation}
\phi=\exp\left\{  -\frac{i}{\hbar\mathrm{v}}\int_{-\infty}^{z}\,V(\mathbf{b}%
,z^{\prime})dz^{\prime}\right\}  .\label{phisol}%
\end{equation}
That is,
\begin{equation}
\psi(\mathbf{r})=\exp\left\{  ikz+i\chi(\mathbf{b},z)\right\}
,\label{psieiksol}%
\end{equation}
where
\begin{equation}
\chi(\mathbf{b},z)=-\frac{1}{\hbar\mathrm{v}}\int_{-\infty}^{z}\,V(\mathbf{b}%
,z^{\prime})dz^{\prime}\label{chieik}%
\end{equation}
is the \textit{eikonal phase}. Given $\,V(\mathbf{r}\,)\,$ one needs a single
integral to determine the wavefunction: a great simplification of the problem.

The {\it eikonal approximation}, in the same form as given by eqs.~(\ref{psieiksol}%
), can be obtained from the Klein-Gordon equation with a (scalar)
potential$\,V$. We will use this approximation in several discussions later on this review.

\subsection{Quantum relativistic Coulomb excitation}

Defining \textbf{r} as the separation between the center of mass of the two
nuclei and \textbf{r$^{\prime}$} as the intrinsic coordinate of the target
nucleus, the inelastic scattering amplitude to first-order is given by [BD04]
\begin{eqnarray}
f(\theta)&=&{\frac{ik}{2\pi\hbar v}}\ \int d^{3}r\ d^{3}r^{\prime}\nonumber \\
&&\left<\Phi_{\mathbf{k^{\prime}}}^{(-)}(\mathbf{r})\ \phi_{f}(\mathbf{r}^{\prime
})\ \left| \ {\cal H}_{int}(\mathbf{r},\ \mathbf{r}^{\prime})\ \right| \ \Phi_{\mathbf{k}}%
^{(+)}(\mathbf{r})\ \phi_{i}(\mathbf{r}^{\prime})\right>,\nonumber \\
\label{fth}%
\end{eqnarray}
where $\Phi_{\mathbf{k^{\prime}}}^{(-)}(\mathbf{r})$ and $\Phi_{\mathbf{k}%
}^{(+)}(\mathbf{r})$ are the incoming and outgoing distorted waves,
respectively, for the scattering of the center of mass of the nuclei, and
$\phi(\mathbf{r}^{\prime})$ is the intrinsic nuclear wavefunction of the
target nucleus.

At intermediate energies, $\Delta E/E_{lab}\ll1$, and forward angles,
$\theta\ll1$, we can use eikonal wavefunctions for the distorted waves; i.e.,
\begin{equation}
\Phi_{\mathbf{k^{\prime}}}^{(-)\ast}(\mathbf{r})\ \Phi_{\mathbf{k}}%
^{(+)}(\mathbf{r})=\exp\left\{  -i\mathbf{q.r}+i\chi(b)\right\}
\;,\label{peik}%
\end{equation}
where
\begin{equation}
\chi(b)=-{\frac{1}{\hbar v}}\int_{-\infty}^{\infty}U_{N}^{opt}(z^{\prime
},\ b)\ dz^{\prime}+i\chi_{C}(b)\label{peikb}%
\end{equation}
is the eikonal-phase, $\mathbf{q}=\mathbf{k}^{\prime}-\mathbf{k}$,
$U_{N}^{opt}$ is the nuclear optical potential, and $\chi_{C}(b)$ is the
Coulomb eikonal phase,
\begin{equation}
\chi_{C}(b)=\frac{2Z_{p}Z_{t}e^{2}}{\hbar\mathrm{v}}\,\ln(kb)\ ,\label{chiC}%
\end{equation} 
where $Z_p$ ($Z_t$) is the projectile (target) nuclear charges.
The Coulomb phase, as given by the above formula, reproduces
the Coulomb amplitude for the scattering of point particles in the eikonal approximation 
for elastic scattering [BD04]. Corrections due to the extended nuclear charges can be easily
incorporated [BD04].
Here we have defined the impact parameter \textbf{b} as
$\mathbf{b}=\mathbf{r}\times\hat{\mathbf{z}}$.

In Eq.  (\ref{fth}) the interaction potential, assumed to be purely
Coulomb, is given by Eqs. \eqref{(higho1)} and \eqref{higho2}. Performing the multipole
expansion as in Eq. \eqref{3.15b},  one finds [BN93]
\begin{align}
f_C({\theta})  & =i{\frac{Zek}{\gamma\hbar v}}\sum_{\pi L M}%
i^{M}\left(  {\frac{\omega}{c}}\right)  ^{L}\sqrt{2L
+1}\ e^{-iM\phi}\nonumber\\
& \times\Omega_{M}(q)\ G_{\pi L M}\left(  {\frac{c}{v}}\right)
\left <I_{f}\ M_{f}\ |\ \mathcal{M}(\pi L,\ -M)\ |I_{i}\ M_{i}\right>\label{fth2}%
\end{align}
where  the functions $G_{\pi L M}$ are given in Eq. \eqref{Glm}. 
The function $\Omega_{m}(q)$ is given by [BN93]
\begin{equation}
\Omega_{M}(q)=\int_{0}^{\infty}db\ b\ J_{M}(qb)\ K_{M}\left(  {\frac{\omega
b}{\gamma v}}\right)  \ \exp\left\{  i\chi(b)\right\}  \;,\label{Omeg}%
\end{equation}
where $q=2k\sin(\theta/2)$ is the momentum transfer, $\theta$ and $\phi$ are
the polar and azimuthal scattering angles, respectively.

In the sharp-cutoff approximation, one assumes
\begin{eqnarray}
 \exp\left\{  i\chi(b)\right\} &=&0, \ \ \  {\rm if} \ \ b\le R,
\nonumber \\
&=& 1, \ \ \ {\rm if} \ \ b>R, \label{sharpct}
\end{eqnarray}
where $R$ is the strong interaction radius, or minimum impact parameter.
Then the integral \eqref{Omeg} can be performed analytically [BB85].

Using techniques
similar to those discussed in previous sections, one obtains
\begin{eqnarray}
{\frac{d^{2}\sigma_{C}}{d\Omega\,dE_{\gamma}}}\ \left(  E_{\gamma}\right)
={\frac{1}{E_{\gamma}}}\ \sum_{\pi L}{\frac{dn_{\pi L}}{d\Omega}%
}\ \sigma_{\gamma}^{\pi L}\ \left(  E_{\gamma}\right) \label{dsig}%
\end{eqnarray}
where d $dn_{\pi L}/d\Omega$ is the
virtual photon number given by [BN93]
\begin{eqnarray}
{\frac{dn_{\pi L}}{d\Omega}}&=&Z^{2}\alpha\ \left(  {\frac{\omega
k}{\gamma v}}\right)  ^{2}{\frac{L \left[  (2L+1)!!\right]  ^{2}%
}{(2\pi)^{3}\ (L+1)}} \nonumber \\ 
&\times& \sum_{M} |G_{\pi LM}|^{2}\ |\Omega
_{M}(q)|^{2}\label{dn}.
\end{eqnarray}

The total cross section for Coulomb excitation can be obtained
from eqs. (\ref{dsig}) and (\ref{dn}), using the approximation
$d\Omega\simeq2\pi qdq/k^{2}$, valid for small scattering angles
and small energy losses. Using the closure relation for the Bessel
functions, 
\begin{equation}\int dq \ q J_M{qx} J_M(qx')={1\over x}\delta(x-x'),
\end{equation}
one obtains
\begin{equation}
{\frac{d\sigma_{C}}{dE_{\gamma}}}\ \left(  E_{\gamma}\right)  ={\frac
{1}{E_{\gamma}}}\ \sum_{\pi L}N_{\pi L}\left(  E_{\gamma}\right)
\ \sigma_{\gamma}^{\pi L}\ \left(  E_{\gamma}\right)  \;,\label{sig}%
\end{equation}
where the total number of virtual photon with energy $\hbar\omega$ is given
by
\begin{equation}
N_{\pi L}(\omega)=Z^{2}\alpha\ {\frac{ L \left[  (2L
+1)!!\right]  ^{2}}{(2\pi)^{3}\ (L+1)}}\ \sum_{m}\ |G_{\pi L
M}|^{2}\ g_{M}(\omega)\;,\label{n}%
\end{equation}
and
\begin{equation}
g_{M}(\omega)=2\pi\ \left(  {\frac{\omega}{\gamma v}}\right)  ^{2}\ \int
db\ b\ K_{M}^{2}\left(  {\frac{\omega b}{\gamma v}}\right)  \ \exp\left\{
-2\ \chi_{I}(b)\right\}  \;,\label{g}%
\end{equation}
where $\chi_{I}(b)$ is the imaginary part of $\chi(b)$, which is
obtained from Eq. ~(\ref{peikb}) and the imaginary part of the
optical potential.

If one uses the sharp-cutoff approximation, Eq. \eqref{sharpct}, one recovers the
result in Eq. \eqref{relcoul7b} [BB85].

We point out that for very light heavy ion partners, the distortion of the
scattering wavefunctions caused by the nuclear field is not important. This
distortion is manifested in the diffraction peaks of the angular
distributions, characteristic of strong absorption processes. If $Z_{p}%
Z_{t}\alpha\gg1$, one can neglect the diffraction peaks in the inelastic
scattering cross sections and a purely Coulomb excitation process emerges. One
can gain insight into the excitation mechanism by looking at how the
semiclassical limit of the excitation amplitudes emerges from the general
result (\ref{dn}).

\subsection{Semiclassical limit of the Coulomb excitation amplitudes}

If we assume that Coulomb scattering is dominant and neglect the
nuclear phase in Eq. ~(\ref{peikb}), we get
\begin{equation}
\Omega_{M}(q)\simeq\int_{0}^{\infty}db\ b\ J_{M}(qb)\ K_{M}\left(
{\frac{\omega b}{\gamma v}}\right)  \ \exp\left\{  i\chi_{C}(b)\right\}
\;.\label{Omegc}%
\end{equation}

This integral can be done analytically by rewriting it as
\begin{equation}
\Omega_{M}(q)=\int_{0}^{\infty}db\ b^{1+i2\eta}\ J_{M}(qb)\ K_{M}\left(
{\frac{\omega b}{\gamma v}}\right)  \;,\label{Oma1}%
\end{equation}
where we used  $\chi_{C}(b)=2\eta\ \ln(kb)$, with
$\eta =Z_{1}Z_{2}e^{2}/\hbar v$. Using standard techniques found
in Ref. [GR80], we find
\begin{align}
\Omega_{M}(q)  & =2^{2i\eta}\ {\frac{1}{M!}}\ \Gamma(1+M+i\eta)\Gamma
(1+i\eta) \Lambda^{M}\ \left(  {\frac{\gamma v}{\omega}}\right)  ^{2+2i\eta
}\nonumber\\
& \times\  F\left(  1+M+i\eta;\ 1+i\eta;\ 1+M;-\Lambda^{2}\right)  \;,\label{Oma2}%
\end{align}
where
\begin{equation}
\Lambda={\frac{q\gamma v}{\omega}}\;,\label{Lamb}%
\end{equation}
and $F$ is the hypergeometric function [GR80].

The connection with the semiclassical results may be obtained by using the low
momentum transfer limit
\begin{align}
J_{M}(qb)  & \simeq\sqrt{\frac{2}{\pi qb}}\ \cos\left(  qb-{\frac{\pi M}{2}%
}-{\frac{\pi}{4}}\right) \nonumber\\
& ={\frac{1}{\sqrt{2\pi qb}}}\ \left\{   e^{iqb-i{\pi\over 2}(M+{1\over 2})/2}%
+e^{-iqb+i{\pi\over 2}(M+{1\over 2})}\right\}  \;,\label{Bes}%
\end{align}
and using the stationary phase method, i.e.,
\begin{equation}
\int G(x)\ e^{i\phi(x)}\ dx\simeq\left(  {\frac{2\pi i}{\phi^{\prime\prime
}(x_{0})}}\right)  ^{1/2}\ G(x_{0})\ e^{i\phi(x_{0})}\;,\label{statb}%
\end{equation}
where
\begin{equation}
{\frac{d\phi}{dx}}(x_{0})=0\qquad\mathrm{and}\qquad\phi^{\prime\prime}%
(x_{0})={\frac{d^{2}\phi}{dx^{2}}}(x_{0})\;.\label{stat2}%
\end{equation}
This result is valid for a slowly varying function $G(x)$.

Only the second term in brackets of Eq. ~(\ref{Bes}) will have a
positive
($b=b_{0}>0$) stationary point. Thus,%
\begin{eqnarray}
\Omega_{M}(q)&\simeq &\frac{1}{\sqrt{2\pi q}}\ \left(  {\frac{2\pi i}%
{\phi^{\prime\prime}(b_{0})}}\right)  ^{1/2}\ \sqrt{b_{0}}\ K_{M}\left(
{\frac{\omega b_{0}}{\gamma v}}\right)\nonumber \\
&  \times& \exp\left\{  i\phi(b_{0}%
)+i{\frac{\pi(M+1/2)}{2}}\right\}  \;,\label{Oma7}%
\end{eqnarray}
where
\begin{equation}
\phi(b)=-qb+2\eta\ \ln(kb)\;.\label{phi}%
\end{equation}
The condition $\phi^{\prime}(b_{0})=0$ implies
\begin{equation}
b_{0}={\frac{2\eta}{q}}={\frac{a_{0}}{\sin(\theta/2)}}\;,\label{b0}%
\end{equation}
where $a_{0}=Z_{p}Z_{t}e^{2}/\mu v^{2}$.

We observe that the relation (\ref{b0}) is the same [with $\cot(\theta
/2)\sim\sin^{-1}(\theta/2)$] as that between impact parameter and deflection
angle of a particle following a classical Rutherford trajectory. Also,
\begin{equation}
\phi^{\prime\prime}(b_{0})=-{\frac{2\eta}{b_{0}^{2}}}=-{\frac{q^{2}}{2\eta}%
}\;,\label{phi2}%
\end{equation}
which implies that in the semiclassical limit
\begin{align}
|\Omega_{M}(q)|_{s.c.}^{2}  & ={\frac{4\eta^{2}}{q^{4}}}\ K_{M}^{2}\left(
{\frac{2\omega\eta}{\gamma vq}}\right) \nonumber\\
& ={\frac{1}{k^{2}}}\ \left(  {\frac{d\sigma}{d\Omega}}\right)  _{Ruth}%
\ K_{M}^{2}\left(  {\frac{\omega a_{0}}{\gamma v\sin(\theta/2)}}\right)
\;.\label{phi3}%
\end{align}
Using the above results, Eq. ~(\ref{dn}) becomes
\begin{eqnarray}
{\frac{dn_{\pi L}}{d\Omega}}&=&\left(  {\frac{d\sigma}{d\Omega}}\right)
_{Ruth}Z^{2}\alpha\ \left(  {\frac{\omega}{\gamma v}}\right)  ^{2}%
{\frac{L\left[  (2L+1)!!\right]  ^{2}}{(2\pi)^{3}\ (L+1)}%
}\nonumber \\ &\times& \sum_{M}\ |G_{\pi LM}|^{2}\ K_{M}^{2}\left(  {\frac{\omega a_{0}%
}{\gamma v\sin(\theta/2)}}\right).\label{dn2}%
\end{eqnarray}

If strong absorption is not relevant, the above formula can be
used to calculate the equivalent photon numbers. The stationary
value given by Eq.  (\ref{b0}) means that the important values of
$b$ which contribute to $\Omega_{m}(q)$ are those close to the
classical impact parameter. Dropping the index $0$ from Eq. 
(\ref{b0}), we can also rewrite Eq.  (\ref{dn2}) as
\begin{eqnarray}
{\frac{dn_{\pi L}}{2\pi b\ db}}&=&Z^{2}\alpha\ \left(  {\frac{\omega
}{\gamma v}}\right)  ^{2}{\frac{L \left[  (2L+1)!!\right]  ^{2}%
}{(2\pi)^{3}\ (L+1)}}\nonumber \\ &\times& \sum_{m}\ |G_{\pi L M}|^{2}\ K_{M}%
^{2}\left(  {\frac{\omega b}{\gamma v}}\right)  \;,\label{dn3}%
\end{eqnarray}
which leads to the semi-classical expressions given in Eqs. \eqref{(A.3a)}-\eqref{(A.3c)}.%

For very forward scattering angles, such that $\Lambda \ll 1$, a further
approximation can be made by setting the hypergeometric function in Eq. 
(\ref{Oma2}) equal to unity [GR80], and one obtains%
\begin{equation}
\Omega_{M}(q)=2^{2i\eta}\ {\frac{1}{M!}}\ \Gamma(1+M+i\eta)\ \Gamma
(1+i\eta)\ \Lambda^{M}\ \left(  {\frac{\gamma v}{\omega}}\right)  ^{2+2i\eta
}\;.\label{Oma3}%
\end{equation}

The main value of $M$ in this case will be $M=0$, for which one gets
\begin{align}
\Omega_{0}(q)  & \simeq2^{2i\eta}\ \Gamma(1+i\eta)\ \Gamma(1+i\eta)\ \left(
{\frac{\gamma v}{\omega}}\right)  ^{2+2i\eta}\nonumber\\
& =-\eta^{2}\ 2^{2i\eta}\ \Gamma(i\eta)\ \Gamma(i\eta)\ \left(  {\frac{\gamma
v}{\omega}}\right)  ^{2+2i\eta}\;,\label{Oma4}%
\end{align}
and
\begin{equation}
|\Omega_{0}(q)|^{2}=\eta^{4}\ \left(  {\frac{\gamma v}{\omega}}\right)
^{4}\ {\frac{\pi^{2}}{\eta^{2}\ \sinh^{2}(\pi\eta)}}\;,\label{Oma5}%
\end{equation}
which, for $\eta\gg1$, results in
\begin{equation}
|\Omega_{0}(q)|^{2}=4\pi^{2}\ \eta^{2}\ \left(  {\frac{\gamma v}{\omega}%
}\right)  ^{4}\ e^{-2\pi\eta}\;.\label{Oma6}%
\end{equation}

This result shows that in the absence of strong absorption and for $\eta\gg1$,
Coulomb excitation is strongly suppressed at $\theta=0$. This also follows
from semiclassical arguments, since $\theta\rightarrow0$ means large impact
parameters, $b\gg1$, for which the action of the Coulomb field is weak.

The results discussed in this section will be useful to study the corrections for
Coulomb excitation at intermediate energy collisions, which we shall discuss later.
But, before that, let us remind ourselves what are giant resonances.

\section{Excitation of giant resonances}

\subsection{What are giant resonances?}

\begin{figure}[t]
\begin{center}
\includegraphics[
natheight=6.364200in,
natwidth=9.968700in,
height=2.5189in,
width=3.3002in
]%
{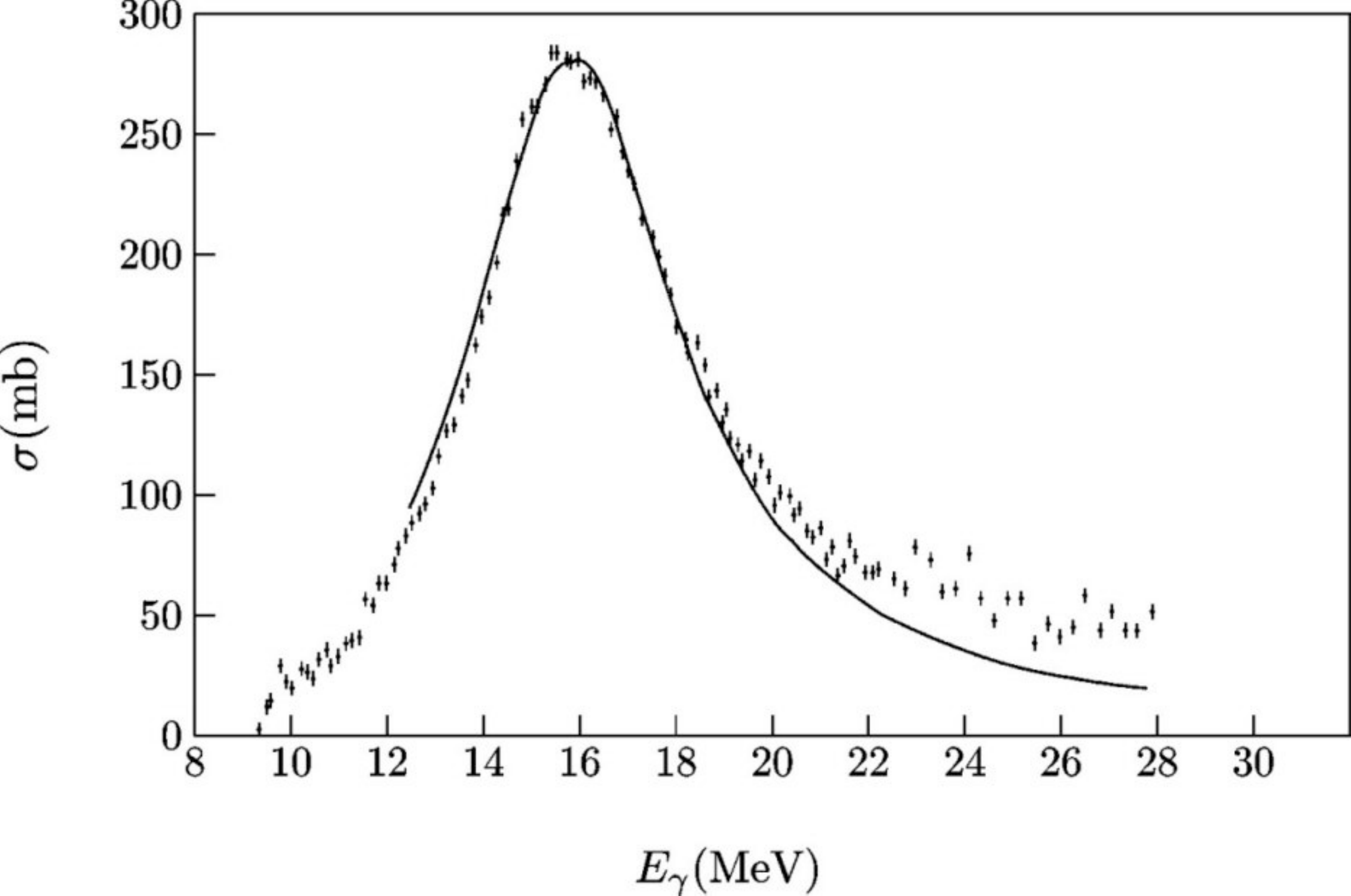}
\caption{\sl Giant resonance in the absorption of photons by
$^{120}$Sn [Le74].}%
\label{gr71}%
\end{center}
\end{figure}

Figure
\ref{gr71} exhibits the excitation function of photoabsorption of $^{120}$Sn
around the electric dipole giant resonance at 15 MeV.
The giant resonance happens in nuclei along the whole periodic table, with the
resonance energy decreasing with $A$ without large oscillations (see Figure
\ref{gr72}) starting at $A=20$. This shows that the giant resonance is a
property of the nuclear matter and not a characteristic phenomenon of nuclei
of a certain type. The widths of the resonances are almost all in the range
between 3.5 MeV and 5 MeV. It can reach 7~MeV in a few cases.

In the Goldhaber-Teller model [GT48], the photon, through the action of its electric
field on the protons, takes the nucleus to an excited state where a group of
protons oscillates in opposite phase against a group of neutrons. In such an
oscillation, those groups interpenetrate, keeping constant the
incompressibility of each group separately. A classic calculation using this
hypothesis leads to a vibration frequency that varies with the inverse of the
squared root of the nuclear radius, i.e., the resonance energy varies with
$A^{-1/6}$.

In the Steinwedel-Jensen model [SJ50] developed a classic study
of the oscillation in another way, already suggested by Goldhaber and Teller,
in which the incompressibility is abandoned. The nucleons move inside of a
fixed spherical cavity with the proton and neutron densities being a function
of the position and time. The nucleons at the surface have fixed position with
respect to each other and the density is written in such a way that, at a
given instant, the excess of protons on one side of the nucleus coincides with
the lack of neutrons on that same side, and vice-versa. Such a model leads to
a variation of the resonance energy with $A^{-1/3}$.

\begin{figure}[t]
\begin{center}
\includegraphics[
width=3.3in
]%
{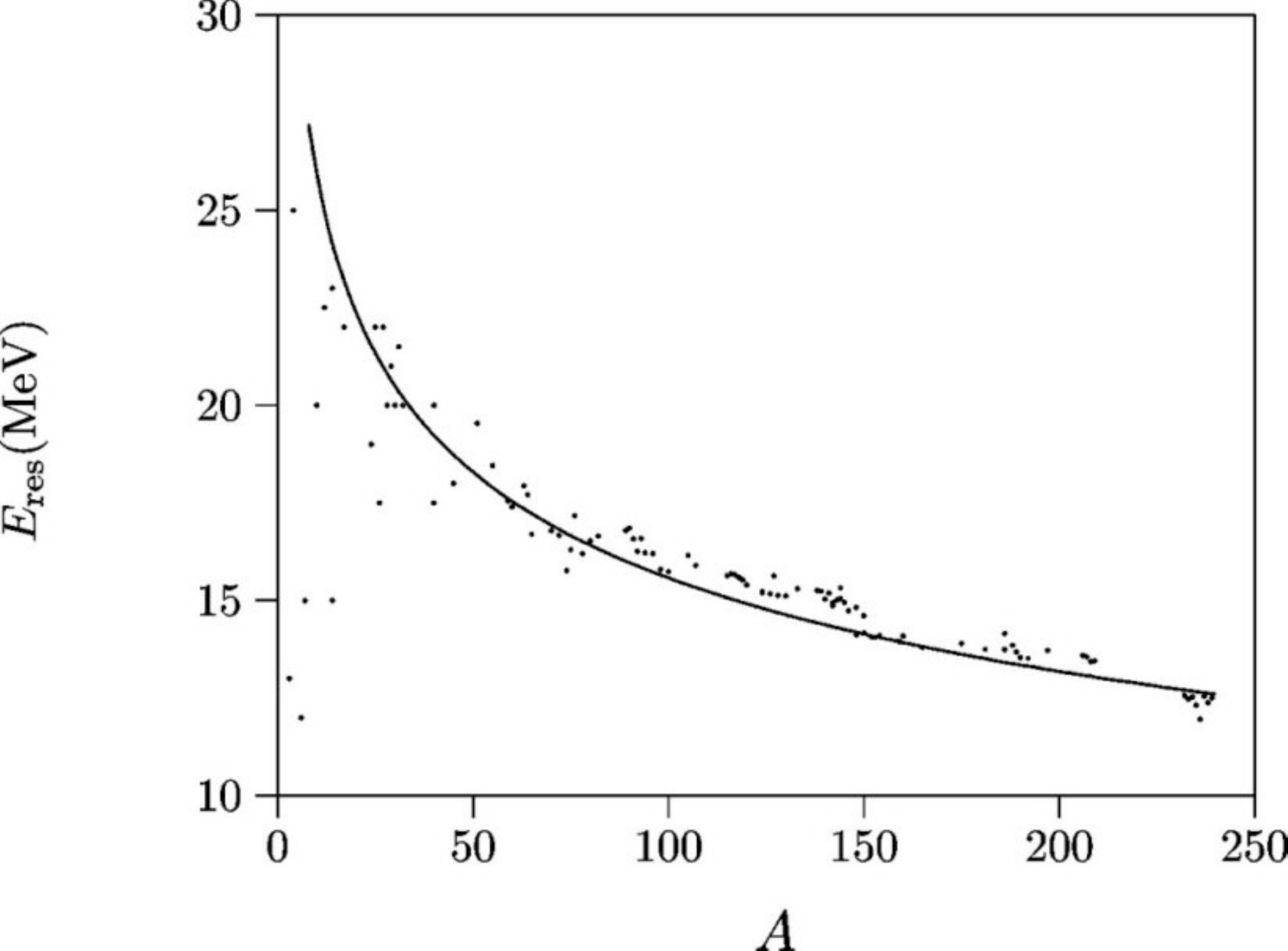}
\caption{\sl Location of the energy of the giant electric dipole
resonance given by (\ref{10.103}) (continuous curve), compared with
experimental points [DB88].}%
\label{gr72}%
\end{center}
\end{figure}

If one assumes  a mixed contribution
of the two models, obtains an expression for $E_{GDR}$ as function of the mass
number $A$ [Mye77],
\begin{equation}
E_{GDR}(\text{MeV})=112\times\left[  A^{2/3}+(A_{0}A)^{1/3}\right]
^{-1/2},\label{10.103}%
\end{equation}
where $A_{0}\cong274$. This expression, with the exception of some very light
nuclei, reproduces the behavior of the experimental values very well, as we
can see in Figure \ref{gr72}. An examination of Equation (\ref{10.103}) shows
that the Gamow-Teller mode prevails broadly in light nuclei, while the
contribution of the Steinwedel-Jensen mode is negligible. The latter mode
increases with $A$ but it only becomes predominant at the end of the periodic
table, at $A=A_{0}$.

{The giant electric dipole resonance arises from an excitation that transmits
1 unit of angular momentum to the nucleus ($\Delta l=1$). If the nucleus is
even-even it is taken to a $1^{-}$ state. What one verifies is that the
transition also changes the isospin of 1 unit ($\Delta T=1$) and, due to that,
it is also named an \textit{isovector resonance}.  Giant isoscalar resonances
($\Delta T=0$) of electric quadrupole ($\Delta l=2$) [PW71] and electric
monopole ($\Delta l=0$) [Ma75] were observed in reactions with charged
particles. The first is similar to the vibrational quadrupole state created by
the absorption of a phonon of $\lambda=2$, since both are, in even-even
nuclei, states of $2^{+}$ vibration. But the giant quadrupole resonance has a
much larger energy. This resonance energy, in the same way argued for the
dipole, decreases smoothly with $A$, obeying the approximate formula }
\begin{equation}
{E_{GQR}(}\text{{MeV}}{)\cong62A^{-1/3}.}\label{10.104}%
\end{equation}
{\hfill}%

\begin{figure}[t]
\begin{center}
\includegraphics[
width=2.2096in
]%
{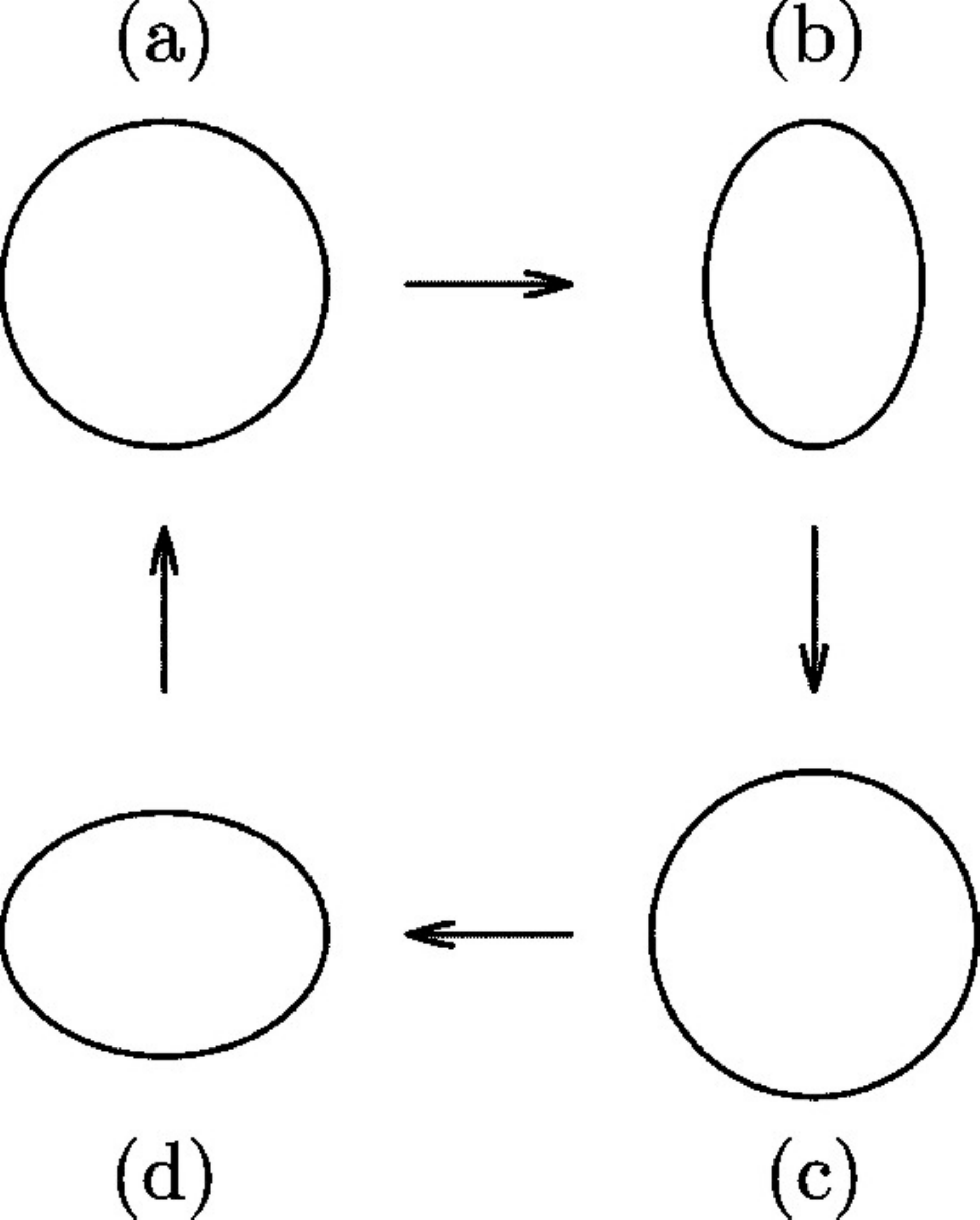}
\caption{\sl Four stages in the vibration cycle of a giant quadrupole
resonance. In an isoscalar resonance, protons and neutrons vibrate in phase,
while in an isovector resonance, the vibrations occur in opposite phase.
For opposite phases, when the
protons are at the stage (b), the neutrons will be at the stage (d), and
vice-versa. }%
\label{gr73}%
\end{center}
\end{figure}

{\ In the state of giant electric quadrupole resonance the nucleus
oscillates between the spherical (supposing that this is the form of
the ground state) and ellipsoidal form. If protons and neutrons act
in phase, we have an isoscalar resonance ($\Delta T=0$) and if they
oscillate in opposite phase the resonance is isovector ($\Delta
T=1$). Figure \ref{gr73} illustrates these two possible vibration
modes. 

The giant monopole resonance is a very special way of nuclear excitation where
the nucleus contracts and expands radially, maintaining its original form but
changing its volume. It is also called the \textit{breathing mode}. It can also
happen in the isoscalar and isovector forms. It is an important way to study
the compressibility of nuclear matter. Again here, the isoscalar form has a
reasonable number of measured cases, the location of the resonance energy
being given by the approximate expression
\begin{equation}
E_{GMR}(\text{MeV})\cong80A^{-1/3}.\label{10.105}%
\end{equation}

Besides the electric giant resonances, associated to a variation in the form
of the nucleus, magnetic giant resonances exist, involving what one calls by
\textit{spin vibrations}. In these, nucleons with spin upward move out of
phase with nucleons with spin downward. The number of nucleons involved in the
process cannot be very large because it is limited by the Pauli principle.

Another important aspect of the study of the giant resonances is the
possibility that they can be induced in already excited nuclei. This
possibility was analyzed theoretically by D. M. Brink and P. Axel [Ax62]
for giant resonances excited \textquotedblleft on top\textquotedblright\ of
nuclei rotating with high angular momentum, resulting in the suggestion that the
frequency and other properties of the giant resonances are not affected by the
excitation. A series of experiences in the decade of the 1980's (see Reference
[BB86a]) gave support to this hypothesis.

A special case happens when the giant resonance is excited on top of another
giant resonance. Understanding the excitation of a giant resonance as the
result of the absorption of one phonon, we can view these double giant
resonances as states of excitation with two vibrational phonons. The double
giant dipole resonance was observed for the first time in reactions with
double charge exchange induced by pions in $^{32}$S [Mo88]. As first
shown in reference [BB86] a much better possibility to study multiple
giant resonances is by means of Coulomb excitation with relativistic heavy
projectiles. Later on, this was indeed verified experimentally and several
properties of multiple giant resonances have been studied theoretically (for a
theoretical review, see [BP99]).

\begin{figure}
[t]
\begin{center}
\includegraphics[
natheight=6.396100in, natwidth=8.364500in, height=2.309in,
width=3.0101in
]%
{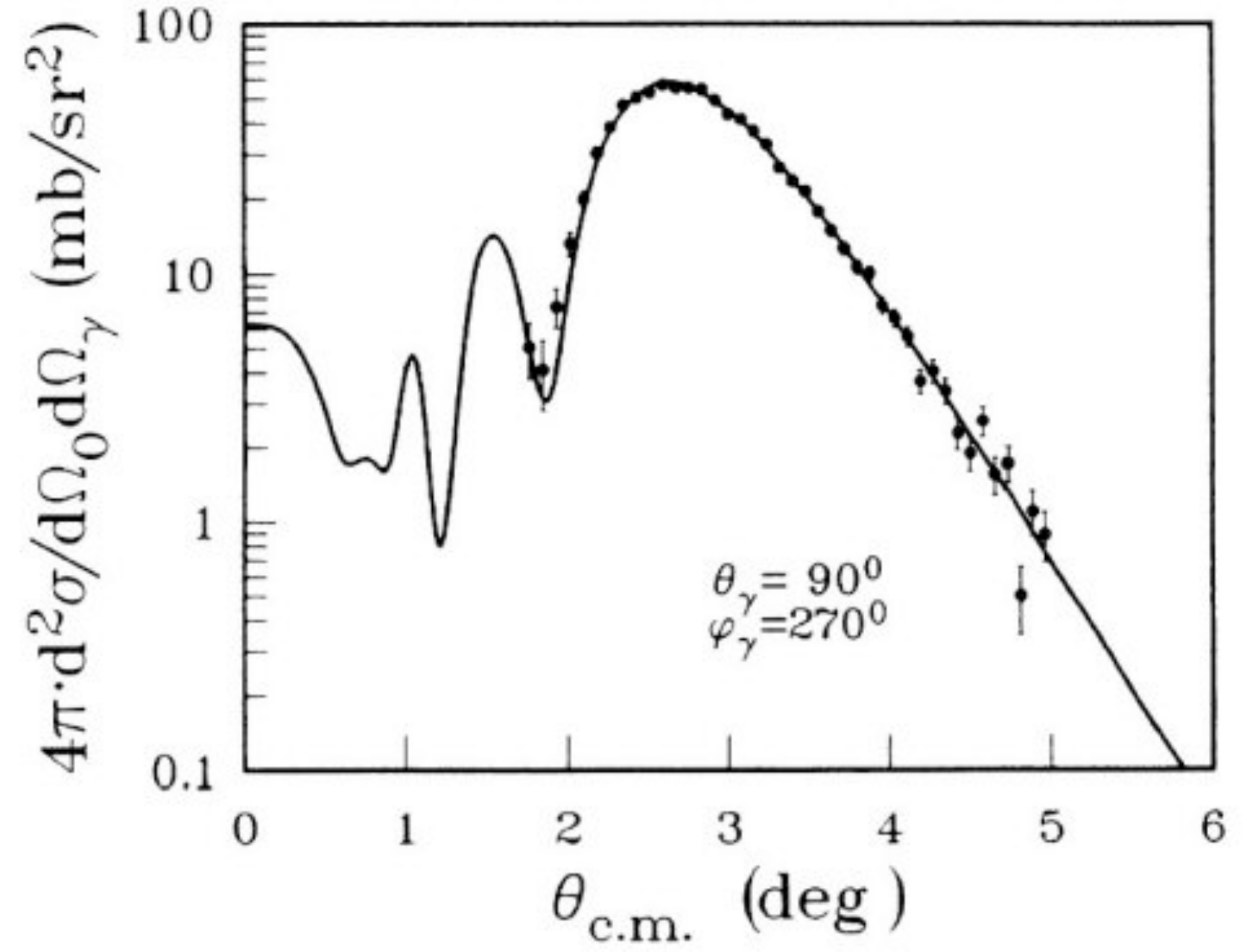}%
\caption{\sl  Cross section for the excitation of the giant dipole
resonance
followed by $\gamma$-decay to the ground state in the reaction $^{208}%
$Pb($^{17}$O,$^{17}$O'$\gamma_{0}$) at 84 MeV/nucleon for fixed $\gamma$ angle
$\theta_{\gamma}=90^{\mathrm{o}}$ and $\phi_{\gamma}=270^{\mathrm{o}}$. The
points are experimental [Bee90]. The curve is predicted for Coulomb
excitation.}%
\label{beene}%
\end{center}
\end{figure}

Figure \ref{beene} shows the cross section for the excitation of the giant
dipole resonance followed by $\gamma$-decay to the ground state in the
reaction $^{208}$Pb($^{17}$O,$^{17}$O'$\gamma_{0}$) at 84 MeV/nucleon for
fixed $\gamma$ angle $\theta_{\gamma}=90^{\mathrm{o}}$ and $\phi_{\gamma
}
=270^{\mathrm{o}}$. The points are experimental [Bee90]. The curve is
predicted for Coulomb excitation using the eikonal wavefunction, as described in the previous section, and a reduced transition strength
calculated according the deformed potential model [BN93]. The agreement with the data is excellent.

\subsection{Sum Rules}

It is useful to be able to estimate the total photoabsorption
cross section summed over all transitions $|i\rangle\rightarrow
|f\rangle$ from the initial, for instance ground, state. Such estimates are
given by the {\it sum rules} (SR) which approximately determine quantities of
the following type:
\begin{equation}
{\cal S}^{(n)}_{i}[F]=\frac{1}{2}\sum_{f}(E_{f}-E_{i})^{n}\Bigl\{\Bigl|
\langle f|F|i\rangle\Bigr|^{2}+\Bigl|\langle f|F^{\dagger}|i\rangle
\Bigr|^{2}\Bigr\}.                                          \label{sra}
\end{equation}
Here the transition probabilities for an arbitrary pair of mutually conjugate
operators $F$ and $F^{\dagger}$ are
weighted with a certain (positive, negative or equal to zero) power $n$ of the
transition energy. For a hermitian operator $F=F^{\dagger}$ the two terms in
(\ref{sra}) are equal and the factor 1/2 is cancelled.

The exact result follows immediately for non-energy-weighted SR, $n=0$,
based on the completeness of the set of the states $|f\rangle$,
\begin{equation}
{\cal S}^{(0)}_{i}[F]=\frac{1}{2}\langle i|F^{\dagger}F+FF^{\dagger}|i\rangle.
                                                               \label{srb}
\end{equation}
Often it turns out to be possible to get a good estimate for the expectation
value in the right hand side of (\ref{srb}), or to extract it from data.

For the {\it energy-weighted} sum rules (EWSR),
${\cal S}^{(1)}$, a reasonable estimate can
be derived for many operators under certain assumptions about the interactions
in the system. First, using again the completeness of the intermediate states
$|f\rangle$, we can identically write down ${\cal S}^{(1)}$ as an
expectation value in the initial state $|i\rangle$ of the double commutator
\begin{equation}
{\cal S}^{(1)}_{i}[F]=\frac{1}{2}\langle i|\Bigl[[F,H],F^{\dagger}\Bigr]|
i\rangle,                                             \label{src}
\end{equation}
where $H$ is the total hamiltonian which has energies $E_{i}$ and $E_{f}$ as
its eigenvalues. Thus, we again need to know the properties of the initial
state only.  Now we choose the operator
$F$ as a one-body quantity depending on coordinates ${\bf r}_{a}$ of the
particles,
\begin{equation}
F=\sum_{a}f_{a}, \quad f_{a}=f({\bf r}_{a}).             \label{srd}
\end{equation}
Apart from that we assume that the hamiltonian does not contain
momentum-dependent interactions. Then only the kinetic part of the
hamiltonian contributes to $[F,H]$, and the result can be found explicitly,
\begin{equation}
[F,H]=\Bigl[\sum_{a}f_{a},\sum_{b}\frac{{\bf p}_{b}^{2}}{2m_{b}}\Bigr]=
\sum_{a}\frac{i\hbar}{2m_{a}}[(\nabla_{a}f_{a}),{\bf p}_{a}]_{+}, \label{sre}
\end{equation}
where $[...\,,\,...]_{+}$ denotes the anticommutator. The outer commutator in
(\ref{src}) leads now to the simple result
\begin{equation}
{\cal S}^{(1)}_{i}[F]=\sum_{a}\frac{\hbar^{2}}{2m_{a}}\langle
i|\Bigl|\nabla_{a} f_{a}\Bigr|^{2}|i\rangle.              \label{srf}
\end{equation}

As an example we take the charge form factor
\begin{equation}
F=\sum_{a}e_{a}e^{i({\bf k}\cdot{\bf r}_{a})}.
                                                            \label{srg}
\end{equation}
The sum rule in this case is universal for any initial state $|i\rangle$,
\begin{equation}
{\cal S}^{(1)}[F]=\hbar^{2}{\bf k}^{2}\sum_{a}\frac{e_{a}^{2}}
{2m_{a}}.                                                     \label{srh}
\end{equation}
Taking ${\bf k}$ along an
(arbitrary) $z$-axis and considering the long wavelength
limit, $kR\ll 1$, we get from the first nonvanishing term in the
expansion of the exponent the EWSR for the dipole operator, $d_z=rY_{10}(\hat{\bf r})$,
\begin{equation}
{\cal S}^{(1)}[d_{z}]=
\sum_{a}\frac{\hbar^{2}e_{a}^{2}}{2m_{a}}.                 \label{sri}
\end{equation}
This is an extension of the old {\it Thomas-Reiche-Kuhn} (TRK)
dipole SR in atomic physics. For a neutral atom, in the center-of-mass
frame attached to the nucleus of charge $Z$ (here $m$ is the electron mass),
\begin{equation}
S^{(1)}[d_{z}]=\frac{\hbar^{2}e^{2}}{2m}Z.                 \label{sri1}
\end{equation}
The atomic TRK SR is essentially exact (up to relativistic velocity-dependent
corrections).

In (\ref{srg}) $e_{a}$ are in fact arbitrary numbers. For intrinsic dipole
excitations we have to exclude the
center-of-mass motion. Therefore our $z$-coordinates should be intrinsic
coordinates, $z_{a}\Rightarrow z_{a}-R_{z}$, where $R_{z}=\sum_{a}z_{a}/A$.
Hence, the intrinsic dipole moment is
\begin{equation}
d_{z}= \sum_{a}e_{a}(z_{a}-R_{z})=e\sum_{p}z_{p}-\frac{Ze}{A}\Bigl(\sum_{p}
z_{p} +\sum_{n}z_{n}\Bigr).                               \label{srj}
\end{equation}
This operator can be rewritten as
\begin{equation}
d_{z}=e_{p}\sum_{p}z_{p}+e_{n}\sum_{n}z_{n}                      \label{srk}
\end{equation}
where protons and neutrons carry {\it effective charges}
\begin{equation}
e_{p}=\frac{N}{A}e, \quad e_{n}=-\frac{Z}{A}e.                \label{srl}
\end{equation}
Now (\ref{sri}) gives the {\it dipole EWSR}
\begin{eqnarray}
{\cal S}^{(1)}_{i}[d_{z}]&\equiv&\sum_{f}E_{fi}|d^{z}_{fi}|^{2}\\
&=&\frac{\hbar^{2}e^{2}}{2m_N}\left[Z\left(
\frac{N}{A}\right)^{2}+N\left(\frac{-Z}{A}\right)^{2}\right]\\
&=&\frac{\hbar^{2}
e^{2}}{2m_N}\frac{NZ}{A},                              \label{srm}
\end{eqnarray}
where $m_N$ is the nucleon mass.

The factor $(NZ/A)$ is connected to the reduced mass for relative motion
of neutrons against protons as required at the fixed center of mass. This
result does not include the dipole strength related to nuclear motion as a
whole. According to the classical SR (\ref{sri1}), this contribution is
$(m\rightarrow Am_N, \,e\rightarrow Ze)$
\begin{equation}
{\cal S}^{(1)}[{\rm c.m.}] =\frac{\hbar^{2}(Ze)^{2}}{2Am_N}.    \label{srn}
\end{equation}
The sum of the global (\ref{srn}) and intrinsic (\ref{srm}) dipole strength
recovers the full TRK SR (\ref{sri1}),
\begin{equation}
{\cal S}^{(1)}_{i}[{\rm tot.\; dip}]=\frac{\hbar^{2}e^{2}}{2m_N}\left(
\frac{NZ}{A}+\frac{Z^{2}}{A}\right)=\frac{\hbar^{2}e^{2}}{2m_N}Z.   \label{sro}
\end{equation}
In contrast to the atomic
TRK case, the nuclear dipole EWSR (\ref{srm}) cannot be
exact. Velocity-dependent and exchange forces are certainly present in nuclear
interactions. Nevertheless, Eq.  (\ref{srm}) gives a surprisingly good estimate
of the realistic dipole strength which is not fully understood.
In a similar way one can consider SR for other multipoles but the results are
not universal and in general depend on the initial state. 

The EWSR (\ref{srm}) is what we need to evaluate the sum of integral
dipole cross sections for real photons over all possible final states $|f\rangle$.
Taking the photon polarization vector along the $z$-axis, we come to the
total dipole photoabsorption cross section
\begin{equation}
\sigma_{tot}^\gamma=\sum_{f}\int dE_{\gamma}\sigma_{fi}^\gamma=2\pi^{2}
\frac{e^{2}\hbar}{m_Nc}\frac{ZN}{A}.                    \label{sr4}
\end{equation}
This universal prediction,
\begin{equation}
\sigma_{tot}^\gamma=0.06 \frac{ZN}{A}{\rm barn}\cdot{\rm MeV}, \label{sr5}
\end{equation}
on average agrees well with experiments in spite of crudeness of
approximations made in the derivation. One should remember that it
includes only dipole absorption.

For the E2 isoscalar giant quadrupole resonances one
has the approximate sum rule [BM75]  
\begin{equation}
\int {dE_{\gamma}\over E_\gamma^2}\sigma_{GQR}^\gamma (E_\gamma)\simeq 0.22 ZA^{2/3}
\ \mu{\rm b\  MeV}^{-1}.                    \label{sr5b}
\end{equation}

\subsection{Coulomb excitation of giant 
resonances}

A simple estimate of Coulomb excitation of giant resonances based on sum rules can be made by
assuming that the virtual photon numbers vary slowly compared to the photonuclear cross sections 
around the resonance peak. Then
\begin{eqnarray}
\sigma_C &\simeq& {N_{E1} (E_{GDR}) 
\over E_{GDR}} \int dE_{\gamma}\sigma_{GDR}^\gamma (E_\gamma)\nonumber \\
&+& N_{E2} (E_{GQR}) E_{GQR} \int {dE_{\gamma}\over E_\gamma^2}\sigma_{GQR}^\gamma (E_\gamma).
    \label{sr6}
\end{eqnarray}

In figure \ref{grs} we show the Coulomb excitation cross section of giant resonances in $^{40}$Ca projectiles hitting a $^{238}$U target as a function of the laboratory energy per nucleon. The dashed line corresponds to the excitation of the giant electric dipole resonance, the dotted to the electric quadrupole, and the lower line to the magnetic dipole which was also obtained using a sum-rule for M1 excitations [BB88]. The solid curve is the sum of these contributions.

\begin{figure}[t]
\begin{center}
\includegraphics[
width=3.4in
]%
{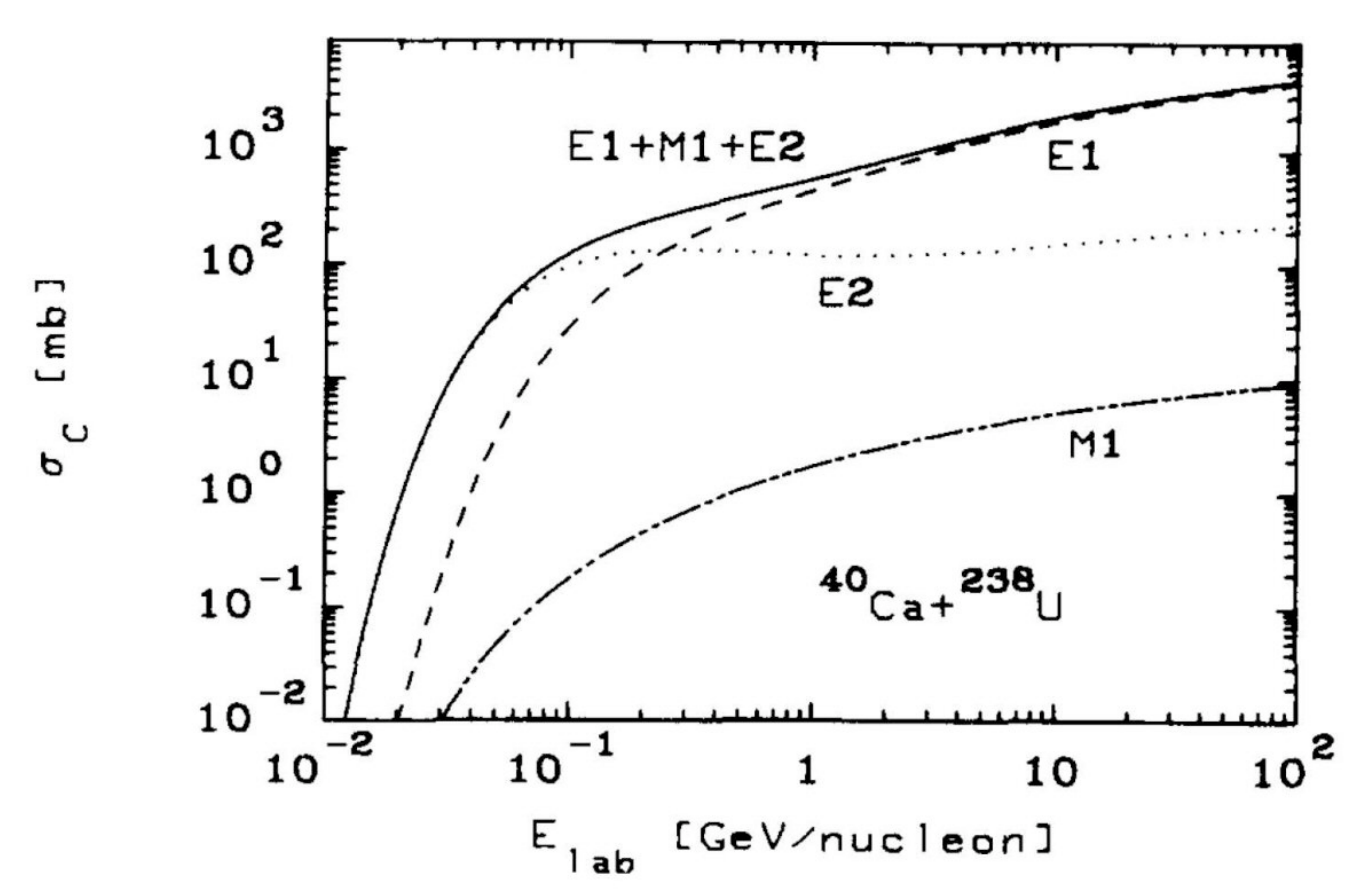}
\caption{\sl   Coulomb excitation cross section of giant resonances in $^{40}$Ca projectiles hitting a $^{238}$U target as a function of the laboratory energy per nucleon. The dashed line corresponds to the excitation of the giant electric dipole resonance, the dotted to the electric quadrupole, and the lower
line to the magnetic dipole. The solid curve is the sum of these contributions. }%
\label{grs}%
\end{center}
\end{figure}

The cross sections increase very rapidly
to large values, which are already attained at intermediate energies.
A salient feature is that the cross section for the excitation of
giant quadrupole
modes is very large at low and intermediate energies, decreasing in
importance (about $10$\% of the $E1$ cross section) as the energy increases
above 1 GeV/nucleon. This occurs because the equivalent photon
number for the $E2$ multipolarity is much larger than that for the $E1$
multipolarity at low collision energies. That is, $n_{E2} \gg n_{E1}$,
for $v\ll c$.
This has a simple explanation, as we already discussed in connection with figure \ref{WW2}.
Pictorially, as seen from an observator at rest,
when a charged particle moves at low energies
the lines of force of its corresponding electric field are
isotropic, diverging from its center in all directions. This means that
the field carries a large amount of tidal ($E2$) components.
On the other hand, when the particle
moves very fast its lines of force appear contracted in the
direction perpendicular to its motion due to Lorentz contraction.
For the observator this field looks like a pulse of plane waves of light.
But plane waves contain all multipolarities with the same weight, and
the equivalent photon numbers become all approximately equal, i.e.,
$n_{E1} \simeq n_{E2} \simeq n_{M1}$, and increase logarithmically
with the energy for $\gamma \gg 1$. The difference in the cross sections
when $\gamma \gg 1$ are then approximately equal to the difference
in the relative strength of the two giant resonances $\sigma_\gamma^{E2}
/\sigma^{E1}_\gamma < 0.1$. The excitation of giant magnetic monopole
resonances is of less importance, since for low energies $n_{M1} \ll
n_{E1}$ ($n_{M1} \simeq (v/c)^2 n_{E1}$), whereas for high energies, where
$n_{M1} \simeq n_{E1}$, it will be also much smaller than the excitation
of electric dipole resonances since their relative strength
$\sigma^{M1}_\gamma/\sigma^{E1}_\gamma$ is much smaller than unity.

At very large energies the cross sections for the Coulomb excitation
of giant resonances overcome the nuclear geometrical cross sections.
Since these resonances decay mostly through particle emission or fission,
this indicates that Coulomb excitation of giant resonances is
a very important process to be considered in relativistic heavy
ion collisions
and fragmentation processes,
especially in heavy ion colliders.
At intermediate energies the cross sections are also
large and this offers good possibilities to establish
and study the properties of giant resonances.

The formalism described in the previous sections has also been used  in the analysis of the data of
Ref. [BB90], in which a projectile of $^{17}$O with an
energy of $E_{lab}=84$ MeV/nucleon excites the target nucleus $^{208}$Pb
to the Giant Dipole Resonance (GDR). The optical potential has a
standard Woods-Saxon form with parameters given in Ref. [BB90]. In order
to calculate the inelastic cross section for the excitation of the
GDR, one can use a Lorentzian parameterization for the
photoabsorption cross section of $^{208}$Pb [Ve70],
\begin{equation}\sigma_{\pi\lambda}^\gamma =\sigma_{m} \;
{E_\gamma^2 \Gamma^2 \over (E_\gamma^2 - E_{R}^2 )^2
+ E_\gamma^2 \Gamma^2} . \label{5.3}\end{equation}
with $\sigma_{m}$ chosen to reproduce the Thomas-Reiche-Kuhn
sum rule for $E1$ excitations, and Eq. \eqref{sr5b} for isoscalar giant
quadrupole resonances.
We use the
widths $\Gamma_{\mbox{\tiny GDR}} = 4 \; $MeV and $\Gamma_{\mbox{\tiny
GQR}} = 2.2 \;$MeV for $^{208}$Pb.

At $ \sim 84$ MeV/nucleon the maximum
scattering angle which still leads to a pure Coulomb scattering
(assuming a sharp cut-off at an impact parameter $b=R_P +R_T$)
is $\sim 4^\circ$. 
Inserting this form into Eq.  (\ref{sig}) and doing the
calculations implicit in Eq.  (\ref{dn}) for $dn_{E1}/d\Omega$, one
obtains the angular distribution which is compared to the data in
Fig. \ref{fig_coul_5_1}. The agreement with the data is excellent,
provided one adjusts the overall normalization to a value
corresponding to $93\,\%$ of the energy weighted sum rule (EWSR)
in the energy interval $7-18.9\;$MeV (see section 6.10). Taking
into account the $\pm10\%$ uncertainty in the absolute cross
sections quoted in Ref.~ [Bar88],
this is consistent with photoabsorption cross section in that energy range.%
\begin{figure}
[ptb]
\begin{center}
\includegraphics[
width=3.5in
]%
{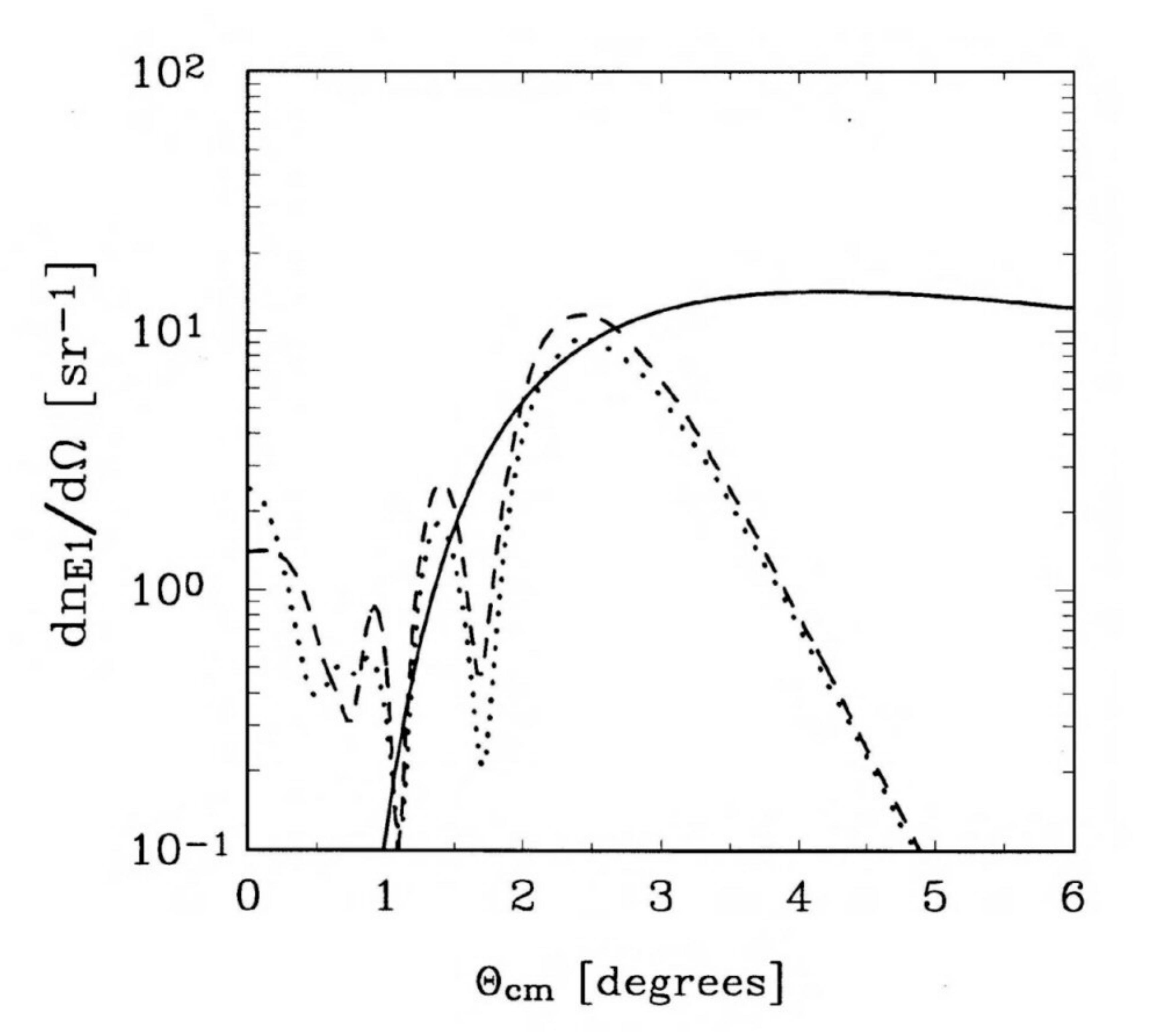}%
\caption{\sl  Virtual photon numbers for the electric dipole multipolarity
generated by 84A MeV $^{17}$O projectiles incident on $^{208}$Pb, as a
function of the center-of-mass scattering angle. The solid curve is a
semiclassical calculation. The dashed and dotted curves are eikonal
calculations with and without relativistic corrections, respectively.}%
\label{fig_coul_4_2}%
\end{center}
\end{figure}

To unravel the effects of relativistic corrections, one \ can
repeat the previous calculations unplugging the factor
$\gamma=(1-v^{2}/c^{2})^{-1}$ which appears in the expressions
(\ref{n}) and (\ref{g}) and using the non-relativistic limit of
the functions $G_{E1m}$ of Eq.  (\ref{Glm}). These modifications
eliminate the relativistic corrections on the interaction
potential. The result of this calculation is shown in fig.
\ref{fig_coul_4_2} (dotted curve). For comparison, 
the result of a full calculation, keeping the relativistic
corrections (dashed curve), is also shown. One observes that the two results have
approximately the same pattern, except that the non-relativistic
result is slightly smaller than the relativistic one. In fact, if
one repeats the calculation for the excitation of GDR using the
non-relativistic limit of eqs. (\ref{n}) and (\ref{g}), one finds
that the best fit to the data is obtained by exhausting $113\,\%$
of the EWSR. This value is very close to the
$110\,\%$ obtained in Ref. [Bar88].%
\begin{figure}
[ptb]
\begin{center}
\includegraphics[
width=3.6in
]%
{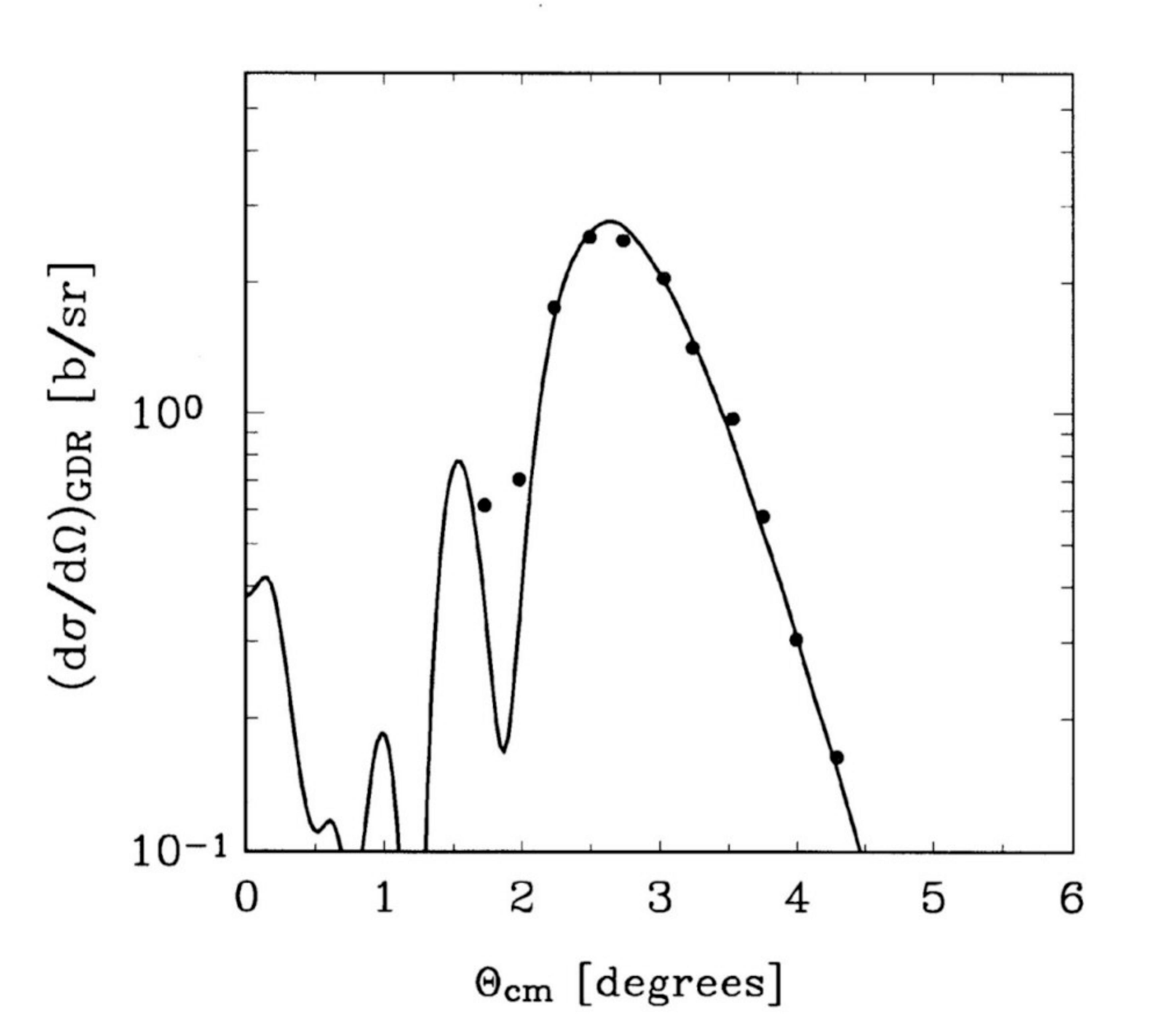}%
\caption{\sl  Differential cross section for the excitation of the isovector giant dipole resonance in Ô$^{208}$Pb by means of $^{17}$0 projectiles at 84 MeV/nucleon, as a function of the center-of-mass scattering angle. Data are from Ref. [BB90]. The theoretical calculation is from Ref. [BN93].}%
\label{fig_coul_5_1}%
\end{center}
\end{figure}

In fig. \ref{fig_coul_4_2} we also show the result of a
semiclassical
calculation (solid curve) for the GDR excitation in lead, using Eq. ~(\ref{dn2}%
) for the virtual photon numbers. The semiclassical curve is
not able to fit the experimental data in figure \ref{fig_coul_5_1}, which shows a perfect agreement with the Coulomb excitation with the eikonal approximation [BN93]. This is mainly because diffraction
effects and strong absorption are not included. But the semiclassical
calculation displays the region of relevance for Coulomb excitation. At small
angles the scattering is dominated by large impact parameters, for which the
Coulomb field is weak. Therefore the Coulomb excitation is small and the
semiclassical approximation fails. It also fails in describing the large angle
data (dark-side of the rainbow angle), since absorption is not treated
properly. One sees that there is a \textquotedblleft window\textquotedblright%
\ in the inelastic scattering data near $\theta=2-3^{\circ}$ in which the
semiclassical and full calculations give approximately the same cross section.%

\begin{figure}
[ptb]
\begin{center}
\includegraphics[
width=3.3in
]%
{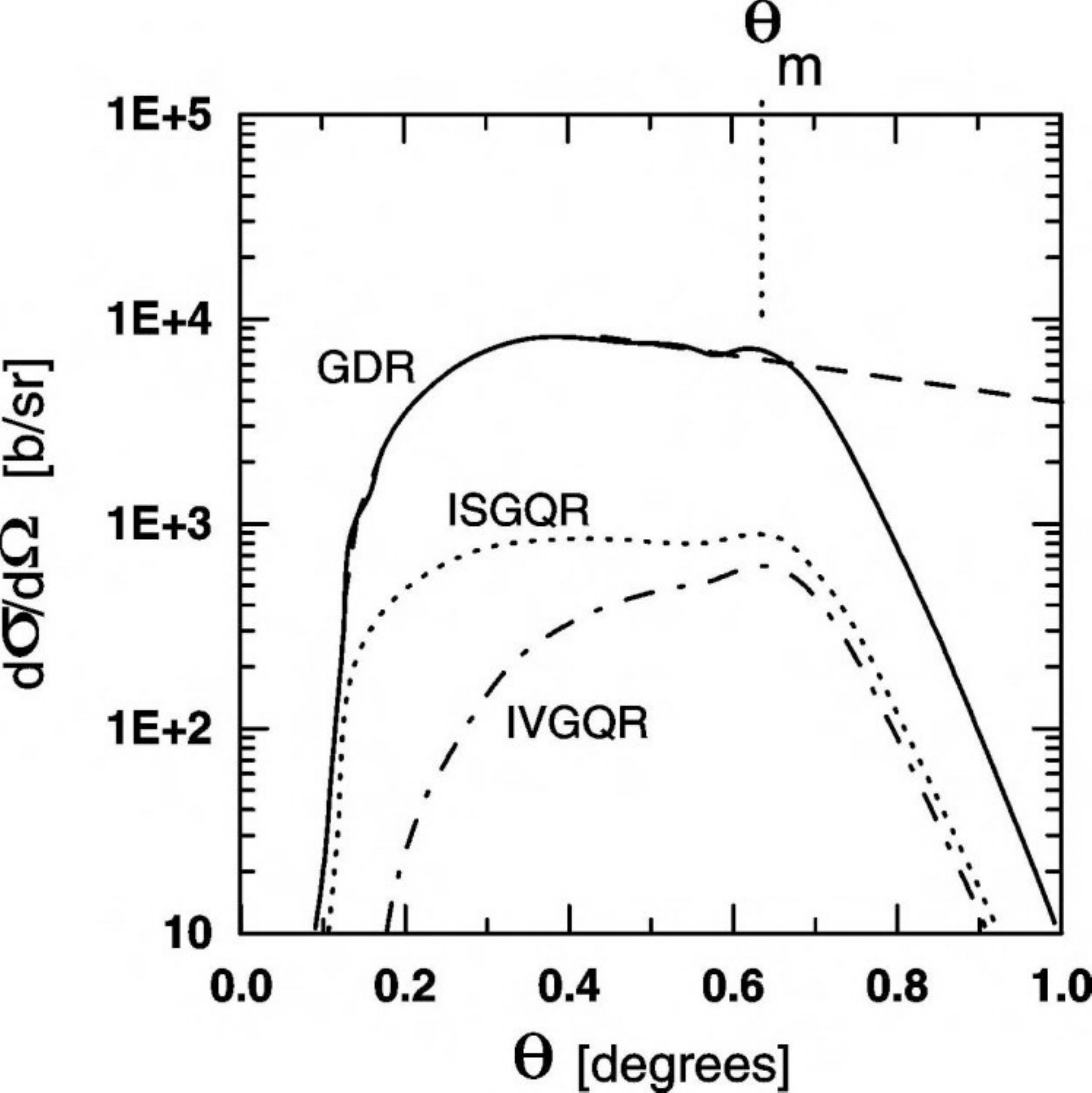}%
\caption{\sl  Differential cross section for the excitation of a giant dipole
resonance in Pb+Pb collisions at 640 MeV/nucleon, as a function of the
center of mass scattering angle. }%
\label{fig_coul_5_2}%
\end{center}
\end{figure}

In fig. \ref{fig_coul_5_2} we show a similar calculation, but for
the excitation of the GDR in Pb for the collision $^{208}$Pb +
$^{208}$Pb at 640 MeV/nucleon. The dashed line is the result of a
semiclassical calculation. Here we see that a purely semiclassical
calculation, is able to reproduce the quantum results up to a
maximum scattering angle $\theta_{m}$, at which strong absorption
sets in. This justifies the use of semiclassical calculations for
heavy systems, even to calculate angular distributions. The cross
sections increase rapidly with increasing scattering angle, up to
an approximately constant value as the maximum Coulomb scattering
angle is approached. This is explained as follows. Very forward angles
correspond to large impact parameter collisions in which case
$\omega b/\gamma v>1$, the virtual photon numbers in Eq.  (\ref{n})
drop quickly to zero, and the excitation of giant resonances in
the nuclei is not achieved. As the impact parameter decreases,
increasing the scattering angle, $\omega b/\gamma v\lesssim1$ and
excitation occurs. 

\subsection{Excitation and photon decay of the GDR}

We now consider the excitation of the target nucleus to the giant dipole
resonance and the subsequent photon decay of that excited nucleus, leaving the
target in the ground state. Experimentally, one detects the inelastically
scattered projectile in coincidence with the decay photon and demands that the
energy lost by the projectile is equal to the energy of the detected photon.
To the extent that the excitation mechanism is dominated by Coulomb
excitation, with the exchange of a single virtual photon, this reaction is
very similar to the photon scattering reaction, except that in the present
case the incident photon is virtual rather than real. In this section, we
investigate whether the connection between these two reactions can be formalized.

We have seen that, under the conditions $\Delta E/E\ll1$, the cross section
for excitation of the target nucleus factorizes into the following expression
(we assume that the contribution of the $E1$-multipolarity is dominant):
\begin{equation}
{\frac{d^{2}\sigma_{C}}{d\Omega dE_{\gamma}}}\left(  E_{\gamma}\right)
={\frac{1}{E_{\gamma}}}\ {\frac{dn_{\gamma}}{d\Omega}}\left(  E_{\gamma
}\right)  \ \sigma_{\gamma}\left(  E_{\gamma}\right).\label{dsig1}%
\end{equation}

Figure \ref{fig_coul_6_1} the cross section for the excitation of the GDR without the detection of
the decay photon. Data are from Ref. ~[Bee90]. The solid line is a calculation from Ref. [BN93] using the eikonal wavefunction
for the Coulomb excitation.

\begin{figure}
[ptb]
\begin{center}
\includegraphics[
width=3.5in
]%
{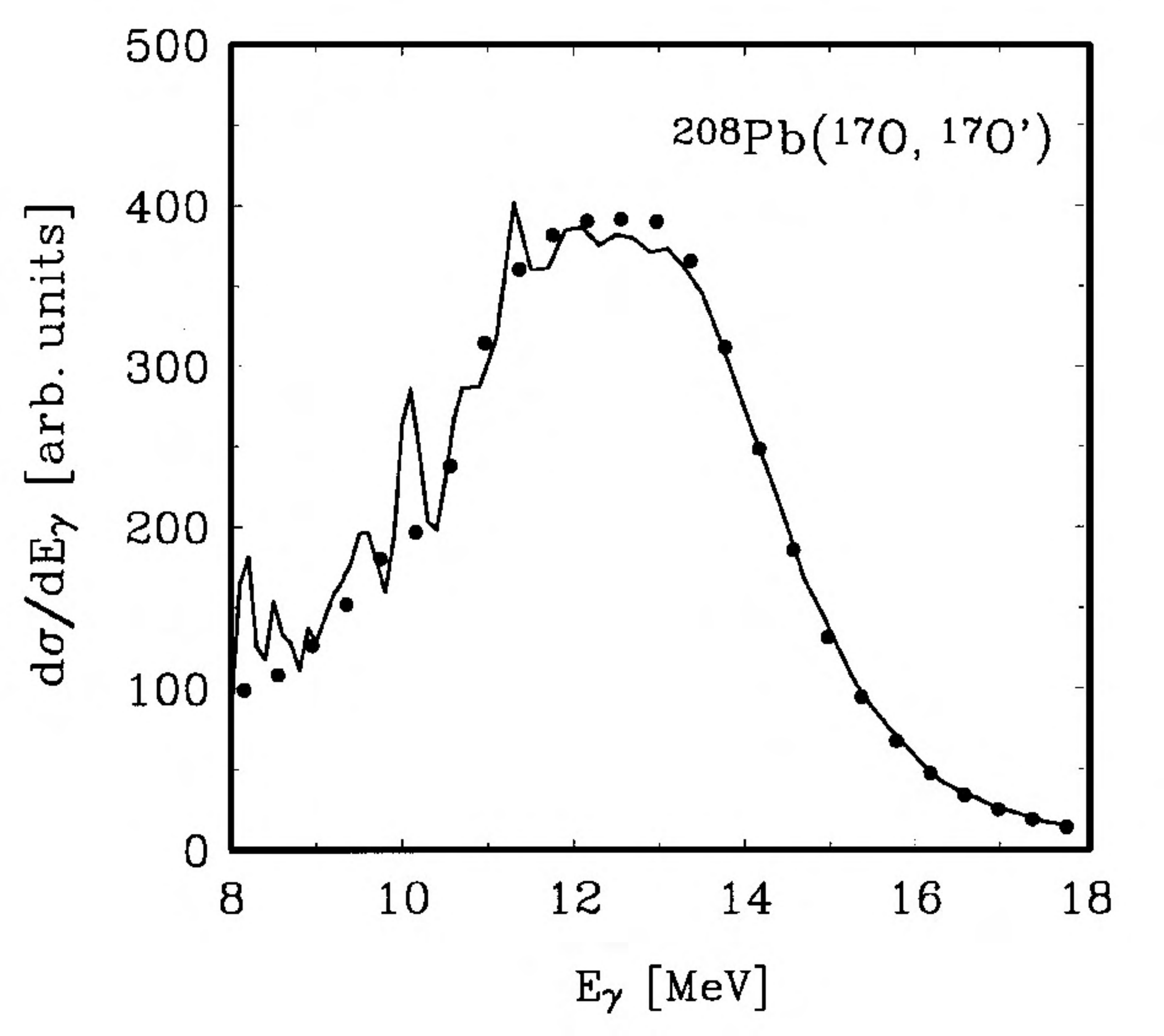}%
\caption{\sl  Cross section for the excitation of the GDR without the detection of
the decay photon. Data are from Ref. ~[Bee90]. The solid line is a calculation from Ref. [BN93].}%
\label{fig_coul_6_1}%
\end{center}
\end{figure}

The usual way to write the cross section $d^{2}\sigma_{C\gamma}/d\Omega
dE_{\gamma}$ for the excitation of the target followed by photon decay to the
ground state is simply to multiply the above expression by a branching ratio
$R_{\gamma}$, which represents the probability that the nucleus excited to an
energy $E_{\gamma}$ will emit a photon leaving it in the ground state
[Bee90]:
\begin{equation}
{\frac{d^{2}\sigma_{C\gamma}}{d\Omega dE_{\gamma}}}\left(  E_{\gamma}\right)
={\frac{1}{E_{\gamma}}}\ {\frac{dn_{\gamma}}{d\Omega}}\left(  E_{\gamma
}\right)  \ \sigma_{\gamma}\left(  E_{\gamma}\right)  \ R_{\gamma}\left(
E_{\gamma}\right)  .\label{dsig2}%
\end{equation}
Instead, one can also use the following expression, in complete analogy with
Eq. ~\eqref{dsig1}:
\begin{equation}
{\frac{d^{2}\sigma_{C\gamma}}{d\Omega dE_{\gamma}}}\left(  E_{\gamma}\right)
={\frac{1}{E_{\gamma}}}\ {\frac{dn_{\gamma}}{d\Omega}}\left(  E_{\gamma
}\right)  \ \sigma_{\gamma\gamma}\left(  E_{\gamma}\right)  ,\label{dsig3}%
\end{equation}
where $\sigma_{\gamma\gamma}\left(  E_{\gamma}\right)  $ is the
cross section for the elastic scattering of photons with energy
$E_{\gamma}$. Formally, these expressions are equivalent in that
they simply define the quantity $R_{\gamma}$. However, if one
treats $R_{\gamma}$ literally as a branching ratio, then these
expressions are equivalent only if it were true that the photon
scattering cross section is just product of the photoabsorption
cross section and the branching ratio. In fact, it is well-known
from the theory of photon scattering that the relationship between
the photoabsorption cross section and the photon scattering cross
section is more complicated. In particular, it is
not correct to think of photon scattering as a two-step process
consisting of absorption, in which the target nucleus is excited
to an intermediate state of energy $E_{\gamma}$, followed by
emission, in which the emitted photon has the same energy
$E_{\gamma}$. Since the intermediate state is not observable, one
must sum over all possible intermediate states and not just those
allowed by conservation of energy. Now, if the energy $E_{\gamma}$
happens to coincide with a narrow level, then that level will
completely dominate in the sum over intermediate states. In that
case, it is proper to regard the scattering as a two-step process
in the manner described above, and the two expressions for the
cross section will be equal. However, for $E_{\gamma}$ in the
nuclear continuum region (e.g., in the region of the GDR), this
will not be the case.

\begin{figure}
[ptb]
\begin{center}
\includegraphics[
width=3.4in
]%
{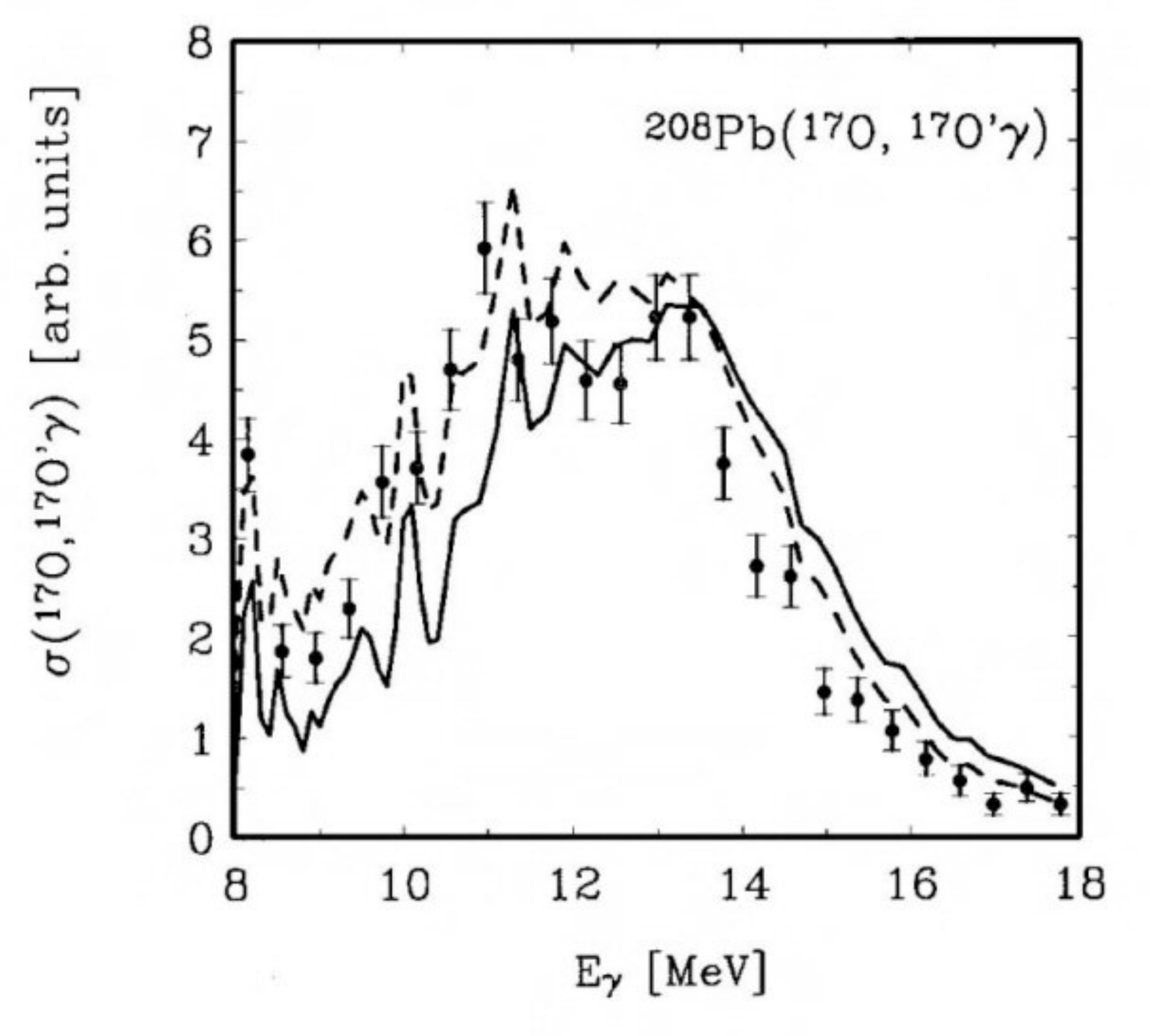}%
\caption{\sl  Cross section for excitation followed by $\gamma$-decay
of $^{208}$Pb by $^{17}$O projectiles at 84A MeV. The solid
(dashed) line includes (excludes) the Thompson scattering
amplitude. Data are from Ref. ~
[Bee90].}%
\label{fig_coul_6_2}%
\end{center}
\end{figure}

The calculation [BN93] is compared to the data in fig.
\ref{fig_coul_6_2}. The left part shows the cross section for the
excitation of the GDR without the detection of the decay photon.
The agreement with the data is excellent. The right part of fig.
\ref{fig_coul_6_2} shows the cross section for the
excitation-decay process as a function of $E_{\gamma}$. Although
the qualitative trend of the data are well described, the
calculation systematically overpredicts the cross section on the
high-energy side of the GDR (solid curve). If the Thompson
amplitude is not included in $\sigma _{\gamma\gamma}$, the
calculation is in significantly better agreement with the data
(dashed curve).

\subsection{Multiphonon resonances}
The Coulomb excitation of harmonic vibrations is formulated in Appendix F. The formalism can be applied to excitation of multiple giant resonances, i.e. giant resonances excited on top of other giant resonances, also called {\it multiphonon giant resonances}. Coulomb excitation of multiphonon giant resonances was formulated in the harmonic vibrator model in Ref. [BB86]. The excitation of multiphonon states may be viewed as the absorption by the target (projectile) of several photons from the pulse of equivalent photons generated by the relativistic projectile (target); or it can be also described as multiple excitation of a harmonic oscillator with GDR quantum energy.  

As explained in Appendix F, the probability for the excitation of a $N$-phonon state, $P_N(b)$ can be obtained from the first-order theory  for the probability, $P_{f.o.}(b)$, 
\begin{equation}
P_{N}=\frac{1}{N!}\;\left| P_{1st} (b)\right| ^{N}\;\exp \left[ -P_{1st}(b)\right] .
\end{equation}
$P_{1st}(b)$ is calculated as in explained in the previous sections.

\begin{figure}
[ptb]
\begin{center}
\includegraphics[
width=3.4in
]%
{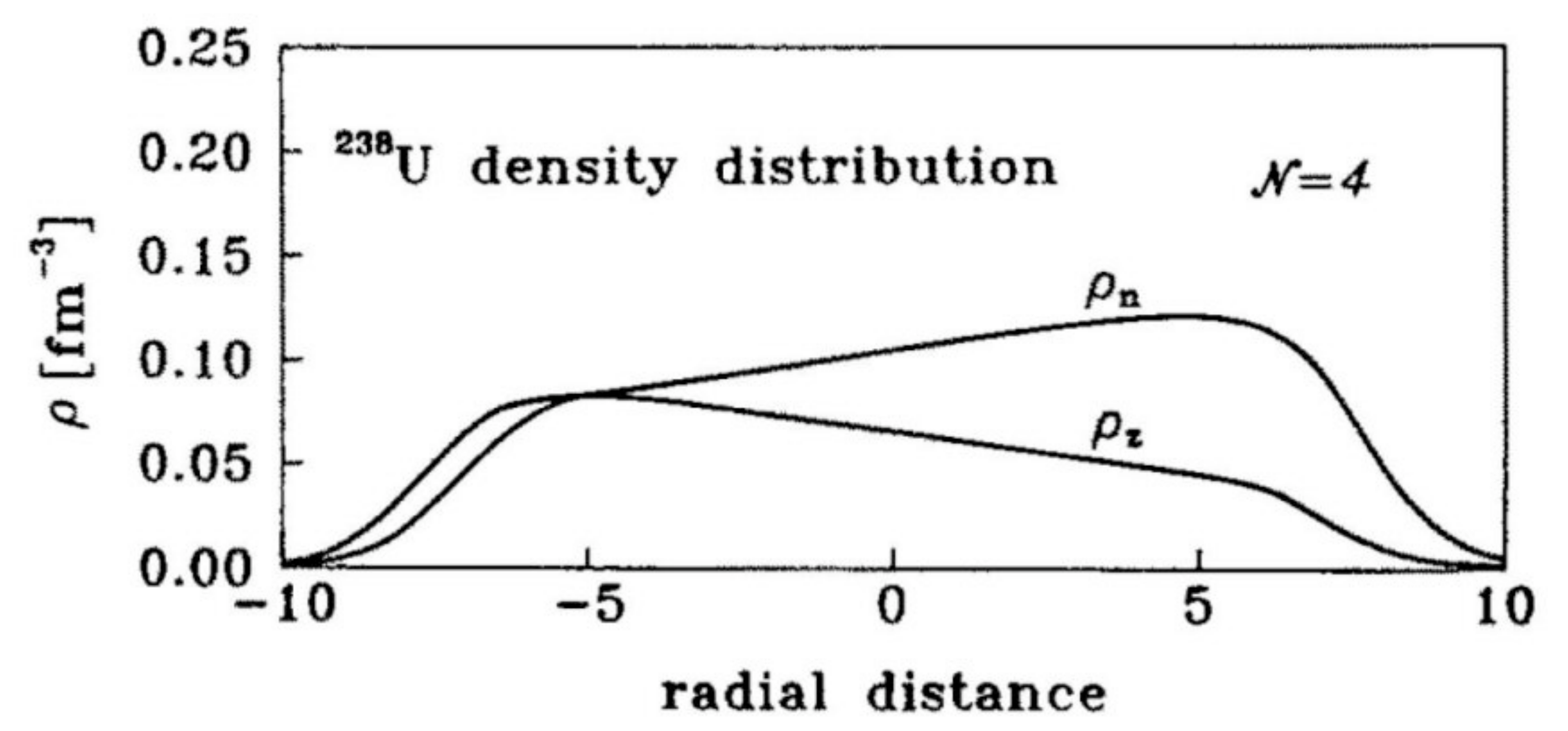}%
\caption{\sl  Proton and neutron density distributions at the moment of largest separation between them in a
$N = 4$ multiphonon state in $^{238}$U [VGB90].}%
\label{mpgrd}%
\end{center}
\end{figure}

The multiphonon states correspond to large-amplitude vibrations of neutrons against protons in nuclei. Although the energy deposit is small (multiples of the energy of a GDR state), such large collective motion may lead to exotic decays of the nuclei. This is a relatively cold fragmentation process, in contrast to the violent fragmentation following central collisions of relativistic heavy ion collisions, in which high pressures and temperatures are achieved. Figure \ref{mpgrd} shows the neutron and proton matter distribution in $^{238}$U for a $N=4$ multiphonon GDR state at the moment of maximum displacement [VGB90].

The first Coulomb excitation
experiments for the excitation of the double giant dipole resonance (DGDR) were performed at the GSI facility in Darmstadt/Germany [Rit93,Sch93]. In Fig. \ref{emling} we show the result of one of these experiments [Sch93], which looked for the
neutron decay channels of giant resonances excited in relativistic projectiles. The excitation
spectrum of relativistic $^{136}$Xe projectiles incident on Pb are compared with the spectrum obtained
in C targets. A comparison of the two spectra immediately proofs that nuclear contribution to the
excitation is very small. Another experiment [Rit93] dealt with the photon decay of the double giant
resonance. A clear bump in the spectra of coincident photon pairs was observed around the energy
of two times the GDR centroid energy in $^{208}$Pb targets excited with relativistic $^{209}$Bi projectiles.

\begin{figure}
[ptb]
\begin{center}
\includegraphics[
width=3.4in
]%
{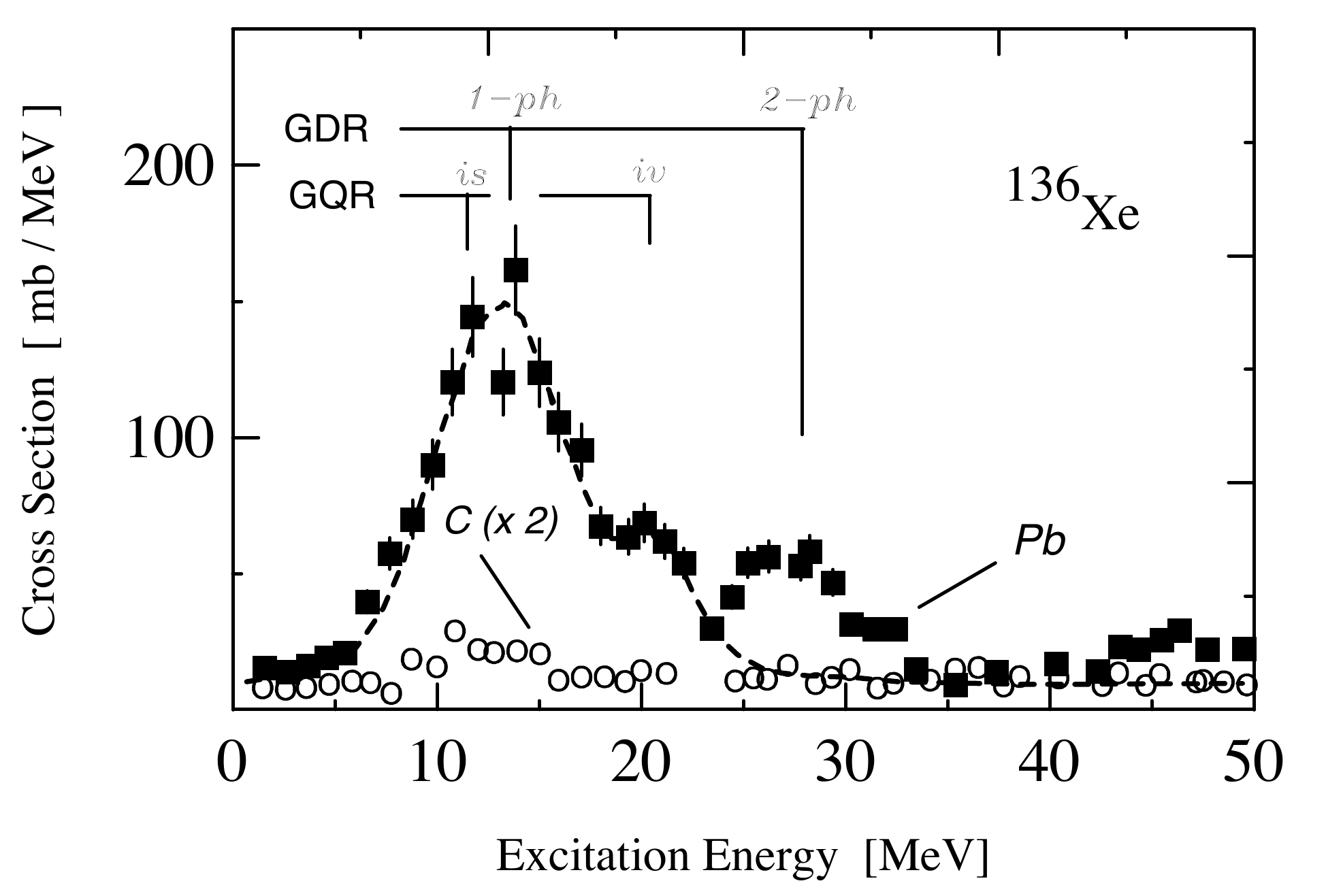}%
\caption{\sl  Experimental results for $^{136}$Xe projectile excitation (at 690 MeV/nucleon) on a Pb target (squares) and a C target
(circles). The spectrum for the C target is multiplied by a factor 2 for better presentation. The resonance energies for one and
two-phonon giant resonances are indicated. The dashed curve reflects the results of a first-order calculation for the
Pb target. The figure is taken from Ref. [Sch93].}%
\label{emling}%
\end{center}
\end{figure}

Much of the interest on multiphonon resonances relies on the possibility of looking at exotic
particle decay of these states. Since there is more energy deposit in the nuclei, other decay channels are open for the
multiphonon states. Generally, the GRs in heavy nuclei decay by neutron emission. One expects
that the double, or triple, GDR decays mainly in the 2n and 3n decay channel. In fact, such a picture
has been adopted in [ABS93] with success to explain the total cross sections for the neutron removal
in peripheral collisions. 

Coulomb excitation of multiphonon resonances is a rich playground for theories of nuclear response. There are still open issues related to the magnitude of the cross sections, the widths of the multiphonon states and the excitation mechanisms in general. Some of the reviews on the subject are Refs.  [ABE98,BP99].

\section{Intermediate energy collisions}

The semiclassical theory of Coulomb excitation in low energy collisions
accounts for the Rutherford bending of the trajectory, but relativistic
retardation effects are neglected [AW75]. On the other hand, in the
theory of relativistic Coulomb excitation [WA79] recoil effects on
the trajectory are neglected (one assumes straight-line motion) but
retardation is handled correctly. In fact, the onset of retardation brings
new important effects such as the steady increase of the excitation cross
sections with bombarding energy. In a heavy ion collision around 100A MeV
the Lorentz factor $\gamma $ is about 1.1. Since this factor enters the
excitation cross sections in many ways, such as in the adiabacity
parameter, $\xi (R)={\omega _{fi}R/}\gamma v$, one expects that some
sizable modifications in the theory of relativistic Coulomb excitation
should occur [AB89]. Recoil corrections are not negligible either,
and the relativistic calculations based on the straight-line parametrization
should not be completely appropriate to describe the excitation
probabilities and cross sections. The Coulomb recoil in a single collision
is of the order of $
a_{0}={{Z_{1}Z_{2}e^{2}}/{m_{0}v^{2}},}$
which is half-distance of closest approach in a head-on collision,
with m$_{0}$ equal to the reduced mass of the colliding nuclei. Although
this recoil is small for intermediate energy collisions, the excitation
probabilities are quite sensitive to it. This is important for example in
the excitation of giant resonances because the adiabacity parameter is of
the order of one. When $\xi (b)\ll 1$, the excitation probabilities depends
on $b$ approximately like $1/b^{2}$, while when $\xi (b)$ becomes greater
than one they decrease approximately as $e^{-2\pi \xi (b)}/b^{2}$.
Therefore, when $\xi \simeq 1$ a slight change of $b$ may vary appreciably
the excitation probabilities.

\subsection{Recoil and retardation corrections}

In intermediate
energy collisions, where one wants to account for recoil and retardation
simultaneously, one should solve the general classical problem of the motion
of two relativistic charged particles. This has been studied in Refs. [AB89,AAB90]. But, even if radiation is neglected,
this problem can only be solved if one particle has infinite mass 
[LL79]. This approximation should be sufficient if we take, e.g., the
collision $^{16}$O + $^{208}$Pb as our system. 

In the classical one-body problem, one starts with the relativistic
Lagrangian 
\begin{equation}
\mathcal{L}=-m_{o}c^{2}\left\{ 1-{\frac{1}{c^{2}}}\,({\dot{r}}^{2}+r^{2}{%
\dot{\phi}}^{2})\right\} ^{1/2}-{\frac{Z_{1}Z_{2}e^{2}}{r}}\ ,  \label{3.6}
\end{equation}
where ${\dot{r}}$ and ${\dot{\phi}}$ are the radial and the angular velocity
of the particle, respectively. Using the
Euler-Lagrange equations one finds three kinds of solutions, depending on
the sign of the charges and the angular momentum in the collision. In the
case of our interest, the appropriate solution relating the collisional
angle $\phi $ and the distance r between the nuclei is [LL79]
\begin{equation}
{\frac{1}{r}}=A\left[ \epsilon \;\cos (W\phi )-1\right]  \label{3.7}
\end{equation}
where 
\begin{eqnarray}
W &=&\left[ 1-({\frac{Z_{1}Z_{2}e^{2}}{c{L}_{0}}})^{2}\right]
^{1/2},\;\;\;\;\;\;A={\frac{Z_{1}Z_{2}e^{2}E}{c^{2}{L}_{0}^{2}W^{2}}}%
\,,  \label{3.8a} \\
\epsilon &=&{\frac{c{L}_{0}}{Z_{1}Z_{2}e^{2}E}}\,\left[
E^{2}-m_{0}^{2}c^{4}+({\frac{m_{0}cZ_{1}Z_{2}e^{2}}{{L}_{0}}})^{2}\right]
^{1/2}\,.  \label{3.8c}
\end{eqnarray}
$E$ is the total bombarding energy in MeV, $m_{0}$ is the mass of the
particle and ${L}_{0}$ its angular momentum. In terms of the Lorentz factor $%
\gamma $ and of the impact parameter $b$, $E=\gamma m_{o}c^{2}$ and ${L}%
_{0}=\gamma m_{0}vb$. The above solution is valid if ${L}%
_{0}>Z_{1}Z_{2}e^{2}/c$. In heavy ion collisions at intermediate energies
one has $L_{0}\gg Z_{1}Z_{2}e^{2}/c$ for impact parameters that do not lead
to strong interactions. It is also easy to show that, from the magnitudes of
the parameters involved in heavy ion collisions at intermediate energies,
the trajectory \eqref{3.7} can be very well described by approximating 
\begin{equation}
W=1\,,\qquad A=\frac{a_{0}}{\gamma b^{2}}\,,\qquad \epsilon =\sqrt{{\frac{%
b^{2}\gamma ^{2}}{a_{0}^{2}}}+1}\ ,  \label{3.9}
\end{equation}
where $a_{0}$ is half the distance of closest approach in a head on
collision (if the nuclei were pointlike and if non-relativistic kinematics
were used), and $\epsilon $ is the eccentricity parameter. In the
approximation \eqref{3.9} $\epsilon $ is related to the deflection angle $%
\vartheta $ by $\epsilon =(a_{0}/\gamma )\cot \vartheta $.

The time dependence for a particle moving along the trajectory, Eq. \eqref{3.7}, may be directly obtained by solving the equation of angular momentum
conservation. Introducing the parametrization 
\begin{equation}
r(\chi )={\frac{a_{0}}{\gamma }}\,\left[ \epsilon \,\cosh \chi +1\right]
\label{3.10}
\end{equation}
we find 
\begin{equation}
t={\frac{a_{0}}{\gamma v}}\,\left[ \chi +\epsilon \,\sinh \chi \right] \,.
\label{3.11b}
\end{equation}
Using the scattering plane perpendicular to the z-axis, one finds that the
corresponding components of $\mathbf{r}$ may be written as 
\begin{eqnarray}
x &=&a\,\left[ \cosh \chi +\epsilon \right] \ ,  \nonumber \\
y &=&a\,\sqrt{\epsilon ^{2}-1}\,\sinh \chi \ ,  \nonumber \\
z &=&0\ ,  \label{3.12cd}
\end{eqnarray}
where $a=a_{0}/\gamma $. This parametrization is of the same form as
in the non-relativistic case, Eq. \eqref{3.12c}, except that $a_{0}$
replaced by $a_{0}/\gamma \equiv a$.

Non-relativistically
the two-body problem is solvable by introduction of center of mass and
relative motion coordinates. Then, the result is equivalent to that of a
particle with reduced mass $m_{0}=m_{P}m_{T}/(m_{P}+m_{T})$ under the action
of the same potential. The particle with reduced mass m$_{0}$ is lighter
than those with mass m$_{P}$ and m$_{T}$, and this accounts for the
simultaneous recoil of them. An exact relativistic solution should reproduce
this behavior as the relative motion energy is lowered. We can use the
reduced mass definition of m$_{0}$ as usual in the parametrization of the
classical trajectory of Coulomb excitation in intermediate energy
collisions, as outlined above. For a $^{16}$O + $^{208}$Pb collision this is
not a bad approximation. For heavier systems like U+U it would be the
simplest way to overcome this difficulty. But, as energy increases, this
approximation is again unimportant since the trajectories will be
straight-lines parametrized by an impact parameter b. An exact result
was obtained numerically in Ref. [AAB90]
using the Darwin Lagrangian to determine the classical trajectory in
collisions at intermediate energies. But, the parametrization of the
classical trajectory as given by Eqs. \eqref{3.12cd} with a reduced mass particle, besides reproducing both the
non-relativistic and the relativistic energies, gives a reasonable solution
to the kind of collisions we want to study.

From here on the procedure is the same as in the derivation of low-energy
excitation amplitudes. Whenever $a_0$ appears in those equations it will have
to be replaced by $a_0/\gamma$. The orbital integrals in Eqs. \eqref{3.19a} and
\eqref{3.19b}, will be done along the trajectory defined by Eqs. \eqref{3.12cd}. Explicitly,
one finds that the orbital integrals in Eq. \eqref{Obtnr} becomes [AB89]
\begin{eqnarray}
I(E \lambda, \mu)
&=& -i ({v \eta \over c}) ^{\lambda+1} \, {1 \over \lambda (2\lambda-1)!!}
\, e^{- \pi \eta / 2  }\nonumber \\
&\times& \int^\infty _ {-\infty} d\chi \; e^{-\eta \epsilon \cosh
\chi}\, e^{i \eta \chi} \nonumber \\
& \times& {(\epsilon + i \sinh
\chi - \sqrt {\epsilon^2 -1} \cosh   \chi)^ \mu \over
(i \epsilon \sinh   \chi + 1) ^{ \mu -1}}\nonumber \\
& \times& \left\lbrack ( \lambda + 1) \, h_ \lambda -
z \, h_{\lambda + 1}
-({v \over c})^2 \epsilon \, \eta \,  \cosh
\chi \cdot h_ \lambda \right\rbrack \nonumber \\
\label{3.27}
\end{eqnarray}
where  $h_\lambda$'s are spherical Hankel functions, and 
\begin{equation}
z = {v \over c} \, \eta \, (i\epsilon \sinh
\chi +1) \, .\label{3.28}
\end{equation}

For magnetic excitations [AB89],
\begin{eqnarray}
I(M\lambda , \mu ) &=&
{i(v \eta /c)^{\lambda +1} \over (2\lambda
-1)!!} \, e^{-\pi \eta /2} \,
\, \int^\infty _{-\infty} d \chi \; h_\lambda (z) \;
e^{- \eta \epsilon \, \cosh   \chi }\nonumber \\
& \times&  e^{i \eta \chi}\,
{ (\epsilon + i \sinh   \chi - \sqrt { \epsilon ^2 -1} \, \cosh
\chi)^\mu
\over (i \epsilon \sinh   \chi + 1)^\mu } \, .
\label{3.19}
\end{eqnarray}

The excitation amplitudes are still given by Eq. \eqref{a1st}, with the
expressions for $S(\pi LM)$ obtained from Eqs. (\ref{3.19a}-\ref{3.19b}) by
replacing $a_0$ by $a_0/\gamma$ and using $\epsilon =(a_{0}/\gamma )\cot \vartheta $.
The excitation cross sections are given by Eq. \eqref{4.5a} and the virtual photon numbers by
Eqs. (\ref{4.7.6}-\ref{4.7.6b}).

\begin{figure}[tb]
\includegraphics[
width=3.4in
]%
{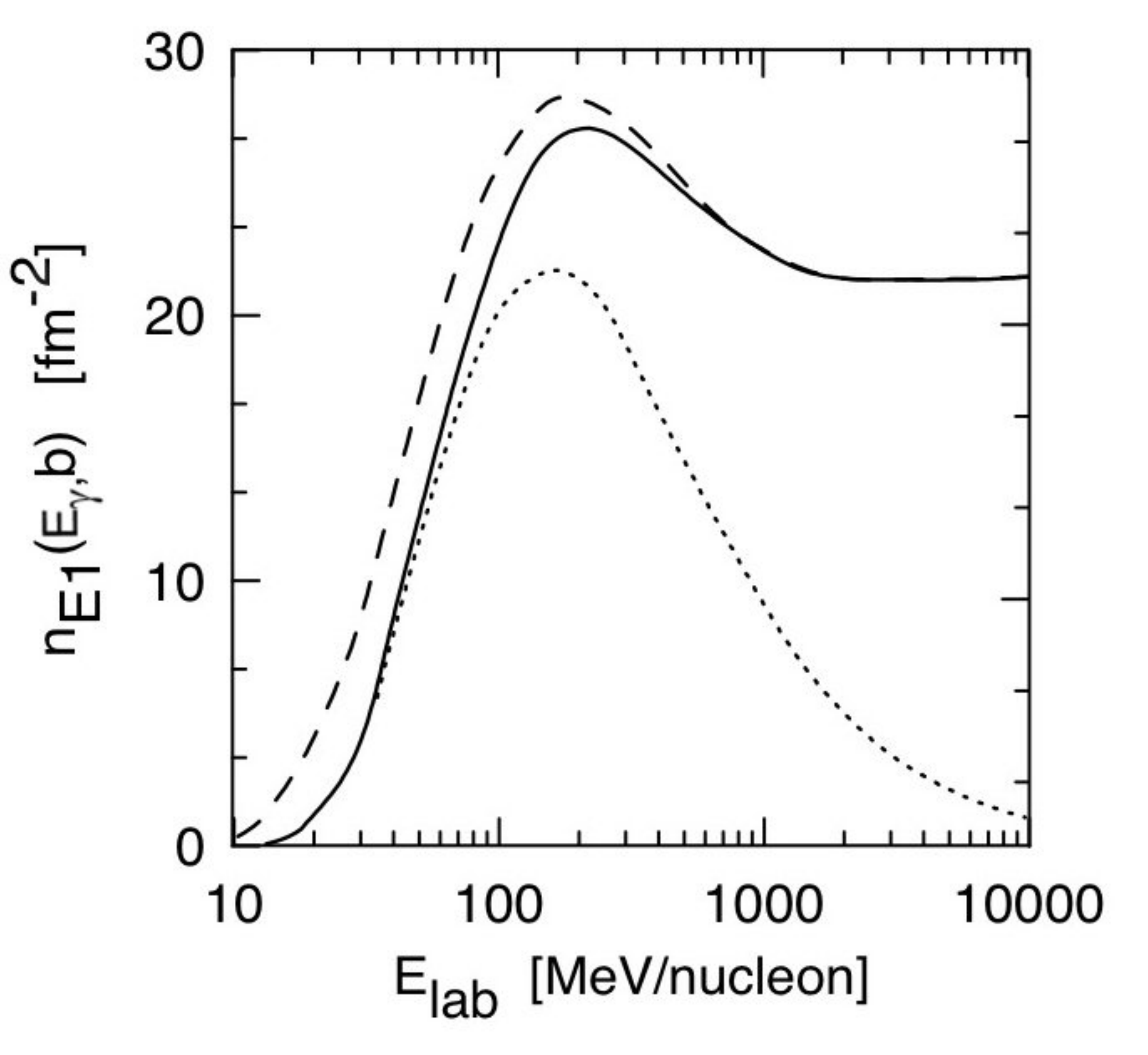}%
\caption
{\sl Electric dipole number of equivalent photons per unit
area $d^2b \equiv 2 \pi
b \, db$, with energy of 10~MeV, incident on $^{208}$Pb in a
collision
with $^{16}$O at impact parameter $b= 15$~fm, and as a function
of the bombarding energy in MeV per nucleon. The dotted line and the
dashed line correspond to calculations performed with the non-relativistic
and with the relativistic approaches, respectively. The solid line represents
a more correct calculation, as described in the text.
\label{a-f3}}
\end{figure}

\begin{figure}[tb]
\includegraphics[
width=3.4in
]%
{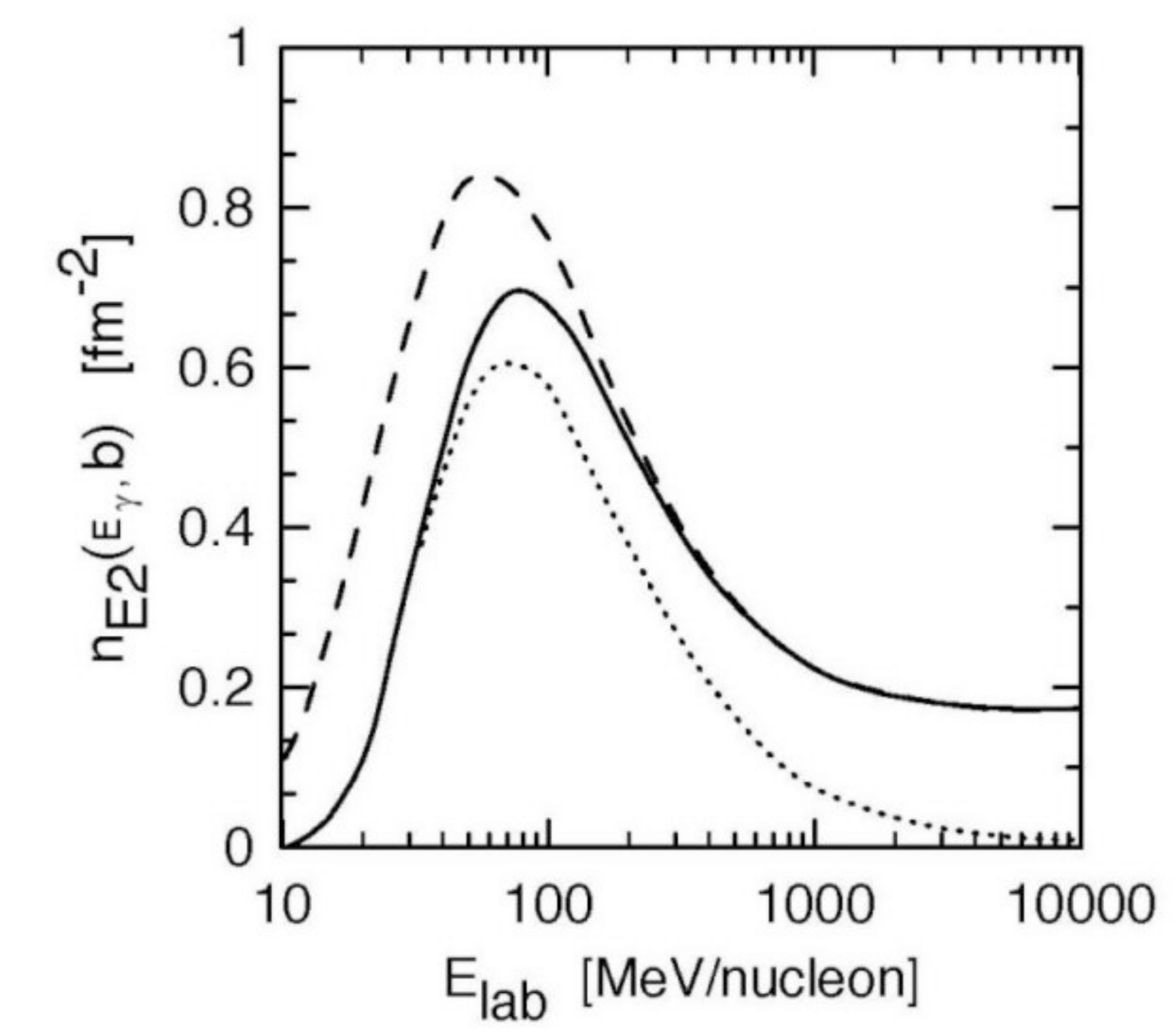}%
\caption{\sl  Same as Fig.~\protect\ref{a-f3}, but for the
$E2$ multipolarity.
\label{a-f4}}
\end{figure}

\begin{figure}[tb]
\includegraphics[
width=3.4in
]%
{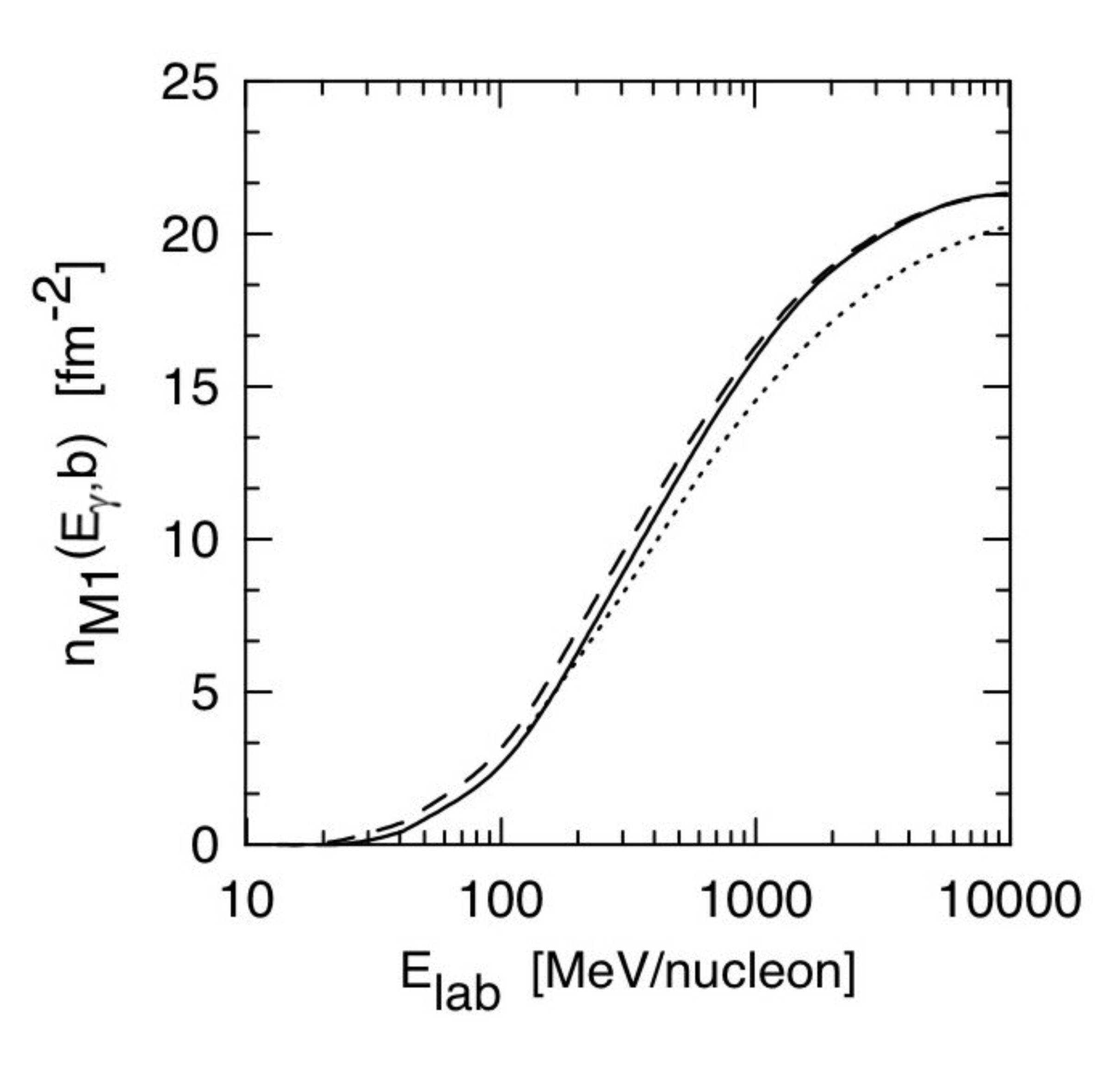}%
\caption
{\sl Same as Fig.~\protect\ref{a-f3}, but for the $M1$ multipolarity.
\label{a-f5}}
\end{figure}

As an example, we consider the excitations induced in $^{208}$Pb in
$^{16}$O + $^{208}$Pb collisions. Since the expression
(\ref{4.5a}), with Eqs. (\ref{4.7.6}-\ref{4.7.6b}),  is quite general, valid for all energies, 
the equivalent photon numbers contain all information
about the differences among the low and the high energy scattering.
In Figs. \ref{a-f3}, \ref{a-f4} and \ref{a-f5} we
show $dn_{\pi \lambda, \varepsilon}$, for the $E1$ (Fig. \ref{a-f3}),
$E2$ (Fig. \ref{a-f4}), and $M1$ (Fig. \ref{a-f5}) multipolarities, and
for the collision
$^{16}$O + $^{208}$Pb with an impact parameter $b=15 \;$fm. They are
the equivalent photon numbers with frequency $\omega = 10 \;$MeV$/\hbar$
incident on $^{208}$Pb.
The dotted lines are obtained by using the non-relativistic
Eq. (\ref{Obtnr}), while the dashed lines correspond
to the relativistic expressions (\ref{(A.3a)},\ref{(A.3b)},\ref{(A.3c)}).
One observes that the relativistic expressions
overestimate the equivalent photon numbers
at low energies, while the non-relativistic
expressions underestimate them at high energies.
The correct values are  given by the solid lines, calculated according to
Eqs. (\ref{3.27}) and (\ref{3.19}). They reproduce the low and the high energy
limits, giving an improved interpolation between these limits at
intermediate energies. These discrepancies are more apparent  for the $E1$
and the $E2$ multipolarities. In the energy interval around 100A MeV
neither the low energy theory nor the high energy one can reproduce well the
correct values. This energy interval is indeed very sensitive to the
effects of retardation and of Coulomb recoil.

At these bombarding energies, the differences between the
magnitude of the
non-relativis\-tic and the correct
relativistic  virtual photon numbers
are kept at a constant value, of about 20\%, for excitation energies
$E_x=\hbar \omega < 10$ MeV. However, they increase sharply
when one reaches the excitation energy of
$E_x=\hbar \omega > 10 \; $MeV.
The reason is that, for such excitation energies,
the adiabaticity factor  becomes greater than
unity ($\xi >1$). This means that excitation energies of order
of $10 \; $MeV (like in the case of giant resonance excitation)
are in the transition region from a constant behavior of the
equivalent photon numbers to that of an exponential ($\sim e^{-\pi \xi}$)
decay. This is more transparent in Fig. \ref{wwphi} where we plot the
equivalent
photon numbers for $E_{lab} = $100 MeV/nucleon, $b=15 \; $fm, and
as a function of $\hbar \omega$. One also
observes from  this figure that the $E2$ multipolarity component
of the electromagnetic field dominates at low frequencies. Nonetheless,
over the range of $\hbar \omega$ up to some tens of MeV, the $E2$
matrix elements
of excitation are much smaller than the $E1$ elements for most nuclei, and the
$E2$ effects become unimportant. However, such effects are relevant
for the excitation of the isoscalar $E2$ giant resonance (GQR$_{is}$)
which have large matrix elements.

Only for the $E1$ multipolarity the orbital integral in Eq. \eqref{3.27} can be done
analytically, 
\begin{eqnarray}
{\frac{dn_{E1}}{d\Omega }}&=&{\frac{Z_{1}^{2}\alpha }{4\pi ^{2}}}\;\bigl({%
\frac{c}{v}}\bigr)^{2}\;{\epsilon ^{4}\;\zeta ^{2}}\;\;e^{-\pi \zeta }\;%
\biggl\{{\frac{1}{\gamma ^{2}}}\;{\frac{\epsilon ^{2}-1}{\epsilon ^{2}}}\;%
\bigl[K_{i\zeta }({\epsilon \zeta })\bigr]^{2}\nonumber \\
&+&\bigl[K_{i\zeta }^{\prime }({%
\epsilon \zeta })\bigr]^{2}\biggr\}\,,  \label{nE1ann}
\end{eqnarray}
where $\epsilon =1/\sin (\vartheta /2)$, $\alpha =1/137$, $\zeta =\omega
a_{0}/\gamma v$. $K_{i\zeta }$ is the
modified Bessel function with imaginary index, $K_{i\zeta }^{\prime }$ is
the derivative with respect to its argument. If one compares this equation with
Eq. \eqref{nE1ann1} one notes that besides the appearance of $\gamma$ factors
in $a=a_0/\gamma$, the first term inside square brackets is also reduced by a $1/\gamma^2$,
as compared to the non-relativistic expression. For relativistic energies, $\zeta\rightarrow 0$,
$\epsilon\rightarrow \infty$, with $\zeta\epsilon\rightarrow {\rm const.}$. It is then easy to see that
the above expression reduces to Eq. \eqref{(A.3a)}, in terms of the impact parameter dependence.

The integration over impact parameter can also be doen analytically for the $E1$ case. One obtains [BB88]
\begin{eqnarray}
N_{E1} &=&
{2\over \pi}
 \ Z_1^2\alpha \;e^{-\pi \zeta }\;\left({c\over v}\right)^2\;\Bigg\{ -\xi
K_{i\zeta }K_{i\zeta }^{\prime }- {1 \over 2}
\left({c\over v}\right)^2\xi ^2 \nonumber \\
&&\times  \Bigg[
\left({\zeta\over \xi}\right)^2K_{i\zeta}^2+K_{i\zeta}^{\prime 2}
-K_{i\zeta }^2\nonumber \\
&-&\frac i{\epsilon _0}
\left( K_{i\zeta }\left(
{{\partial K_\mu ^{\prime }}\over
{\partial \mu }}
\right) _{\mu =i\zeta }-
K_{i\zeta }^{\prime }
\left( {{\partial K_\mu }\over
{\partial \mu }}
\right) _{\mu =i\zeta }\right)
\Bigg] \Bigg\} \ ,\nonumber \\ \label{Ne1an}
\end{eqnarray}
where $\zeta =\omega
a_{0}/\gamma v$,
\begin{equation}
\epsilon_0 =
\left\{ \begin{array}{ll}
1,     & \mbox{for    $ 2a>b_{mim} $ ,}\\
R/a-1, & \mbox{for    $ 2a<b_{mim} $ ,}
\end{array}
\right.
\end{equation}
and $\xi=\epsilon_0 \zeta=\omega b_{mim}/\gamma v$.

It is easy to see that this equation reduces to Eq. (\ref{(A.6a)})
in the relativistic limit, when $\zeta \longrightarrow 0$,
$\epsilon_0 \longrightarrow \infty$.
In practice, it is much easier to numerically calculate the orbital integrals in Eqs. (\ref{3.27}-\ref{3.19}) directly. 
A numerical code (COULINT) is available for the purposes [Ber04].

\section{E\&M response of weakly-bound systems}

Coulomb excitation has been one of the main tools to study the structure of unstable isotopes in radioactive beam facilities. This is a relatively new field and many new aspects of Coulomb excitation have been developed in the last two decades. We will discuss few of these aspects and, in particular, the role of transitions in/to the continuum of loosely-bound systems.

\subsection{Dominance of one resonance}

{It often occurs that the Coulomb excitation mechanism is dominated
by transitions between the ground state and a continuum resonant state. This part of the
excitation mechanism can be described exactly in a coupled-channels
approach, while the other excitations can be described perturbatively [Ca96].  Fig. \ref{a-f23} represents the procedure. One
resonance is coupled to the ground state while the remaining resonances are
fed by these two states according to first order perturbation theory. The
coupling matrix elements involves the ground state and a set of doorway
states }$|D_{\lambda \mu }^{(n)}>${ , where }$n${ \ specifies
the kind of resonance and }$\lambda \mu ${ \ are angular momentum
quantum numbers. The amplitudes of these resonances in real continuum states
are } 
\begin{equation}
\alpha ^{(n)}(\epsilon )=<\phi (\epsilon )|\mathcal{D}_{\lambda \mu }^{(n)}>,
\label{aeps}
\end{equation}
{where }$\phi (\epsilon )${ \ denotes the wavefunction of one
of the numerous states which are responsible for the broad structure of the
resonance. In this equation }$\epsilon =E_{x}-E_{n}${ , where }$E_{x}$%
{ \ is the excitation energy and }$E_{n}${ \ is the centroid of
the resonance considered.}

\begin{figure}[tb]
\includegraphics[
width=3.4in
]%
{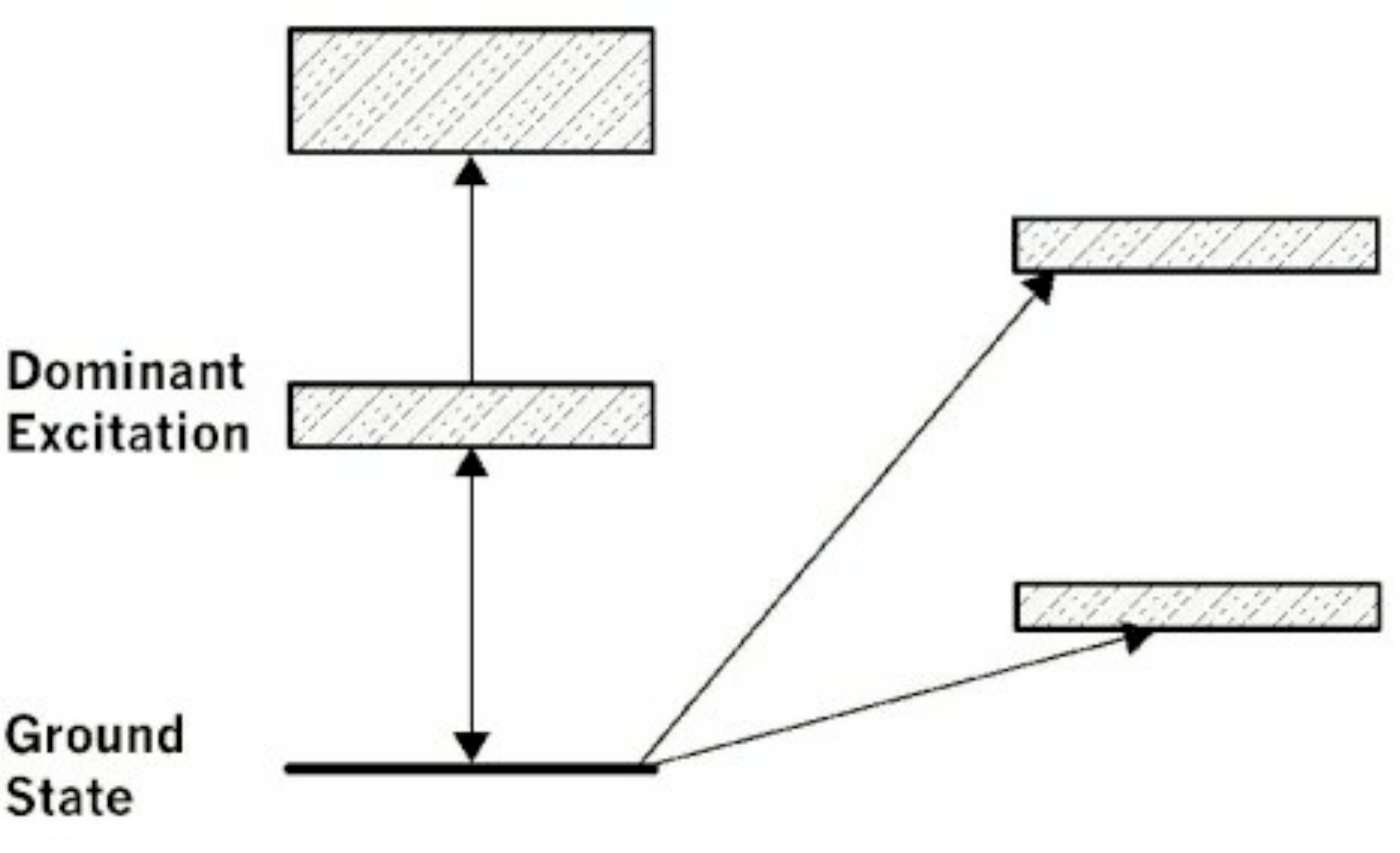}%
\caption
{\sl Schematic representation of
Coulomb excitation of a dominant resonant state and of other states
weakly-coupled to the ground state.
\label{a-f23}}
\end{figure}

{Let us assume for simplicity that the dominant resonant state has
spin and parity }$J^{P}=1^{-}${ . However, the following results can
be easily generalized to all spin-parity types. The following
coupled-channels equations are obtained [Ca96]:} 
\begin{eqnarray}
i\hbar \frac{d{a}_{0}}{dt}(t) &=&\sum_{\mu }\int d\epsilon \left<\phi (\epsilon )|%
\mathcal{D}_{1\mu }^{(1)}\right>\left<\mathcal{D}_{1\mu }^{(1)}|V_{E1,\mu }(t)|0\right>\nonumber \\
&\times&\exp
\left\{ -{\frac{i}{\hbar }}(E_{1}+\epsilon )t\right\} a_{\epsilon ,1\mu
}^{(1)}(t)  \notag \\
&=&\sum_{\mu }\int d\epsilon \ \alpha ^{(1)}(\epsilon )\ V_{\mu }^{(01)}(t)\
\nonumber \\
&\times&
\exp \left\{ -{\frac{i}{\hbar }}(E+\epsilon )t\right\} \ a_{\epsilon ,1\mu
}^{(1)}(t),  \label{da0}
\end{eqnarray}
{and } 
\begin{eqnarray}
i\hbar \ \frac{d{a}_{\epsilon ,1\mu }^{(1)}}{dt}(t)&=&\left[ (\alpha
^{(1)}(\epsilon )\ V_{\mu }^{(01)}(t)\right] ^{\ast }\nonumber \\
&\times& \exp \left\{
{i\over \hbar}(E_{1}+\epsilon )t\right\} \ a_{0}(t).  \label{da1}
\end{eqnarray}
{Above, }$(n=1)${ \ stands for the dominant resonance, }$a_{0}$%
{denotes the occupation amplitude of the ground state and }$%
a_{\epsilon ,1\mu }^{(1)}${ \ the occupation amplitude of a state
located at an energy }$\epsilon ${ \ away from the centroid of the
resonance, and with magnetic quantum number }$\mu ${ \ (}$\mu =-1,0,1$%
{ ). We used the short hand notation }$V_{\mu }^{(01)}(t)=\left<D_{1\mu
}^{(1)}|V_{E1,\mu }(t)|0\right>$.

{Integrating Eq.~\eqref{da1} and inserting the result in Eq.~\eqref{da0},
we get the integro-differential equation for the ground state occupation
amplitude } 
\begin{eqnarray}
&&\frac{d^{2}{a}_{0}}{dt^{2}}(t) =-\frac{1}{\hbar ^{2}} \sum_{\mu }\ V_{\mu
}^{(01)}(t) \int d\epsilon  |\alpha ^{(1)}(\epsilon )|^{2}  \notag \\
&\times &\int_{-\infty }^{t}dt^{\prime }\left[ V_{\mu }^{(01)}(t^{\prime })%
\right] ^{\ast } \exp \left\{ -{i\over \hbar}(E_{1}+\epsilon )(t-t^{\prime })
\right\}  a_{0}(t^{\prime }), \nonumber \\ \label{da01}
\end{eqnarray}
{where we used that }$a_{\epsilon ,1\mu }^{(1)}(t=-\infty )=0${. 
To carry out the integration over }$\epsilon ${ , we should use an
appropriate parametrization for the doorway amplitude }$\alpha
^{(1)}(\epsilon )${ . A convenient choice is the Breit-Wigner (BW)
form which yields the square amplitude } 
\begin{equation}
|\alpha ^{(1)}(\epsilon )|^{2}={\frac{1}{2\pi }}\ \left[ {\frac{\Gamma _{1}}{%
\epsilon ^{2}+\Gamma _{1}^{2}/4}}\right] ,  \label{aeps2}
\end{equation}
{where }$\Gamma _{1}${ \ is chosen to fit the experimental
width. In this case, this integral will be the simple exponential } 
\begin{eqnarray}
&&\int d\epsilon \ |\alpha ^{(1)}(\epsilon )|^{2}\ \exp \left\{ -i{\frac{%
(E_{1}+\epsilon )t}{\hbar }}\right\} \nonumber \\
&=&\exp \left\{ -i{\frac{(E_{1}-i\Gamma
_{1}/2)t}{\hbar }}\right\} .
\end{eqnarray}

{A better agreement with the experimental line shapes of the giant
resonances is obtained by using a Lorentzian (L) parametrization for }$%
|\alpha ^{(1)}(\epsilon )|^{2}${ , i.e., } 
\begin{equation}
|\alpha ^{(1)}(\epsilon )|^{2}={\frac{2}{\pi }}\ \left[ {\frac{\Gamma
_{1}E_{x}^{2}}{(E_{x}^{2}-E_{1}^{2})^{2}+\Gamma _{1}^{2}E_{x}^{2}}}\right] ,
\label{L}
\end{equation}
{where }$E_{x}=E_{1}+\epsilon ${. The energy integral can
still be performed exactly, but now it leads to the more
complicated result 
\begin{eqnarray}
&&\int d\epsilon \ |\alpha ^{(1)}(\epsilon )|^{2}\ \exp \left\{ -i{\frac{%
(E_{1}+\epsilon )t}{\hbar }}\right\} 
\nonumber \\
&=&\left( 1-i{\frac{\Gamma _{1}}{2E_{1}}%
}\right) \ \exp \left\{ -i{\frac{(E_{1}-i\Gamma _{1}/2)t}{\hbar }}\right\} \ 
\notag \\
&+&\ \Delta C(t)\ ,
\end{eqnarray}
{where }$\Delta C(t)${ \ is a non-exponential correction to the
decay. This correction can be shown numerically to be negligible. It
will therefore be ignored. After integration over }$\epsilon ${ , Eq.~%
\eqref{da01} reduces to} 
\begin{eqnarray}
\frac{d^{2}{a}_{0}}{dt^{2}}(t)&=&-\mathcal{S}_{1}\sum_{\mu }V_{\mu
}^{(01)}(t)\int_{-\infty }^{t}dt^{\prime }\left[ V_{\mu }^{(01)}(t^{\prime })%
\right] ^{\ast }\nonumber \\
&\times&\exp \left\{ -i{\frac{(E_{1}-i\Gamma _{1}/2)(t-t^{\prime })}{%
\hbar }}\right\} a_{0}(t^{\prime })\nonumber \\
\end{eqnarray}
{where the factor }$S_{1}${ \ is }$S_{1}=1${ \ for
BW-shape and }$S_{1}=1-i\Gamma _{1}/2E_{1}${ \ for L-shape.}

{We can take advantage of the exponential time-dependence in the
integral of the above equation, to reduce it to a set of second order
differential equations. Introducing the auxiliary amplitudes }$A_{\mu }(t)$%
{, given by the relation } 
\begin{equation}
a_{0}(t)=1\ +\ \sum_{\mu }A_{\mu }(t),  \label{A}
\end{equation}
{ with initial conditions }$A_{\mu }(t=-\infty )=0${ , and
taking the derivative of Eq.~(\ref{A}), one obtains } 
\begin{eqnarray}
&&{\ddot{A}}_{\mu }(t)-\left[ {\frac{{\dot{V}}_{\mu }^{(01)}(t)}{V_{\mu
}^{(01)}(t)}}\ -\ {\frac{i}{\hbar }}\left( E_{1}-i{\frac{\Gamma _{1}}{2}}%
\right) \right] {\dot{A}}_{\mu }(t)\ \nonumber \\
&+&\ \mathcal{S}_{1}\ {\frac{|V_{\mu
}^{(01)}(t)|^{2}}{\hbar ^{2}}}\left[ 1+\sum_{\mu ^{\prime }}A_{\mu ^{\prime
}}(t)\right] =0.
\end{eqnarray}

{Solving the above equation, we get }$a_{0}(t)${. Using this
amplitude and integrating Eq.~(\ref{da1}), one can evaluate }$a_{\epsilon
,1\mu }^{(1)}(t)${ . The probability density for the population of a
dominant continuum state with energy }$E_{x}${ \ in a collision with
impact parameter }$b${, }$P_{1}(b,E_{x})${ , is obtained trough
the summation over the asymptotic (}$t\rightarrow \infty ${ )
contribution from each magnetic substate. This yields} 
\begin{eqnarray}
P_{1}(b,\ E_{x})&=&|\alpha ^{(1)}(E_{x}-E_{1})|^{2}\ \sum_{\mu }\Bigg|
\int_{-\infty }^{\infty }dt^{\prime }\nonumber\\
&\times& \exp \left\{ iE_{x}t^{\prime
}\right\} \ \left[ V_{\mu }^{(01)}(t^{\prime })\right] ^{\ast }\
a_{0}(t^{\prime })\Bigg| ^{2},  \label{pro1}
\end{eqnarray}
{where }$|\alpha ^{(1)}(E_{x}-E_{1})|^{2}${ \ is given by Eq.~%
\eqref{aeps2} or by Eq.~\eqref{L}, depending on the choice of the resonance
shape.}

{To first order, the contribution to the excitation of another
resonance (the one on the top of Fig. \eqref{a-f23}) from the dominant one is
given by} 
\begin{eqnarray}
P_{2}(b,E_{x}) &=&|\alpha ^{(2)}(E_{x}-E_{2})|^{2}\,\mathcal{S}_{1}
\nonumber \\
&\times&\sum_{\nu
}\Bigg| \int_{-\infty }^{\infty }dt^{\prime }
\exp \big\{ iE_{x}t^{\prime
}\big\} \ \sum_{\mu }\ \Big[ V_{\nu \mu }^{(12)}(t^{\prime })\Big]
^{\ast }  \notag \\
&&\times \int_{-\infty }^{t^{\prime }}dt^{\prime \prime }V_{\mu
}^{(01)}(t^{\prime \prime })\nonumber \\
&\times&\exp \Bigg\{ -i{\frac{(E_{1}-i\Gamma
_{1}/2)(t-t^{\prime })}{\hbar }}\Bigg\} \ a_{0}(t^{\prime \prime })\Bigg|
^{2}.  \nonumber \\ \label{P2}
\end{eqnarray}
{We should point out that Eq.~\eqref{P2} is not equivalent to
second-order perturbation theory. This would be true only in the limit }$%
a_{0}(t)\longrightarrow 1${ . In this approach, }$a_{0}(t)\neq 1$%
{, since it is modified by the time-dependent coupling to the
dominant state. This coupling is treated exactly by means of the
coupled-channels equations. This approach is justified due to the (assumed)
small excitation amplitude for the transition }$1\longrightarrow 2${ ,
since }$a_{1}(t)\ll a_{0}(t)${.}

{Equations similar to (\ref{pro1}) can also be used to calculate the
excitation probabilities directly from the ground-state, with the proper
choice of energies, widths, and transition potentials.} The above formalism has been used
to study the excitation of giant resonances multiphonon resonances [Ca96].  It can be easily
applied to description of pigmy resonances in nuclei far from the stability. 

\subsection{Cluster-like light nuclei}

In first-order perturbation theory, the photo-nuclear cross sections are related to the reduced matrix elements,
for the excitation energy $E_{x}$  through the relation 
\begin{equation}
\frac{d\sigma _{C} }{dE_{\gamma}}\left( E_{\gamma}\right)=\frac{(2\pi )^{3}(L +1)%
}{L \left[ (2L +1)!!\right] ^{2}}\left( \frac{E_{\gamma}}{\hbar c}%
\right) ^{2L -1}\frac{dB}{dE_{x}}\left( \pi L ,E_{\gamma}\right)  \label{(1.2)}
\end{equation}
where $dB/dE_{x}$, are the reduced matrix elements, or \textit{Coulomb response
functions}.

The response function for electric transitions, $dB\left(
EL\right)  /dE_{\gamma},$ in Eq.  \eqref{(1.2)} is given by
\begin{eqnarray}
&&\frac{dB\left(  EL\right)  }{dE_{\gamma}}=\frac{\left\vert \left\langle
J_{f}\left\Vert Y_{L}\left(  \widehat{\mathbf{r}}\right)  \right\Vert
J_{i}\right\rangle \right\vert ^{2}}{2J_{i}+1}\nonumber \\
&\times&\left[  \int_{0}^{\infty
}dr\ r^{2+L}\ \ \delta\rho_{if}^{\left(  EL\right)  }\left(  r\right)
\right]  ^{2}w\left(  E_{\gamma}\right)  , \label{photo2}%
\end{eqnarray}
\newline where $w\left(  E_{\gamma}\right)  $ is the density of final
states (for nuclear excitations into the continuum) with energy $E_{\gamma
}=E_{f}-E_{i}$. 

The transition density $\delta\rho_{if}^{\left(  EL\right)  }\left(  r\right)
$ will depend upon the nuclear model adopted.
We will discuss here very simple models to calculate this response. To get a better reproduction of experimental
data, more microscopic calculations such as the Random Phase Approximation (RPA) are often necessary.

\subsubsection{Cluster response function}

Due to the small
binding energy of nuclear halo systems composed of two clusters ($a=b+c$), the most important part of the
wavefunction is beyond the range of the nuclear potential. The external part
of the wavefunction is well described by an Yukawa of the form 
\begin{equation}
\psi _{bc}(\mathbf{r})=N_{0}\sqrt{\frac{\eta }{2\pi }}\frac{e^{-\eta r}}{r}
\label{(2.1)}
\end{equation}
where $\eta =\sqrt{2\mu _{bc}S/\hbar },$ $\mu _{bc}$ is the reduced mass of (%
$b+c$), $S$ is the separation energy, and $N_{0}=e^{\eta r_{0}}/\sqrt{1+\eta
r_{0}}$ is a normalization factor which corrects the wavefunction to account
for the finite range, $r_{0}$, of the ($b+c$) potential. There is no
resonance structure in the $b+c$ continuum. This is clearly a good
assumption for the deuteron and also for other neutron {\it halo} systems.

If we further assume that the final state is a plane-wave state (i.e., we
neglect final state interactions) $\psi _{f}\equiv \left\langle \mathbf{q}|%
\mathbf{r}\right\rangle =e^{i\mathbf{q.r}}.$ The response functions for
electric multipole transitions
\begin{equation}
\frac{dB}{dE_{x}}(E\lambda ,E_{x})=SF\sum_{\mu }\Big| \left\langle \mathbf{q}%
\left| {\cal M}(ELM )\right| \psi _{bc}(\mathbf{r})\right\rangle
\Big| ^{2}\frac{d^{3}q}{(2\pi )^{3}}  \label{(2.2)}
\end{equation}
where $SF$ is a spectroscopic factor (i.e., probability to find the system in
the state ($b+c$)). ${\cal M}(ELM )$ is the electric multipole operator
 corrected to exclude c.m. motion, i.e.,
\begin{eqnarray}
{\cal M}(EL \mu )&=&e\sum\limits_{i=p}\left[ \left( 1-\frac{1}{A}%
\right) ^{L }+(-1)^{L }\frac{\left( Z-1\right) }{A^{L }}%
\right] r_{i}^{L }Y_{L \mu }\nonumber \\
&+&\;e\sum_{i=n}\;Z\left( -%
\frac{1}{A^{L }}\right) r_{i}^{L }Y_{L M }
\label{(2.3)}
\end{eqnarray}
where the sum runs over all protons (p) and neutrons (n) in the nucleus.\ 

For a cluster-system the sum runs over the (effective) charges of the
clusters. For example, 
\begin{eqnarray}
{\cal M}(E1M ) &=&e\left( \frac{Z_{b}A_{c}-Z_{c}A_{b}}{A_{a}}\right)
r_{bc}Y_{1M }(\widehat{\mathbf{r}}_{bc})  \notag \\
{\cal M}(E2M ) &=&e\left( \frac{Z_{b}A_{c}^{2}+Z_{c}A_{b}^{2}}{A_{a}^{2}%
}\right) r_{bc}Y_{2M }(\widehat{\mathbf{r}}_{bc})  \label{(2.4}
\end{eqnarray}
where $\widehat{\mathbf{r}}_{bc}\;$is the relative position of b-c. 

Using $
d^{3}q=q^{2}dqd\Omega _{q}=\sqrt{2E_{x}}(\mu _{bc}/\hbar ^{2})^{3/2}dE
d\Omega _{q}$ in Eq. \eqref{(2.2)} and integrating over $\Omega ,$
one finds [BS92] 
\begin{eqnarray}
\frac{dB(EL ;E_{\gamma})}{dE_{\gamma}}&=&\frac{2^{L -1}}{\pi ^{2}}(L
!)^{2}(2L +1)(SF) N_{0}^{2}\left[ O(L )\right] ^{2}\nonumber \\
&\times& \left( \frac{%
\hbar ^{2}}{\mu _{bc}}\right) ^{L }\frac{\sqrt{S}(E_{\gamma}-S)^{L
+1/2}}{E_{\gamma}^{2L +2}}  \label{(2.5)}
\end{eqnarray}
where 
\begin{equation}
O(L )=\frac{Z_{b}A_{c}^{L }-(-1)^{L }Z_{c}A_{b}^{L }%
}{A_{a}^{L }}\;e.  \label{(2.6)}
\end{equation}

The maximum of this function occurs at 
\begin{equation}
E_{0}^{EL}=\frac{L +1/2}{L +3/2}\;S.  \label{(2.7)}
\end{equation}

In Figure \ref{softm} we show the photonuclear cross sections obtained by
using Eq. \eqref{(2.5)} in the definition \eqref{(1.2)}. The $E2$ and $E3$
photonuclear cross sections are smaller than the $E1$ by factors of order of
10$^{5}$ and $10^{9}$, respectively.
The spectroscopic factor and the normalization constant ($S$, and $N_{o}$,
in Eq.  \eqref{(2.5)} were set to unity, for simplicity. The have a longer
tail than the $E1$ photonuclear cross section due to the factor $%
E_{x}^{2L -1}$ in Eq. \eqref{(1.2)}. Thus, we conclude that a
cluster-like correlation is not as effective in
producing an enhancement of the low energy part of the photonuclear cross
sections for higher multipolarities, as in the dipole case.

\begin{figure}[tb]
\includegraphics[
width=3.4in
]%
{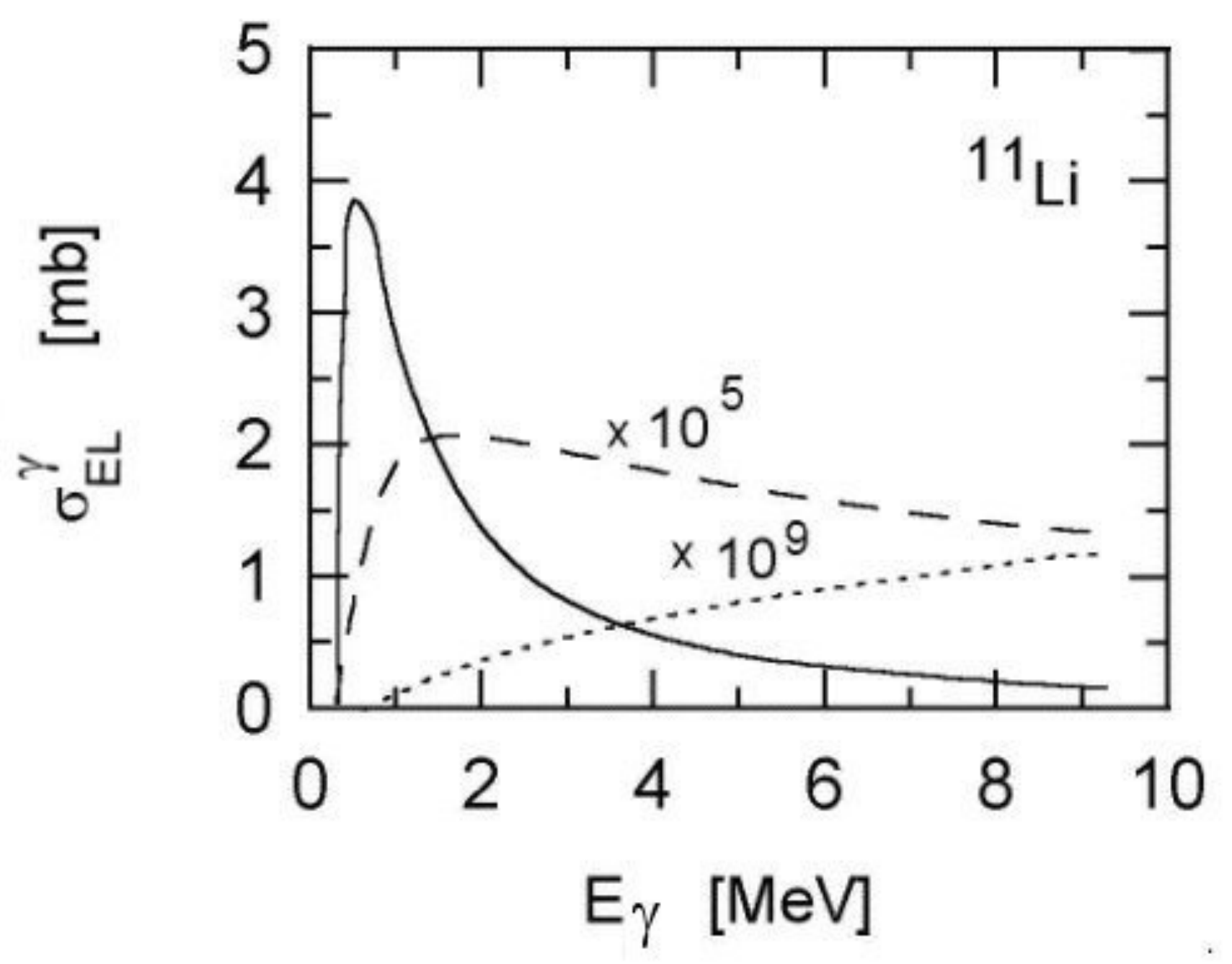}%
\caption
{\sl
Photo absorption cross
section of $^{11}$Li\ in the cluster model. The solid (dashed) [dotted]
curve corresponds to dipole (quadrupole) [octupole] multipolarity. The
quadrupole (octupole) values were multiplied by a factor of $10^{5}$\ (10$%
^{9}$) [BS92].
\label{softm}}
\end{figure}

\subsubsection{One-neutron halo}

We will now discuss the effect of final state interactions. We start by considering a one-neutron halo nucleus.
The initial wavefunction can be written as $\Psi_{JM}=r^{-1}u_{ljJ}%
(r)\mathcal{Y}_{lJM}$, where $R_{ljJ}(r)$ is the radial wavefunction and
$\mathcal{Y}_{lJM}$ is a spin-angle function [BD04]. 

As before, we use a simple Yukawa form for an s-wave
initial wavefunction, $u_{0}(r)=N_{0}\exp(-\eta r)$, and a p-wave final
wavefunction, $\ u_{1}(r)=j_{1}(kr)\cos\delta_{1}-\ n_{1}(kr)\sin\delta_{1}$.
In these equations $\eta$\ is related to the neutron separation energy
$S_{n}=\hbar^{2}\eta^{2}/2\mu$, $\mu$\ is the reduced mass of the neutron +
core system, and $\hbar k=\sqrt{2\mu E_{r}}$, with $E_{r}$ being the final
energy of relative motion between the neutron and the core nucleus. $N_{0}$ is
the normalization constant of the initial wavefunction. 

The transition density
is given by $r^{2}\delta\rho_{if}\left(  r\right)  =e_{eff}A_{i}u_{i}%
(r)u_{f}(r)$, where $i$ and $f$ indices include angular momentum dependence
and $e_{eff}=-eZ_{c}/A$ is the effective charge of a neutron+core nucleus with
charge $Z_{c}$. The $E1$ transition integral $\mathcal{I}_{l_{i}l_{f}}%
=\int_{0}^{\infty}dr\ r^{3}\ \ \delta\rho_{if}\left(  r\right)  $ for the
wavefunctions described above yields%
\begin{align}
\mathcal{I}_{s\rightarrow p}  &  =e_{eff}\frac{2k^{2}}{\left(  \eta^{2}%
+k^{2}\right)  ^{2}}\left[  \cos\delta_{1}+\sin\delta_{1}\frac{\eta\left(
\eta^{2}+3k^{2}\right)  }{2k^{3}}\right] \nonumber\\
&  \simeq\frac{e_{eff}\hbar^{2}}{2\mu}\frac{2E_{r}}{\left(  S_{n}%
+E_{r}\right)  ^{2}}\Bigg[  1+\nonumber\\
&\left(  \frac{\mu}{2\hbar^{2}}\right)
^{3/2}\frac{\sqrt{S_{n}}\left(  S_{n}+3E_{r}\right)  }{-1/a_{1}+\mu r_{1}%
E_{r}/\hbar^{2}}\Biggr]  , \label{isp}%
\end{align}
where the effective range expansion of the phase shift, $k^{2l+1}\cot
\delta\simeq-1/a_{l}+r_{l}k^{2}/2,$ was used in the second line of the above
equation. For $l=1$, $a_{1}$ is the \textquotedblleft scattering
volume\textquotedblright\ (units of length$^{3}$) and $r_{1}$ is the
\textquotedblleft effective momentum\textquotedblright\ (units of 1/length).
Their interpretation is not as simple as the $l=0$ effective range parameters.
Typical values are, e.g. $a_{1}=-13.82$ fm$^{-3}$ and $r_{1}=-0.419$ fm$^{-1}$
for n+$^{4}$He $p_{1/2}$-wave scattering and $a_{1}=-62.95$ fm$^{-3}$ and
$r_{1}=-0.882$ fm$^{-1}$ for n+$^{4}$He $p_{3/2}$-wave scattering [Arn73].

The energy dependence of Eq.  \eqref{isp} has some unique features. As
shown in the previous section, the matrix elements
for electromagnetic response of weakly-bound nuclei present a small
peak at low energies, due to the proximity of the bound state to the
continuum. This peak is manifest in the response function as
\begin{equation}
\frac{dB(EL)}{dE}\propto\left\vert \mathcal{I}_{s\rightarrow p}\right\vert
^{2}\propto\frac{E_{r}^{L+1/2}}{\left(  S_{n}+E_{r}\right)  ^{2L+2}}.
\label{deel}%
\end{equation}
It appears centered at the energy [BS92]
$E_{0}^{(EL)}\simeq\frac{L+1/2}{L+3/2}S_{n}$ for a generic electric
response of multipolarity $L$. For $E1$ excitations, the peak occurs
at $E_{0}\simeq3S_{n}/5$.

\begin{figure}[ptb]
\begin{center}
\includegraphics[
height=2.9464in,
width=3.1298in
]{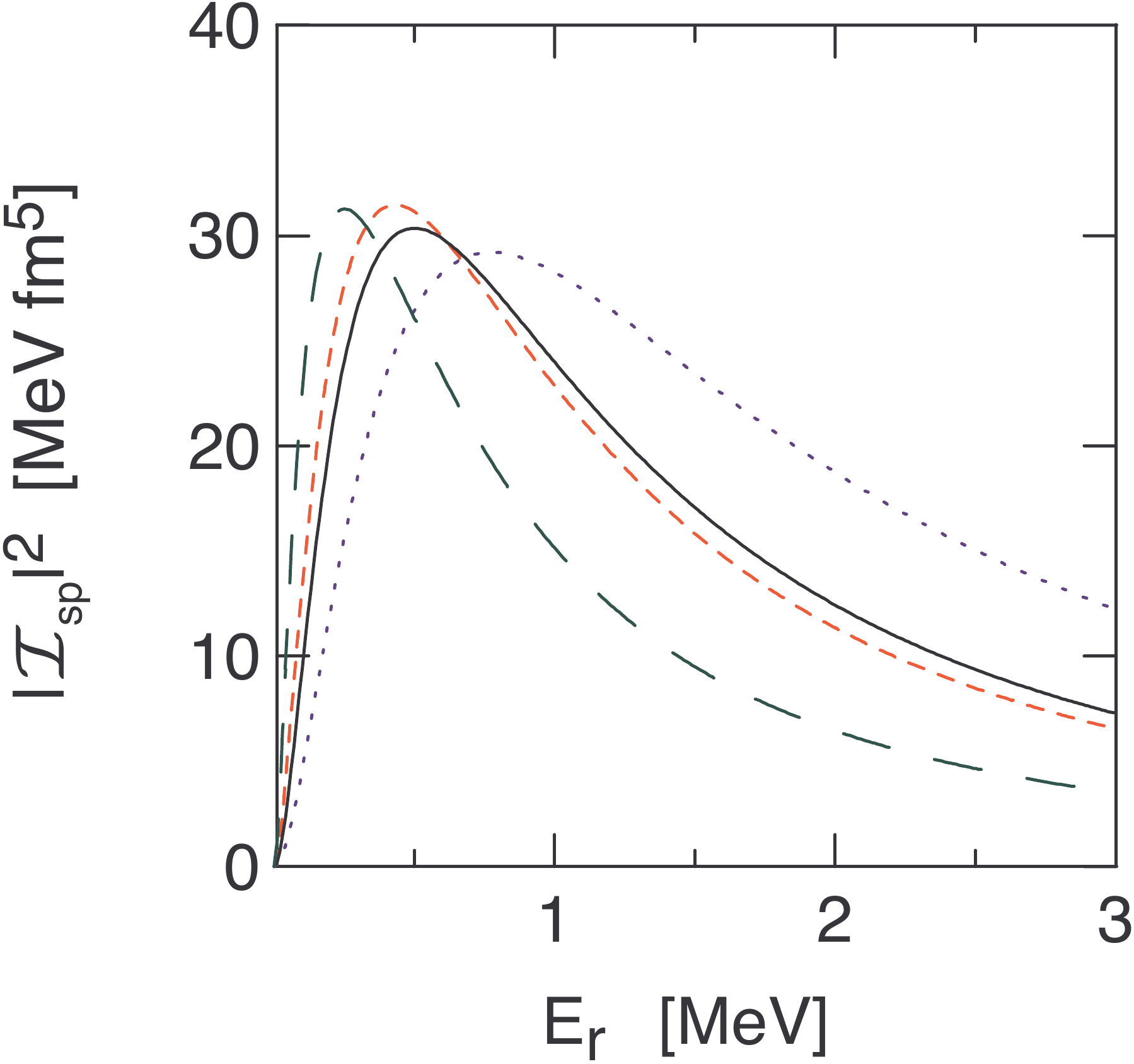}
\end{center}
\caption{\sl  $\left\vert \mathcal{I}_{s\rightarrow
p}\right\vert ^{2}$ \ calculated using Eq.  \eqref{isp}, assuming
$e_{eff}=e$, $A=11$ and $S_{n}=0.5$ MeV, as a function of $E_{r}$, the relative energy of the fragments.
The long dashed curve corresponds to \ $a_{1}=-10$ fm$^{-3}$ and
$r_{1}=-0.5$ fm$^{-1}$, the dashed curve corresponds to \
$a_{1}=-50$ fm$^{-3}$ and $r_{1}=1$ fm$^{-1},$ the solid curve
corresponds to\ $a_{1}=$ $r_{1}=0$, and finally, the dotted curve
corresponds to\ $a_{1}=-10$ fm$^{-3}$ and $r_{1}=0.5$ fm$^{-1}$.}%
\label{fig3}%
\end{figure}

The second term inside brackets in Eq. \eqref{isp} is a modification
due to final state interactions. This modification may become
important, as shown in figure \ref{fig3}, where $\left\vert
\mathcal{I}_{s\rightarrow p}\right\vert ^{2}$ calculated with Eq.
\eqref{isp} is plotted as a function of $E_{r}$. Here, for simplicity,
we have assumed the values $e_{eff}=e$, $A=11$ and $S_{n}=0.5$ MeV.
This does not correspond to any known nucleus and is used to assess
the effect of the scattering length and effective range in the
transition matrix element. The long dashed curve corresponds to \
$a_{1}=-10$ fm$^{-3}$ and $r_{1}=-0.5$ fm$^{-1}$, the dashed curve
corresponds to \ $a_{1}=-50$ fm$^{-3}$ and $r_{1}=1$ fm$^{-1},$ the
solid curve corresponds to\ $a_{1}=$ $r_{1}=0$, and finally, the
dotted curve corresponds to\ $a_{1}=-10$ fm$^{-3}$ and $r_{1}=0.5$
fm$^{-1}.$ Although the effective range expansion is only valid for
small values of $E_{r}$, it is evident from the figure that the
matrix element is very sensitive to the effective range expansion
parameters.

The strong dependence of the response function on the effective
range expansion parameters makes it an ideal tool to study the
scattering properties of light nuclei which are of interest for
nuclear astrophysics. It is important to notice that the one-halo
has been studied in many experiments, e.g. for the case of $^{11}$Be
for which there are many data available. In the literature one can find a detailed
analysis of how the nuclear shell-model can explain the experimental
data, by fitting the spectroscopic factors for several
single-particle configurations. 

\subsubsection{Two-neutron halo}

Weakly-bound nuclei, such $^{6}$He or $^{11}$Li, require a three--body
treatment in order to reproduce the electromagnetic response more accurately.
In a popular three-body model, the bound--state wavefunction in the center of
mass system is written as an expansion over {\it hyperspherical harmonics} (HH), see
e.g.~[Zhu93],
\begin{equation}
\Psi\left(  \mathbf{x},\mathbf{y}\right)  =\frac{1}{\rho^{5/2}}\sum
_{KLSl_{x}l_{y}}\Phi_{KLS}^{l_{x}l_{y}}\left(  \rho\right)  \left[
\mathcal{J}_{KL}^{l_{x}l_{y}}\left(  \Omega_{5}\right)  \otimes\chi
_{S}\right]  _{JM}. \label{hhwf}%
\end{equation}
Here $\mathbf{x}$ and $\mathbf{y}$ are Jacobi vectors where (see
figure \ref{threeb}) $\mathbf{x}=\frac{1}{\sqrt{2}}\left(
\mathbf{r}_{1}-\mathbf{r}_{2}\right)$ and $y=\sqrt{\frac{2\left(
A-2\right)  }{A}}\left(
\frac{\mathbf{r}_{1}+\mathbf{r}_{2}}{2}-\mathbf{r}_{c}\right)$,
where A is the nuclear mass, $\mathbf{r}_{1}$ and $\mathbf{r}_{2}$
are the position of the nucleons, and $\mathbf{r}_{c}$ is the
position of the core. The hyperradius \ $\rho$ determines the size
of a three-body state: $\rho ^{2}=x^{2}+y^{2}$. The five angles
$\left\{ \Omega_{5}\right\}  $ include usual angles
$(\theta_{x},\phi_{x})$, $(\theta_{y},\phi_{y})$ which parametrize
the direction of the unit vectors $\widehat{\mathbf{x}}$ and
$\widehat{\mathbf{y}}$ and the hyperangle $\theta$, related by
$x=\rho \sin\theta$ and $y=\rho\cos\theta$, where
$0\leq\theta\leq\pi/2$.

\begin{figure}[ptb]
\begin{center}
\includegraphics[
height=2.2485in,
width=2.6117in
]{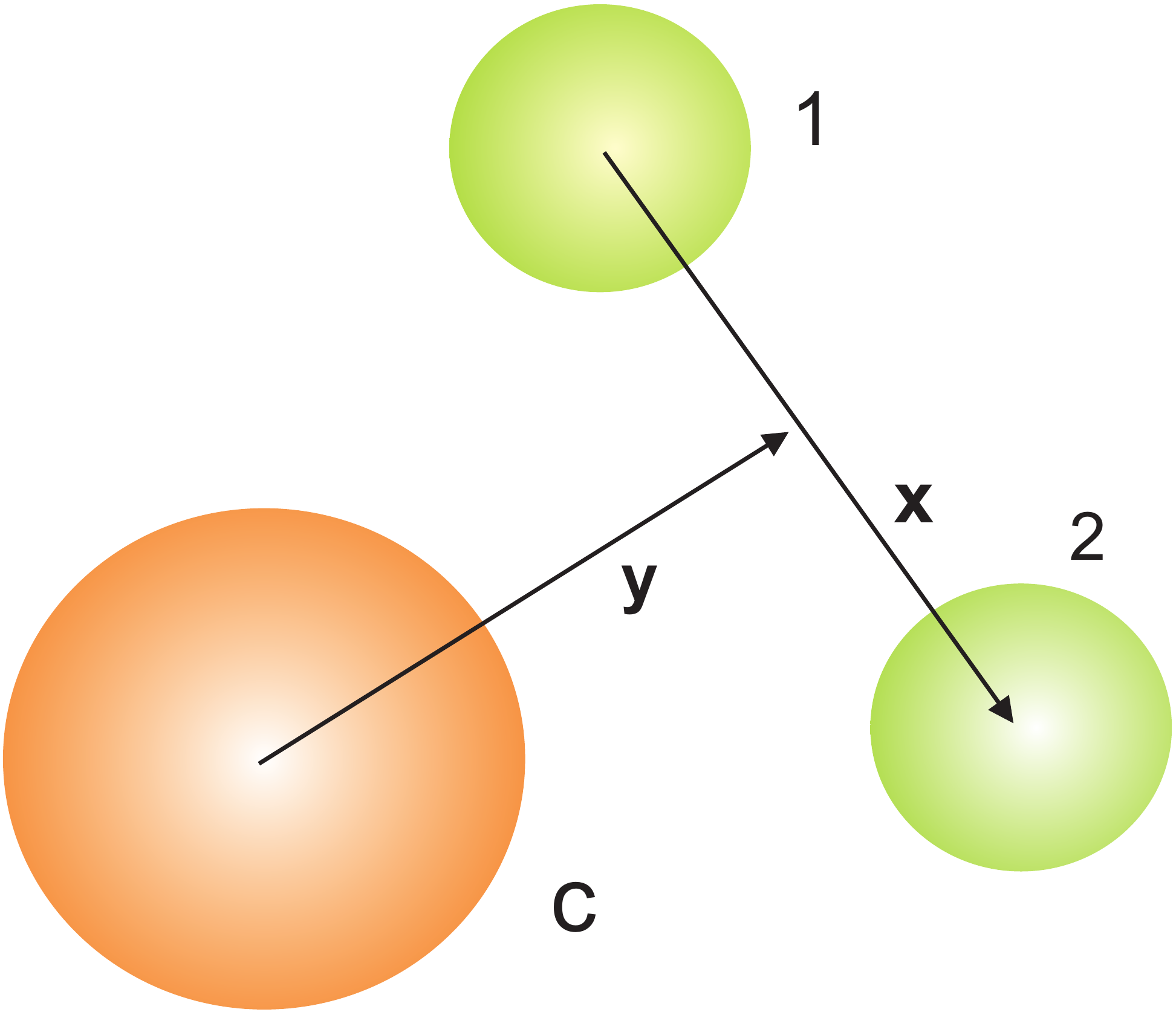}
\end{center}
\caption{\sl  Jacobian coordinates (\textbf{x} and
\textbf{y}) for a three-body
system consisting of a core (c) and two nucleons (1 and 2). }%
\label{threeb}%
\end{figure}

The insertion of the three-body wavefunction, Eq.  \eqref{hhwf}, into
the Schr\"odinger equation yields a set of coupled differential
equations for the hyperradial wavefunction
$\Phi_{KLS}^{l_{x}l_{y}}\left(  \rho\right) $. Assuming that the
nuclear potentials between the three particles are known, this
procedure yields the bound-state wavefunction for a three-body
system with angular momentum $J$.

In order to calculate the electric response we need the scattering
wavefunctions in the three-body model to calculate the integrals in
Eq.  \eqref{photo2}. One would have to use final wavefunctions with
given momenta, including their angular information. When the final
state interaction is disregarded these wavefunctions are three-body
plane waves [Pus96]. To carry out the calculations, the
plane waves can be expanded in products of hyperspherical harmonics
in coordinate and momentum spaces. However, since we are only
interested in the energy dependence of the response function, we do
not need directions of the momenta. Thus, instead of using plane
waves, we will use a set of final states which just include the
coordinate space and energy dependence.

For weakly-bound systems having no bound subsystems the hyperradial
functions entering the expansion \eqref{hhwf} behave asymptotically as [Mer74]
$ \Phi_{a}\left(  \rho\right)
\longrightarrow\text{constant}\times\ \exp\left( -\eta\rho\right)$
as $\rho\longrightarrow \infty$,
where the two-nucleon separation energy is related to $\eta$ by $S_{2n}%
=\hbar^{2}\eta^{2}/\left(  2m_{N}\right)  $. This wavefunction has
similarities with the two-body case, when $\rho$ is interpreted as the
distance $r$ between the core and the two nucleons, treated as one single
particle. But notice that the mass $m_{N}$ would have to be replaced by
$2m_{N}$ if a simple two-body (the {\it dineutron-model} [BB88]) were
used for $^{11}$Li or $^{6}$He.

\begin{figure}[ptb]
\begin{center}
\includegraphics[
height=2.6187in,
width=3.3347in
]{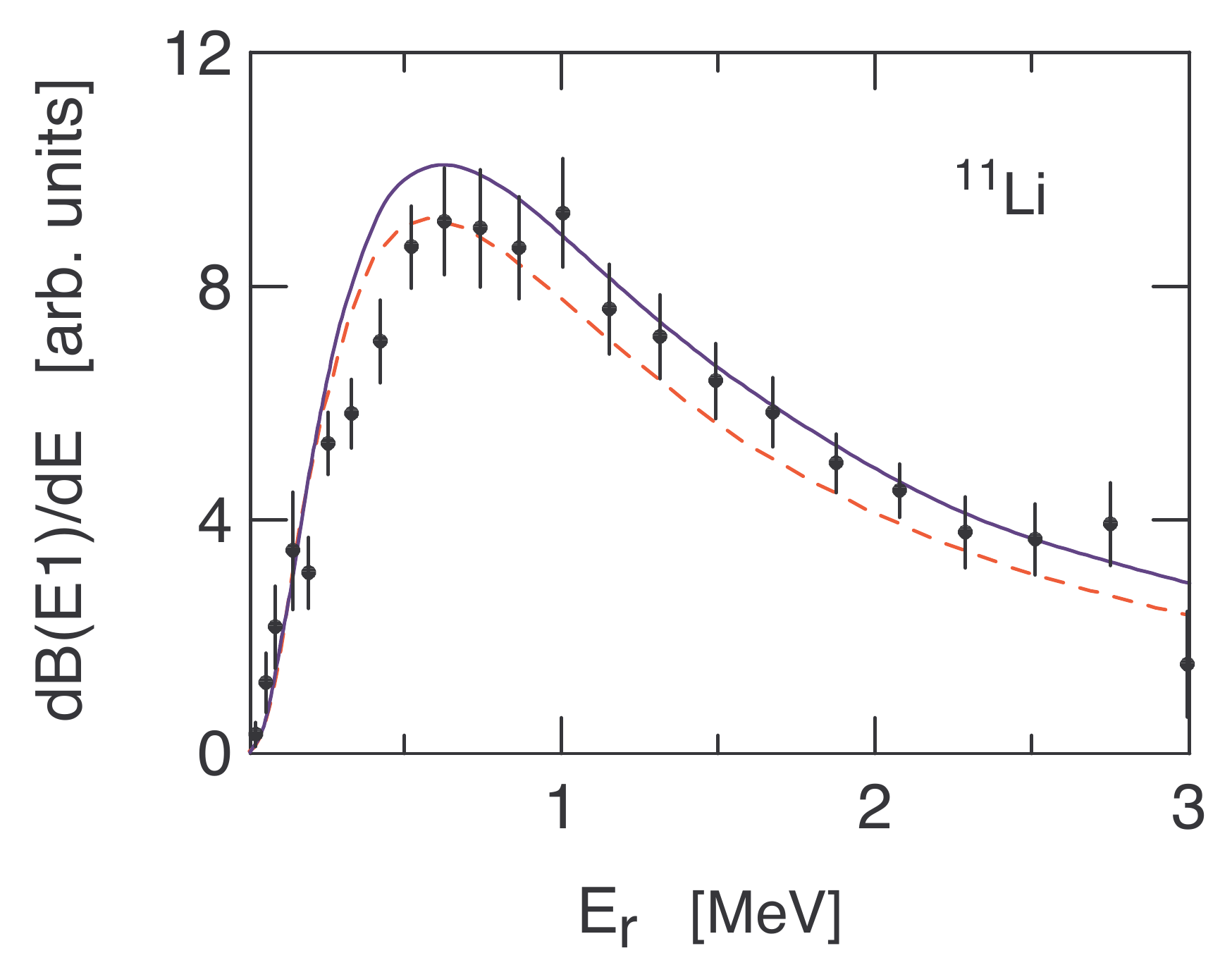}
\end{center}
\caption{\sl  Comparison between the calculation of the
response function (in arbitrary units) with eqs. \eqref{ie1} and
\eqref{dbde3b}, using $\delta_{nn}=0$ and $\delta_{nc}=0$, (dashed
line), or including the effects of final state interactions
(continuous line). The experimental data are from Ref.
[Shi95]. }%
\label{fig4}%
\end{figure}

Since only the core carries charge, in a three-body model the $E1$
transition operator is given by $M\sim yY_{1M}\left(
\widehat{\mathbf{y}}\right)  $ for the final state. The $E1$ transition matrix element is obtained by a
sandwich of this operator between $\Phi_{a}\left(  \rho\right)
/\rho^{5/2}$ and scattering wavefunctions. In Ref. [Pus96] the
scattering states were taken as plane waves. We will use distorted
scattering states,
leading to the expression%
\begin{equation}
\mathcal{I}\left(  E1\right)  =\int dxdy\frac{\Phi_{a}\left(  \rho\right)
}{\rho^{5/2}}\ y^{2}x u_{p}\left(  y\right)  u_{q}\left(  x\right)  ,
\label{ie1}%
\end{equation}
where $u_{p}\left(  y\right)  =j_{1}\left(  py\right)  \cos\delta_{nc}%
-n_{1}\left(  py\right)  \sin\delta_{nc}$ is the core-neutron
asymptotic continuum wavefunction, assumed to be a $p$-wave, and
$u_{q}\left(  x\right) =j_{0}\left(  qx\right)
\cos\delta_{nn}-n_{0}\left(  qx\right)  \sin \delta_{nn}$ is the
neutron-neutron asymptotic continuum wavefunction, assumed to be an
$s$-wave. The relative momenta are given by
$\mathbf{q}=\frac{1}{\sqrt{2}}\left(
\mathbf{q}_{1}-\mathbf{q}_{2}\right)$, and
$\mathbf{p}=\sqrt{\frac{2\left( A-2\right)  }{A} }\left(
\frac{\mathbf{k}_{1}+\mathbf{k}_{2}}{2}-\mathbf{k}_{c}\right)$.

The $E1$ strength function is proportional to the square of the matrix element
in Eq.  \eqref{ie1} integrated over all momentum variables, except for the total
continuum energy $E_{r}=\hbar^{2}\left(  q^{2}+p^{2}\right)  /2m_{N}$. This
procedure gives%
\begin{eqnarray}
\frac{dB\left(  E1\right)  }{dE_{r}}&=&\text{constant }\cdot \int\left\vert
\mathcal{I}\left(  E1\right)  \right\vert ^{2}E_{r}^{2}\cos^{2}\Theta\nonumber\\
&\times&\sin
^{2}\Theta d\Theta d\Omega_{q}d\Omega_{p}, \label{dbde3b}%
\end{eqnarray}
where $\Theta=\tan^{-1}\left(  q/p\right)  $.

The $^{1}$S$_{0}$ phase shift in neutron-neutron scattering is
remarkably well reproduced up to center of mass energy of order of 5
MeV by the first two
terms in the effective-range expansion $k\cot\delta_{nn}\simeq-1/a_{nn}%
+r_{nn}k^{2}/2.$ Experimentally these parameters are determined to be
$a_{nn}=-23.7$ fm and $r_{nn}=2.7$ fm$.$ The extremely large (negative) value
of the scattering length implies that there is a virtual bound state in this
channel very near zero energy. The p-wave scattering in the n-$^{9}$Li
($^{10}$Li) system appears to have resonances at low energies [Thoe99].
We assume that this phase-shift can be described by the resonance relation%
$\sin\delta_{nc}=({\Gamma/2})/{\sqrt{\left(  E_{r}-E_{R}\right)  ^{2}%
+\Gamma^{2}/4}}$, with $E_{R}=0.53$ MeV and $\Gamma=0.5$ MeV
[Thoe99].

Most integrals in eqs. \eqref{ie1} and \eqref{dbde3b} can be done analytically,
leaving two remaining integrals which can only be performed numerically. The
result of the calculation is shown in figure \eqref{fig4}. The dashed line was
obtained using $\delta_{nn}=0$ and $\delta_{nc}=0$, that is, by neglecting
final state interactions. The continuous curve includes the effects of final
state interactions, with $\delta_{nn}$ and $\delta_{nc}$ parametrized as
described above. The experimental data are from Ref. [Shi95]. The data
and theoretical curves are given in arbitrary units. Although the experimental
data is not perfectly described by either one of the results, it is clear that
final state interactions are of extreme relevance.

As pointed out in Ref. [Pus96], the $E1$ three-body response
function of $^{11}$Li can still be described by an expression
similar to Eq.  \eqref{deel}, but with different powers. Explicitly,
${dB\left(  E1\right)  }/{dE_{r}}\propto{E_{r}^{3}}/{\left(
S_{2n}^{eff}+E_{r}\right)  ^{11/2}}$. Instead of $S_{2n}$, one has
to use an effective $S_{2n}^{eff}=aS_{2n}$, with $a\simeq1.5$. With
this approximation, the peak of the strength function in the
three-body case is situated at about three times higher energy than
for the two-body case, Eq.  \eqref{deel}. In the
three-body model, the maximum is thus predicted at $E_{0}^{(E1)}%
\simeq1.8S_{2n}$, which fits the experimentally determined peak position for
the $^{11}$Li $E1$ strength function very well [Pus96]. It is thus
apparent that the effect of three-body configurations is to widen and to shift
the strength function $dB\left(  E1\right)  /dE$ to higher energies.

It is worthwhile mentioning that the data presented in figure
\ref{fig4}, and of other experiments, is different
in form and magnitude of the more recent experiment of Ref.
[Nak06]. The reason for the discrepancy is attributed to an
enhanced sensitivity in the experiment of Ref. [Nak06] to low
relative energies below $E_{rel}=0.5$ MeV compared to previous
experiments. Also, this recent experiment agrees very well with the
nn-correlated model of Ref. [EB92]. This
theoretical model is different than the model presented in this
section in many aspects. In principle, the three-body models should
be superior, as they include the interactions between the
three-particles without any approximation. For example,  Ref. 
[EB92] use a simplified interaction between the two-neutrons.
On the other hand, they include the many-body effects, e.g. the
Pauli blocking of the occupied states in the core. It is not well
known the reason why the data of Ref. [Nak06] is better
described with the model of Ref.  [EB92] than traditional 3-body
models.

\subsection{Pigmy resonances}

We have seen that the energy position where the soft dipole response
peaks depends upon the few body model adopted. Except for a two-body
resonance in $^{10}$Li, there was no reference to a resonance in the
continuum. The peak in the response function can be simply explained
by the fact that it has to grow from zero at low energies and return
to zero at large energies. In few-body, or cluster, models, the form
of the bound-state wavefunctions and the phase space in the
continuum determine the position of the peak in the response
function. Few-body resonances will lead to more peaks.

Now we shall consider the case in which a collective resonance is
present. As with giant dipole resonances (GDR) in stable nuclei, one
believes that {\it pygmy resonances} at energies close to the threshold
are present in halo, or neutron-rich, nuclei (in English, pigmy = pygmy). This was proposed in
Ref. [SIS90] using the hydrodynamical model for
collective vibrations. The possibility to explain the soft dipole
modes (figure \ref{fig4}) in terms of direct breakup, has made it
very difficult to clearly identify the signature of pygmy resonances
in light exotic nuclei.

The hydrodynamical model needs
adjustments to explain collective response in light, neutron-rich,
nuclei. Because clusterization in light nuclei exists, not all
neutrons and protons can be treated equally. The necessary
modifications are straight-forward and discussed next. 

\begin{figure}[ptb]
\begin{center}
\includegraphics[
height=2.3082in, width=2.8911in ]{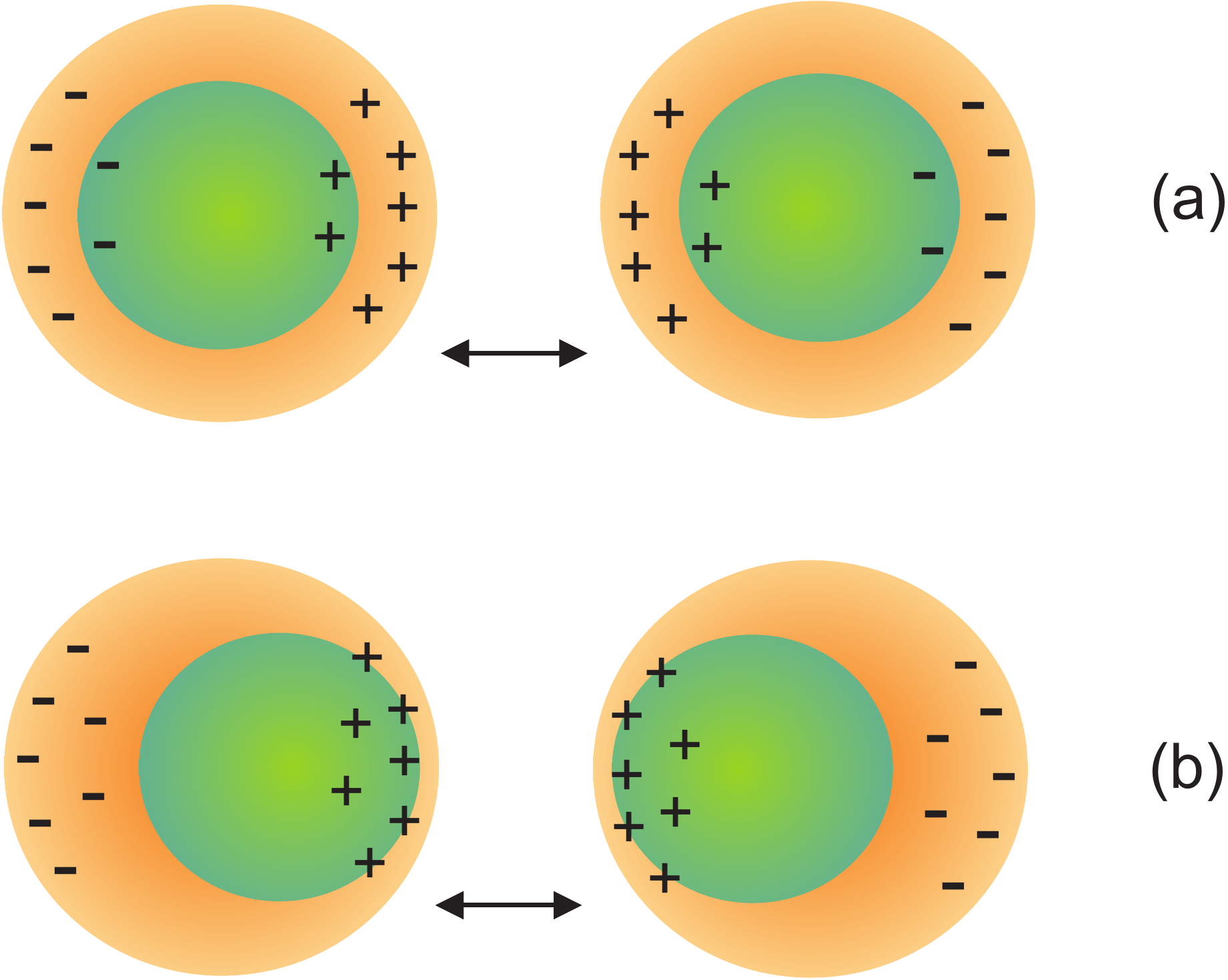}
\end{center}
\caption{\sl  Hydrodynamical model for collective nuclear
vibrations in halo nuclei. The (a) Steinwedel-Jensen (SJ) mode and
the (b) Goldhaber-Teller (GT)
mode are shown separately.}%
\label{gtsj}%
\end{figure}

When a collective vibration of protons against neutrons is present
in a nucleus with charge (neutron) number $Z$ ($N$), the neutron and
proton fluids are displaced with respect to each other by
$d_{1}=\alpha_{1}R$ and each of the fluids are displaced from the
origin (center of mass of the system) by $d_{p}=Nd_{1}/A$ and $d_{n}%
=-Zd_{1}/A$. This leaves the center of mass fixed and one gets for
the dipole moment $D_{1}=Zed_{p}=\alpha _{1}NZeR/A$. The GT (Goldhaber-Teller) model [Gt48]
assumes that the restoring force is due to the increase of the
nuclear surface which leads to an extra energy proportional to
$A^{2/3}$. In this model, the inertia is proportional to $A$ and the
excitation energy is consequently given by $E_{x}\propto\sqrt{A^{2/3}%
/A}=A^{-1/6}$.

For light, weakly-bound nuclei, it is more appropriate to assume
that the neutrons inside the core ($A_{c},Z_{c}$) vibrate in phase
with the protons. The neutrons and protons in the core are tightly
bound. An overall displacement among them requires energies of the
order of 10-20 MeV, well above that of the soft dipole modes. The
dipole moment becomes $\mathbf{D}_{1} =e\mathbf{d}_{1}{\left(  Z_{c}%
A-ZA_{c}\right)  }/{A}  =Z_{eff}^{(1)}e\mathbf{d}_{1}$, where
$\mathbf{d}_{1}$ is a vector connecting the center of mass of the
two fluids (core and excess neutrons).  We see that the dipole
moment is now smaller than before because the
effective charge changes from $NZ/A$ in the case of the GDR to $Z_{eff}%
^{(1)}=$ $\left(  Z_{c}A-ZA_{c}\right)  /A$. This effective charge is zero if
$Z_{c}A=ZA_{c}$ and no pigmy resonance is possible in this model, only the
usual GDR.

Figure \ref{gtsj} shows a schematic representation of the hydrodynamical model
for collective nuclear vibrations in a halo nucleus, as considered here. Part
(a) of the figure shows the Steinwedel-Jensen (SJ) mode [SJ50] in which the total
matter density of both the core and the halo nucleons do not change locally.
Only the local ratio of the neutrons and protons changes. Part (b) of the
figure shows a particular case of the Goldhaber-Teller (GT) mode, in which the
core as a whole moves with respect to the halo nucleons.

\begin{figure}[ptb]
\begin{center}
\includegraphics[
natheight=3.589000in,
natwidth=4.678600in,
height=3.0597in,
width=3.9825in
]{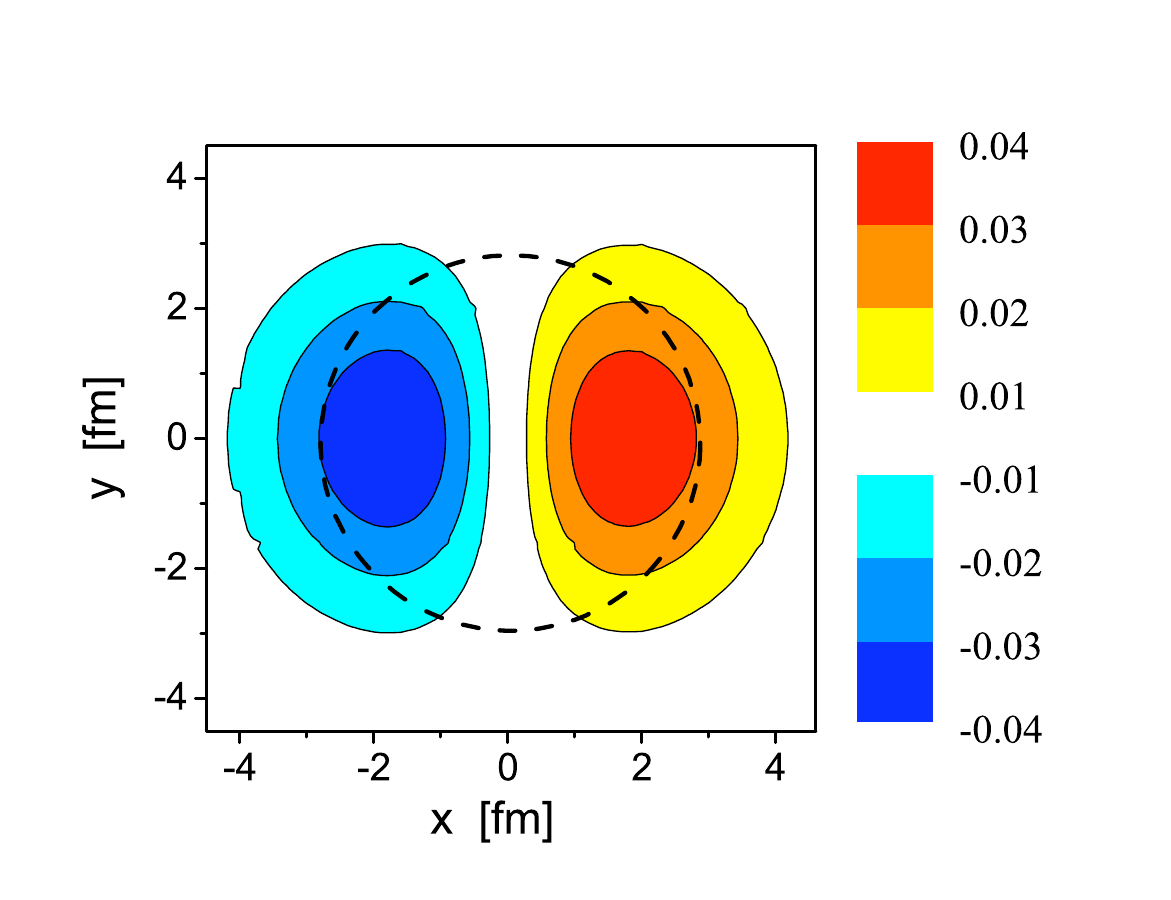}
\end{center}
\caption{\sl  Contour plot of the nuclear transition
density in the hydrodynamical model consisting of a mixture of GT
and SJ vibrations. The darker areas represent the larger values of
the transition density in a nucleus which has an average radius
represented by the dashed circle. The legend on the right
displays the values of the transition density within each contour limit. }%
\label{pygmy}%
\end{figure}

For spherically symmetric densities, the transition density in the GT mode can
be calculated from $\delta\rho_{p}=\rho_{p}\left(  \left\vert \mathbf{r}%
-\mathbf{d}_{p}\right\vert \right)  -\rho_p\left(  \mathbf{r}\right)
$, where $\rho_p$ is the charge density. Using $d_{1}\ll R$, it is
straight-forward to show that  $ \delta\rho_{p}^{(1)}\left(
\mathbf{r}\right) =\delta\rho_{p}^{(1)}\left(  r\right) Y_{10}\left(
\widehat{\mathbf{r}}\right)$ , where
\begin{equation}
\delta\rho_{p}^{(1)}\left(  r\right)
=\sqrt{\frac{4\pi}{3}}Z_{eff}\alpha _{1}R\frac{d\rho_{0}}{dr},
\end{equation}
and $\rho_0$ is the ground state matter density.

In the Steinwedel-Jensen (SJ) mode, the local variation of the
density of protons is found to be $\delta\rho_{p}^{(2)}\left(
\mathbf{r}\right)  =\delta \rho_{p}^{(2)}\left(  r\right)
Y_{10}\left( \widehat{\mathbf{r}}\right)$,
where%
\begin{equation}
\delta\rho_{p}^{(2)}\left(  r\right)  =\sqrt{\frac{4\pi}{3}}Z_{eff}%
^{(2)}\alpha_{2}Kj_{1}\left(  kr\right)  \rho_{0}\left(  r\right)  ,
\end{equation}
where $K=9.93$. If the proton and neutron content of the core does
not change [SIS90],
the effective charge number in the SJ mode is given by $Z_{eff}^{(2)}%
=Z^{2}(N-N_{c})/A(Z+N_{c})$.

The transition density at a point $\mathbf{r}$ from the
center-of-mass of the nucleus is a combination of the SJ and GT
distributions and is given by $ \delta\rho_{p}\left(
\mathbf{r}\right)     =\delta\rho_{p}\left(  r\right)  Y_{10}\left(
\widehat{\mathbf{r}}\right)$,
where%
\begin{equation}
\delta\rho_{p}\left(  r\right)  =\sqrt{\frac{4\pi}{3}}R\left\{  Z_{eff}%
^{(1)}\alpha_{1}\frac{d}{dr}+Z_{eff}^{(2)}\alpha_{2}\frac{K}{R}j_{1}\left(
kr\right)  \right\}  \rho_{0}(r). \label{transdsjgt}%
\end{equation}
Changes can be accommodated in these expressions to account for the different
radii of the proton and neutron densities.

Figure \ref{pygmy} shows the contour plot, in arbitrary units, of
the nuclear transition density in the hydrodynamical model,
consisting of a mixture of GT and SJ vibrations. The darker areas
represent the larger values of the transition density in a nucleus
which has an average radius represented by the dashed circle. In
this particular case,  a Hartree-Fock density calculation
for $^{11}$Li, and a radius $R=3.1$ fm, was used. The parameters $\alpha_{1}$
and $\alpha_{2}$ were chosen so that $Z_{eff}^{(1)}\alpha
_{1}=Z_{eff}^{(2)}\alpha_{2}$, i.e$.$ a symmetric mixture of the SJ
and GT modes.

\begin{figure}[ptb]
\begin{center}
\includegraphics[
height=2.6333in,
width=3.4688in
]{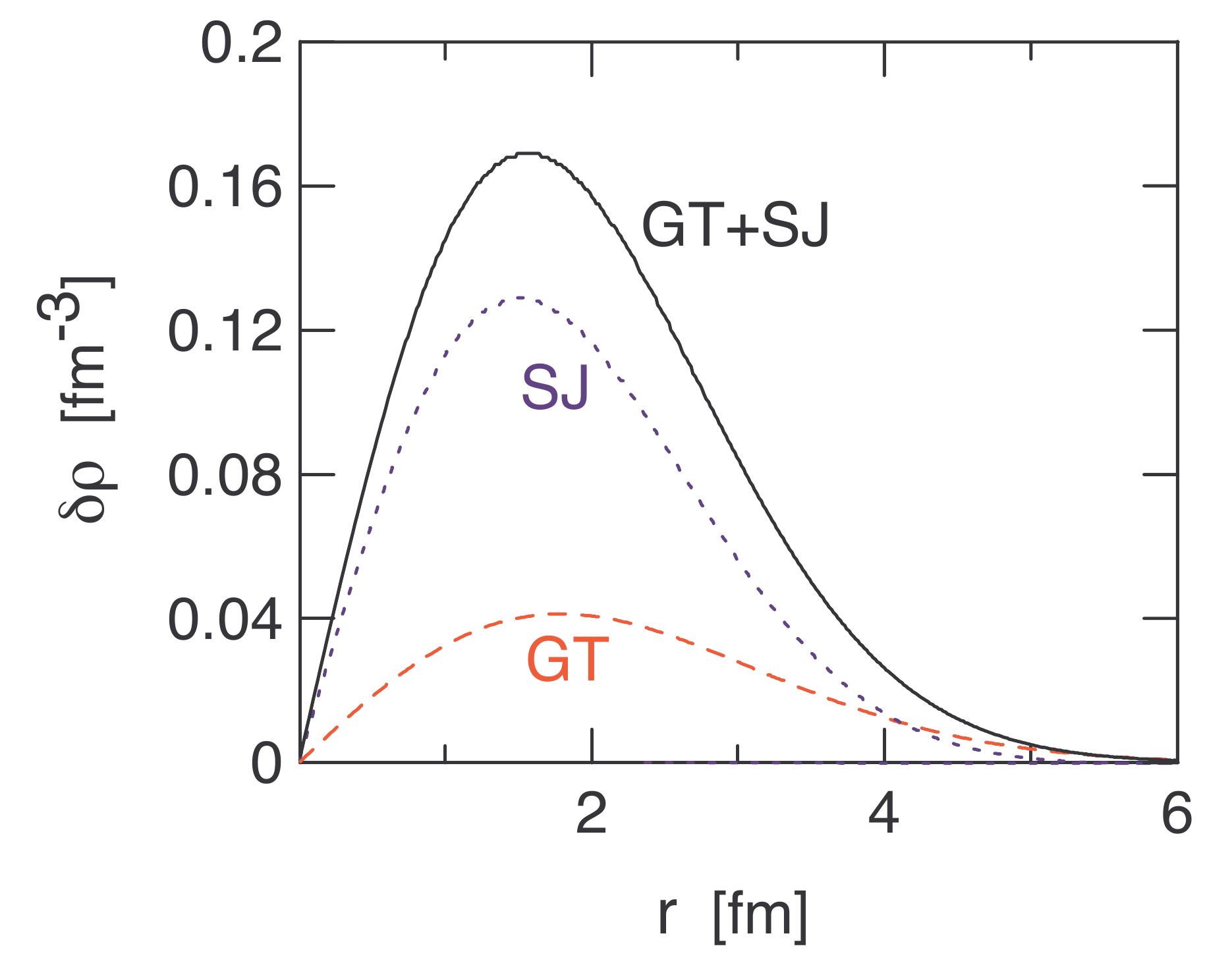}
\end{center}
\caption{\sl  Hydrodynamical transition densities for
$^{11}$Li and three different assumptions for the SJ+GT admixtures,
according to Eq.  \eqref{transdsjgt}. The dashed curve is for a GT
oscillation mode, with the core vibrating against the halo neutrons,
with effective charge number $Z_{eff}^{(1)}=6/11$, radius $R=3.1$
fm, and $\alpha_{1}=1$. The dotted curve is for an SJ oscillation
mode, with effective charge number $Z_{eff}^{(2)}=2/11$, and
$\alpha_{2}=1$.
The solid curve is their sum.}%
\label{trand}%
\end{figure}

Figure \ref{trand} shows the transition densities for $^{11}$Li for
three different assumptions of the SJ+GT admixtures, according to
Eq. \eqref{transdsjgt}. The dashed curve is for a GT oscillation mode,
with the core vibrating against the halo neutrons, with effective
charge number $Z_{eff}^{(1)}=6/11$, radius $R=3.1$ fm, and
$\alpha_{1}=1$. The dotted curve
is for an SJ oscillation mode, with effective charge number $Z_{eff}%
^{(2)}=2/11$, and $\alpha_{2}=1$. The solid curve is their sum.
Notice that the transition densities are peaked at the surface, but
at a
radius smaller than the adopted \textquotedblleft rms\textquotedblright%
\ radius $R=3.1$ fm.

The liquid drop model predicts an equal admixture of SJ+GT
oscillation modes for large nuclei [Mye77]. The contribution
of the SJ oscillation mode decreases with decreasing mass number,
i.e. $\alpha_2 \longrightarrow 0  $ as $A\longrightarrow0$. This is
even more probable in the case of halo nuclei, where a special type
of GT mode (oscillations of the core against the halo nucleons) is
likely to be dominant. For this special collective motion an
approach different than that used in Ref. 
[SIS90] has to be considered.  The resonance energy formula
derived by Goldhaber and Teller  [GT48] changes to%
\begin{equation}
E_{PR}=\left(  \frac{3\varphi\hbar^{2}}{2aRm_{N}A_{r}}\right)  ^{1/2},
\label{EPGDR}%
\end{equation}
where $A_{r}=A_{c}\left(  A-A_{c}\right)  /A$ and $a$ is the length within
which the interaction between a neutron and a nucleus changes from a
zero-value outside the nucleus to a high value inside, i.e. $a$ is the size of
the nuclear surface. $\varphi$ is the energy needed to extract one neutron
from the proton environment.

Goldhaber and Teller [GT48] argued that in a heavy stable
nucleus $\varphi$ is not the binding energy of the nucleus, but the
part of the potential energy due to the neutron-proton interaction.
It is proportional to the asymmetry energy. In the case of
weakly-bound nuclei this picture changes and it is more reasonable
to associate $\varphi$ to the separation energy of the valence
neutrons, $S$. I will use $\varphi=\beta S$, with a parameter
$\beta$ which is expected to be of order of one. Since for halo
nuclei the product $aR$ \ is proportional to $S^{-1},$ we obtain the
proportionality $E_{PR}\propto S$. Using Eq.  \eqref{EPGDR} for
$^{11}$Li , with \ $a=1$ fm, $R=3$ fm and $\varphi=S_{2n}=0.3$ MeV,
we get $E_{PR}=1.3$ MeV. Considering that the pygmy resonance will
most probably decay by particle emission, one gets $E_{r}\simeq1$
MeV for the kinetic energy of the fragments, which is within the
right ballpark (see figure \ref{fig4}).

Both the direct dissociation model and the hydrodynamical model
yield a bump in the response function proportional to $S$, the
valence nucleon(s) separation energy. In the direct dissociation
model the width of the response function obviously depends on the
separation energy. But it also depends on the nature of the model,
i.e. if it is a two-body model, like the model often adopted for
$^{11}$Be or $^{8}$B, or a three-body model, appropriate for
$^{11}$Li and $^{6}$He. In the two-body model the phase-space
depends on energy as $\rho\left( E\right)  \propto
d^{3}p/dE\propto\sqrt{E}$, while in the three-body model $\rho\left(
E\right)  \sim E^{2}$. This explains why the peak of figure
\ref{fig4} is pushed toward higher energy values, as compared to the
prediction of Eq.  \eqref{deel}. It also explains the larger width of
$dB/dE$ obtained in three-body models. In the case of the pigmy
resonance model, this question is completely open.

The {\it hydrodynamical model} predicts [Mye77] for the width of the
collective mode $\Gamma=\hbar\overline{\mathrm{v}}/R$, where
$\overline{\mathrm{v}}$ \ is the average velocity of the nucleons
inside the nucleus. This relation can be derived by assuming that
the collective vibration is damped by the incoherent collisions of
the nucleons with the walls of the nuclear potential well during the
vibration cycles (piston model). Using
$\overline{\mathrm{v}}=3\mathrm{v}_{F}/4$, where
$\mathrm{v}_{F}=\sqrt{2E_{F}/m_{N}}$ is the Fermi velocity, with
$E_{F}=35$ MeV and $R=6$ fm, one gets $\Gamma\simeq6$ MeV. This is
the typical energy width a giant dipole resonance state in a heavy
nucleus. In the case of neutron-rich light nuclei
$\overline{\mathrm{v}}$ is not well defined. There are two average
velocities: one for the nucleons in the core, $\overline
{\mathrm{v}}_{c}$, and another for the nucleons in the skin, or
halo, of the nucleus, $\overline{\mathrm{v}}_{h}$. One is thus
tempted to use a
substitution in the form $\overline{\mathrm{v}}=\sqrt{\overline{\mathrm{v}%
}_{c}\overline{\mathrm{v}}_{h}}.$ Following Ref. [BM93], the width of
momentum distributions of core fragments in knockout reactions, $\sigma_{c}$,
is related to the Fermi velocity of halo nucleons by $\mathrm{v}_{F}%
=\sqrt{5\sigma_{c}^{2}}/m_{N}$. Using this expression with $\sigma_{c}%
\simeq20$ MeV/c, we get $\Gamma=5$ \ MeV (with $R=3$ fm). This value
is also not in discordance with experiments (see figure \ref{fig4}).

Better microscopic models, e.g. those based on random phase
approximation (RPA) calculations  are necessary to
study pigmy resonances. The halo nucleons have to be treated in an
special way to get the response at the right energy position, and
with approximately the right width. 

\subsection{Comparison between cluster breakup and pigmy resonance excitation}

One might argue that the total breakup cross section would be a good
signature for discerning direct dissociation versus the dissociation
through the excitation of a pigmy collective vibration. The trouble
is that the energy weighted sum rule for both cases are
approximately of the same magnitude. This can be
shown by using the electric dipole strength function in the 
cluster breakup
model, namely%
\begin{equation}
\frac{dB(E1)}{dE}=\mathcal{C}\frac{3\hbar e^{2}Z_{eff}^{2}}{\pi^{2}\mu}%
\frac{\sqrt{S_{n}}\left(  E-S_{n}\right)  ^{3/2}}{E^{4}},\label{dbe1de}%
\end{equation}
where $E=E_{r}+S_{n}$ is the total excitation energy. $\mathcal{C}$
is a constant of the order of one, accounting for the corrections to
the wavefunction.

The sum rule for dipole excitations, $S_{1}\left(  E1\right)
=\int_{S_n}^\infty dE\ E\ \frac{dB\left( E1\right)}{dE}$, is
\begin{equation}
S_{1}\left(  E1\right)  =\mathcal{C}\left(  \frac{9}{8\pi}\right)  \frac
{\hbar^{2}e^{2}}{\mu}Z_{eff}^{2},\label{srule}%
\end{equation}
with $Z_{eff}^2=\left(  Z_{c}A-ZA_{c}\right)  ^{2}/[AA_{c}\left(  A-A_{c}%
\right)  ]$. This is the same (with $\mathcal{C}=1$) as Eq.  1 of
Ref. [AGB82], which is often quoted as the standard value to
which models for the nuclear response in the region of pigmy
resonance should be compared to. The response function in Eq. 
\eqref{dbe1de}, with $\mathcal{C}=1$, therefore exhausts 100\% of the
so-called cluster sum rule  [AGB82]. The total cross section
for electron breakup of weakly-bound systems is roughly proportional
to $S_{1}$. This assertion can be easily verified by using eqs.
\eqref{EPA} and \eqref{dbe1de}, assuming that the logarithmic dependence
of the virtual photon numbers on the energy $E\equiv E_{\gamma}$ can
be factored out of the integral in Eq.  \eqref{EPA}.

The dipole strength of the pigmy dipole resonance is given by the
same equation \eqref{srule}. The constant $\mathcal{C}$ is still of
the order of unity, but not necessarily the same as in Eq. 
\eqref{dbe1de} and the effective charges are also different. For the
Goldhaber-Teller pigmy dipole model  the effective charge is given
by $Z_{eff}^{(1)}=$ $\left(  Z_{c}A-ZA_{c}\right)  /A$, whereas for
the Steinwedel-Jensen it is
$Z_{eff}^{(2)}=(Z^{2}/A)(N-N_{c})/(Z+N_{c})$. Assuming that the
Goldhaber-Teller mode prevails, one gets the simple prediction for
the ratio between the cross sections for
direct breakup versus excitation of a pigmy collective mode:%
\begin{equation}
\frac{\sigma^{direct}}{\sigma^{pigmy}}=\mathcal{C}\frac{A
}{A_{c}\left(  A-A_{c}\right)  }.
\end{equation}
For $^{11}$Li this ratio is 11$\mathcal{C}$/18 while for $^{11}$Be
it is 11$\mathcal{C}$/10. One thus concludes that it is very difficult to identify if the
nuclear response at low energies is due to
collective excitations of pigmy resonances, or a trivial phase-space effect due to the cluster dissociation, 
which is a basic manifestation of the low binding energy. Other observables are necessary to make firm conclusions, 
such as correlations of emitted fragments.

\subsection{Higher-order effects}

Breakup processes in nucleus-nucleus collisions are complicated, in whatever
way they are treated. They constitute at least a three-body problem, which
is further complicated due to the long range Coulomb force. Exact treatments
(such as the Fadeev-approach) are therefore prohibitively cumbersome. On the
other hand, many approximate schemes have been developed and these approaches have been used with
considerable success.

Higher-order effects in Coulomb breakup were first observed in  Ref. [Ie93]. In Fig. \ref{break4} we see that the velocity of $%
^{9}$Li fragments are faster, in average, than those of the two detached
neutrons in a Coulomb break-up of 28 MeV/nucleon $^{11}$Li incident on lead
targets.

\begin{figure}[ptb]
\begin{center}
\includegraphics[
height=2.6333in,
width=3.4688in
]{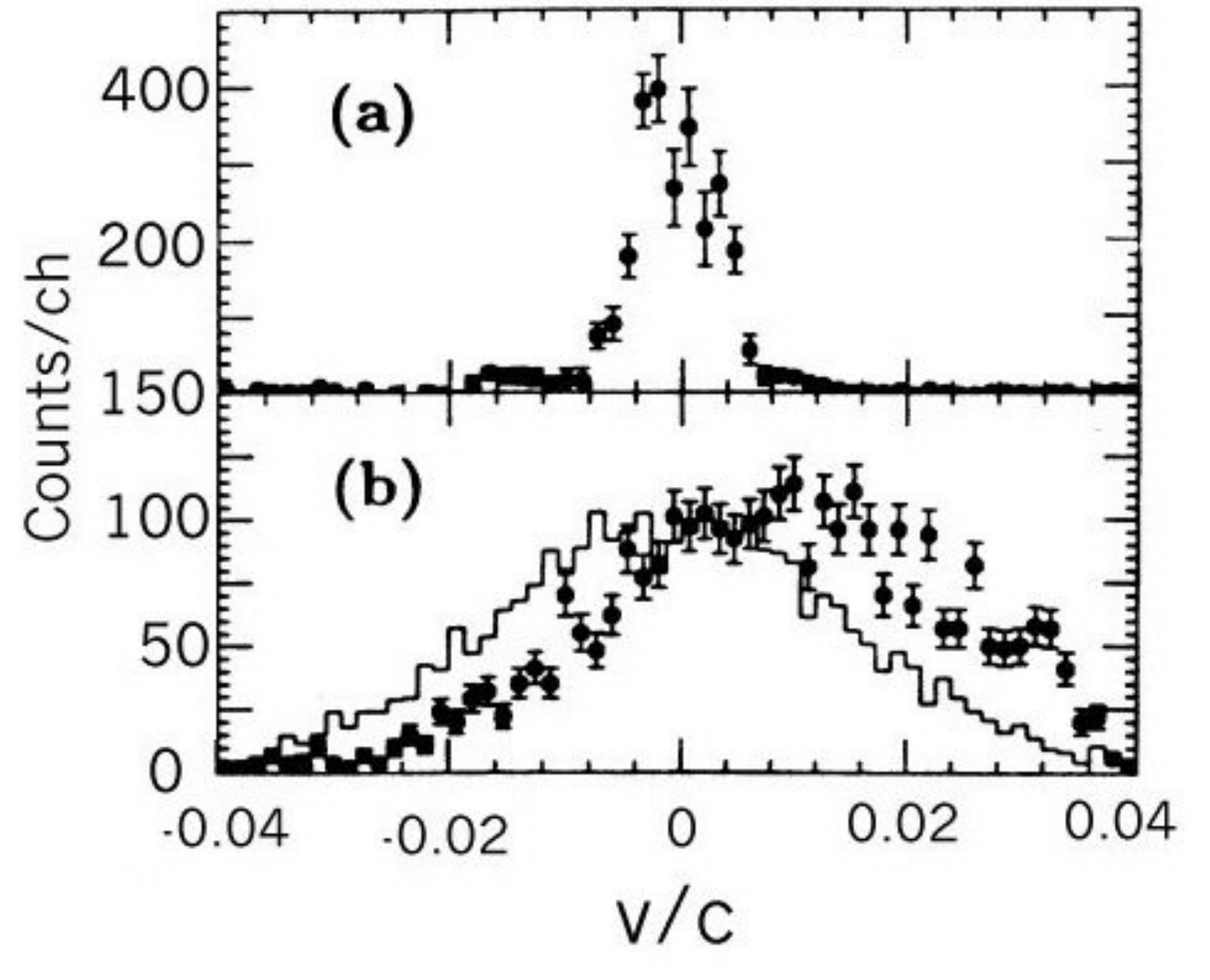}
\end{center}
\caption{\sl  Spectrum of the
longitudinal component of the center-of-mass velocity of $^{9}$Li\ and two
neutrons in the frame of the incident $^{11}$Li. (b) Spectrum of the
longitudinal component of the relative velocity V$_{9}$\ - V$_{2n}.$\ The
histogram shows the result of a Monte Carlo simulation assuming no Coulomb
acceleration effects. }%
\label{break4}%
\end{figure}

This effect can be
understood qualitatively as follows. The $^{11}$Li is deccelerated up to a
point where it dissociates. This occurs around the distance of closest
approach. Afterwards the $^{9}$Li fragments is accelerated, whereas the
neutrons are not. Since $^{9}$Li is lighter than $^{11}$Li its final
velocity is greater than the incoming beam velocity. The neutrons are
consequently slower.

One should expect that perturbation theory fails in describing the breakup
process when the cross sections attain very high values. In fact, the Coulomb  break-up probability for weakly-bound
nuclei calculated with first-order perturbation theory is close to unity.
This can be understood with use of simple arguments. The energy transferred
by the Coulomb field to the excitation of a projectile nucleus, with $N$
neutrons and $Z$ protons, incident with velocity $v$ on a target nucleus
with charge $eZ_{T}$ at an impact parameter $b$ is approximately given by 
\begin{equation}E^{\ast
}={2(NZ/A)(Z_{T}e^{2})^{2}\over m_{N}b^{2}v^{2}}\end{equation} where $m_{N}$ is the nucleon
mass. For $^{11}$Be projectiles ($N=7,\;Z=4$) incident on lead at $b=15\;$fm
and $v\approx 0.5c$, one gets $E^{\ast }\approx 1\;$MeV. This energy is more
than sufficient to break $^{11}$Be apart, since the separation energy a
neutron from this nucleus is about $0.5\;$MeV. This means that, at small
impact parameters the break-up probability is of order of unity and a
non-perturbative treatment of the break-up process should be carried out.

A higher-order treatment of Coulomb dissociation is therefore
desirable. However, one faces the difficulty that the final states are in
the continuum  and
the coupling matrix elements present divergency problems, caused by the
non-localized behavior of the continuum wavefunctions. This difficulty is
avoided by a discretization of the continuum along the lines discussed below, introduced in
Ref. [BC92].
The method is a useful starting point to understand the main features of a
coupled-channels calculation with states in the continuum. What I will describe next is a simple example of
what is commonly known as the {\it Continuum-Discretized Coupled-Channels} (CDCC) method, which can include several levels of
sophistication.

\subsubsection{Discretization of the continuum}

Our basis of time-dependent discrete states are defined as  ($B$ s the binding energy)
\begin{eqnarray}
&&\left| \phi _{0}\right\rangle =e^{-iE_{0}t/\hbar }\left| 0\right\rangle
,\;\;\;\mathrm{with}\;\;E_{0}=-B\nonumber\\
&&\mathrm{and}\;\;\left| \phi _{j\ell
m}\right\rangle =e^{-iE_{j}t/\hbar }\int \Gamma _{j}(E)\;\left| E,\ell
m\right\rangle  \label{BC9231}
\end{eqnarray}
where $|E,\ell m>$ are continuum wavefunctions of the projectile fragments
(without the interaction with the target), with good energy and angular
momentum quantum numbers $E,\,\ell ,\,m$. The functions $\Gamma _{j}(E)$ are
assumed to be strongly peaked around an energy $E_{j}$ in the continuum.
Therefore, the discrete character of the states $|\phi _{j\ell m}>$
(together with $|\phi _{0}>$) allows an easy implementation of the
coupled-states calculations. 

We assume that the projectile has no bound
excited states. This assumption is often the rule for very loosely-bound
systems. The orthogonality of the discrete states \eqref{BC9231} is guaranteed
if 
\begin{equation}
\int dE\;\Gamma _{i}(E)\;\Gamma _{j}(E)=\delta _{ij}.  \label{BC9232}
\end{equation}

For the continuum set $|E\ell m>$ we use, for the sake of simplicity, the
plane wave basis 
\begin{eqnarray}
<\mathbf{r}|E\ell m>&=&u_{\ell ,\,E}(r)\,Y_{\ell m}(\mathbf{r})\nonumber \\
&=&\left( {\frac{%
2\mu }{\hbar ^{2}}}\right) ^{3/4}\;{\frac{E^{1/4}}{\sqrt{\pi }}}\;j_{\ell
}(qr)\;Y_{\ell m}(\mathbf{\hat{r}})  \label{BC9233}
\end{eqnarray}
which obey the normalization condition ($E=\hbar ^{2}q^{2}/2\mu $) 
\begin{equation}
<E\ell m|E^{\prime }\ell ^{\prime }m^{\prime }>=\delta _{\ell \ell ^{\prime
}}\delta _{mm^{\prime }}\delta (E-E^{\prime }).  \label{BC9234}
\end{equation}
These states arise from the partial wave expansion of the plane wave $\exp (i%
\mathbf{q.r})$. Writing the time-dependent Schr\"{o}dinger equation for $%
\Psi (t)=\sum_{j\ell m}a_{j\ell m}\,\phi _{j\ell m}$, taking the scalar
product with the basis states and using orthonormality relations, we get the
equations  (see Appendix B)
\begin{equation}
i\hbar {\frac{da_{j\ell m}}{dt}}=\sum_{j^{\prime }\ell ^{\prime }m^{\prime
}}V_{j\ell m;j^{\prime }\ell ^{\prime }m^{\prime }}\;a_{j^{\prime }\ell
^{\prime }m^{\prime }}\;e^{-i(E_{j}^{\prime }-E_{j})t/\hbar }.
\label{BC9235}
\end{equation}
We use the index $j=0$ for the ground state $|0>$ and $j=1,2,\dots $for the
discrete continuum states. $V_{j\ell m;j^{\prime }\ell ^{\prime }m^{\prime
}} $ are the matrix elements $<\phi _{j\ell m}|V|\phi _{j^{\prime }\ell
^{\prime }m^{\prime }}>$.

For $\Gamma _{j}(E)$ we consider two different sets of functions. Firstly
the set $\Gamma _{1}(E),\dots $ $,\Gamma _{N}(E);$%
\begin{eqnarray}
\Gamma _{j}(E)&=&{\frac{1}{\sqrt{\sigma }}}\,,\;\;\;\;\mathrm{for}%
\;\;(j-1)\sigma <E<j\sigma, \nonumber\\
\Gamma _{j}(E)&=&0\,,\qquad 
\mathrm{otherwise.}  \label{BC9236}
\end{eqnarray}

\begin{figure}[ptb]
\begin{center}
\includegraphics[
height=3.6333in,
width=3.1688in
]{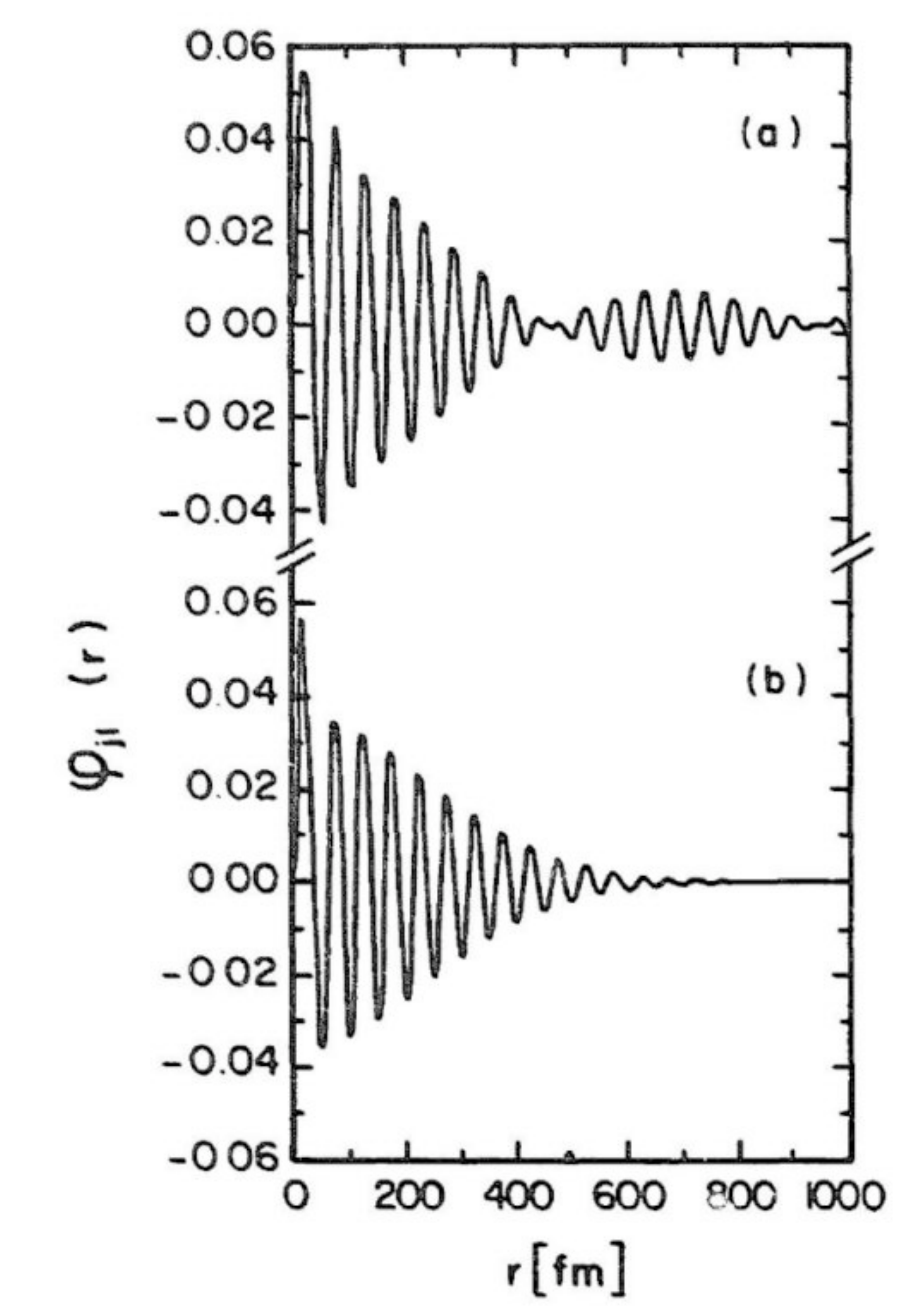}
\end{center}
\caption{\sl  Radial wave functions for the
discretized continuum using the histogram set (a) and the continuous set
(b). From [BC92].}%
\label{felipe3}%
\end{figure}

This set corresponds to histograms of constant height $1/\sqrt{\sigma }$ and
width $\sigma $. The states $\Gamma _{j}(E)$ trivially satisfy the
orthonormalization condition of Eq. \eqref{BC9232}. They present the advantage
of leading to simple analytical expressions for the coupling matrix
elements. On the other hand they have discontinuities at the edges, which
lead to numerical difficulties. The second set consists of the functions 
\begin{equation}
\chi _{j}(E)=N_{n_{j}}\;\left( {\frac{E}{\sigma }}\right)
^{n_{j}^{2}}\;e^{-n_{j}(E/\sigma )}.  \label{BC9237a}
\end{equation}
The normalization constant 
\begin{equation}
N_{n_{j}}={\frac{1}{\sqrt{\sigma }}}\;\left[ {\frac{(2n_{j})^{2n_{j}^{2}+1}}{%
(2n_{j}^{2})!}}\right] ^{1/2},  \label{BC9237b}
\end{equation}
guarantees that $\int \chi _{j}(E)\,\chi _{j}(E)\,dE=1$. The functions $\chi
_{j}$ are peaked at $E=n_{j}\,\sigma $ and have width $\approx \sigma $. The
integer $n_{j}=K.j$ is proportional to the index-j and the proportionality
constant, a small integer $K$, is a parameter of the set which determines
the overlap of two consecutive functions $\chi _{j}$ and $\chi _{j+1}$.
However, this
set fails to satisfy the orthogonality condition of Eq. \eqref{BC9235}. This
shortcoming can be fixed by the definition of a new set $\Gamma _{j}(E)$ of
linear combinations 
\begin{equation}
\Gamma _{j}(E)=\sum_{k=1}^{N}\,C_{jk}\,\chi _{k}(E),  \label{BC9238}
\end{equation}
with the coefficients $C_{ij}$ determined so that the resulting combinations
be orthogonal. These coefficients can be found by means of an
orthogonalization procedure as, e.g., the \textit{Gram-Schmidt method}. The set of Eq. 
\eqref{BC9238} has the advantages of being continuously derivable and of
leading to reasonably simple coupling matrix elements.

A comparison between basis states $\phi _{j\ell m}(r)$ generated with each
of these sets [through Eq. \eqref{BC9231}] is made in Fig. \ref{felipe3}.  One observes that the discrete wavefunctions $\phi _{j\ell m}$
decrease rapidly enough with $r$, so that the matrix elements $<\phi _{j\ell
m}|rY_{1\mu }|\phi _{j^{\prime }\ell ^{\prime }m^{\prime }}>$ are finite.
The use of the histograms \eqref{BC9236} for $\Gamma _{j}(E)$ leads to beats
in $\phi _{j\ell m}$ as displayed in Fig. \ref{felipe3}(a). These beats are
the result of the discontinuous nature of $\Gamma _{j}(E)$ and arise from
the interference from the borders of the histograms. Due to this behavior,
the numerical evaluation of $<\phi _{j\ell m}|rY_{1\mu }|\phi _{j^{\prime
}\ell ^{\prime }m^{\prime }}>$ is more involved than with the second set of $%
\Gamma _{j}$-functions, \eqref{BC9238}. Indeed, as we see from Fig. \ref
{felipe3}(b) the beats disappear with the use of the basis set \eqref{BC9238}%
. Although the use of plane-wave basis allows the derivation of simple
results with both sets, this fact is of relevance for improved
coupled-channels calculations in the continuum.

Using Eq. \eqref{BC9231} and the properties of the spherical harmonics one finds 
\begin{eqnarray}
V_{j\ell m;j^{\prime }\ell ^{\prime }m^{\prime }}&=&{\frac{(-1)^{m}}{\sqrt{2}%
}}\gamma Z_{T}e^{2}\left( Z_{c}\,{\frac{m_{b}}{m_{a}}}-Z_{b}\,{\frac{m_{c}}{%
m_{a}}}\right)\nonumber \\
&\times& {\frac{\sqrt{(2\ell +1)(2\ell ^{\prime }+1)}}{(b^{2}+\gamma
^{2}v^{2}t^{2})^{3/2}}}\left( {{\ell }\atop{0}}{{1}\atop{0}}{{\ell
^{\prime }}\atop{0}}\right)  \notag \\
&\times&\Big\{ ib\Big[ \left( {{\ell }\atop{-m}}{{1}\atop{1}}{{%
\ell ^{\prime }}\atop{m^{\prime }}}\right) +\left( {{\ell }\atop{-m}}{{1}\atop{%
-1}}{{\ell ^{\prime }}\atop{m^{\prime }}}\right) \Big] \nonumber \\
&+&\sqrt{2}\,\gamma
vt\left( {{\ell }\atop{-m}}{{1}\atop{0}}{{\ell ^{\prime }}\atop{m^{\prime
}}}\right) \Big\} I_{j\ell ;j\prime \ell \prime }  \label{BC-9a}
\end{eqnarray}
where 
\begin{eqnarray}
I_{j\ell ;j^{\prime }\ell ^{\prime }}&=&\int r^{3}dr\int dE\;\Gamma
_{j}(E)\int dE^{\prime }\;\Gamma _{j^{\prime }}(E^{\prime })\nonumber\\
&\times&u_{\ell
,\,E}^{\ast }(r)u_{\ell ^{\prime },\,E^{\prime }}(r).  \label{BC-9b}
\end{eqnarray}
From \eqref{BC-9a} one deduces that the interaction potential is different
from zero only if $|\ell -\ell ^{\prime }|=1$, as expected.

The use of the plane wave basis is especially useful because, exploiting the
recursion and closure relations of the spherical Bessel functions, one
obtains the general result 
\begin{equation}
I_{j\ell ;j^{\prime }\ell ^{\prime }}={\frac{\hbar ^{2}}{\mu }}\left\{ {%
\frac{\ell +\ell ^{\prime }+2}{2}}F_{jj^{\prime }}+\delta _{\ell ,\ell
^{\prime }+1}G_{j,j^{\prime }}+\delta _{\ell +1,\ell ^{\prime
}}G_{j^{\prime }j}\right\} ,  \label{BC92310}
\end{equation}
where 
\begin{eqnarray}
F_{jj^{\prime }}&=&\int dq\;\Gamma _{j}(E)\;\Gamma _{j^{\prime
}}(E)\nonumber\\
G_{jj^{\prime }}&=&\int dq\;q\;\Gamma
_{j}(E)\;{\frac{d}{dq}}\Gamma _{j^{\prime }}(E)  \label{BC92311}
\end{eqnarray}
with $E=\hbar ^{2}q^{2}/2\mu $. Explicit forms can be found for each basis
set:

\textit{(a) Histogram - } Applying this relation to the histogram set \eqref
{BC9236}, one can show that for $j,\;j^{\prime }\neq 0$%
\begin{equation}
I_{j\ell ;j^{\prime }\ell ^{\prime }}=\hbar \sqrt{\frac{2}{\mu \sigma }}%
\left\{ 
\begin{array}{l}
{\frac{\ell +\ell ^{\prime }+1}{2}}\;\left[ \sqrt{j}-\sqrt{j-1}\right]
\;\;\mathrm{if\;}\;j=j^{\prime } \\ 
-(-1)^{(j+\ell -j^{\prime }-\ell ^{\prime })\over 2}\sqrt{\frac{j+j^{\prime }-1}{2%
}}\;\mathrm{if}\;|j-j^{\prime }|=1 \\ 
0\;\;\;\;\mathrm{otherwise.}
\end{array}
\right. \;  \label{BC92312}
\end{equation}
For $j=0$ or $j^{\prime }=0$, only the integral with $\ell $, or $\ell
^{\prime }=1$ is necessary, and the result is 
\begin{equation}
I_{00;\;j1}=I_{j1;\;00}={\frac{\sqrt{2\eta \sigma }}{\pi }}\;{\frac{%
E_{j}^{3/4}}{(E_{0}+E_{j})^{2}}}\;\left( {\frac{\hbar ^{2}}{2\mu }}\right)
^{3/4}
\end{equation}
where $E_{j}=(j-1/2)\,\sigma $.

\textit{(b) Continuous basis - } For the set of continuous energy functions 
\eqref{BC9238} one finds, for $j,\,j^{\prime }\neq 0$ 
\begin{eqnarray}
\left\{ 
\begin{array}{c}
F_{jj^{\prime }} \\ 
G_{jj^{\prime }}
\end{array}
\right\} &=&\sqrt{\frac{\mu \sigma }{2\hbar ^{2}}}\sum_{n,n^{\prime
}}C_{jn}\,C_{j^{\prime }n^{\prime }}N_{n}\,N_{n^{\prime }}{\frac{\Gamma
(n^{2}+n^{\prime }{}^{2}+1/2)}{(n+n^{\prime })^{n^{2}+n^{\prime }{}^{2}+1/2}}%
}\nonumber \\ 
&\times&\left\{ 
\begin{array}{c}
1 \\ 
2n^{\prime }{}^{2}-\frac{n^{\prime }(2n^{2}+2n^{\prime }{}^{2}+1)}{%
n+n^{\prime }}
\end{array}
\right\}  \label{BC92314}
\end{eqnarray}
where $\Gamma (z)$ is the gamma-function and we simplified the notation
using $n\equiv n_{j}$. For $j=0$, or $j^{\prime }=0$, one finds 
\begin{eqnarray}
I_{00;\;j1}&=&I_{j1;\;00}={\frac{\sqrt{2\eta \sigma }}{\pi }}\;{\frac{%
E_{j}^{3/4}}{(E_{0}+E_{j})^{2}}}\;\left( {\frac{\hbar ^{2}}{2\mu }}\right)
^{3/4}\nonumber \\
&\times&\sum_{n}\,{\frac{n^{2}!}{n^{n^{2}+1}}}\;\sqrt{\frac{(2n)^{2n^{2}+1}}{%
(2n^{2})!}}\;C_{jn}\,.  \label{BC92315}
\end{eqnarray}

\begin{figure}[ptb]
\begin{center}
\includegraphics[
height=1.6333in,
width=3.3688in
]{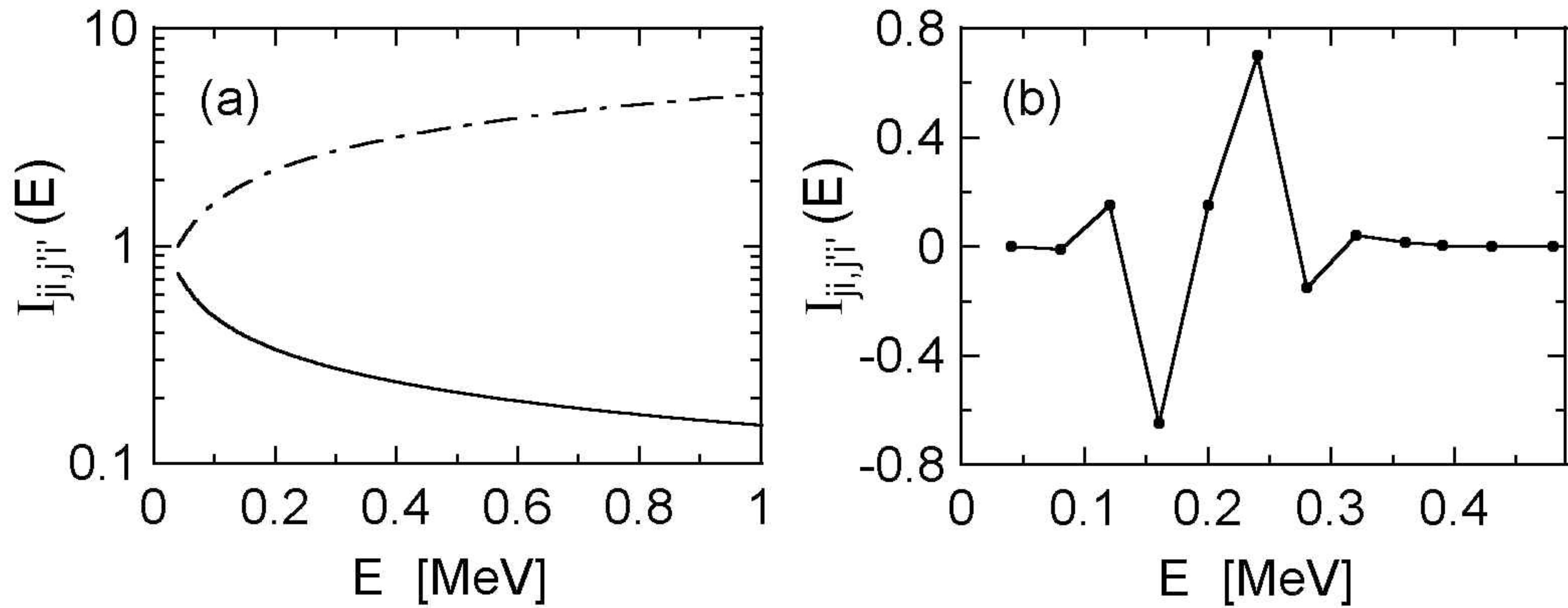}
\end{center}
\caption{\sl  (a) Radial matrix elements,
Eqs. \eqref{BC-9a} and \eqref{BC-9b} for the transition $j\rightarrow j+1$
(dashed-line), and for the $j\rightarrow j$ one (solid line). We used $\ell
=0$ and $\ell ^{\prime }=1$. (b) Radial matrix elements for the transition $%
j\rightarrow j^{\prime }$, keeping $E_{j}= const.$ and varying $%
E_{j^{\prime }}$ [BC92]. }%
\label{felipe2}%
\end{figure}

As we have seen above, the use of the plane wave basis \eqref{BC9233} results
in the elegant derivation of $I_{j\ell ;\;j^{\prime }\ell ^{\prime }}$
presented by Eqs. \eqref{BC92310} and \eqref{BC92311}. Nonetheless, the s-wave ($%
\ell =0$) state of Eq. \eqref{BC9233} is not orthogonal to the bound-state
wave function. To restore orthogonality one has to add an extra piece to
this function. One expects however that this approximation does not affect
the results appreciably since to access this state one needs at least two
transitions: the $0\rightarrow j1$ followed by the $j^{\prime }1\rightarrow
j^{\prime }0$ one. But the later transition competes with the transition to
the ground state, $j1\rightarrow 00$, which is the dominant one. A more
severe restriction is the use of plane waves to describe the continuum. A
realistic calculation would have to use outgoing waves for $u_{\ell
,\,E}^{(+)}(r)$ which would carry information about the final state
interactions of the ($b+c$)-system.

The break-up probability per unit energy interval, $P_{(E)}^{BU}$, is given
by 
\begin{equation}
P_{(E)}^{BU}=\sum_{ij}\Gamma _{i}(E)\;\Gamma _{j}(E)\;Q_{ij}  \label{BC9214a}
\end{equation}
where 
\begin{equation}
Q_{ij}=\mathcal{R}e\;\left[ \sum_{\ell m}a_{i\ell m}^{\ast }\,a_{j\ell m}%
\right] \,.  \label{BC9241b}
\end{equation}

The validity of the dipole approximation for the interaction potential \eqref
{BC-9a} to calculate the continuum-continuum coupling can only be justified
for $qr\ll 1$. But, as shown in Fig. \eqref{felipe3}, the discretized
wavefunctions extend up to $400\;$\textrm{fm}. Thus, unless the matrix
elements for the continuum-continuum coupling, Eq. \eqref{BC-9b}, have its
main contribution from $r\ll 20$\textrm{fm}, the dipole approximation is not
valid. The $jj$-coupling do satisfy this requirement. In this case the wave
functions have equal energies, but different angular momenta. This causes an
asymptotically ($r\gg 1/q$) constant phase difference between the wave
functions entering in $I_{jl;jl^{\prime }}$. This leads to cancellations in
the integrand of Eq. \eqref{BC-9b} for large $r$. The situation is different
for the ($j,j^{\prime }\neq j$)-coupling. In this case the integrand has
contributions from larger values of $r$ and these contributions increase
with the energy. With a correct treatment of the multipole expansion of the
interaction potential  the integrals $I_{j\ell ;j^{\prime }\neq
j,\ell ^{\prime }}$ would decrease with $E$. We expect that the transitions
between $00\rightarrow j^{\prime },\ell =1$ and $j^{\prime },\ell
=1\rightarrow 00$ dominate the excitation process, so that the matrix
elements between states with $j\neq j^{\prime }\neq 0$ do not play an
important role. Also, to minimize the consequence of the breaking down the
dipole approximation in the continuum-continuum coupling at $j\neq j^{\prime
}$, one can use a large parameter $K$ (e.g., here $K=4$ was used). This
leads to small $I_{j\neq j^{\prime }}$.

\begin{figure}[ptb]
\begin{center}
\includegraphics[
height=1.6333in,
width=3.3688in
]{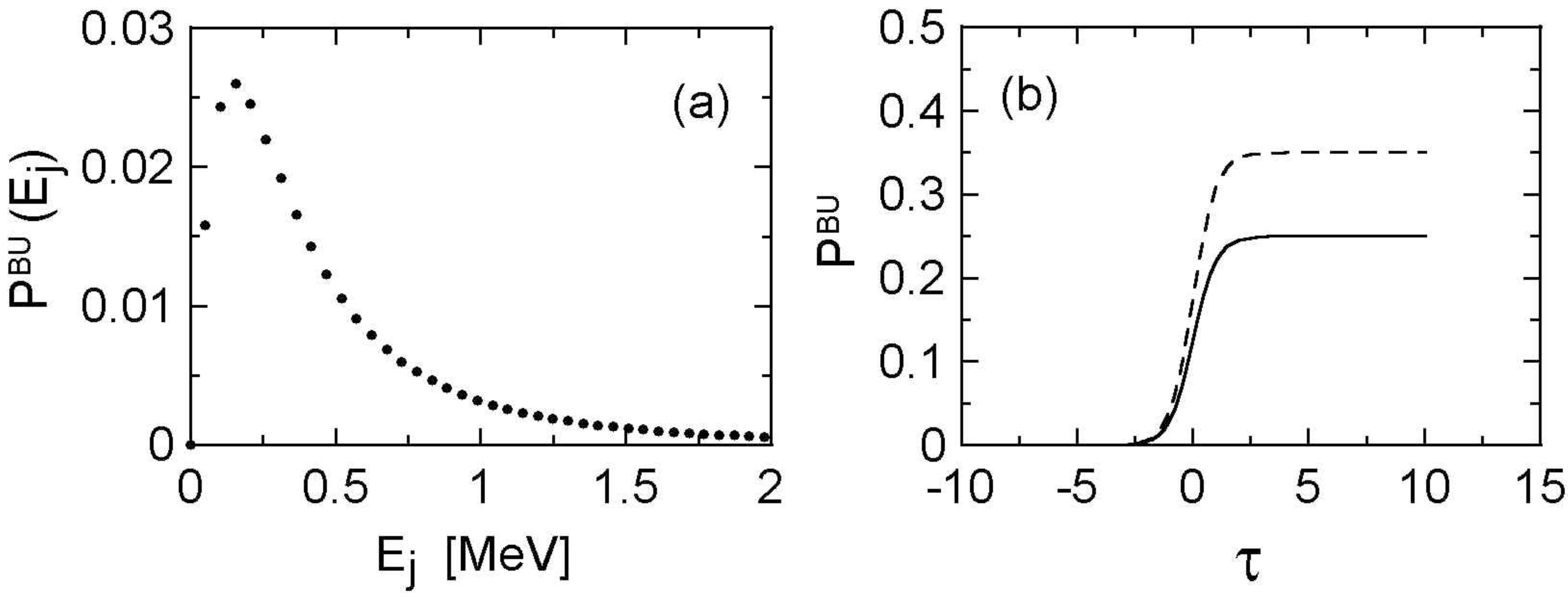}
\end{center}
\caption{\sl  (a) Coulomb break-up
probability, per unit energy interval (MeV$^{-1}$), of $^{11}$Li projectiles
incident on lead at 100 MeV/nucleon and $b=15\;$fm, as a function of the
final total kinetic energy of the fragments. (b) Breakup probability as a
function of the time, $\protect\tau =vt/c$ [BC92]. }%
\label{felipe4}%
\end{figure}

In Figure \ref{felipe4}(a) we show the break-up probability per unit energy
interval for the reaction $^{11}\mathrm{Li}+\mathrm{Pb}$ at $100$
MeV/nucleon and $b=15\;$fm, calculated from Eq. \eqref{BC9214a} by solving the
coupled-differential Eqs. \eqref{BC9235} for $a_{i\ell m}$. We see that the
energy distribution of the fragments is peaked at $E\sim 0.3\;$MeV.
Therefore, the most relevant momentum transfer to the $^{11}$Li nucleus
occurs at $q=\sqrt{2\mu _{bc}B}/\hbar \sim 20\;$\textrm{fm}$^{-1}$.

In Figure \ref{felipe4}(b) the solid line represents $P^{BU}$, the total
breakup probability [Eq. \eqref{BC9214a} integrated over energy], as a
function of the adimensional parameter $\tau =vt/b$, for $b=15\;$\textrm{fm}%
. This is obtained by solving the coupled-channels Eqs. \eqref{BC9235} for a
time $t$ and calculating the sum $P^{BU}(t)=\sum_{j\ell m}|a_{j\ell m}|^{2}$%
. The dashed-line corresponds to the neglect of all transitions, except for
the $0\rightarrow j\ell $ ones.  The solid-line includes all
possible transitions. The break-up probability occurs in a time scale of $%
\Delta t\sim b/v$. As $t\rightarrow \infty $ the break-up probability is $%
40\%$ smaller than that calculated by first order perturbation theory.

\subsubsection{Dynamical model in space-time lattice}

Another method for the treatment of higher-order effects in Coulomb breakup is the solution of the 
Schr\"odinger equation in a discretized space-time lattice [BB93,BB94,EBB95]. 
This should yield the same results as the CDCC model described in the previous section, if the model space is
large enough. By that we mean that the wavefunction expansion in the CDCC method, and the lattice discretization in the present model, 
are robust, convergent, and accurate. With present computer power, this can only be achieved with rather simplified structure models. 

One assumes a potential model for a cluster-like ($c+x$)
halo nucleus. For example, a Woods-Saxon potential may be used for the
nuclear interaction. To this potential a time-dependent
non-relativistic Coulomb interaction is added, 
\begin{equation}
V_{coul}=Z_{\tau }e^{2}[\frac{Z_{c}}{\mid \mathbf{r}_{c}-\mathbf{R}(t)\mid }+%
\frac{Z_{x}}{\mid \mathbf{r}_{x}-\mathbf{R}(t)\mid }-\frac{Z_{a}}{R(t)}]
\label{(3.8)}
\end{equation}
where $\;a=c+x$ \ is the projectile, $\mathbf{r}_{c}=-\mathbf{r}A_{x}/A_{a}$%
, and $\mathbf{r}_{c}=\mathbf{r}(1-A_{x}/A_{a}).$

The potential above can be expanded into multipoles. Normally, the dipole
part of the expansion is the most relevant one. The time dependent
wavefunction for the relative motion of $c+x$ is given by 
\begin{equation}
\Psi (\mathbf{r},t)=\frac{1}{r}\sum_{\ell m}u_{\ell m}(r,t)Y_{\ell m}(%
\widehat{\mathbf{r}}).  \tag{(3.9) }
\end{equation}

Inserting this expansion in the Schr\"{o}dinger equation and retaining only
the dipole expansion of the potential \eqref{(3.8)} we get 
\begin{eqnarray}
&&\left[ \frac{d^{2}}{dr^{2}}-\frac{\ell (\ell +1)}{r^{2}}-\frac{2\mu _{cx}}{%
\hbar }V_{N}(r)\right] \;u_{\ell m}(r,t)\nonumber \\ &+&\sum\limits_{\ell ^{\prime
}m^{\prime }}S_{\ell ^{\prime }m^{\prime }}^{(\ell m)}u_{\ell ^{\prime
}m^{\prime }}(r,t)=
-\frac{2\mu _{cx}}{\hbar }\frac{\partial u_{\ell m}}{%
\partial t}  \label{(3.10b)}
\end{eqnarray}
where 
\begin{eqnarray}
S_{\ell ^{\prime }m^{\prime }}^{(\ell m)}&=&-\frac{2\mu _{cx}}{\hbar }\frac{%
(-1)^{m}}{\sqrt{2}}\left[ b^{2}+v^{2}t^{2}\right] ^{-3/2}\nonumber \\
&\times& \sqrt{(2\ell
+1)(2\ell ^{\prime }+1)}\left( 
\begin{array}{lll}
\ell & 1 & \ell ^{\prime } \\ 
0 & 0 & 0
\end{array}
\right)  \notag \\
&\times &\Bigg\{ ib\left[ \left( 
\begin{array}{lll}
\ell & 1 & \ell ^{\prime } \\ 
-m & 1 & -m^{\prime }
\end{array}
\right) +\left( 
\begin{array}{lll}
\ell & 1 & \ell ^{\prime } \\ 
-m & -1 & -m^{\prime }
\end{array}
\right) \right] \nonumber \\
&+&\sqrt{2}vt\left( 
\begin{array}{lll}
\ell & 1 & \ell ^{\prime } \\ 
-m & 0 & -m^{\prime }
\end{array}
\right) \Bigg\} r.  \notag \\
&&  \label{(3.11c)}
\end{eqnarray}
This equation can be solved by a finite difference method. A truncation on
the $\ell ,m$ values is needed. Only a few angular momentum states are
needed. Denoting $\alpha \equiv (\ell ,m),$ the wave
function $u_{\alpha }$ at time $t+\Delta t$ is obtained from the wave
function at time $t$, according to the algorithm [BB93] 
\begin{eqnarray}
u_{\alpha }(t+\Delta t)&=&\left[ \frac{1}{i\tau }-\Delta ^{\left( 2\right) }+%
\frac{\Delta t}{2\hbar \tau }V_{\alpha }\right] ^{-1}\nonumber\\
&\times&\left[ \frac{1}{i\tau }%
+\Delta ^{(2)}-\frac{\Delta t}{2\hbar \tau }V_{\alpha }+\frac{\Delta t}{%
\hbar \tau }\;\widehat{S}\right] u_{\alpha }(t).  \nonumber\\
\label{(3.12)}
\end{eqnarray}
In this equation $\tau =\hbar \Delta t/4\mu _{bx}(\Delta r)^{2}$ and\ $%
\widehat{S}u_{\alpha }(t)=\sum\limits_{\alpha ^{\prime }}S_{\alpha ^{\prime
}}^{(\alpha )}u_{\alpha ^{\prime }}(t).$ Also, 
\begin{equation}
V_{\alpha }(r)=V_{N}(r)+\frac{\hbar ^{2}\ell (\ell +1)}{2\mu _{bc}r^{2}}.
\label{(3.13a)}
\end{equation}

The wave functions $u_{\alpha }(r,t)$ are discretized in a mesh in space,
with a mesh-size $\Delta r$. The second difference operator $\Delta ^{(2)}$
is defined as 
\begin{eqnarray}
&&\Delta ^{(2)}u_{\alpha }^{(j)}=u_{\alpha }^{(j+1)}(t)+u_{\alpha
}^{(j-1)}(t)-2u_{\alpha }^{(j)}(t),\nonumber \\
&&\text{with\ \ }u_{\alpha
}^{(j)}\equiv u_{\alpha }(r_{j},t).  \label{(3.13b)}
\end{eqnarray}

The 1$^{st}$ operation on the right side of Eq. \eqref{(3.12)} is trivial. The
second one needs special attention (see Appendix H).

The wave function calculated numerically at a very large time will not be
influenced by the Coulomb field. The numerical integration can be stopped
there. The continuum part of the wave function is extracted by means of the
relation (and normalized to unity) 
\begin{eqnarray}
\Psi _{c}(\mathbf{r},t)&=&\left[ \Psi -\Psi _{gs}<\Psi _{gs}\mid \Psi >\right]\nonumber\\
&\times&\left[ 1-\mid <\Psi _{gs}\mid \Psi >\mid ^{2}\right] ^{-1/2}
\label{(3.13)}
\end{eqnarray}
where $\Psi _{gs}$ is the initial wave function.

This wave function can be projected onto continuum eigenstates of the
potential $V_{N}(r)$ in order to obtain the excitation probability of that
state. For illustration lets us use the simplifying assumption that the
final states are plane waves. A projection onto these states is equivalent
to a Fourier transform of the time dependent wave function. One gets 
\begin{equation}
\Psi _{c}(\mathbf{p)}=\sum\limits_{\ell m}C_{\ell m}(p)Y_{\ell m}(\widehat{%
\mathbf{p}})  \label{(3.14)}
\end{equation}
where 
\begin{equation}
C_{\ell m}(p)=\sqrt{\frac{2}{\pi }}\;i^{\ell }\;\int dr\;r\;j_{\ell
}(pr)u_{\ell m}^{(c)}(r,t)
\end{equation}
where $j_{\ell }$ is the spherical Bessel function. The probability density
for an excitation to a final state with energy $E$ is 
\begin{equation}
P(b,E,\Omega )=\frac{1}{2}\left( \frac{2\mu bx}{\hbar ^{2}}\right) ^{3/2}%
\sqrt{E}\mid \Psi _{c}(\widehat{\mathbf{p}})\mid ^{2}.  \label{(3.16)}
\end{equation}
Integrating over $\Omega $ : 
\begin{equation}
P(b,E)=\frac{1}{2}\left( \frac{2\mu _{bx}}{\hbar ^{2}}\right) ^{3/2}\sqrt{E}%
\sum\limits_{\ell m}\mid C_{\ell m}(p)\mid ^{2}.  \label{(3.17)}
\end{equation}

In first-order perturbation theory this spectrum would be given by 
\begin{eqnarray}
P^{(1)}(b,E)&=&\frac{4\pi }{9}\left( \frac{2Z_{T}e^{2}}{\hbar v}\right)
^{2}\left( \frac{E}{\hbar v}\right) ^{2}\;\;\frac{dB(E1)}{dE}\nonumber \\
&\times& \left[
K_{0}^{2}(x)+K_{1}^{2}(x)\right]  \label{(3.18)}
\end{eqnarray}
where the $K^{\prime }s$ are the modified Bessel functions and $\;x=Eb/\hbar
v.$ In this case $dB(E1)/dE$ is calculated by using the ground state and
continuum states of the same potential  for $c+x$. 

As shown in figure \ref{reacc},  the first-order perturbation theory
does not work well
for the break-up of $^{11}$Li at 28 MeV/nucleon. 

\begin{figure}[ptb]
\begin{center}
\includegraphics[
height=1.6333in,
width=3.4688in
]{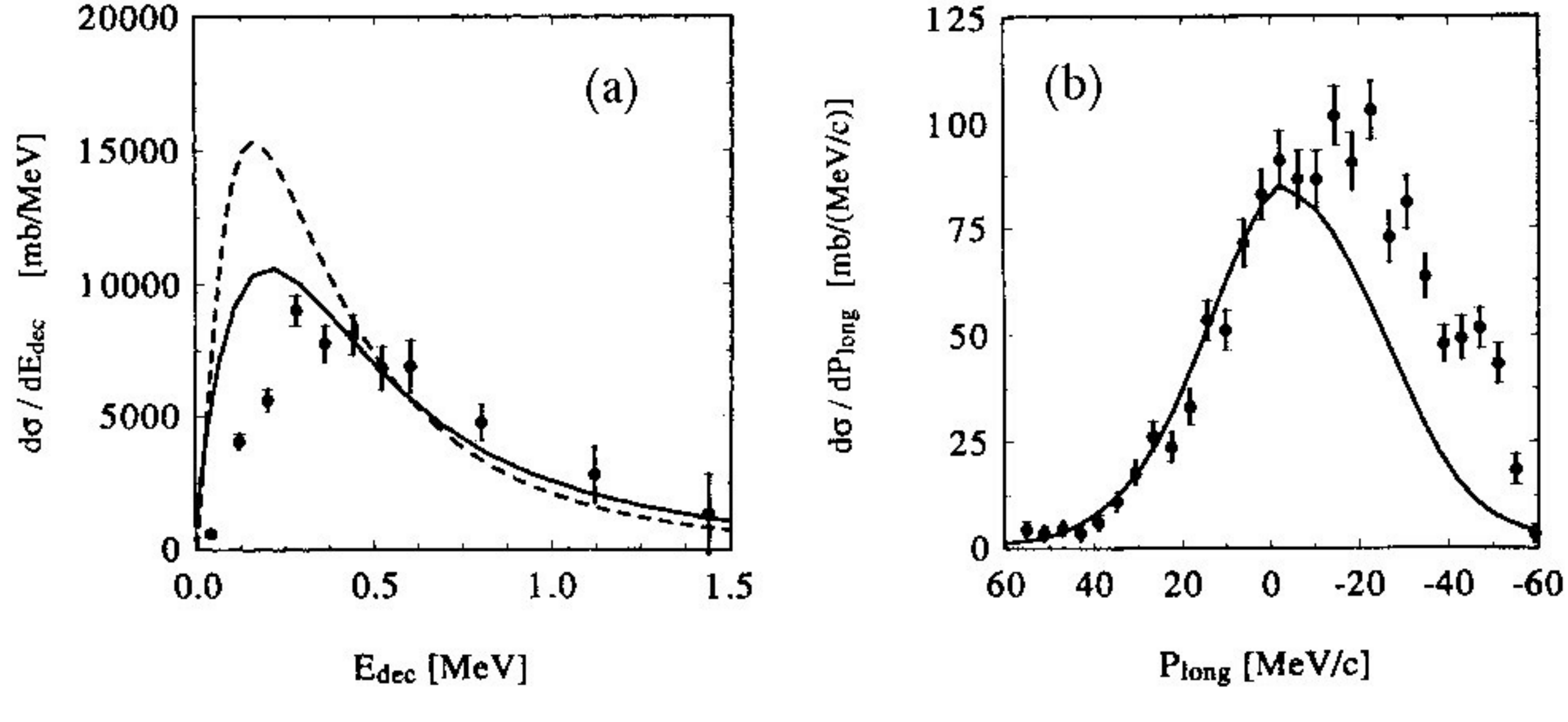}
\end{center}
\caption{\sl  (a) Energy spectrum for the
breakup of $^{11}$Li\ at 28 MeV/nucleon. Data points are from 
[Iek93]. Solid (dashed) line includes (does not include) the reacceleration
effect [EBB95]. (b) Longitudinal relative momentum of the
neutrons and the $^{9}$Li\ fragments in the breakup of $^{11}$Li\
projectiles at 28 MeV/nucleon. Data are from Ref. [Iek93]. Solid
curve is a calculation from Ref. [EBB95]. }%
\label{reacc}%
\end{figure}

The reacceleration effect is very important in this case due to the very
large break-up probabilities at small impact parameters. This induces higher
order dynamical continuum-continuum interactions which distort the spectrum
appreciably. One also observes that although the dynamical corrections
modify the spectrum in the right direction the discrepancy with the
experiment is appreciable. This might be a deficiency of the cluster model
and has to be studied more carefully. Also, the spectrum of the relative
longitudinal momentum of the fragments is not in a good agreement with the
dynamical calculations of Ref. [EBB95], at least for the higher part
of the spectrum. This is shown in the figure \ref{reacc}(b).

Higher-order effects are manifest
in published theoretical calculations, and seem to have some experimental support.
However, in most cases the effect is not relevant. There are also some
theoretical results which does not show  post-acceleration
effects. 

\section{Radiative capture in stars}

If you have read to this point, you are now convinced that there must be numerous situations where
Coulomb excitation and dissociation can be used for accessing precious information on nuclear structure. Obviously, any nucleus is part of a process in stars (perhaps the superheavies are not). Nuclear physics is therefore ultimately linked to astrophysics. We also expect that Coulomb dissociation plays a role in stars. In fact, it also plays a role in Cosmology. The  Greisen-Zatsepin-Kuzmin (GZK) effect, for example, is nothing more that Coulomb excitation by  (3 K) real photons  remnant from the Big Bang and by virtual photons in intergalactic magnetic fields [GZK66].

As we have seen before, Coulomb excitation probes the same matrix elements as real photons. It is thus natural that many astrophysical reactions in stars involving photons in the initial or final channels can be studied on earth via Coulomb  excitation. In particular, Coulomb dissociation is directly related to radiative capture in stars. In the next sections we discuss the {\it Coulomb dissociation method}, its experimental applications to some astrophysical problems and some of the newest theoretical developments and difficulties associated to it.

\subsection{The Coulomb dissociation method}\label{sec:CD}

Coulomb dissociation is a process  analogous to what happened to the comet
Shoemaker-Levy as it disintegrated during its approximation  
to Jupiter in 1994 (see Figure \ref{comet}).  Approximately 1.5 to 2.2 hours
after closest approach, the comet (which was presumably a  
single body at the time) was broken apart by tidal forces into at least 21
pieces. The pieces continued to orbit Jupiter with a period of approximately 2
years. Due to gravitational forces from the Sun, which changed the orbits
slightly on the next approach to Jupiter, the pieces impacted the planet. 
Much stronger tidal forces occur when nuclei come close to each other due to
their mutual electromagnetic field. This leads to their dissociation,  
especially for weakly-bound nuclei.

\begin{figure}[t]
\begin{center}
{\includegraphics[width=6cm]{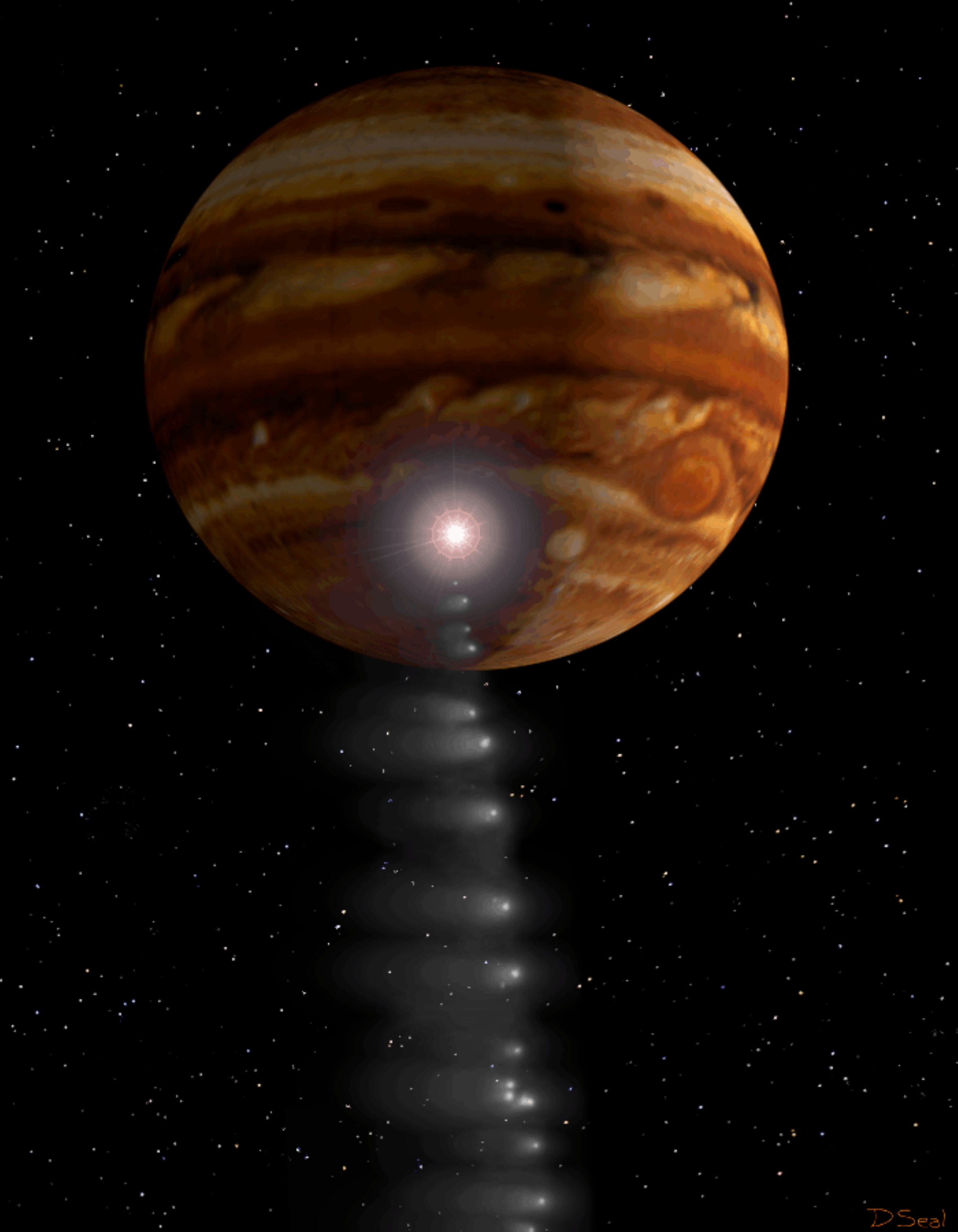}}
\end{center}
\caption{\sl  \label{comet}
Jupiter and comet Shoemaker-Levy 9, as imaged by the
Hubble Space Telescope (HST), on May 18, 1994, when the giant planet was at a
distance of 420 million miles (670 million km) from Earth. The gravitational interaction
of Jupiter with the comet has broken it up into many pieces (picture from
NASA). Much stronger tidal forces 
occur when nuclei come close to each other due to their mutual electromagnetic
fields. }
\end{figure}

The idea behind the Coulomb dissociation method is relatively simple. The
(differential, or angle-integrated) Coulomb breakup cross 
section for $a+A\rightarrow b+c+A$ follows from 
\begin{equation}
{d\sigma_{C}^{\pi L }(E_\gamma)\over
d\Omega}=N^{\pi L}(E_\gamma;\theta;\phi)\ .\ \sigma_{\gamma+a\
\rightarrow\ b+c}^{\pi L}(E_\gamma),\label{CDmeth}
\end{equation}
where $E_\gamma$ is the energy transferred from the relative motion to
the breakup, and $\sigma_{\gamma+a\ \rightarrow\
b+c}^{\pi L}(E_\gamma)$ is the photo-dissociation cross section for
the multipolarity ${\pi L}$ and photon energy $E_\gamma$.  Time reversal
allows one to deduce the radiative
capture cross section $b+c\rightarrow a+\gamma$ from $\sigma_{\gamma+a\ \rightarrow\ b+c}%
^{\pi L}(E_\gamma)$, i.e.,
\begin{equation}
\sigma_{b+c \rightarrow\ \gamma+a}={2(2j_a+1)\over (2j_b+1)(2j_c+1)} {k_\gamma^2\over k^2}\sigma_{\gamma+a\ \rightarrow\ b+c},
\end{equation}
where $k_\gamma=E_\gamma/\hbar c$ is the  photon wavenumber, and $k=\sqrt{2\mu(E_\gamma-B)}/\hbar$ is  the wavenumber
for the relative motion of b+c. Except for the extreme case very close to the threshold ($k \rightarrow 0$), we have
$k_\gamma \ll k$, so that the phase space favors the photodisintegration cross section as
compared to the radiative capture. Direct measurements of the photodisintegration
near the break-up threshold do hardly provide experimental advantages
and seem presently impracticable. On the other hand the copious source
of virtual photons  acting on a fast charged nuclear projectile when passing the
Coulomb field of a (large Z) nucleus offers a more promising way to study the
photodisintegration process as Coulomb dissociation. 

 This method was introduced in Ref. [BBR86] and has
been tested 
successfully in a number of reactions of interest to astrophysics.
The most celebrated case is the reaction
$^{7}$Be$(p,\gamma)^{8}$B, first studied in Ref. [Mot94],
followed by 
numerous experiments in the last decade. This reaction is important because it produces $^8$B in the core of
our sun. These nuclei decay by emitting a high energy neutrinos which are one of the best probes of the sun's interior.  The measurement
of such neutrinos is very useful to test our theoretical solar models.

\begin{figure}[t]
\begin{center}
{\includegraphics[width=8cm]{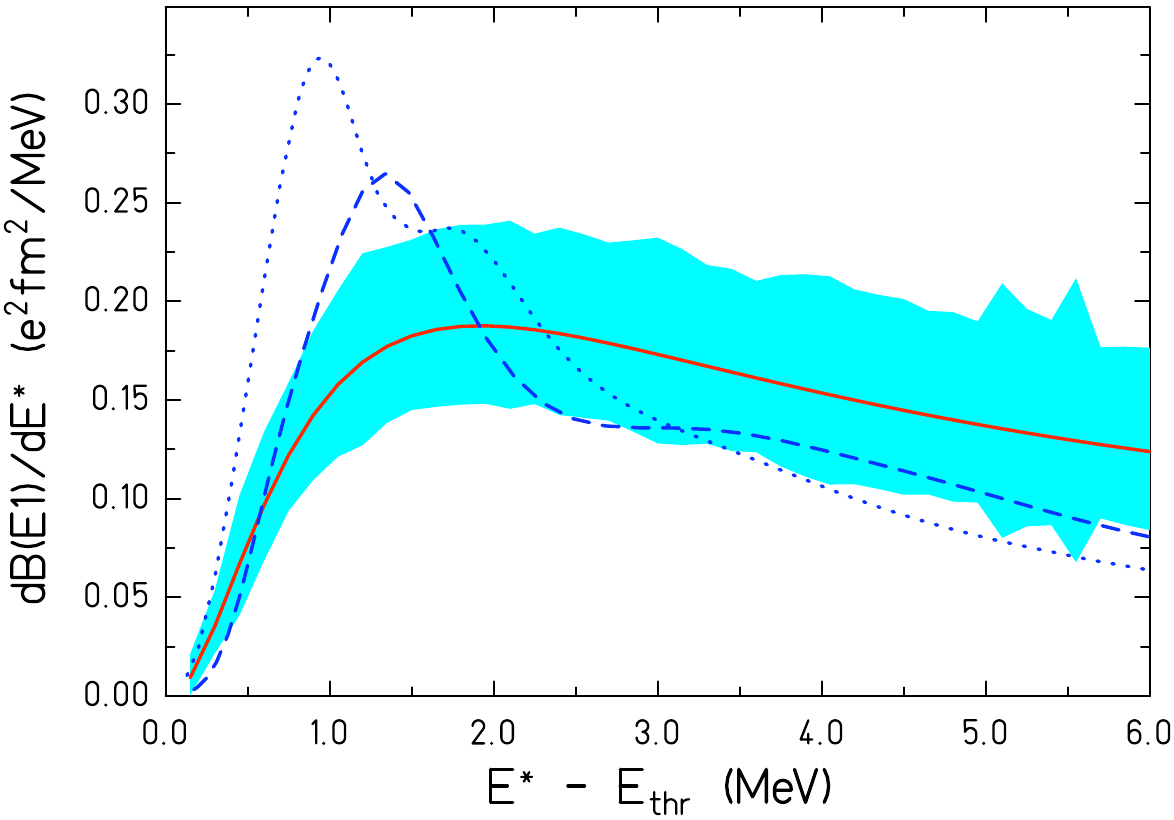}}
\end{center}
\vspace{-0.5cm}
\caption{\sl  
Electric dipole response function for $^6$He. The shaded area
represents the experimental results from a Coulomb dissociation experiment
[Aum99]. The dashed and dotted lines correspond to results from  
three-body decay models from Refs. [Da98,Cob97].  (Courtesy of
T. Aumann).}  
\label{aumannfig1}
\end{figure}

Another example is the two-neutron capture on $^4$He could perhaps play a role in the post-collapse
phase in type-II supernovae. The bottleneck in this nucleosynthesis scenario
is the formation of nuclei with $A \ge 9$ from nucleons and
$\alpha$-particles. In principle, the reaction $^4$He$(2n,\gamma)^6$He could  be
relevant in bridging the instability gap at $A=5$, although it is believed that this
reaction cannot compete with the ($\alpha n,\gamma$) process in a type-II
supernova scenario. Experiments with Coulomb dissociation have been used to
study this question, as shown in the example presented in Figure
\ref{aumannfig1}. The  figure  displays the  electric dipole
response function for $^6$He. The shaded areas represent the experimental
results from a Coulomb dissociation experiment [Aum99]. The dashed and
dotted lines correspond to results from  
three-body decay models from Refs. [Dan98,Cob97]. 

In  figure  \ref{aumannfig2}
we show the measurement of two-body correlations in the three-body decay
of $^6$He. The lower panels display the ratio between the measured $\alpha$-n and
$n$-$n$ relative-energy spectra (upper panels) and the spectra simulated
(histograms) according to standard phasespace distributions
[Aum99]. From the analysis of this experiment it was found that 10\% of
the dissociation cross section proceeds via the formation of $^5$He. A rough
estimate yields 1.6 mb MeV for the photoabsorption cross section for
$^6$He$(\gamma,n)^5$He, which agrees with theoretical calculations
[Efr96]. From this experiment one concludes that the cross sections for
formation of $^5$He and $^6$He via one (two) neutron capture by $^4$He are not
large enough to compete with the  ($\alpha$n, $\gamma$) capture process (for
more details, see Ref. [Aum06]). Nonetheless, this and the previously
mentioned examples, show the relevance of the Coulomb dissociation method to
assess some of the basic questions of relevance for nuclear astrophysics.  

\begin{figure}[t]
\begin{center}
{\includegraphics[width=8.5cm]{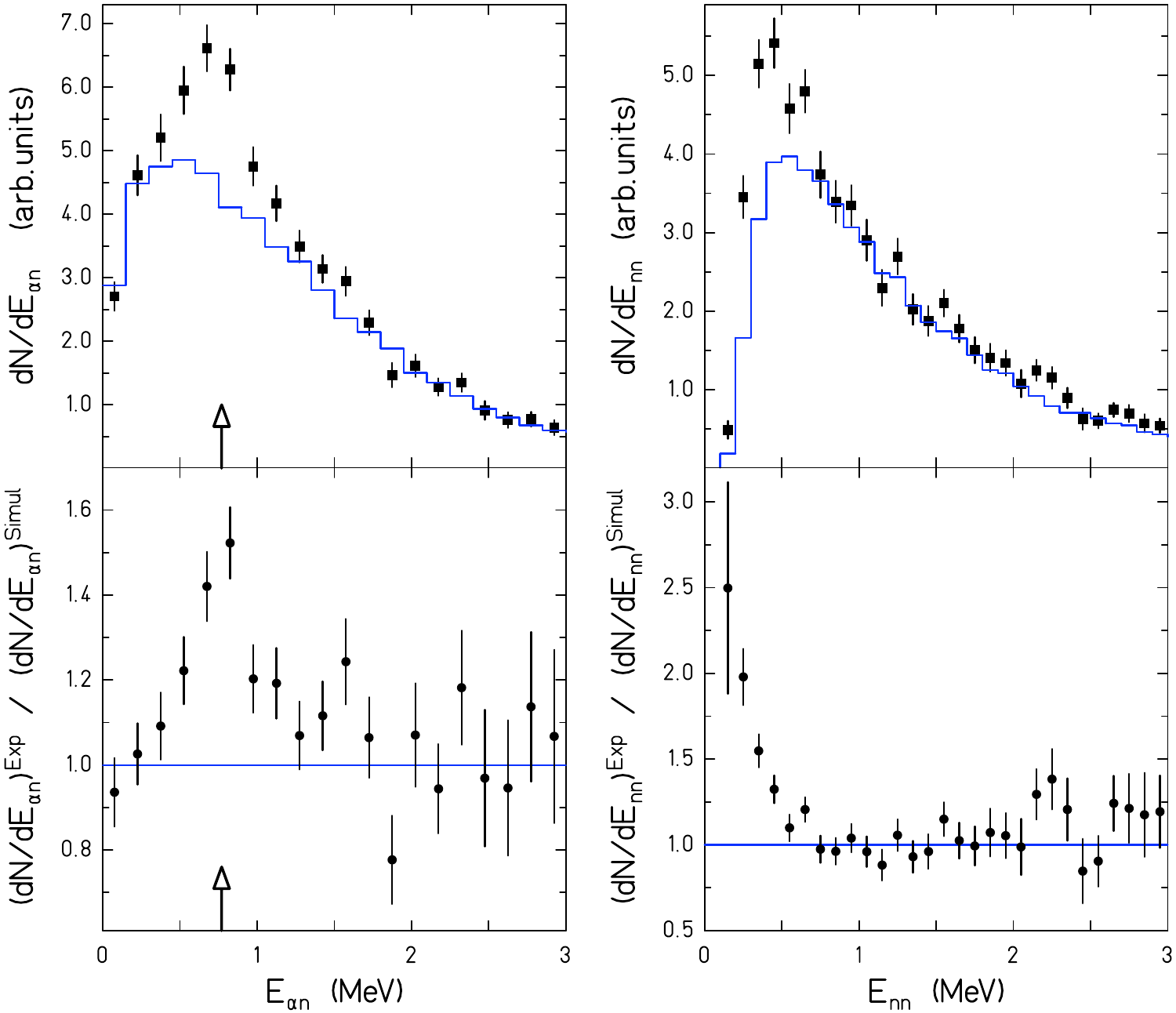}}
\end{center}
\vspace{-0.5cm}
\caption{\sl  
Measurement of two-body correlations in the three-body decay of $^6$He. The
lower panels show the ratio between the measured $\alpha-n$ and $n$-$n$
relative-energy spectra (upper panels) and the spectra simulated (histograms)
according to standard phasespace distributions [Aum99]. (Courtesy of
T. Aumann).}  
\label{aumannfig2}
\end{figure}

\subsection{Higher-order corrections}

The Coulomb dissociation method is specially useful if first-order perturbation theory is valid.
If not, one can still extract the electromagnetic matrix elements involved in radiative capture reactions.
But a much more careful analysis of the high-order effects need to be done (see figure \ref{figcdcc}. We will discuss briefly a continuum-discretized
coupled-channels calculation, appropriate for relativistic collision energies.

Coupled-channels calculations are a complication that we could  live better without. It helps some people to get jobs, but they do not
make life easier, as they have very often unphysical (or better said, unpredictable) results. The reason for the uncertain behavior is that, in each interaction step, plus and minus signs of matrix elements accumulate in one or the other direction (like a brownian motion). In seminars you might ask why a coupled-channels calculation reduced, or increased, the cross section. The plain answer is invariably: ``it is a coupled-channels effect".  Well, who said that life is easy?

\subsubsection{Continuum-Discretized Coupled-Channels}

The continuum-discretized coupled-channels method (CDCC)
[Kam87] is one of the most accurate models to describe
the breakup of halo nuclei taking account of higher-order couplings
explicitly. The eikonal CDCC method (E-CDCC) [Oga03,Oga06], is a derivation of CDCC
that enables one to efficiently treat the nuclear and Coulomb
breakup reactions at $E_{lab} \ge 50$ MeV/nucleon. An essential
prescription described in Refs.~ [Oga03,Oga06] is the
construction of hybrid (quantum and eikonal) scattering amplitudes,
with which one can make quantum-mechanical (QM) corrections to the
pure eikonal wavefunctions with a minimum task. These corrections
are, however, expected to become less important as the incident
energy increases.

First I will describe a simplified CDCC method presented in Ref. {Ber05] where the 
inclusion of relativistic continuum-continuum effects in CDCC calculations were introduced.
Let us consider the Klein-Gordon (KG) equation with a potential $V_{0}$ which
transforms as the time-like component of a four-vector [AC79]. For a
system with total energy $E$ (including the rest mass $M$), the KG equation
can be cast into the form of a Schr\"{o}dinger equation (with $\hbar=c=1$),
\begin{equation}\left(  \nabla^{2}+k^{2}-U\right)  \Psi=0,\end{equation} where $k^{2}=\left(  E^{2}%
-M^{2}\right)  $ and $U=2V_{0}(2E-V_{0})$. When $V_{0}\ll M$, and $E\simeq M$,
one gets $U=2MV_{0}$, as in the non-relativistic case. The condition $V_{0}\ll
M$ is met in peripheral collisions between nuclei at all collision energies.
Thus, one can always write $U=2EV_{0}$. A further simplification is to assume
that the center of mass motion of the incoming projectile and outgoing
fragments is only weakly modulated by the potential $V_{0}$. 

To get the
dynamical equations, one discretizes the wavefunction in terms of the
longitudinal center-of-mass momentum $k_{z}$, using the ansatz%
\begin{equation}
\Psi=\sum_{\alpha}\mathcal{S}_{\alpha}\left(  z,\mathbf{b}\right)
\ \exp\left(  ik_{\alpha}z\right)  \ \phi_{k_{\alpha}}\left(
\mathbf{\mbox{\boldmath$\xi$}}\right). \label{eq1}%
\end{equation}
In this equation, $\left(  z,\mathbf{b}\right)  $ is the projectile's
center-of-mass coordinate, with \textbf{b} equal to the impact parameter.
$\ \phi\left(  \mathbf{\mbox{\boldmath$\xi$}}\right)  $ is the projectile
intrinsic wavefunction and $\left(  k,\mathbf{K}\right)  $ is the projectile's
center-of mass momentum with longitudinal momentum $k$\ and transverse
momentum $\mathbf{K}$\textbf{.} There are hidden, uncomfortable, assumptions
in Eq.  \eqref{eq1}. The center of mass of a relativistic system of interacting
particles is not a well defined quantity. Also, the separation between the
center of mass and intrinsic coordinates is not permissible under strict
relativistic treatments. For high energy collisions we can at best justify Eq. 
\eqref{eq1} for the scattering of light projectiles on heavy targets. Eq.
\eqref{eq1} is only reasonable if the projectile and target closely maintain
their integrity during the collision, as in the case of very peripheral collisions.

\begin{figure}[t]
\begin{center}
{\includegraphics[width=7.5cm]{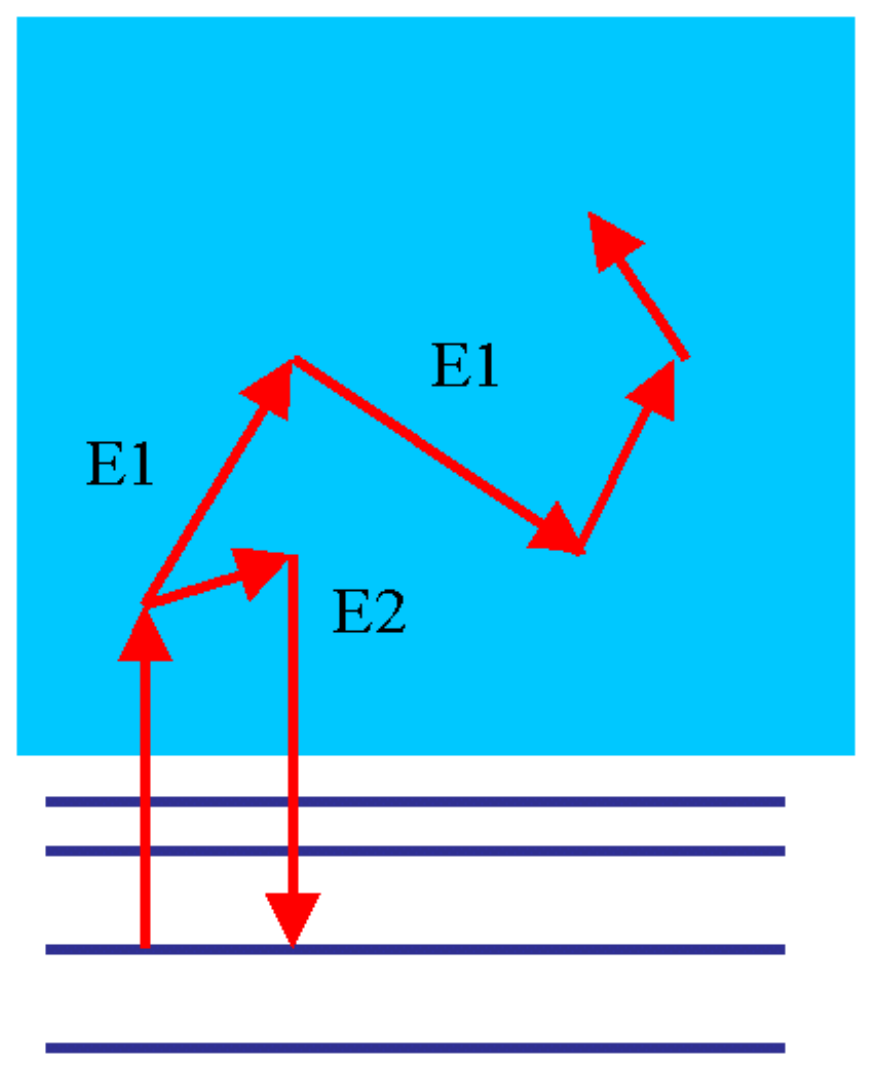}}
\end{center}
\vspace{-0.5cm}
\caption{\sl  
Schematic description of a multi-step excitation, with excursions in the continuum.}  
\label{figcdcc}
\end{figure}

Neglecting the internal structure means $\phi_{k_{\alpha}}\left(
\mathbf{\mbox{\boldmath$\xi$}}\right)  =1$ and the sum in Eq.  \eqref{eq1}
reduces to a single term with $\alpha=0$, the projectile remaining in its
ground-state. It is straightforward to show that inserting Eq.  \eqref{eq1} in
the KG equation \begin{equation}\left(  \nabla^{2}+k^{2}-2EV_{0}\right)  \Psi=0,\end{equation} and
neglecting $\nabla^{2}\mathcal{S}_{0}\left(  z,\mathbf{b}\right)  $ relative
to $ik\partial_{Z}\mathcal{S}_{0}\left(  z,\mathbf{b}\right)  $ one gets
$ik\partial_{Z}\mathcal{S}_{0}\left(  z,\mathbf{b}\right)  =EV_{0}%
\mathcal{S}_{0}\left(  z,\mathbf{b}\right)  $, which leads to the center of
mass scattering solution \begin{equation}\mathcal{S}_{0}\left(  z,\mathbf{b}\right)
=\exp\left[  -iv^{-1}\int_{-\infty}^{z}dz^{\prime}\ V_{0}\left(  z^{\prime
},\mathbf{b}\right)  \right]  ,\end{equation} with $v=k/E$. Using this result in the
Lippmann-Schwinger equation, one gets the familiar result for the eikonal
elastic scattering amplitude, i.e. \[ f_{0}=-i { k\over 2\pi}  \int
d\mathbf{b}\ \exp\left(  i\mathbf{Q\cdot b}\right)  \ \left\{  \exp\left[
i\chi(\mathbf{b})\right]  -1\right\}, \] where the eikonal phase is given by
$\chi(\mathbf{b})=\mathcal{S}_{0}\left(  \infty,\mathbf{b}\right)  $, and
$\mathbf{Q}=\mathbf{K}^{\prime}-\mathbf{K}$ is the transverse momentum
transfer. \ Therefore, the elastic scattering amplitude in the eikonal
approximation has the same form as that derived from the Schr\"odinger
equation in the non-relativistic case.

For inelastic collisions we insert Eq.  \eqref{eq1} in the KG equation and use
the orthogonality of the intrinsic wavefunctions $\phi_{k_{\alpha}}\left(
\mathbf{\mbox{\boldmath$\xi$}}\right)  $. This leads to a set of
coupled-channels equations for $\mathcal{S}_{\alpha}$:%
\begin{equation}
\left(  \nabla^{2}+k^{2}\right)  \mathcal{S}_{\alpha}\mathrm{e}^{ik_{\alpha}
z}=\sum_{\alpha}\left\langle \alpha\left\vert U\right\vert \alpha^{\prime
}\right\rangle \ \mathcal{S}_{\alpha^{\prime}}\ \mathrm{e}^{i k_{\alpha
^{\prime}} z}, \label{eq2}%
\end{equation}
\bigskip with the notation $\left\vert \alpha\right\rangle =\left\vert
\phi_{k_{\alpha}}\right\rangle $. Neglecting terms of the form $\nabla
^{2}\mathcal{S}_{\alpha}\left(  z,\mathbf{b}\right)  $ relative to
$ik\partial_{Z}\mathcal{S}_{\alpha}\left(  z,\mathbf{b}\right)  $, Eq. 
\eqref{eq2} reduces to%
\begin{equation}
iv\frac{\partial\mathcal{S}_{\alpha}\left(  z,\mathbf{b}\right)  }{\partial
z}=\sum_{\alpha^{\prime}}\left\langle \alpha\left\vert V_{0}\right\vert
\alpha^{\prime}\right\rangle \ \mathcal{S}_{\alpha^{\prime}}\left(
z,\mathbf{b}\right)  \ \mathrm{e}^{i\left(  k_{\alpha^{\prime}}-k_{\alpha
}\right)  z}. \label{eq3}%
\end{equation}
The scattering amplitude for the transition $0\rightarrow\alpha$ is given by%
\begin{equation}
f_{\alpha}\left(  \mathbf{Q}\right)  =-\frac{ik}{2\pi}\int d\mathbf{b}%
\ \ \exp\left(  i\mathbf{Q\cdot b}\right)  \ \left[  S_{\alpha}\left(
\mathbf{b}\right)  -\delta_{\alpha,0}\right]  , \label{eq4}%
\end{equation}
with $S_{\alpha}\left(  \mathbf{b}\right)  =\mathcal{S}_{\alpha}\left(
z=\infty,\mathbf{b}\right)  $. The set of equations \eqref{eq3} and \eqref{eq4}
are the eikonal-CDCC equations (E-CDCC). They are much simpler to solve
than the low-energy CDCC equations because the $z$ and $b$ coordinates
decouple and only the evolution on the $z$ coordinate needs to be treated non-perturbatively.

The E-CDCC equations can be used to study the dissociation of $^{8}$B
projectiles at high energies. The energies transferred to the projectile are
small, so that the wavefunctions can be treated non-relativistically in the
projectile frame of reference. In this frame the wavefunctions will be
described in spherical coordinates, i.e. $\left\vert \alpha\right\rangle
=\left\vert jlJM\right\rangle, $ where $j$, $l$, $J$ and $M$ denote the
angular momentum numbers characterizing the projectile state. It is important
to notice that the matrix element $\left\langle \alpha\left\vert
V_{S}\right\vert \alpha^{\prime}\right\rangle $ is Lorentz invariant. Boosting
a volume element from the projectile to the laboratory frame means $d^{3}%
\xi\rightarrow d^{3}\xi/\gamma$, where $\gamma=\left(  1-v^{2}\right)
^{-1/2}$ is the Lorentz contraction factor. The intrinsic projectile
wavefunction is a scalar and transforms according to $\phi_{\alpha}\left(
\xi_{x},\xi_{y},\xi_{z}\right)  \rightarrow\phi_{\alpha}\left(  \xi_{x}%
,\xi_{y},\gamma\xi_{z}\right)  $, while $V_{0}$, being the time-like component
of a four-vector, transforms as $V_{0}\left(  b,z;\xi_{x},\xi_{y},\xi
_{z}\right)  \rightarrow\gamma V_{0}\left(  b,z;\xi_{x},\xi_{y},\gamma\xi
_{z}\right)  $. Thus, redefining the integration variable $z$ in the
laboratory as $\mathbf{\xi}_{z}^{\prime}=\gamma\mathbf{\xi}_{z}$ leads to the
afore mentioned invariance. We can therefore calculate $\left\langle
\alpha\left\vert V_{0}\right\vert \alpha^{\prime}\right\rangle $\ in the
projectile frame.

The longitudinal wavenumber $k_{\alpha}\simeq(E^{2}-M^{2})^{1/2}$\ also
defines how much energy is gone into projectile excitation, since for small
energy and momentum transfers $k_{\alpha}^{\prime}-k_{\alpha}=\left(
E_{\alpha}^{\prime}-E_{\alpha}\right)  /v$. In this limit, eqs. \eqref{eq3} and
\eqref{eq4} reduce to semiclassical coupled-channels equations, if one uses
$z=vt$ for a projectile moving along a straight-line classical trajectory, and
changing to the notation $\mathcal{S}_{\alpha}\left(  z,b\right)  =a_{\alpha
}(t,b)$, where $a_{\alpha}(t,b)$ is the time-dependent excitation amplitude
for a collision wit impact parameter $b$.
With these changes, equation \eqref{eq4} also agrees with the equivalent
semiclassical equation for the inelastic scattering amplitude. Therefore, for non-relativistic collisions and small momentum
and energy transfers, the above derivation reduces to previous methods used in
the literature.

If the state $\left\vert \alpha\right\rangle $\ is in the continuum (positive
proton+$^{7}$Be energy) the wavefunction is discretized according to
\begin{equation}\left\vert \alpha;E_{\alpha}\right\rangle =\int dE_{\alpha}^{\prime}%
\ \Gamma(E_{\alpha}^{\prime})\ \left\vert \alpha;E_{\alpha}^{\prime
}\right\rangle ,\end{equation} where the functions $\Gamma(E_{\alpha})$ are assumed to be
strongly peaked around the energy $E_{\alpha}$ with width $\Delta E$. For
convenience the histogram set is chosen. The
inelastic cross section is obtained by solving the E-CDCC equations and using
\begin{equation}{d\sigma\over d\Omega dE_{\alpha}}=\left\vert f_{\alpha}\left(  \mathbf{Q}\right)
\right\vert ^{2}\ \Gamma^{2}(E_{\alpha}).\end{equation}

The potential $V_{0}$ contains contributions from the nuclear and the Coulomb
interaction. The nuclear potentials are constructed along traditional lines of
non-relativistic theory.  The  potentials are also
expanded into  multipolarities. These potentials are then transformed
as the time-like component of a four-vector, as described above. The multipole expansion of the Coulomb interaction in the
projectile frame including retardation. The first term (monopole) of the expansion
is the retarded Lienard-Wiechert potential which does not contribute to the
excitation, but to the center of mass scattering. Due to its long range, it is
hopeless to solve Eq.  \eqref{eq3} with the Coulomb monopole potential, as
$\mathcal{S}_{\alpha}\left(  z,\mathbf{b}\right)  $\ will always diverge. This
can be rectified by using the regularization scheme \begin{equation}S_{\alpha}\left(
\mathbf{b}\right)  \rightarrow\exp\left[  i\left(  2\eta\ln\left(  kb\right)
+\chi_{a}\left(  b\right)  \right)  \right]  \ S_{\alpha}\left(
\mathbf{b}\right), \end{equation} where $\eta=Z_{P}Z_{T}/\hbar v$ and the $S_{\alpha
}\left(  \mathbf{b}\right)  $\ on the right-hand side is now calculated
without inclusion of the Coulomb monopole potential in Eq.  \eqref{eq3}. The
purely imaginary absorption phase, $\chi_{a}\left(  b\right)  $, was
introduced to account for absorption at small impact parameters.  

The multipole-expansion of the relativistic Coulomb potential
between the target nucleus (T) with the atomic number $Z_{\rm T}$
and the projectile (P), consisting of C and v clusters, are given in
Ref.~[Ber05]:
\begin{equation}
V_{{\rm E1}\mu}=\sqrt{\frac{2\pi}{3}}\xi Y_{1\mu}\left(  \mathbf{\hat
{\mbox{\boldmath$\xi$}}}\right)  \frac{\gamma Z_{\rm T}ee_{\rm E1}}{\left(
b^{2}+\gamma ^{2}z^{2}\right)^{3/2}}\left\{
\begin{array}
[c]{c}%
\mp b\ \ (\mathrm{if}\ \ \ \mu=\pm1)\\
\sqrt{2}z\ \ (\mathrm{if}\ \ \ \mu=0)\
\end{array}
\right.  \label{eq6}%
\end{equation}
for the E1 (electric dipole) field and%
\begin{align}
V_{{\rm E2}\mu}= &  \sqrt{\frac{3\pi}{10}}\xi^{2}Y_{2\mu}\left(
\mathbf{\hat {\mbox{\boldmath$\xi$}}}\right)  \frac{\gamma
Z_{\rm T}ee_{\rm E2}}{\left(  b^{2}+\gamma
^{2}z^{2}\right)  ^{5/2}}\nonumber\\
&  \times\left\{
\begin{array}
[c]{c}%
b^{2}\ \ \ \ (\mathrm{if}\ \ \ \mu=\pm2)\\
\mp2\gamma^{2}bz\ \ \ \ (\mathrm{if}\ \ \ \mu=\pm1)\\
\sqrt{2/3}\left(  2\gamma^{2}z^{2}-b^{2}\right)  \ \ \ \ (\mathrm{if}%
\ \ \ \mu=0)\
\end{array}
\right.  \label{eq7}%
\end{align}
for the E2 (electric quadrupole) field. In
Eqs.~(\ref{eq6})--(\ref{eq7}), $e_{{\rm E}\lambda}=[Z_{\rm v}(A_{\rm
C}/A_{\rm P})^\lambda +Z_{\rm C}(-A_{\rm v}/A_{\rm P})^\lambda]e$
are effective charges for $\lambda=1$ and 2 multipolarities for the
breakup of ${\rm P}\rightarrow{\rm C}+{\rm v}$. The intrinsic
coordinate of v with respect to C is denoted by
$\mathbf{\mbox{\boldmath$\xi$}}$ and $b$ is the impact parameter (or
transverse coordinate) in the collision of P and T, which is defined
by $b=\sqrt{x^2+y^2}$ with ${\bf R} = (x,y,z)$, the relative
coordinate of P from T in the Cartesian representation.  Note
that these relations are obtained with so-called far-field
approximation [EB05], i.e. $R$ is assumed to be always larger
than $\xi$. The Coulomb coupling potentials in E-CDCC are obtained
with Eqs.~(\ref{eq6})--(\ref{eq7}) as shown below.

The full E-CDCC equations, including relativistic effects and recoil corrections for the three-body reaction 
are given by [OB09]:
\begin{eqnarray}
\dfrac{i\hbar^2}{E_c}K_c^{(b)}(z)
\dfrac{d}{d z}\psi_{c}^{(b)}(z) &=& \sum_{c'}
{\mathfrak{F}}^{(b)}_{cc'}(z) \; {\cal R}^{(b)}_{cc'}(z) \nonumber\\
&\times&\psi_{c'}^{(b)}(z) \ e^{i\left(K_{c'}-K_c \right) z}, \label{cceq4}
\end{eqnarray}
where $c$ denotes the channel indices \{$i$, $\ell$, $m$\}; $i>0$
($i=0$) stands for the $i$th discretized-continuum (ground) state
and $\ell$ and $m$\ are respectively the orbital angular momentum
between the constituents (C and v) of the projectile and its
projection on the $z$-axis taken to be parallel to the incident
beam. Note that we neglect the internal spins of C and v for
simplicity. The impact parameter $b$ is relegated to a superscript
since it is not  a dynamical variable. The total energy and the
asymptotic wave number of P are denoted by $E_c$ and $K_c$,
respectively, and \begin{equation}{\cal R}_{cc'}^{(b)}(z)={(K_{c'} R-K_{c'}
z)^{i\eta_{c'}}\over  (K_c R-K_c z)^{i\eta_c}}\end{equation} with $\eta_c$ the
Sommerfeld parameter. The local wave number $K_c^{(b)}(z)$ of P is
defined by energy conservation as
\begin{equation}
E_c
=
\sqrt{(m_{\rm P}c^2)^2+\left\{\hbar c K_c^{(b)}(z)\right\}^2}
+
\dfrac{Z_{\rm P}Z_{\rm T}e^2}{R},
\label{klcl}
\end{equation}
where $m_{\rm P}$ is the mass of P and $Z_{\rm P}e$ ($Z_{\rm T}e$)
is the charge of P (T).
The reduced coupling-potential ${\mathfrak{F}}^{(b)}_{cc'}(z)$ is given by
\begin{equation}
{\mathfrak{F}}^{(b)}_{cc'}(z)
=
{\cal F}^{(b)}_{cc'}(z)
-\dfrac{Z_{\rm P}Z_{\rm T}e^2}{R}\delta_{cc'},
\label{FF1}
\end{equation}
where
\begin{equation}
{\cal F}^{(b)}_{cc'}(z)
=
\left\langle
\Phi_{c}
|
U_{\rm CT}+U_{\rm vT}
|
\Phi_{c'}
\right\rangle_{\bf \xi}
e^{-i(m'-m) \phi_R}.
\label{FF2}
\end{equation}
The $\Phi$ denotes the internal wave functions of P, $\phi_R$ is the
azimuthal angle of ${\bf b}$ and $U_{\rm CT}$ ($U_{\rm vT}$) is the
potential between C (v) and T consisting of nuclear and Coulomb
parts. In actual calculations we use the multipole expansion ${\cal
F}^{(b)}_{cc'}(z)=\sum_\lambda {\cal F}^{\lambda (b)}_{cc'}(z)$, the
explicit form of which is shown in Ref.~[Oga06].

In order to include the dynamical relativistic effects described above,
we make the replacement
\begin{equation}
{\cal F}^{\lambda (b)}_{cc'}(z) \rightarrow \gamma f_{\lambda,m-m'}
{\cal F}^{\lambda (b)}_{cc'}(\gamma z) .\label{FF3}
\end{equation}
The factor $f_{\lambda,\mu}$ is set to unity for nuclear couplings,
while for Coulomb couplings we take
\begin{equation}
f_{\lambda,\mu}
=
\left\{
\begin{array}{cl}
1/\gamma & \quad (\lambda=1, \mu=0) \\
\gamma   & \quad (\lambda=2, \mu=\pm1) \\
1        & \quad ({\rm otherwise}) \\
\end{array}
\right.
\label{FF4}
\end{equation}
following Eqs.~(\ref{eq6}) and (\ref{eq7}). Correspondingly, we use
\begin{equation}
\dfrac{Z_{\rm P}Z_{\rm T}e^2}{R}\delta_{cc'}
\rightarrow
\gamma\dfrac{Z_{\rm P}Z_{\rm T}e^2}{\sqrt{b^2+(\gamma z)^2}}\delta_{cc'}
\label{FF5}
\end{equation}
in Eqs.~(\ref{klcl}) and (\ref{FF1}). The Lorentz contraction factor
$\gamma$ may have channel-dependence, i.e., $\gamma=E_c/(m_{\rm
P}c^2)$, which we approximate by the value in the incident channel,
i.e., $E_0/(m_{\rm P}c^2)$.

It should be remarked that we neglect the recoil motion of T in
Eq.~(\ref{cceq4}); this can be justified because we consider
reactions in which T is significantly heavier than P and we only
treat forward-angle scattering in the present study [Ber05], as
shown below. Note also that in the high incident-energy limit ${\cal
R}_{cc'}^{(b)}(z)\to 1$ and $K_c^{(b)}(z) \to K_c$, unless the
energy transfer is extremely large. Thus, in this limit
Eq.~(\ref{cceq4}) becomes Lorentz covariant, as desired.

Using Eqs.~(\ref{cceq4})--(\ref{FF5}), the dissociation
observables in reactions of loosely bound nuclei $^8$B and $^{11}$Be
on $^{208}$Pb targets was calculated [OB09]. 

\begin{figure}[htpb]
\begin{center}
\includegraphics[width=0.45\textwidth,clip]{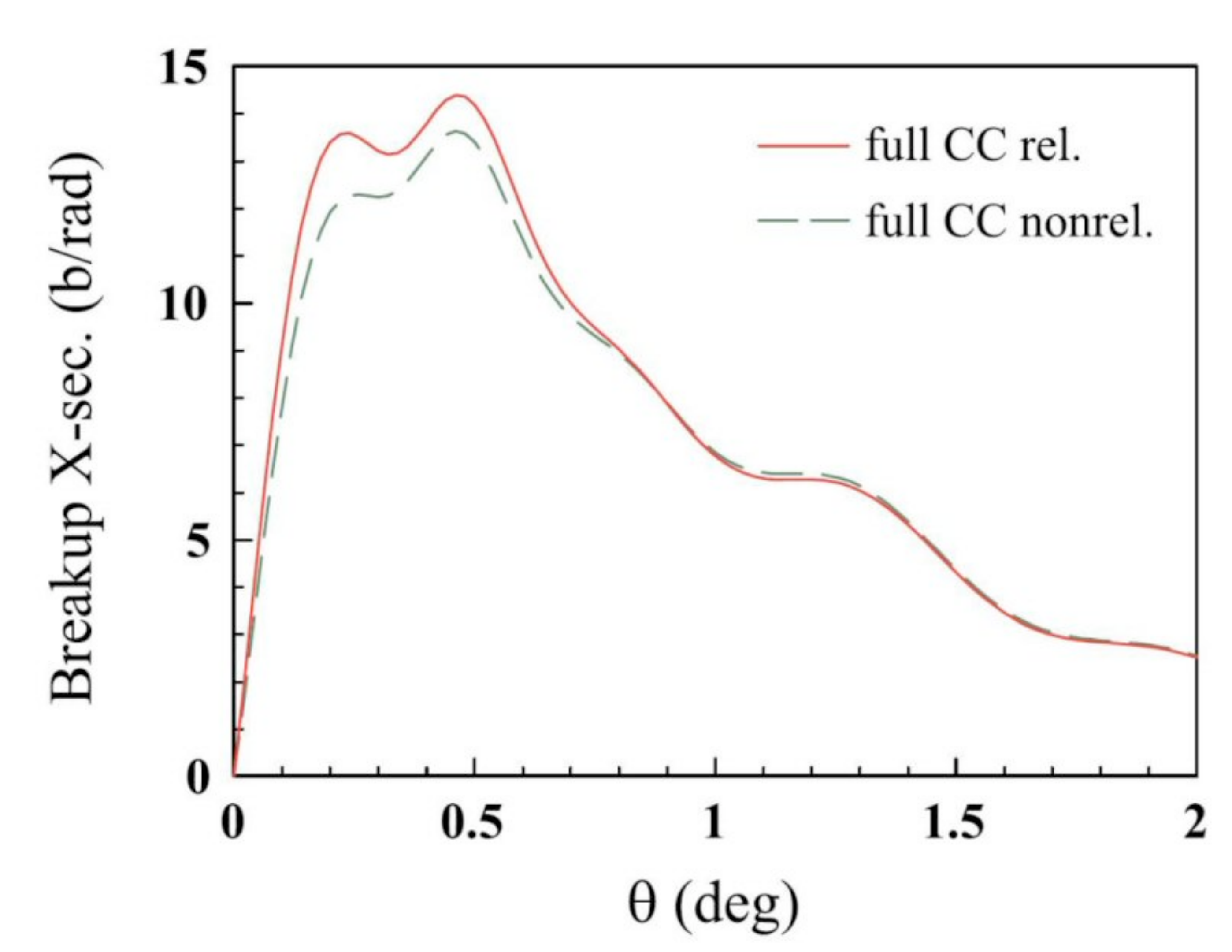}
\end{center}
\caption{\sl  
The total breakup cross section for $^8$B$+^{208}$Pb at
250 MeV/nucleon, as a function of the scattering angle
of the c.m. of the projectile after breakup.
The solid and dashed lines show the results of the full CC
calculation with and without the dynamical relativistic effects,
respectively.
}
\label{ogata1}
\end{figure}
Figure \ref{ogata1} shows the total breakup cross section of $^8$B by
$^{208}$Pb at 250 MeV/nucleon, as a function of the scattering angle
$\theta$ of the center-of-mass (c.m.) of the projectile after
breakup. The solid and dashed lines represent the results of the
E-CDCC calculation with and without the dynamical relativistic
effects, respectively; in the latter we set $\gamma=1$ instead of
the proper value 1.268 in Eqs.~(\ref{FF3})--(\ref{FF5}). Note that
in all calculations shown in this work we use relativistic
kinematics. One sees that the dynamical relativistic correction
gives significantly larger breakup cross sections for $\theta <
1.3$ degrees; the difference between the two around the peak is
sizable, i.e. of the order of 10--15 \%.

We show in Fig.~\ref{ogata2} how the coupled-channel calculations affect the
role of the dynamical relativistic correction. The left (right)
panel corresponds to the calculation with both nuclear and Coulomb
breakup (only Coulomb breakup).

\begin{figure}[htpb]
\begin{center}
\includegraphics[width=0.48\textwidth]{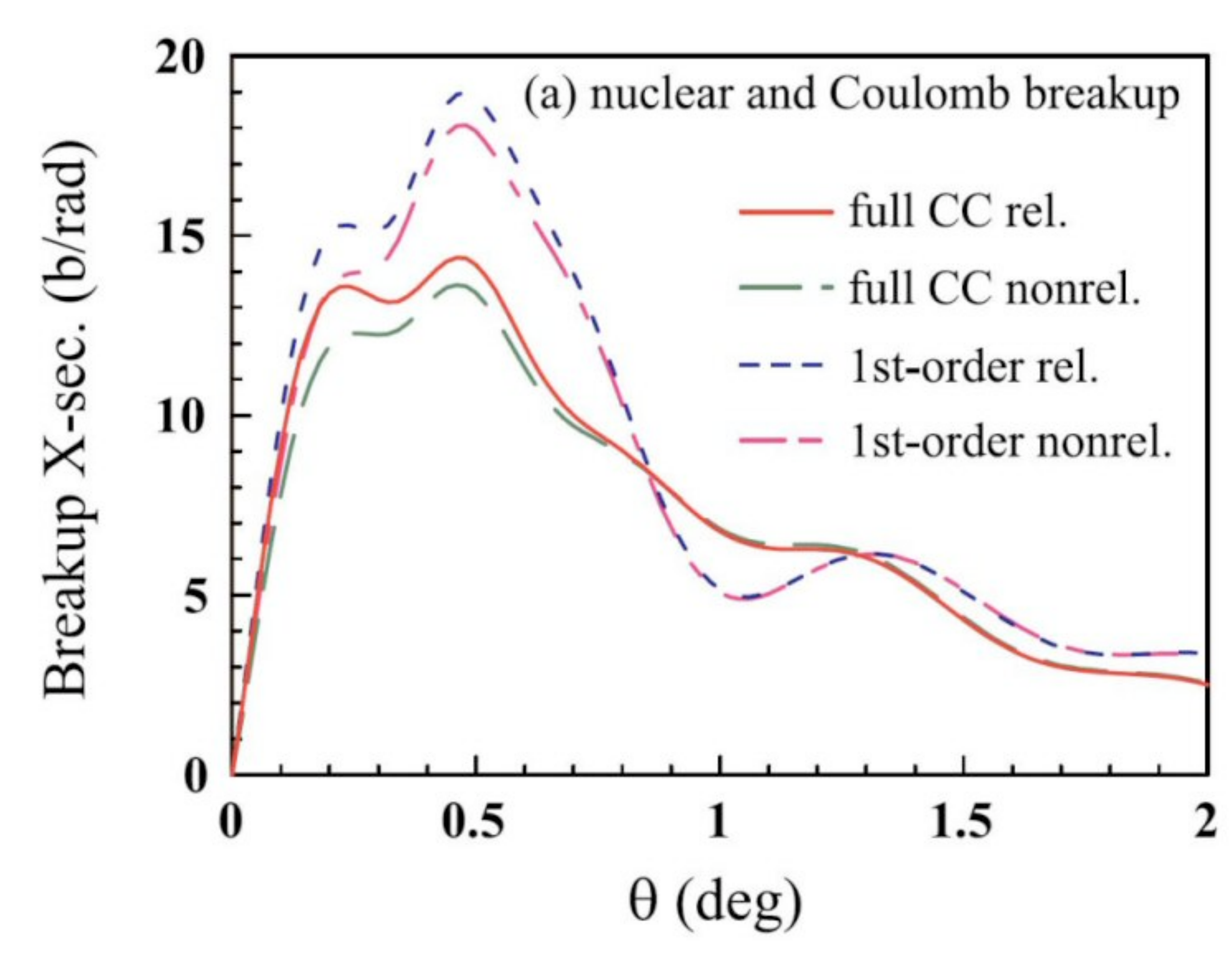}
\includegraphics[width=0.45\textwidth]{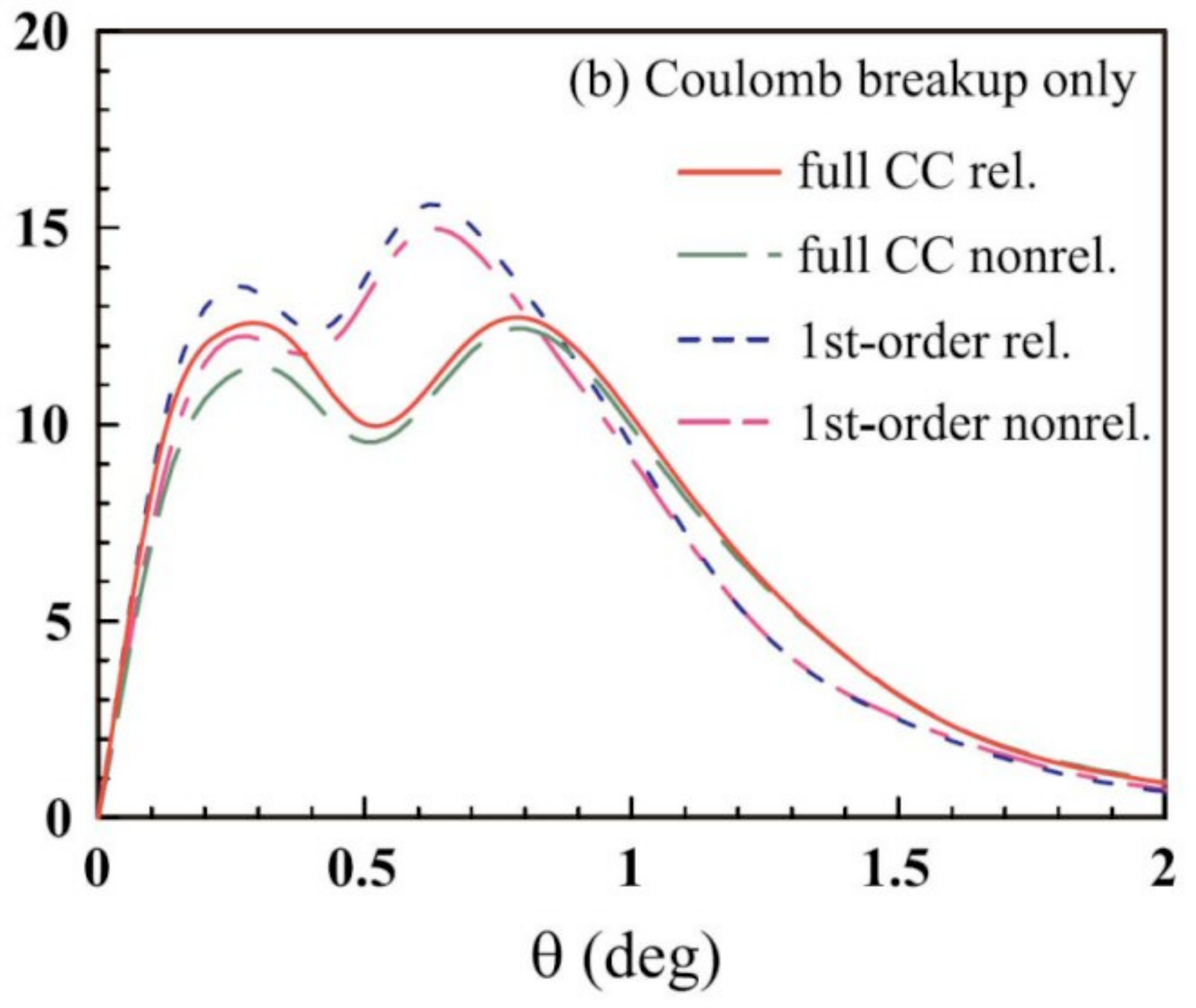}
\caption{\sl  
The total breakup cross sections for $^8$B$+^{208}$Pb at
250 MeV/nucleon with nuclear and Coulomb breakup (top panel)
and only Coulomb breakup (bottom panel).
The solid (dotted) and dashed (dash-dotted) lines show
the results of the full CC (first-order perturbative)
calculation with and without the relativistic correction, respectively.
}\label{ogata2}
\end{center}
\end{figure}

In each panel the solid (dotted) and dashed (dash-dotted) lines show
the results of the full CC (first-order perturbative) calculation
with and without the dynamical relativistic correction,
respectively. One sees from the left panel that relativistic
corrections are slightly quenched when the first-order calculation
is adopted. This is also the case when only Coulomb breakup is
included, as shown in the right panel of Fig.~\ref{ogata2}. We stress here that
due to continuum-continuum couplings relativistic effects are
non-linear (and non-trivial to interpret). Thus, as shown in Fig.~\ref{ogata2},
one cannot infer the effect of relativistic corrections by simply
carrying out first-order calculations. Moreover, it should be
remarked that the full CC and first-order calculations give quite
different breakup cross sections.

One concludes that the effects of relativistic
corrections of the nuclear and Coulomb coupling potentials on the
breakup cross sections of the weakly bound projectiles $^8$B and
$^{11}$Be by $^{208}$Pb targets at 250 and 100 MeV/nucleon. The
relativistic corrections modify appreciably the breakup cross
sections, at the level of 15 \% (10\%), in collisions at  250 (100)
MeV/nucleon. This change is found to be due mainly to the
modification of the Coulomb potential. We have also shown that
relativistic corrections influence continuum-continuum transitions.

\section{What you wanted to know, but were afraid to ask}

\subsection{Close and far fields}

So far, in all derivations of Coulomb excitation we
restricted to distant collision, where the intrinsic radial coordinates
are smaller than the minimum distance between the
colliding nuclei. In calculations of the Coulomb dissociation
of proton halo nuclei, it is of interest also to consider close
collisions, since the density of the valence proton can extend
to very large distances. Moreover, higher-order processes
may also play a role, as suggested by the nonrelativistic calculations
of the $^8$B breakup. This may be of relevance for the calculation of 
continuum-continuum matrix elements.

We derive a general expression for the multipole expansion of
the electro-magnetic interaction from a fast charge particle, which
can be employed in higher-order dynamical calculations of the Coulomb 
excitation of halo nuclei [EB05]. 

The relativistic expression for the electro-magnetic interaction
from a fast charged particle, moving on a straight line trajectory
in the z-direction and impact parameter {\it b} with respect
to a target nucleus, i.e. ${\bf r}'=(b,0,vt)$, and acting on a proton
with coordinates ${\bf r}$ is
\begin{equation}V_{\rm em}({\bf r},{\bf r}') = 
Z_Te^2\Bigl(\phi-{{\bf v}\over 2mc^2}
[\hat{\bf p}\phi+\phi\hat{\bf p}]\Bigr),\label{EB1}\end{equation}
where $\phi$ is the Li\'enard-Wiechert potential
\begin{equation}\phi={\gamma\over \sqrt{|{\bf b}-{\bf b}'|^2
+\gamma^2(z-z')^2}},\label{EB2}\end{equation}
and $\hat{\bf p}$ is the momentum operator associated with ${\bf r}$. 

We start out with the Fourier representation of $\phi$ which 
can be written as (see appendix of Ref. [BB88])
\begin{eqnarray}
\phi &=& {4\pi\over (2\pi)^3} \int d{\bf q} \ 
{e^{i{\bf q}\cdot ({\bf r}-{\bf r}')} \over q_\perp^2+q_z^2/\gamma^2}\nonumber \\
&=& {4\pi\over (2\pi)^3} \int d{\bf q} \ 
{e^{i{\bf q}\cdot ({\bf r}-{\bf r}')} \over q^2}
\ {1\over 1-\beta^2 \ \cos^2(\theta_q)}.\label{EB3}
\end{eqnarray}
where $\beta=v/c$. Except for the last factor, this is just the
ordinary Coulomb potential. 

To proceed, we introduce the multipole expansion of the last term 
in Eq. \eqref{EB3} (see Eq. 8.825 of Ref. [GR80])
\begin{equation}{1\over 1 - \beta^2 \cos^2(\theta)} = 
\sum_{\lambda \ {\rm even}} (2\lambda+1) \ \beta^{-1} \
Q_\lambda(\beta^{-1}) \ P_\lambda(\cos(\theta)),\label{EB4}\end{equation}
and insert the plane wave expansion
\begin{equation}e^{i{\bf q\cdot r}} = 4\pi \sum_{LM} i^L \ j_L(qr) \
Y_{LM}^*({\hat q}) Y_{LM}({\hat r})\end{equation}
for the two plane waves in Eq. \eqref{EB3}. Thus we obtain
\begin{equation}\phi = 4\pi \sum_{LL'} i^{L-L'} \
R_{LL'}(r,r') \sum_m A_{lL'M}(\beta) \
Y_{LM}({\hat r}) \ Y_{L'M}^*({\hat r}'),\label{EB5}\end{equation}
where
\begin{eqnarray}R_{LL'}(r,r') &=& {2\over \pi}  
\int_0^\infty dq \ j_L(qr) \ j_{L'}(qr')\nonumber \\&=&
{1\over \sqrt{rr'}} \
\int_0^\infty {dq\over q}
J_{L+1/2}(qr) \ J_{L'+1/2}(qr'),\label{EB5a}\nonumber \\ \end{eqnarray}
and
\begin{equation}A_{LL'M}(\beta) =
\langle Y_{LM}| {1\over 1-\beta^2\cos^2(\theta)} | Y_{L'M}\rangle\end{equation}
\begin{equation}= \sum_{\lambda \ {\rm even}} (2\lambda+1) \ \beta^{-1}
Q_\lambda(\beta^{-1}) \ 
\langle L'M\lambda 0|LM\rangle \ 
\langle L0\lambda 0 |L'0\rangle.\label{EB5b}\end{equation}
The last two Clebsch-Gordan coefficients is the matrix element
$\langle Y_{LM}| P_\lambda(\cos(\theta))| Y_{L'M}\rangle.$

We note that different multipoles, $(LM)$ and $(L'M)$,
are mixed in the expansion \eqref{EB5}. Let use therefore first
take a look at the contribution from diagonal terms with $L'=L$.
Here the integral \eqref{EB5a} is particularly simple 
\begin{equation}R_{LL}(r,r') ={1\over 2L+1} \ {r_<^L\over r_>^{L+1}}\label{EB6}\end{equation}
where $r_<=\min(r,r')$ and $r_>=\max(r,r')$. Inserting this into 
Eq. \eqref{EB5} we obtain the diagonal contribution
\begin{equation}\phi_{\rm diag} = \sum_{L} {4\pi\over 2L+1} \
{r_<^L\over r_>^{L+1}}
\sum_m A_{LLM}(\beta) \
Y_{LM}({\hat r}) \ Y_{LM}^*({\hat r}').\label{EB7}\end{equation}
This has a structure similar to the non-relativistic expression,
except for the factor $A_{LLM}(\beta)$.
In fact, in the non-relativistic limit we have 
$A_{LL'M}(\beta=0)=\delta_{LL'}$ and the non-relativistic
expression is recovered.

The off-diagonal (i.~e. $L'\neq L$) contributions to $\phi$ 
are more complicated.
According to Eq. \eqref{EB5b}, and the symmetry in $L$ and $L'$, we only 
need expressions for $L'=L+\Lambda$, $\Lambda=2,4,..$. 
One finds
that  [EB05]
\begin{equation}R_{L,L+\Lambda}(a,b) = R_{L+\Lambda,L}(b,a)\label{EB8a}\end{equation}
\begin{equation}R_{L,L+\Lambda}(a,b) = 
0, \ \ \ {\rm for} \ \ a\geq b, \ \ \ {\rm and} \ \ 
\Lambda=2,4,...\label{EB8b}\end{equation}
The explicit, non-zero expression is [EB05]
\begin{equation}R_{L,L+\Lambda}(a,b) = 
{1\over 2L+1+\Lambda} \ {a^L\over b^{L+1}} \
P_{\Lambda/2}^{(L+1/2,-1)}(1-2(a/b)^2),\label{EB8c}\end{equation}
which is valid when $a<b$, and $\Lambda=0,2,4,...$.

It is convenient to exploit the properties (\ref{EB8a}-\ref{EB8b}) and write
expressions for $\phi$ that are valid for distant and close collisions,
i.~e. $r<r'$ and $r>r'$, respectively. The expression one obtains
for distant collisions is
\begin{eqnarray}\phi_{\rm dist}& =& \sum_{LM} 4\pi Y_{LM}({\hat r})
\sum_{\Lambda=0,2,.}
i^\Lambda \ R_{L,L+\Lambda}(r,r') \nonumber \\
&\times& A_{L,L+\Lambda,M} \ 
Y_{L+\Lambda,M}^*({\hat r}'),\label{EB9a}\end{eqnarray}
whereas the contribution for close collisions is
\begin{eqnarray}\phi_{\rm close} &=& \sum_{LM} 4\pi Y_{LM}({\hat r})
\sum_{\Lambda=0,2,.}
i^\Lambda \ R_{L-\Lambda,L}(r',r) \nonumber \\
&\times& A_{L-\Lambda,L,m} \ 
Y_{L-\Lambda,L,m}^*({\hat r}').\label{EB9b}\end{eqnarray}
The last sum over $\Lambda$ is finite since $L-\Lambda$ must be 
non-negative. The two expressions \eqref{EB9a} and \eqref{EB9b} reduce to Eq. \eqref{EB7} 
for $\Lambda=0$. 

The above results have been recently used in Ref. [OB09b] to study the Coulomb breakup of
$^8$B. It was concluded that the close field contributes very little to the breakup cross section. As there exists very few proton halo 
nuclei as loosely-bound as $^8$B, we conclude that the usual multipole expansion with distant fields describes Coulomb excitation satisfactorily.

\subsection{Comparison to electron scattering}
In the plane wave Born approximation (PWBA) the cross section for inelastic
electron scattering is given by [Ba62,EG88]
\begin{align}
\frac{d\sigma}{d\Omega}  &  =\frac{8\pi e^{2}}{\left(  \hbar c\right)  ^{4}%
}\left(  \frac{p^{\prime}}{p}\right)  \sum_{L}\Bigg\{  \frac{EE^{\prime}%
+c^{2}\mathbf{p\cdot p}^{\prime}+m^{2}c^{4}}{q^{4}}\nonumber \\
&\times\left\vert F_{ij}\left(
q;CL\right)  \right\vert ^{2} \nonumber\\
&    +\frac{EE^{\prime}-c^{2}\left(  \mathbf{p\cdot q}\right)  \left(
\mathbf{p}^{\prime}\cdot\mathbf{q}\right)  -m^{2}c^{4}}{c^{2}\left(
q^{2}-q_{0}^{2}\right)  ^{2}}\nonumber \\
&\times\left[  \left\vert F_{ij}\left(  q;ML\right)
\right\vert ^{2}+\left\vert F_{ij}\left(  q;EL\right)  \right\vert
^{2}\right]  \Bigg\}  \label{PWBA}%
\end{align}
where $J_{i}$ $\left(  J_{f}\right)  $ $\ $is the initial (final)
angular momentum of the nucleus, $\left(  E,\mathbf{p}\right)  $ and
($E^\prime ,\mathbf{p}^{\prime}$) are the initial and final energy
and momentum of the electron, and $\left(  q_{0},\mathbf{q}\right)
=\left(\frac{  (E-E^{\prime})}{\hbar c},\frac{\left(
\mathbf{p-p}^{\prime}\right)} {\hbar}\right)  $ is the energy and
momentum transfer in the reaction. $F_{ij}\left(  q;\Pi L\right)  $
are form factors for momentum transfer $q$ and for Coulomb ($C$),
electric ($E$) and magnetic ($M$) multipolarities, $\Pi=C,E,M$,
respectively.

Here we will only treat electric multipole transitions. Moreover, we will
treat low energy excitations such that $E,E^{\prime}\gg\hbar cq_{0}$, which is
a good approximation for electron energies \ $E\simeq500$ MeV and small
excitation energies $\Delta E=\hbar cq_{0}\simeq1-10$ MeV. These are typical
values involved in the dissociation of nuclei far from the stability line.

Using the Siegert's theorem [Sie37,Sa51], one can show that
the Coulomb and electric form factors in eq.
\eqref{PWBA} are proportional to each other. Moreover, for very forward
scattering angles ($\theta\ll1$) the PWBA cross section, eq.
\ref{PWBA}, can be rewritten as
\begin{equation}
\frac{d\sigma}{d\Omega dE_{\gamma}}=\sum_{L}\frac{dN_{e}^{(EL)}\left(
E,E_{\gamma},\theta\right)  }{d\Omega dE_{\gamma}}\ \sigma_{\gamma}%
^{(EL)}\left(  E_{\gamma}\right)  , \label{EPA}%
\end{equation}
where $\sigma_{\gamma}^{(EL)}\left(  E_{\gamma}\right)  $, with $E_{\gamma
}=\hbar cq_{0}$,\ is the photo-nuclear cross section for the $EL$%
-multipolarity, given by eq. \eqref{photonn}, which we rewrite as [BB88]
\begin{equation}
\sigma_{\gamma}^{(EL)}\left(  E_{\gamma}\right)  =\frac{\left(  2\pi\right)
^{3}\left(  L+1\right)  }{L\left[  \left(  2L+1\right)  !!\right]  ^{2}%
}\left(  \frac{E_{\gamma}}{\hbar c}\right)  ^{2L-1}\frac{dB\left(  EL\right)
}{dE_{\gamma}}. \label{sigphoto}%
\end{equation}
In the long-wavelength approximation, the response function, $dB\left(
EL\right)  /dE_{\gamma},$ in eq. \eqref{sigphoto} is given by
\begin{eqnarray}
\frac{dB\left(  EL\right)  }{dE_{\gamma}}&=&\frac{\left\vert \left\langle
J_{f}\left\Vert Y_{L}\left(  \widehat{\mathbf{r}}\right)  \right\Vert
J_{i}\right\rangle \right\vert ^{2}}{2J_{i}+1}\nonumber \\
&\times&\left[  \int_{0}^{\infty
}dr\ r^{2+L}\ \ \delta\rho_{if}^{\left(  EL\right)  }\left(  r\right)
\right]  ^{2}w\left(  E_{\gamma}\right), \nonumber \\\label{photo3}%
\end{eqnarray}
\newline where $w\left(  E_{\gamma}\right)  $ is the density of final
states (for nuclear excitations into the continuum) with energy $E_{\gamma
}=E_{f}-E_{i}$. The
the transition density $\delta\rho_{if}^{\left(  EL\right)  }\left(  r\right)
$ will depend upon the nuclear model adopted.

For $L\ge 1$ one obtains from Eq. \eqref{PWBA} that
\begin{align}
&\frac{dN_{e}^{(EL)}\left(  E,E_{\gamma},\theta\right)  }{d\Omega dE_{\gamma}}
  =\frac{4L}{L+1}\frac{\alpha}{E}\left[  \frac{2E}{E_{\gamma}}\sin\left(
\frac{\theta}{2}\right)  \right]  ^{2L-1}\nonumber\\
&  \times\frac{\cos^{2}\left(  \theta/2\right)  \sin^{-3}\left(
\theta/2\right)  }{1+\left(  2E/M_{A}c^{2}\right)  \sin^{2}\left(
\theta/2\right)  }\nonumber \\
&\times \left[  \frac{1}{2}+\left(  \frac{2E}{E_{\gamma}}\right)
^{2}\frac{L}{L+1}\sin^{2}\left(  \frac{\theta}{2}\right)  +\tan^{2}\left(
\frac{\theta}{2}\right)  \right]  . \label{EPAE}%
\end{align}

One can also define a differential cross section integrated over angles. Since
$\sigma_{\gamma}^{(EL)}$ does not depend on the scattering
angle, this can be obtained from eq. \eqref{EPAE} by integrating $dN_{e}^{(EL)}
/d\Omega dE_{\gamma}$ over angles, from $\theta_{\min}=E_{\gamma}/E$ to
a maximum value $\theta_{m}$, which depends upon
the experimental setup.

Eqs. (\ref{EPA}-\ref{EPAE}) show that under the conditions of the
proposed electron-ion colliders, electron scattering will offer the
same information as excitations induced by real photons. The
reaction dynamics information is contained in the virtual photon
spectrum of eq. \eqref{EPAE}, while the nuclear response dynamics
information will be contained in Eq. \eqref{photo3}. This is akin to a
method developed long time ago by Fermi [Fe24] and usually
known as the Weizsaecker-Williams method [We34] (see section on historical note}. The quantities
$dN_{e}^{(EL)}/d\Omega dE_{\gamma}$ can be interpreted as the number
of equivalent (real) photons incident on the nucleus per unit
scattering angle $\Omega$ and per unit photon energy $E_{\gamma}.$
Note that $E0$ (monopole) transitions do not appear in this
formalism. As immediately inferred from Eq. \eqref{photo3}, for $L=0$
the response function $dB\left( EL\right) /dE_{\gamma}$ vanishes
because the volume integral of the transition density also vanishes
in the long-wavelength approximation. But for larger scattering
angles the Coulomb multipole matrix elements ($CL$) in Eq.
\eqref{PWBA} are in general larger than the electric ($EL$)
multipoles, and monopole transitions become relevant [Sch54].

\begin{figure}[tb]
\begin{center}
\includegraphics[
height=2.8945in,
width=3.1375in
]{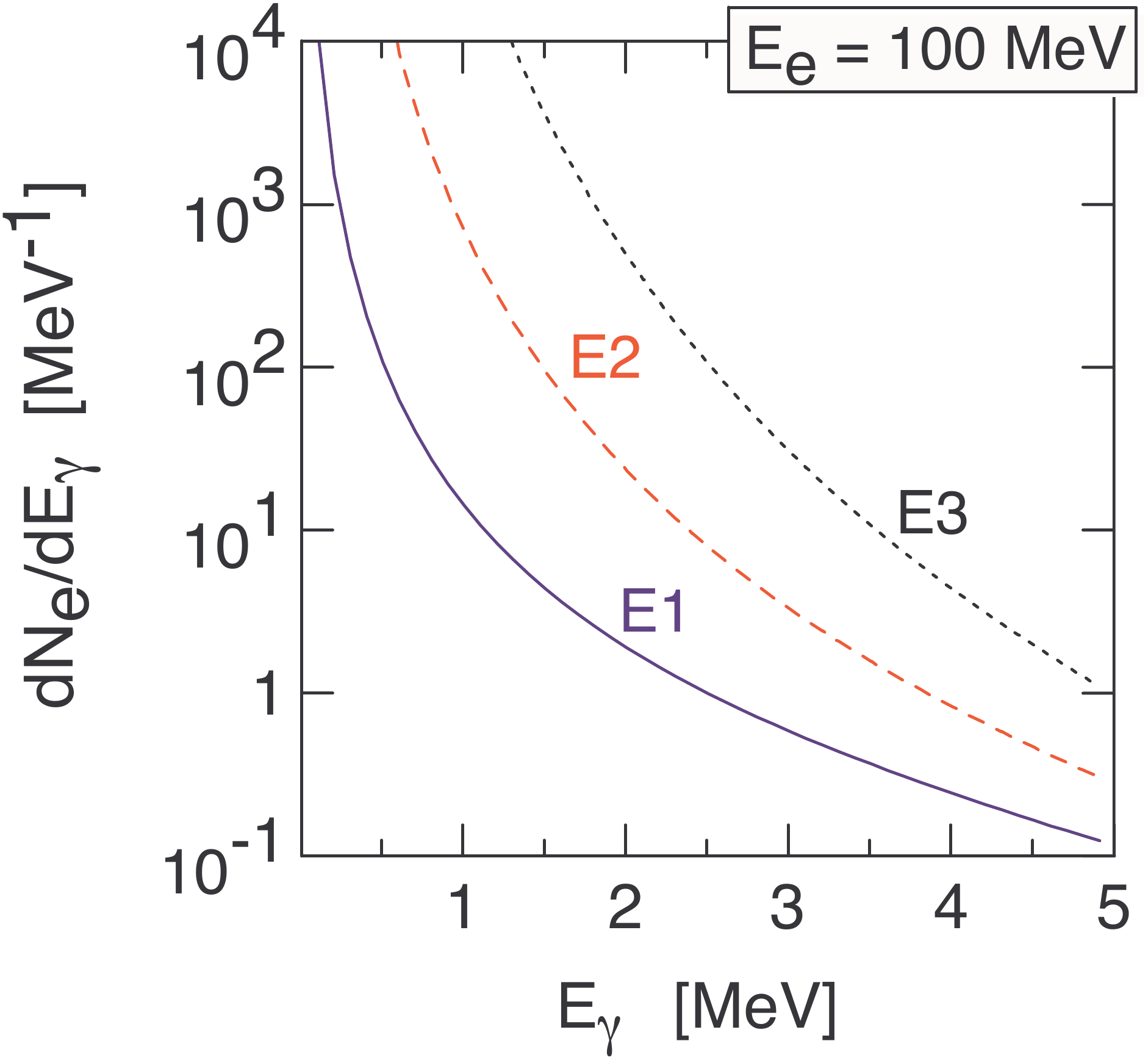}
\end{center}
\caption{\sl Virtual photon spectrum for the $E1$, $E2$
and $E3$ multipolarities in electron scattering off arbitrary nuclei
at $E_{e}=100$ MeV and maximum
scattering angle of 5 degrees.}%
\label{dnde1}%
\end{figure}

In figure \ref{dnde1} we show the virtual photon spectrum for the $E1$, $E2$
and $E3$ multipolarities for electron scattering off arbitrary nuclei at
$E_{e}=100$ MeV. These spectra have been obtained
assuming a maximum scattering angle of 5 degrees. An evident feature
deduced from this
figure is that the spectrum increases rapidly with decreasing energies. Also,
at excitation energies of 1 MeV, the spectrum yields the ratios $dN_{e}%
^{(E2)}/dN_{e}^{(E1)}\simeq500$ and $dN_{e}^{(E3)}/dN_{e}^{(E2)}\simeq100$.
However, although $dN_{e}^{(EL)}/dE_{\gamma}$ increases with the multipolarity
$L$, the nuclear response decreases rapidly with $L$, and $E1$ excitations
tend to dominate the reaction. For larger electron energies the ratios
$N^{(E2)}/N^{(E1)}$ and $N^{(E3)}/N^{(E1)}$ decrease rapidly.

In figure \ref{fig2} we show a comparison between the $E1$ virtual photon
spectrum, $dN_{e}/dE_{\gamma}$, of 1 GeV electrons with the spectrum generated
by 1 GeV/nucleon heavy ion projectiles. In the case of Coulomb excitation, the
virtual photon spectrum was calculated according to Eq. \eqref{(A.6a)}. For
simplicity, we use for the strong interaction distance $R=10$ fm. The spectrum
for the heavy ion case is much larger than that of the electron for large
projectile charges. For $^{208}$Pb projectiles it can be of the order of 1000
times larger than that of an electron of the same energy. As a natural
consequence, reaction rates for Coulomb excitation are larger than for
electron excitation. But electrons have the advantage of being a cleaner
electromagnetic probe, while Coulomb excitation at high energies needs a
detailed theoretical analysis of the data due to contamination by nuclear
excitation. As one observes in figure \ref{fig2}, the virtual spectrum for the
electron contains more hard photons, i.e. the spectrum decreases slower with
photon energy than the heavy ion photon spectrum. This is because, in both
situations, the rate at which the spectrum decreases depends on the ratio of
the projectile kinetic energy to its rest mass, $E/mc^{2},$ which is much
larger for the electron ($m=m_{e}$) than for the heavy ion ($m=$ nuclear mass).

To obtain an effective luminosity per unit energy, the equivalent
photon number is multiplied by the experimental luminosity,
$L_{eA}$, i.e. $\frac{dL_{eff}}{dE_\gamma}=L_{eA}
\frac{dN}{dE_\gamma}$. The number of events per unit time, $N_\tau$,
is given by the integral $N_\tau=\int \sigma(E_\gamma)  dL_{eff} $,
where $\sigma(E_\gamma)$ is the photonuclear cross section. Assuming
that the photonuclear cross section peaks at energy $E_0$ and using
the Thomas-Reiche-Kuhn (TRK) sum rule, Eq. \eqref{sr4}, we can
approximate this integral by $N_\tau=\frac{dL_{eff}}{dE_0} \times
6\times 10^{-26}\frac{NZ}{A}$, where $dL_{eff}/dE$ is expressed in
units of cm$^{-2}$s$^{-1}$MeV$^{-1}$. The giant resonances exhaust
most part of the TRK sum rule and occur in nuclei at energies around
$E_0=15$ MeV. For 1 GeV electrons $dN(E_\gamma=E_0)/dE_\gamma\approx
6\times 10^{-3}$/MeV. With a luminosity of $L_{eA}=10^{25}$
cm$^{-2}$s$^{-1}$, one gets $\frac{dL_{eff}}{dE_0}\approx 6\times
10^{22}$ cm$^{-2}$MeV$^{-1}$s$^{-1}$ and a number of events
$N_\tau\approx4\times 10^{-3}NZ/A\approx10^{-3}A $ s$^{-1}$. Thus,
for medium mass nuclei, one expects thousands of events per day.

\begin{figure}[ptb]
\begin{center}
\includegraphics[
height=2.911in,
width=3.2007in
]{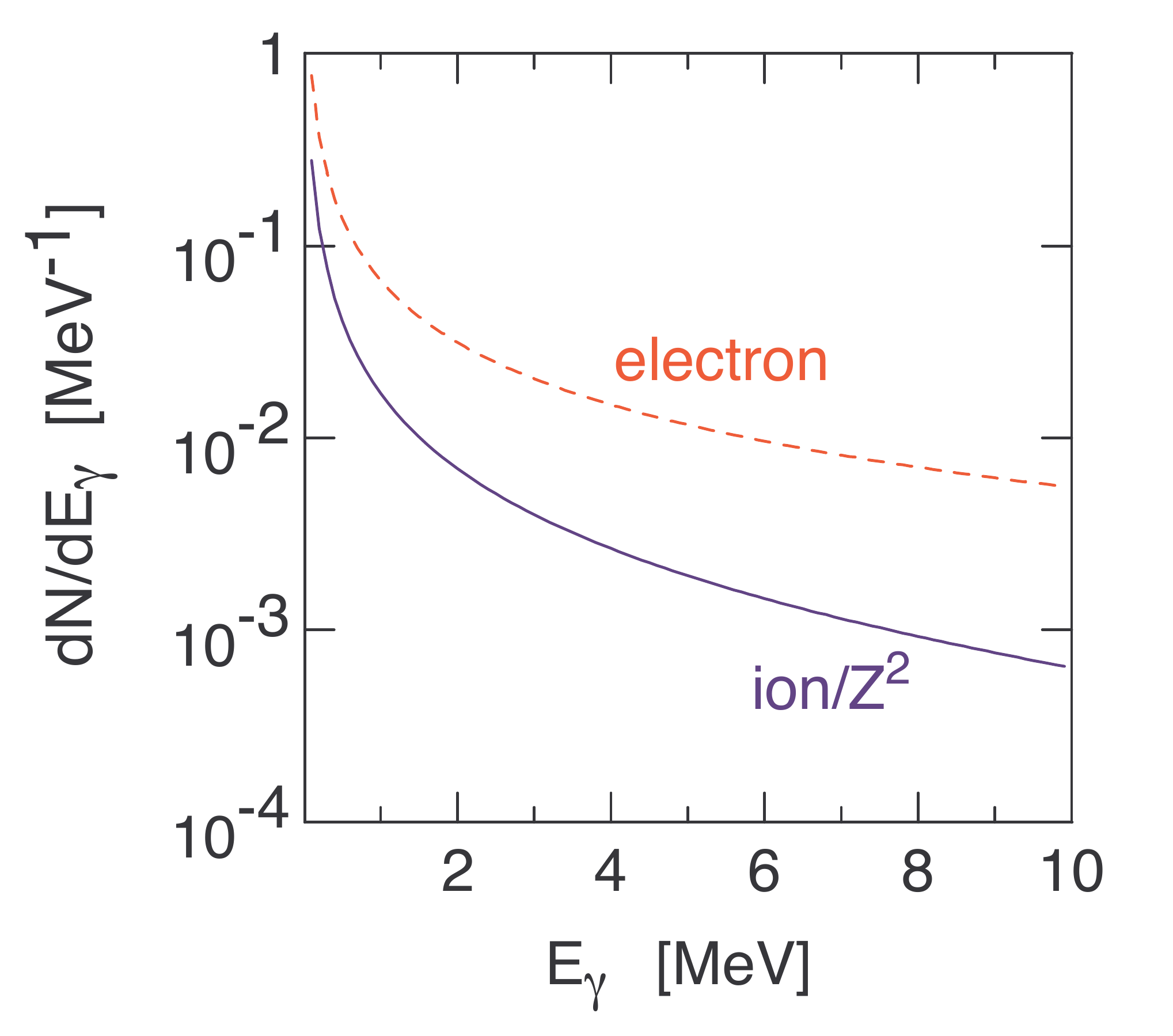}
\end{center}
\caption{\sl Comparison between the virtual photon
spectrum of 1 GeV electrons (dashed line), and the spectrum
generated by a 1 GeV/nucleon heavy ion projectile (solid line) for
the $E1$ multipolarity, as a function of the photon energy. The
virtual photon spectrum for the ion has been divided by the
square of its charge number.}%
\label{fig2}%
\end{figure}

These estimates increase linearly with the accelerator luminosity,
$L_{eA}$, and show that studies of giant resonances in neutron-rich
nuclei is very promising at the proposed facilities
[Haik05,Sud01]. Only a small fraction, of the order of
5\%-10\%, of the TRK sum-rule goes into the excitation of
soft-dipole modes. However, these modes occur at a much
lower energy, $E_r\approx 1$ MeV, where the number of equivalent
photons (see figure \ref{fig2}) is at least one order of magnitude larger
than for giant resonance energies. Therefore, inelastic processes
leading to the excitation of soft dipole modes will be as abundant
as those for excitation of giant resonances. However, one has to
keep in mind that it is not clear if experiments with very
short-lived nuclei will be feasible at the proposed electron-ion
colliders.

\subsection{Non-inertial effects in Coulomb excitation}

Extremely large accelerations occur when atomic nuclei collide. For
instance, two lead nuclei in a head on collision with a center of
mass kinetic energy of 500 MeV, reach a closest distance of 19.4 fm
before they bounce back and move outward. At this distance each
nucleus accelerates with an intriguing $\sim 10^{27}$ m/s$^2$. Very
few other physical situations in the Universe involve nuclei
undergoing such large accelerations, usually related to
astrophysical objects, as in the vicinity of neutron stars and black
holes, where huge gravitational fields exist. Here we will
discuss the effects of large accelerations and large gravitational
fields,  and their possible influence on nuclear reactions in the
laboratory and in astrophysical environments. Nuclear reactions are
crucial for the formation of stellar structures and their rates
could be affected by various factors. 

There are two
systems of reference which are often used to describe the effects of
the collision: (a) the center-of-mass (cm) system of the two nuclei
and (b) the system of reference of the excited nucleus. System (b)
is appropriate to use when the intrinsic properties of the excited
nucleus is described in some nuclear model. A typical example is the
case of Coulomb excitation. One assumes that the nuclei scatter and
their cm wave functions are described by Coulomb waves due to the
Coulomb repulsion between the nuclei. Then one considers the
residual effect of the Coulomb potential on the motion of the
nucleons inside the nuclei. This is done by expanding the  Coulomb
potential in multipoles and using the high order terms (higher than
first order) as a source of the excitation process. In this approach
one illustrates the privileged role of the cm of the nuclear system:
the net effect of the external forces is to (i) accelerate all the
particles together, along with the cm of the system, and (ii) to
change the intrinsic quantum state of the system through the spatial
variation of the interaction within the system. Thus the theoretical
treatment of accelerated many-body systems is well under control in
non-relativistic dynamics.

In the non-relativistic case, the separation of variables into
intrinsic motion and relative motion between the cm of each nucleus
is a simple algebraic procedure. A problem arises when one wants to
extend the method to describe intrinsic excitations of relativistic
many-body systems.  Very few works exist in the literature
addressing this problem. The reason is that for nuclear reactions in
the laboratory, the effect is expected to be very small, a common
belief which must be tested. Another other reason is that in stellar
environments where the gravitational fields are large, huge
pressures develop, "crushing" atoms, stripping them from their
electrons, and ultimately making nuclei dissolve into their
constituents. Effects of nuclear excitation are not relevant in the
process. 

Nuclei participating in nuclear reactions in a gaseous phase of a
star follow inertial trajectories between collisions with other
nuclei. Such trajectories are ``free fall'' trajectories in which
all particles within the nucleus have the same acceleration. That is
surely true in the non-relativistic case, but not in the
relativistic one because retardation effects lead to corrections due
to the nuclear sizes. The central problem here is the question
regarding the definition of the center of mass of a relativistic
many body system. We follow Refs.
[LL89,Pry48,Few84], with few modifications, to show that a
correction term proportional to the square of the acceleration
appears in the frame of reference of the accelerated system. 
The results shown here are from Ref. [BHK09].

\subsubsection{Hamiltonian of an accelerated many-body system}

Starting with a Lagrangian of a free particle in an inertial frame
and introducing a coordinate transformation into an accelerated
frame  with acceleration $\mathcal{A}$, a ``fictitious force" term
appears in the Lagrangian when written in coordinates fixed to the
accelerated frame. Thus, in an accelerated system the Lagrangian $L$
for a free particle can be augmented by a (non-relativistic)
interaction term of the form $-m\mathcal{A}z$, that is
\begin{equation}
L=-mc^{2}+\frac{1}{2} mv^{2}-m\mathcal{A}z, \label{Lnr}
\end{equation}
where $z$
is the particle's coordinate along the direction of acceleration of
the reference frame [LL89].

In the relativistic case, the first step to obtain the Lagrangian of
a many body system in an accelerated frame is to setup an
appropriate measure of space-time in the accelerated frame, i.e. one
needs to find out the proper space-time metric. The free-particle
action $S=-mc\int ds$ requires that $ds=(c-v^2/2c+{\cal A}z)dt$,
which can be used to obtain $ds^2$. To lowest order in $1/c^2$ one
gets
\begin{equation}
ds^{2}=c^{2}\left(1+\frac{\mathcal{A}z}{c^{2}}\right)^{2}dt^{2}
-dx^{2}-dy^{2}-dz^{2}=g_{\mu \nu}d\xi^{\nu}d\xi^{\mu},
\label{metr1}%
\end{equation}
where ${\bf v}dt = d{\bf r}$ was used, with $d\xi^{\mu}=(cdt,dx,dy,dz)$ and
$g_{\mu\nu}=\left(  g_{00}%
,-1,-1,-1\right)  $, $g_{00}=\left( 1+\mathcal{A}z/c^{2}\right)
^{2}$. The indices $\mu$ run from $0$ to $3$. Eq.~(\ref{metr1})
gives a general form for the metric in an accelerated system. This
approach can be found in standard textbooks (see, e.g. Ref.
[LL89], $\S$ 87).

From the definition for the Hamiltonian,
$H=\mathbf{p}\cdot\mathbf{v}-L$, with ${\bf p}=\partial
L/\partial\mathbf{v}=m{\bf v}/\sqrt{g_{00}-v^2/c^2}$, and using the
action with the metric of Eq.~(\ref{metr1}), after a straightforward
algebra one finds
\begin{equation}
H=\frac{g_{00}mc^{2}}{\sqrt{g_{00}-\frac{v^{2}}{c^{2}}}}=c\sqrt{g_{00}\left(
p^{2}+m^{2}c^{2}\right)  }. \label{ham1}%
\end{equation}
Expanding $H$ in powers of $1/c^{2}$, one obtains
\begin{equation}
H    =\frac{p^{2}}{2m}\left(  1-\frac{p^{2}}{4m^{2}c^{2}}\right)
+m\mathcal{A}z\left(  1+\frac{p^{2}}%
{2m^{2}c^{2}}\right)
  +\mathcal{O}\left(  {1\over c^4}\right)  . \label{ham2}%
\end{equation}

This Hamiltonian can be applied to describe a system of particles
with respect to a system of reference moving with acceleration
$\mathcal{A}$, up to order $1/c^{2}$. For an accelerated nucleus the
obvious choice is the cm system of the nucleus. But then the term
carrying the acceleration correction  averages out to zero in the
center of mass, as one has ($\sum_{i}m_{i}\mathcal{A}z_{i}=0$).
There is an additional small contribution of the acceleration due to
the term proportional to $p^2$. Instead of exploring the physics of
this term, one has to account for one more correction as explained
below.

The above derivation of the Hamiltonian for particles in accelerated
frames does not take into account that the definition of the cm of a
collection of particles is also modified by relativity. This is not
a simple task as might seem at first look. There is no consensus in
the literature about the definition of the cm of a system of
relativistic particles. The obvious reason is the role of
simultaneity and retardation. Ref. [Pry48] examines several
possibilities. For a system of particles it is found convenient to
define the coordinates $q^{\mu}$ of the center of mass as the mean
of coordinates of all particles weighted with their dynamical masses
(energies). The relativistic (covariant) generalization of center of
mass is such that the coordinates $q^{\mu}$ must satisfy the
relation~[Pry48]
\begin{equation}
P^{0}q^{\mu}=%
{\displaystyle\sum\limits_{i}}
p_{i}^{0}z_{i}^{\mu}, \label{cm1}%
\end{equation}
where the coordinates of the \textit{i}th particle with respect to
the center of mass are denoted by $z_{i}^{\mu}$ and the total
momentum vector by
$P^{\mu}=%
{\displaystyle\sum\limits_{i}} p_{i}^{\mu}$. Ref. [Pry48]
chooses eq. (\ref{cm1}) as the one that is most qualified to
represent the definition of cm of a relativistic system, which also
reduces to the non-relativistic definition of the center of mass. We
did not find a better discussion of this in the literature and we
could also not find a better way to improve on this definition.

The above definition, Eq.~(\ref{cm1}), leads to the compact form, to
order $1/c^{2}$,
\begin{eqnarray}
{\displaystyle\sum\limits_{i}}
\frac{m_{i}\mathbf{r}_{i}}{\sqrt{g_{00}-\frac{v_{i}^{2}}{c^{2}}}}&=&
{\displaystyle\sum\limits_{i}} m_{i}\mathbf{r}_{i}\left(
1+\frac{v_{i}^{2}}{2c^{2}}-\frac{z_{i}\mathcal{A}
}{c^{2}}+\mathcal{O}({1\over c^4})\right)\nonumber\\ &=&0,
\label{cm2}
\end{eqnarray}
where $\mathbf{r}_{i}=(x_{i},y_i,z_{i})$ is the coordinate and
$v_{i}$ is the velocity of the {\it i}th particle with respect to
the cm.

For a system of non-interacting particles the condition in
Eq.~(\ref{cm2})
implies that, along the direction of motion,%
\begin{equation}%
{\displaystyle\sum\limits_{i}}
\mathcal{A}m_{i}z_{i}=-%
{\displaystyle\sum\limits_{i}}
\mathcal{A}m_{i}z_{i}\left(  \frac{v_{i}^{2}}{2c^{2}}-\frac{z_{i}\mathcal{A}%
}{c^{2}}\right)  . \label{cm3}%
\end{equation}
Hence, the Hamiltonian of Eq.~(\ref{ham2}) for a collection of
particles becomes
\begin{equation}
H    =%
{\displaystyle\sum\limits_{i}}
\frac{p_{i}^{2}}{2m_{i}}\left(  1-\frac{p_{i}^{2}}{4m_{i}^{2}c^{2}}%
\right)  +\frac{\mathcal{A}^{2}}{c^{2}}%
{\displaystyle\sum\limits_{i}} m_{i}z_{i}^{2}
  +U\left(  r_{i}\right) +\mathcal{O}({1\over c^{4}})  , \label{hinert}%
\end{equation}
where we have added a scalar potential $U\left( r_{i}\right)$, which
would represent a (central) potential within an atom, a nucleus, or
any other many-body system.

Notice that the term proportional to $-m\mathcal{A}z$ completely
disappears from the Hamiltonian after the relativistic treatment of
the cm. This was also shown in Ref. [Few84]. It is important
to realize that non-inertial effects will also carry modifications
on the interaction between the particles. For example, if the
particles are charged, there will be relativistic corrections
(magnetic interactions) which need to be added to the scalar
potential $U\left( r_{i}\right)  =\sum_{j\neq
i}Q_{i}Q_{j}/\left\vert \mathbf{r}_{i}-\mathbf{r}_{j}\right\vert $.
As shown in Ref. [Few84], the full treatment of non-inertial
effects together with relativistic corrections will introduce
additional terms proportional to $\mathcal{A}$ and $\mathcal{A}^{2}$
in the Hamiltonian of Eq.~(\ref{hinert}), to order $1/c^{2}$. Thus,
a more detailed account of non-inertial corrections of a many-body
system requires the inclusion of $\mathcal{A}$-corrections in the
interaction terms, too. We refer the reader to Ref. [Few84]
where this is discussed in more details. Here we will only consider
the consequences of the acceleration correction term in
Eq.~(\ref{hinert}),
\begin{equation}
H_{nin}=\frac{\mathcal{A}^{2}}{c^{2}}%
{\displaystyle\sum\limits_{i}} m_{i}z_{i}^{2} .\label{hinert2}
\end{equation}

\subsubsection{Reactions in the laboratory}

For a nuclear fusion reaction, at the Coulomb radius
(distance of closest approach, $R_C$) the nuclear acceleration is given by
\begin{equation}
\mathcal{A}_{C}={Z_{1}Z_{2}e^{2}\over R_{C}^{2}m_{0}}, \label{acc}
\end{equation}
where
$m_{0}=m_{N}A_{1}A_{2}/\left(  A_{1}+A_{2}\right)$ is the reduced
mass of the system and $m_{N}$ is the nucleon mass. For typical
values, $E=1$ MeV, $Z_{1}=Z_{2}=10$, and $A_{1}=A_{2}=20$, one obtains
$R_C=Z_1Z_2e^2/E=144$ fm and $\mathcal{A}%
_{C}=6.2\times10^{25}$ m/s$^{2}$. This is the acceleration that the
cm of each nucleus would have with respect to the laboratory system.

The c.m. of the excited nucleus
is the natural choice for the reference frame. This is because it is
easier to adopt a description of atomic and nuclear properties in
the cm frame of reference. Instead, one could also chose the cm of
the colliding particles. This later (inertial) system makes it
harder to access the acceleration effects, as one would have to
boost the wave functions to an accelerated system, after calculating
it in the inertial frame. This is a more difficult task. Therefore
we adopt the cm reference frame of the excited nucleus, using the
Hamiltonian of the previous section. This Hamiltonian was deduced for a
constant acceleration. If the acceleration is time-dependent, the
metric of Eq.~(\ref{metr1}) also changes. Thus, in the best case
scenario, the Hamiltonian of Eq.~(\ref{hinert}) can be justified in
an adiabatic situation in which the relative velocity between the
many-body systems is much smaller than the velocity of their
constituent particles with respect to their individual center of
masses.

\subsubsection{Reactions involving halo nuclei}

The nuclear wave-function of a (s-wave) loosely-bound, or ``halo",
state can be conveniently parameterized by
\begin{equation}
\Psi\simeq\sqrt{\alpha\over 2\pi} \ {\exp\left(  -\alpha r\right)
\over r}, \end{equation} where the variable $\alpha$ is related to
the nucleon separation energy through
$S=\hbar^{2}\alpha^{2}/2m_{N}$. In first order perturbation theory
the energy shift of a halo state will be given by
\begin{equation}
\Delta E_{nin}^{N}=\left\langle \Psi\left\vert H_{nin}\right\vert
\Psi\right\rangle={1 \over 8S} {\left( Z_{1}Z_{2}e^{2}\hbar\over
R_{C}^{2}m_{0}c\right) ^{2}}.
\end{equation}
Assuming a small separation energy $S=100$ keV, and using the same
numbers in the paragraph after Eq.~(\ref{acc}), we get $\Delta
E_{nin}^{N}= 0.024$ eV, which is very small, except for states very
close to the nuclear threshold, i.e. for $S\rightarrow 0$. But the
effect increases with $Z^2$ for symmetric systems (i.e.
$Z_1=Z_2=A_1/2$). It is thus of the order of $\Delta E_{nin}^{N}=
1-10$ eV for larger nuclear systems.

There might exist situations where this effect could be present. For
instance, the triple-alpha reaction which bridges the mass = 8 gap
and forms carbon nuclei in stars relies on the lifetime of only
10$^{-17}$ s of $^{8}$Be nuclei. It is during this time that another
alpha-particle meets $^{8}$Be nuclei in stars leading to the
formation of carbon nuclei. This lifetime corresponds to an energy
width of only $5.57\pm0.25$ eV~[Ti04]. As the third alpha
particle approaches $^{8}$Be, the effects of linear acceleration
will be felt in the reference frame of $^8$Be. This will likely
broaden the width of the $^{8}$Be resonance (which peaks at
$E_{R}=91.84\pm0.04$ KeV) and consequently its lifetime. However,
this line of thought could be wrong if one assumes that the third
alpha particle interacts individually with each of the two alpha
particles inside $^8$Be, and that the effects of acceleration
internal to the $^8$Be nucleus arise from the different distances
(and thus accelerations) between the third alpha and each of the
first two. This effect has not been discussed
elsewhere and deserves further investigation, if not for
this particular reaction maybe for other reactions of astrophysical
interest involving very shallow nuclear states.

\subsubsection{Nuclear transitions}

Many reactions of astrophysical interest are deduced from
experimental data on nucleus-nucleus scattering. Important
information on the position and widths of resonances, spectroscopic
factors, and numerous other quantities needed as an input for
reaction network calculations in stellar modeling are obtained by
the means of nuclear spectroscopy using nuclear collisions in the
laboratory. During the collision the nuclei undergo huge
acceleration, of the order of $\mathcal{A}\simeq10^{28}$ m/s$^{2}$.
Hence, non-inertial effects will be definitely important.

A simple proof of the statements above can be obtained by studying
the Coulomb excitation. The simplest treatment that one can use in
the problem is  a semi-classical calculation. 
Following the same procedure leading to Eq. \eqref{10.99}, we can calculate the
contribution of the Hamiltonian of Eq.~(\ref{hinert2}). In this
case, ${\cal A}=Z_1Z_2e^2/m_0r^2$ and the pertubative potential $\phi=V_{nin}$
is given by
\begin{equation}
V_{nin}=\left({Z_1Z_2e^2\over m_0}\right)^2{Xm_N\over c^2r^4},
\end{equation}
where we assume that $X$ nucleons participates in the transition.
One then finds  \begin{equation} a_{if}^{nin}=\left({Z_1Z_2e^2 \over
m_0}\right)^2{32Xm_NQ_{if}\over 15is^3\hbar v_0c^2} . \end{equation}
The ratio between the two transition probabilities (the above one and Eq. \eqref{10.99}) is
\begin{equation}
\left\vert {a_{if}^{nin} \over a_{if}}
 \right\vert^2 = \left(
8Xm_N  Z_1Z_2^2e^2\over 5s m_0^2c^2\right)^2 \label{ratV}.
\end{equation}

Applying Eq. \eqref{ratV} to the lead-lead collision at 500 MeV,  we find $\left\vert {a_{if}^{nin} /
a_{if}}
 \right\vert^2 = (0.0093X)^2$.  This yields very small results for
the effect of non-inertial effects in single particle
transitions ($X\simeq 1$), but can become appreciable for the
excitation of collective states such as the giant resonances, for
which $X\gg 1$. 

\section{Codes}
\begin{description}
\item[GOSIA -] A non-relativistic semiclassical coupled-channel Coulomb excitation least-squares search code, GOSIA, has been developed to analyze the large sets of experimental data. Manual and code available from:

http://www.pas.rochester.edu/~cline/Gosia/

\item[DWEIKO -]  A relativistic coupled-channels code for  Coulomb and nuclear excitations of E1, E2, E3,M1, and M2 multipolarities. 

Manual: C.A. Bertulani, C.M. Campbell, T. Glasmacher, Computer Physics Communications 152 (2003) 317. 

Code available from the Computer Physics Communications Program Library: 

http://www.cpc.cs.qub.ac.uk/cpc/

\item[COULINT -] A code which incorporates retardation corrections and recoil at intermediate energy collisions ($\sim 100$ MeV/nucleon), appropriate for Coulomb excitation at rare isotopes facilities, such as GANIL, GSI, NSCL  and RIKEN. Under development. Preliminary versions already available [Ber04].

\end{description}

\section{Appendices}

Here I discuss a few of the basic principles of quantum mechanics which are necessary to understand this review.

\subsection{Decay rates}

A quantum system, described by a wavefunction that is an Hamiltonian
eigenfunction, is in a well defined energy state and, if it does not suffer
external influences, it will remain indefinitely in that state. But this ideal
situation does not prevail in excited nuclei, or in the ground state of an
unstable nuclei. Interactions of several types can add a perturbation to the
Hamiltonian and the pure energy eigenstates no more exist. In this situation a
transition to a lower energy level of the same or of another nucleus can occur.

An unstable state normally lives a long time compared to the fastest nuclear
processes, e. g., the time spent by a particle with velocity near that of the
light to cross a nuclear diameter. In this way we can admit that a nuclear
state is approximately stationary, and to write for its wave function:
\begin{equation}
\Psi(\mathbf{r},t)=\psi(\mathbf{r})e^{-iWt/\hbar}.\label{5.29.6}%
\end{equation}
$|\Psi(\mathbf{r},t)|^{2}dV$\ is the probability to find the nucleus in the
volume $dV$\ and, if the state described by $\Psi$\ decays with a decay
constant $\lambda $, it is reasonable to write{  \ }
\begin{equation}
|\Psi(\mathbf{r},t)|^{2}=|\Psi(\mathbf{r},0)|^{2}e^{-\lambda  t}.\label{5.30.6}%
\end{equation}

To obey simultaneously (\ref{5.29.6}) and (\ref{5.30.6}), $W$\
must be a complex quantity with imaginary part $-\lambda \hbar/2$.
We can write:{  \ }
\begin{equation}
W=E_{0}-{\frac{\hbar \lambda  }{2}}i,\label{5.31}%
\end{equation}
which shows that the wave function (\ref{5.29.6}) does not
represent a well defined stationary state since the exponential
contains a real part $-\lambda  t/2$. However, we can write the
exponent of (\ref{5.29.6}) as a superposition of values
corresponding to well defined energies $E$\ (for $t\geq0$):
\begin{equation}
e^{-(iE_{0}/\hbar+\lambda /2)t}=\int_{-\infty}^{+\infty}A(E)e^{-iEt/\hbar
}\,dE.\label{5.32}%
\end{equation}

Functions connected by a Fourier transform relate to each other as
\begin{align}
f(t)  & ={\frac{1}{\sqrt{2\pi}}}\int_{-\infty}^{+\infty}g(\omega)e^{-i\omega
t}\,d\omega,\label{5.33a}\\
g(\omega)  & ={\frac{1}{\sqrt{2\pi}}}\int_{-\infty}^{+\infty}f(t)e^{i\omega
t}\,dt.\label{5.33b}%
\end{align}
This allows to establish the form of the amplitude $A(E)$:
\begin{equation}
A(E)={\frac{1}{2\pi\hbar}}\int_{0}^{+\infty}e^{[i(E-E_{0})/\hbar-\lambda 
/2]t}\,dt,\label{5.34}%
\end{equation}%
where the lower limit indicates that the stationary sate was
created at the time $t=0$. The integral (\ref{5.34}) is of easy
solution, giving
\begin{equation}
A(E)={\frac{1}{h\lambda /2-2\pi i(E-E_{0})}}.\label{5.35}%
\end{equation}

\begin{figure}
[t]
\begin{center}
\includegraphics[
width=2.9in
]%
{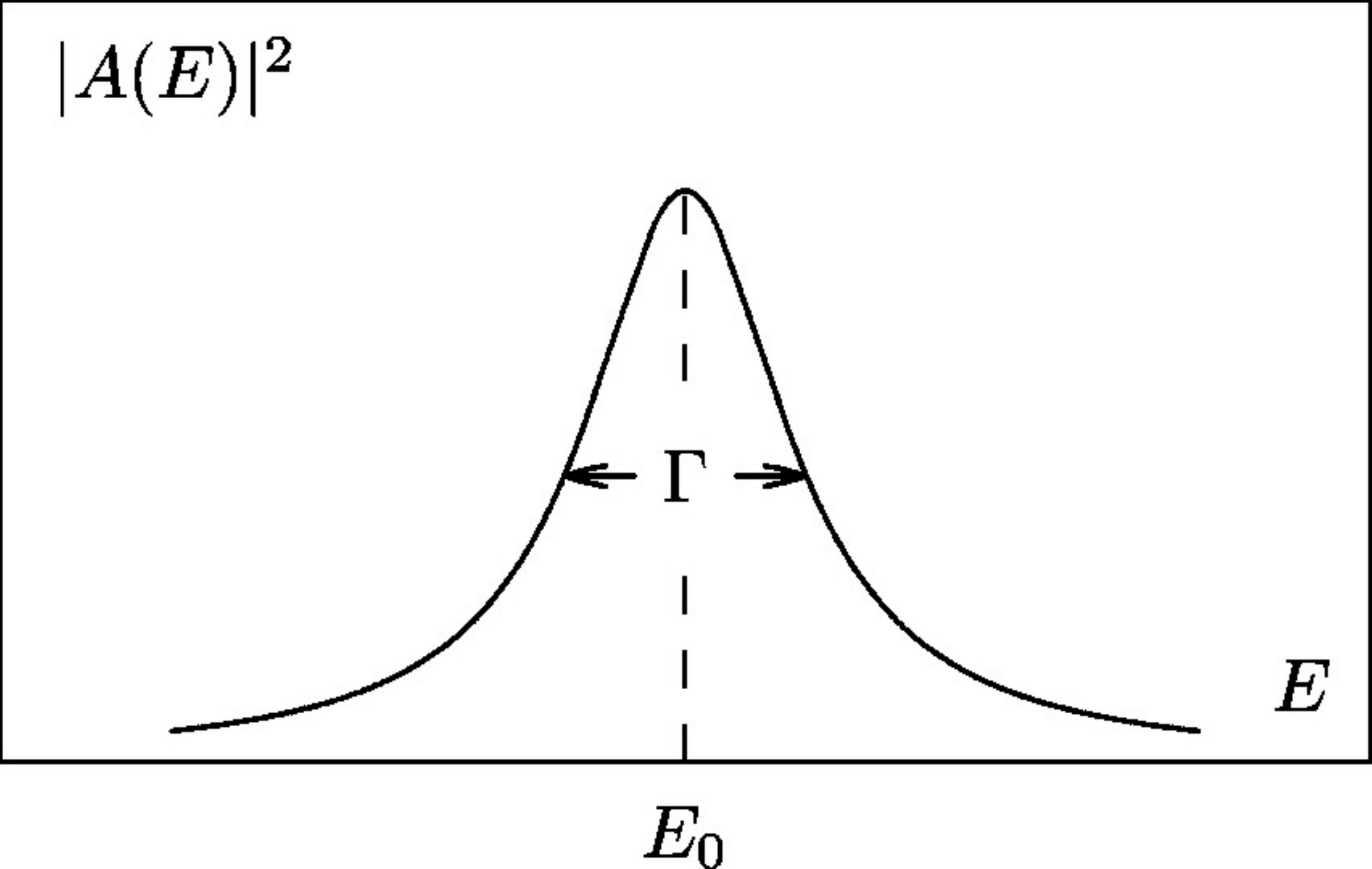}%
\caption{\sl  Form of the distribution (\ref{5.36}).}%
\label{direct3}%
\end{center}
\end{figure}

The probability of finding a value between $E$\ and $E+dE$\ in an energy
measurement is given by the product
\begin{equation}
A^{\ast}(E)A(E)={\frac{1}{h^{2}\lambda ^{2}/4+4\pi^{2}(E-E_{0})^{2}}%
},\label{5.36}%
\end{equation}
and this function of energy has the form of a Lorentzian, with the
aspect shown in figure \ref{direct3}. Its width at half-maximum is
$\Gamma =\hbar \lambda  =\hbar/\tau$. The relationship
\begin{equation}
\tau\Gamma=\hbar\label{5.37}%
\end{equation}
between the half-life and the width of a state is directly connected to the
uncertainty principle and shows that the longer time a state survives the
greater is the precision with which its energy can be determined. In
particular, only to stable states one can attribute a single value for the energy.

The decay constant $\lambda $\ was presented as the probability per unit time
of occurrence of a transition between quantum states, and its values were
supposed to be known experimental data. Now we will show that a formula to
evaluate the decay constant can be obtained from the postulates of
perturbation theory.

\subsection{Coupled-channels}

We can describe an unstable state by the addition of a perturbation to a
stationary state. Formally we can write
\begin{equation}
H=H_{0}+V,\label{5.38}%
\end{equation}
where $H$\ is the Hamiltonian of the unstable state, composed by the
non-perturbed Hamiltonian $H_{0}$\ and a small perturbation $V$. The
Hamiltonian $H_{0}$\ satisfies an eigenvalue equation
\begin{equation}
H_{0}\psi_{n}=E_{n}\psi_{n},\label{5.39}%
\end{equation}
whose eigenfunctions form a complete basis in which the total wavefunction
$\Psi$, that obeys
\begin{equation}
H\Psi=i\hbar\frac{\partial\Psi}{\partial t},\label{5.40}%
\end{equation}
can be expanded:
\begin{equation}
\Psi=\sum_{n}a_{n}(t)\psi_{n}e^{-iE_{n}t/\hbar}.\label{5.41}%
\end{equation}
Using (\ref{5.38}) and (\ref{5.41}) in (\ref{5.40}), and with the aid of
(\ref{5.39}), we obtain:
\begin{equation}
i\hbar\sum_{n}\dot{a}_{n}\psi_{n}e^{-iE_{n}t/\hbar}=\sum_{n}Va_{n}\psi
_{n}e^{-iE_{n}t/\hbar},\label{5.42}%
\end{equation}
with $\dot{a}_{n}\equiv\partial a_{n}(t)/\partial t$. Using the
orthogonalization properties of the $\psi_{n}$, let us multiply (\ref{5.42})
by $\psi_{k}^{\ast}$\ and integrate it in the coordinate space. From this,
results the coupled-channels equations%
\begin{equation}
\dot{a}_{k}\left(  t\right)  =-\frac{i}{\hbar}\sum_{n}a_{n}\left(  t\right)
\ V_{kn}\left(  t\right)  \ e^{i{\frac{E_{k}-E_{n}}{\hbar}}t},\label{5.43}%
\end{equation}
where we introduced the matrix element ($d\tau$\ is the volume element)
\begin{equation}
V_{kn}=\int\psi_{k}^{\ast}V\psi_{n}\,d\tau.\label{5.44}%
\end{equation}

\subsection{Transition rates and cross sections}
Let us make the following assumptions about the perturbation $V$: it begins to
act at time $t=0$, when the unperturbed system is described by an eigenstate
$\psi_{m}$. It stays at a very low value and after a short time interval
becomes zero at $t=T$. These assumptions allow us to say that the conditions%
\begin{equation}
a_{m}=1,\ \ \ \ \ \ \ \ \ \ \ \ \ \text{and}\ \ \ \ \ \ a_{m}%
=0\;\;\;\;\;\text{if}\;\;\;(m\neq n)\label{5.45}%
\end{equation}
are rigorously verified for $t<0$\ and also works approximately for $t>0$.
Thus, (\ref{5.43}) has only one term and the value of the amplitude is
obtained from
\begin{equation}
a_{k}=-\frac{i}{\hbar}\int_{0}^{T}V_{km}e^{i\left(  E_{k}-E_{m}\right)
t/\hbar}\,dt,\label{5.46}%
\end{equation}
whose value must be necessarily small by the assumption that
followed (\ref{5.45}). The above approach is also known as {\it
first order perturbation theory}. The integral of (\ref{5.46})
gives
\begin{equation}
a_{k}=\frac{V_{km}\left[  1-\exp\left(  i\frac{E_{k}-E_{m}}{\hbar}T\right)
\right]  }{E_{k}-E_{m}}.\label{5.47}%
\end{equation}

We need now to interpret the meaning of the amplitude $a_{k}$. The quantity
$a_{k}^{\ast}a_{k}$\ measures the probability of finding the system in the
state $k$. This characterizes a transition occurring from the initial state
$m$\ to the state $k$, and the value of $a_{k}^{\ast}a_{k}$\ divided by the
interval $T$\ should be a measure of the decay constant $\lambda_{k}%
$\ relative to the state $k$. The total decay constant is obtained by the sum
over all states:
\begin{equation}
\lambda =\sum_{k\neq m}L _{k}=\frac{\sum|a_{k}|^{2}}{T}.\label{5.48}%
\end{equation}

Let us now suppose that there is a large number of available states $k$. We
can in this case replace the summation in (\ref{5.48}) by an integral.
Defining $\rho(E)$\ as the density of available states around the energy
$E_{k}$, we write
\begin{eqnarray}
\lambda &=&\frac{1}{T}\int_{-\infty}^{+\infty}|a_{k}|^{2}\rho(E_{k}%
)\,dE_{k}\nonumber \\
&=&\frac{4}{T}\int_{-\infty}^{+\infty}|V_{km}|^{2}\frac{\sin
^{2}[\left[  \left(  {\frac{E_{k}-E_{m}}{2\hbar}}\right)  T\right]  }%
{(E_{k}-E_{m})^{2}}\rho(E_{k})\,dE_{k}.\nonumber \\
\label{5.49}%
\end{eqnarray}

The function sin$^{2}x/x^{2}$\ only has significant amplitude near the origin.
In the case of (\ref{5.49}), if we suppose that $V_{km}$\ and $\rho$\ do not
vary strongly in a small interval of the energy $E_{k}$\ around $E_{m}$, both
these quantities can be taken outside of the integral, and we obtain the final
expression
\begin{equation}
\lambda=\frac{2\pi}{\hbar}|V_{km}|^{2}\rho(E_{k})\ .\label{5.50}%
\end{equation}
Eq. (\ref{5.50}) is known as the golden rule
$n^{\underline{\circ}}\,2$\ (also known as \textit{Fermi golden
rule}), and allows to determine the decay constant if we know the
wavefunctions of the initial and final states.

We now describe how we can obtain a reaction cross section for the system A(a,b)B from Fermi's Golden rule.
For the application of (\ref{5.50}) the entrance channel will be understood as
an initial quantum state, constituted of particles of mass $m_{a}$ hitting a
target $A$ of mass $m_{A}$. The final quantum state is the exit channel, where
particles of mass $m_{b}$ move away from the nucleus $B$ of mass $m_{B}$. The
essential hypothesis for the application of the golden rule is that the
interaction responsible for the direct reaction can be understood as a
perturbation among a group of interactions that describe the system as a whole.

If we designate as $\lambda$ the transition rate of the initial quantum state
to the final quantum state, we can write a relationship between $\lambda$ and
the total cross section:
\begin{equation}
\lambda=v_{a}\sigma=\sigma\frac{k_{a}\hbar}{m_{a}},\label{10.71}%
\end{equation}
where $v_{a}$ is now the velocity of the particles of the incident beam. If
$d\lambda$ is the part relative to the emission in the solid angle $d\Omega$,
then
\begin{equation}
d\sigma=\frac{m_{a}}{k_{a}\hbar}d\lambda\label{10.72}%
\end{equation}
and, using (\ref{5.50}),%
\begin{equation}
d\sigma=\frac{m_{a}}{k_{a}\hbar}\frac{2\pi}{\hbar}|V_{fi}|^{2}d\rho
(E_{f}),\label{10.73}%
\end{equation}
where the indexes $i$ and $f$ refer to the initial and final stages of the reaction.

The density of final states can be easily obtained by counting the available number of states for the projectile in a large box within an energy interval $dE$. For a box with volume $s^3$, one finds [BD04]
\begin{equation}
\rho=\frac{dN}{dE}=\frac{pm_{b}s^{3}}{2\pi^{2}\hbar^{3}}.\label{10.74}%
\end{equation}
If we assume an isotropic distribution of momentum for the final states, the
fraction $d\rho$ that corresponds to the solid angle $d\Omega$ is
\begin{equation}
d\rho=\frac{pm_{b}s^{3}}{2\pi^{2}\hbar^{3}}\frac{d\Omega}{4\pi}=\frac
{k_{b}m_{b}s^{3}}{8\pi^{3}\hbar^{2}}d\Omega\ .\label{10.75}%
\end{equation}
Using (\ref{10.75}) in (\ref{10.73}), we obtain an expression for the
differential cross section:
\begin{equation}
\frac{d\sigma}{d\Omega}=\frac{m_{a}m_{b}k_{b}}{(2\pi\hbar^{2})^{2}k_{a}%
}|V_{fi}|^{2},\label{10.76}%
\end{equation}
involving the matrix element
\begin{equation}
V_{fi}=\int\Psi_{b}^{\ast}\Psi_{B}^{\ast}\chi_{\beta}^{(-)\ast}(\mathbf{r}%
_{\beta})V\Psi_{a}\Psi_{A}\chi_{\alpha}^{(+)}(\mathbf{r}_{\alpha}%
)d\tau.\label{10.77}%
\end{equation}
$\Psi_{a},\Psi_{b},\Psi_{A},\Psi_{B}$ are the internal wavefunctions of the
nuclei $a$, $b$, $A$ and $B$. $\chi_{\alpha}^{(+)},\chi_{\beta}^{(-)} $ are
the scattering wavefunctions of the relative momentum in the entrance channel $\alpha$
and in the exit channel $\beta$. The volume $s^{3}$ was not written in
(\ref{10.76}) because it can be chosen as the unity, since the wavefunctions
that appear in (\ref{10.77}) are normalized to 1 particle per unit volume. $V$
is the perturbation potential that causes the \textquotedblleft
transition\textquotedblright\ from the entrance to the exit channel. It can be
understood as an additional interaction to the average behavior of the
potential and, in this sense, can be written as the difference between the
total potential in the exit channel and the potential of the optical model in
that same channel.

\subsection{Electromagnetic potentials}

The Lorentz equation, or the equation of motion of a charge in an electromagnetic field, is
given by
\begin{equation}
\frac{d}{dt}\frac{m\mathbf{v}}{\sqrt{1-\mathrm{v}^{2}/c^{2}}}=q(\mathbf{E}%
+\frac{\mathbf{v}}{c}\;\times\mathbf{B}\text{\ }).\label{(A.1)}%
\end{equation}
We now introduce the electromagnetic potentials ($\phi,{\bf A}$) by means of the relations
\begin{equation}
\mathbf{B}=\mathbf{\nabla}\;\times\text{ }\mathbf{A}%
\;;\;\;\;\;\;\;\;\;\;\;\;\mathbf{E}=-\mathbf{\nabla}\phi-\frac{1}{c}%
\frac{\partial\mathbf{A}}{\partial t}.\label{(A.2)}%
\end{equation}
In terms of these potentials, we get
\begin{align}
\frac{d}{dt}\frac{m\mathbf{v}}{\sqrt{1-\mathrm{v}^{2}/c^{2}}}  & =q\left[
-\mathbf{\nabla}\phi-\frac{1}{c}\frac{\partial\mathbf{A}}{\partial t}%
+\frac{\mathbf{v}}{c}\cdot(\mathbf{\nabla}\;\times\;\mathbf{A})\right]
\nonumber\\
& =\mathbf{\nabla}\left(  -q\phi+q\frac{\mathbf{v}}{c}.\mathbf{A}\right)
-\frac{q}{c}\left(  \frac{\partial}{\partial t}+\mathbf{v}.\mathbf{\nabla
}\right)  \mathbf{A}\nonumber\\
& =\mathbf{\nabla}\left(  -q\phi+q\frac{\mathbf{v}}{c}.\mathbf{A}\right)
-\frac{q}{c}\frac{d\mathbf{A}}{dt}\label{(A.3)}%
\end{align}
where we have used the convective derivative
\begin{equation}
\frac{d\mathbf{A}\left[  \mathbf{r}(t),t\right]  }{dt}=\left(  \frac{\partial
}{\partial t}+\mathbf{v}.\mathbf{\nabla}\right)  \mathbf{A}\left[
\mathbf{r}(t),t\right]  .\label{(A.4)}%
\end{equation}
{We can rewrite (\eqref{(A.3)}) as}
\begin{equation}
\frac{d}{dt}\left(  \frac{m\mathbf{v}}{\sqrt{1-\mathrm{v}^{2}/c^{2}}}+\frac
{q}{c}\;\mathbf{A}\right)  +\mathbf{\nabla}\left(  q\phi-\frac{q}%
{c}\;\mathbf{v}\cdot\mathbf{A}\right)  =0.\label{(A.5)}%
\end{equation}

This equation has the form of the Euler-Lagrange equation
\[
\frac{d}{dt}\left(  \mathbf{\nabla}_{\mathbf{r}}\mathcal{L}\right)
-\mathbf{\nabla}_{\mathbf{r}}\mathcal{L}=0.
\]

The suitable Lagrangian that reproduces Eq. \eqref{(A.3)} when replaced in Eq. (\ref{(A.5)}) is given by
\begin{equation}
\mathcal{L}\left(  \mathbf{r},\mathbf{v}\right)  {  =}-\;mc^{2}%
\sqrt{1-\mathrm{v}^{2}/c^{2}}+\frac{q}{c}\;\mathbf{v}\cdot\mathbf{A}%
(r,t)-q\phi(r,t).\label{(A.6)}%
\end{equation}

The canonical momentum is
\begin{equation}
\mathbf{p}=\mathbf{\nabla}_{\mathbf{v}}\mathcal{L}=\frac{m\mathbf{v}}%
{\sqrt{1-\mathrm{v}^{2}/c^{2}}}+\frac{q}{c}\;\mathbf{A}(\mathbf{r}%
,t)=\mathbf{P}+\frac{q}{c}\;\mathbf{A}(\mathbf{r},t)\label{(A.7)}%
\end{equation}
{  where}
\begin{equation}
\mathbf{P}\;=m\mathbf{v}/\sqrt{1-\mathrm{v}^{2}/c^{2}}\label{(A.8)}%
\end{equation}
is the kinetic momentum and $q\mathbf{A}/c$   is the momentum
carried by the electromagnetic field.

The Hamiltonian is
\begin{equation}
\mathcal{H}=\mathbf{p}.\;\mathbf{v}-\mathcal{L}=\frac{mc^{2}}{\sqrt
{1-\mathrm{v}^{2}/c^{2}}}+q\phi.\label{(A.9a)}%
\end{equation}
We can rewrite (\ref{(A.9a)}) as
\begin{equation}
\mathcal{H}{  (\mathbf{r,p})=c}\left\{  \left[  \mathbf{p}-\frac{q}%
{c}\;\mathbf{A}(\mathbf{r},t)\right]  ^{2}+(mc)^{2}\right\}  ^{1/2}%
+q\phi(\mathbf{r},t).\label{(A.9b)}%
\end{equation}

For non-relativistic particles,
\begin{equation}
\mid\mathbf{P}\mid\;=\;\mid\mathbf{p}-\;\frac{q}{c}\;\mathbf{A}\mid
\;<<\;mc\label{(A.10)}%
\end{equation}
{  and}
\begin{equation}
{\cal H}\left(  \mathbf{r},\mathbf{p}\right)  =mc^{2}+\frac{(\mathbf{p}%
-\;q\mathbf{A/}c)^{2}}{2m}+q\phi.\label{(A.11)}%
\end{equation}

The second term has as part of its contribution the quantity
\[
\left(  q\mathbf{A}\right)  ^{2}/\;2mc^{2}%
\]
which is relevant only in processes where two photons are involved and
may be ignored. The remaining terms yield the Hamiltonian for the electromagnetic
interaction,
\begin{equation}
\mathcal{H}_{int}=q\phi-\frac{q}{c}\;\mathbf{v}.\mathbf{A}\label{(A.12)}%
\end{equation}
where the rest mass and kinetic energy of the particle was subtracted from Eq. \eqref{(A.11)}.

For systems involving a charge density  $\rho\left(  \mathbf{r}%
,t\right)  $  and current density \textbf{j}(\textbf{r},t), ${\cal H}_{int}$ can be generalized to
\begin{equation}
\mathcal{H}_{int}{  =}\int\left[  \rho\phi-\frac{1}{c}\;\mathbf{j}%
.\mathbf{A}\right]  {  d}^{3}{  r.}%
\end{equation}

\subsection{$\gamma$-ray emission following excitation}

After the excitation, the nuclear state $\left|  I_{f}\right\rangle $ can
decay by gamma emission to another state $\left|  I_{g}\right\rangle $. Complications arise from the fact that the nuclear levels are not only
populated by Coulomb excitation, but also by conversion and $\gamma
$-transitions cascading down from higher states (see figure \ref{cascade1}%
(a)). To compute the angular distributions one must know the parameters
$\Delta_{l}\left(  i\longrightarrow j\right)  $ and $\epsilon_{l}\left(
i\longrightarrow j\right)  $\, for $l\geq1$ [AW66],
\begin{equation}
\epsilon_{l}^{2}\left(  i\longrightarrow j\right)  =\alpha_{l}\left(
i\longrightarrow j\right)  \Delta_{l}^{2}\left(  i\longrightarrow j\right)  ,
\label{casc1}%
\end{equation}
where $\alpha_{l}$ is the total $l$-pole conversion coefficient, and
\begin{eqnarray}
\Delta_{\pi l}&=&\left[  \frac{8\pi\left(  l+1\right)  }{l\left[  \left(
2l+1\right)  !!\right]  ^{2}}\frac{1}{\hbar}\left(  \frac{\omega}{c}\right)
^{2l+1}\right]  ^{1/2}\nonumber \\ 
&\times&\left(  2I_{j}+1\right)  ^{-1/2}\left\langle
I_{j}\left\|  i^{s(l)}\mathcal{M}(\pi l)\right\|  I_{i}\right\rangle \ ,
\label{gad0_1}%
\end{eqnarray}
with $s(l)=l$ for electric $\left(  \pi=E\right)  $ and $s(l)=l+1$ for
magnetic $\left(  \pi=M\right)  $ transitions. The square of $\Delta_{\pi l}$
is the $l$-pole $\gamma$-transition rate (in $\sec^{-1}$).

The angular distributions
of gamma rays following the excitation depend on the frame of reference used.
In our notation, the z-axis corresponds to the beam axis, and the statistical
tensors are given by (we use the notation of [WB65,AW66])
\begin{align}
\alpha_{k\kappa}^{(0)}(f)  &  =\frac{\left(  2I_{f}+1\right)  ^{1/2}}{\left(
2I_{1}+1\right)  }\sum_{M_{f}=-(M_{f}^{^{\prime}}+\kappa),M_{f}^{^{\prime}}%
}\left(  -1\right)  ^{I_{f}+M_{f}} \nonumber\\
&  \times\left(
\begin{tabular}
[c]{lll}%
$I$$_{f}$ & $I$$_{f}$ & $k$\\
$-M$$_{f}$ & $M$$_{f}^{^{\prime}}$ & $\kappa$%
\end{tabular}
\right)\sum_{M_{1}}a_{I_{f}M_{f}^{^{\prime}}}^{\ast}(M_{1})\;a_{I_{f}M_{f}%
}(M_{1})\;, \label{stat}%
\end{align}
where $f$ is the state from which the gamma ray is emitted, and $1$ denotes
the initial state of the nucleus, before the excitation. To calculate the
angular distributions of the gamma rays one needs the statistical tensors for
$k=0,2,4$ and $-k\leq\kappa\leq k$ (see [WB65,AW66]).

Instead of the diagram of figure \ref{cascade1}(a), we will consider here the
much simpler situation in which the $\gamma$-ray is emitted directly from the
final excited state $f$ to a lower state $g$, which is observed experimentally
(see figure \ref{cascade1}(b)). The probability amplitude for this process is
\begin{equation}
a_{i\rightarrow f\rightarrow g}=\sum_{M_{f}}a_{i\rightarrow f}\;\left\langle
I_{g}M_{g}\mathbf{k}\sigma\left|  H_{\gamma}\right|  I_{f}M_{f}\right\rangle ,
\label{gad1}%
\end{equation}
where $\left\langle I_{g}M_{g}\mathbf{k}\sigma\left|  H_{\gamma}\right|
I_{f}M_{f}\right\rangle $ is the matrix element for the transition
$f\rightarrow g$ due to the emission of a photon with momentum \textbf{k} and
polarization $\sigma$. The operator $H_{\gamma}$ accounts for this transition.
The angular dependence of the $\gamma$-rays is given explicitly by the
spherical coordinates $\theta$ and $\phi$ of the vector \textbf{k}.

Since the angular emission probability will be normalized to unity, we can
drop constant factors and write it as (an average over initial spins is
included)
\begin{eqnarray}
W\left(  \theta\right) & =&\sum_{M_{i},\;M_{g},\sigma}\left|  a_{i\rightarrow
g}\right|  ^{2}\nonumber \\
&=&\sum_{M_{i},\;M_{g},\sigma}\left|  \sum_{M_{f}}a_{i\rightarrow
f}\;\left\langle I_{g}M_{g}\mathbf{k}\sigma\left|  H_{\gamma}\right|
I_{f}M_{f}\right\rangle \right|  ^{2}.\nonumber \\ \label{gad2}%
\end{eqnarray}

The transition operator $H_{\gamma}$\ can be written as
\begin{equation}
H_{\gamma}=\sum_{lm^{{}}}H_{\gamma}^{\left(  lm\right)  }=\sum_{lm}%
\hat{\mathcal{O}}_{lm}^{(nuc)}\otimes\hat{\mathcal{O}}_{lm}^{\left(
\gamma\right)  }, \label{gad2_b}%
\end{equation}
where the first operator in the sum acts between nuclear states, whereas the
second operator acts between photon states of well defined angular momentum,
$l,m$.

Expanding the photon state $\left|  \mathbf{k}\sigma\right\rangle $ in a
complete set $\left|  lm\right\rangle $\ of the photon angular momentum, and
using the Wigner-Eckart theorem (angular momentum notation of Ref. 
[Ed60], one gets
\begin{align}
\left\langle I_{g}M_{g}\mathbf{k}\sigma\left|  H_{\gamma}\right|  I_{f}%
M_{f}\right\rangle  &  =\sum_{l,m}\left\langle \mathbf{k}\sigma
|lm\right\rangle \left\langle I_{g}M_{g}\left|  H_{\gamma}^{\left(  lm\right)
}\right|  I_{f}M_{f}\right\rangle \nonumber\\
&  =\left(  -1\right)  ^{I_{f}-M_{f}}\sum_{l,m}\left(
\begin{array}
[c]{ccc}%
I_{f} & l & I_{g}\\
-M_{f} & m & M_{g}%
\end{array}
\right)  \nonumber \\
&\times \left\langle \mathbf{k}\sigma|lm\right\rangle \left\langle
I_{g}\left\|  H_{\gamma}^{\left(  l\right)  }\right\|  I_{f}\right\rangle .
\label{gad3}%
\end{align}

One can rewrite $\left|  \mathbf{k}\sigma\right\rangle $ in terms of $\left|
\mathbf{z}\sigma\right\rangle $, i.e., in terms of a photon propagating in the
z-direction. This is accomplished by rotating $\left|  \mathbf{k}
\sigma\right\rangle $ to the z-axis, using of the rotation matrix [BS94],
$D_{mm^{\prime}}^{l}$, i.e.,
\begin{equation}
\left\langle \mathbf{k}\sigma|lm\right\rangle =\sum_{m^{\prime}}D_{mm^{\prime
}}^{l}\left(  \mathbf{z\longrightarrow k}\right)  \left\langle \mathbf{z}
\sigma|lm^{\prime}\right\rangle . \label{gad4}%
\end{equation}

Expanding the photon field in terms of angular momentum eigenfunctions, one
can show that [FS65,EG88]
\begin{equation}
\left\langle \mathbf{z}\sigma|lm\pi\right\rangle =\left\{
\begin{array}
[c]{c}%
\sqrt{\frac{2l+1}{2}}\delta_{\sigma m}\;\;\;\;\;\mathrm{for}\;\;\;\pi=E\\
\sqrt{\frac{2l+1}{2}}\sigma\delta_{\sigma m}\;\;\;\;\;\mathrm{for}\;\;\;\pi=M.
\end{array}
\right.  \label{gad5}%
\end{equation}

One has now to express the operator $\hat{\mathcal{O}}_{l^{\prime}m^{\prime}%
}^{\left(  \gamma\right)  }$ in Eq.  \eqref{gad2_b} in terms of the electric and
magnetic multipole parts of the photon field. This problem is tedious but
straightforward [BR53]. Inserting eqs. \eqref{gad4} and \eqref{gad5} in Eq. 
\eqref{gad3}, yields (neglecting constant factors)
\begin{align}
&\left\langle I_{g}M_{g}\mathbf{k}\sigma\left|  H_{\gamma}\right|  I_{f}%
M_{f}\right\rangle   =\sum_{l,m}\left(  -1\right)  ^{I_{g}-l+M_{f}}\nonumber \\
&\times \sqrt{\left(  2l+1\right)  \left(  2I_{f}+1\right)  }\left(
\begin{array}
[c]{ccc}%
I_{g} & l & I_{f}\\
M_{f} & m & -M_{f}%
\end{array}
\right) \nonumber\\
&  \times D_{m\sigma}^{l}\left(  \mathbf{z\longrightarrow k}\right)  \left[
\Delta_{El}+\sigma\Delta_{Ml}\right]  , \label{gad6}%
\end{align}
where $\Delta_{\pi l}$ is given by Eq.  \eqref{gad0_1}.

Inserting Eq.  \eqref{gad6} into Eq.  \eqref{gad2} one gets a series of sums over
the intermediate values of the angular momenta
\begin{align}
&W\left(  \theta\right)     = \sum_{ \overset{M_{i},M_{g},\sigma,M_{f},}%
{M_{f}^{\prime}, l,m,l^{\prime},m^{\prime}} } a_{i\longrightarrow
f}a_{i\longrightarrow f^{\prime}}^{\ast}\nonumber\\
&\times \sqrt{\left(  2l+1\right)  \left(
2l^{\prime}+1\right)  \left(  2I_{f}+1\right)  }\left(
\begin{array}
[c]{ccc}%
I_{g} & l & I_{f}\\
M_{g} & m & -M_{f}%
\end{array}
\right) \nonumber\\
&  \times\left(  -1\right)  ^{M_{f}+M_{f}^{\prime}-l-l^{\prime}}\left(
\begin{array}
[c]{ccc}%
I_{g} & l^{\prime} & I_{f}\\
M_{g} & m^{\prime} & -M_{f}^{\prime}%
\end{array}
\right)  \nonumber\\
&\times D_{m\sigma}^{l}\left[  D_{m^{\prime}\sigma}^{l^{\prime}}\right]
^{\ast}\Delta_{l}\Delta^{*}_{l^{\prime}}\;, \label{gad7_b}%
\end{align}
where $\Delta_{l}=\Delta_{El}+l\Delta_{Ml}$. The product $\Delta_{l}\Delta
^{*}_{l^{\prime}}$ is always real since $(-1)^{s(l)}=\Pi$ (the parity).

Assuming that the particles are detected symmetrically around the z-axis one
can integrate over $\phi_{particle}$, what is equivalent to integrating, or
averaging, over $\phi_{\gamma}$. This yields the following integral
\begin{align}
&  \int d\phi D_{m\sigma}^{l}\left[  D_{m^{\prime}\sigma}^{l^{\prime}}\right]
^{\ast}=
\delta_{mm^{\prime}}\left(  -1\right)  ^{m-\sigma}\nonumber\\
&\times \sum_{j}\frac{2j+1}
{\sqrt{4\pi}}\left(
\begin{array}
[c]{ccc}
l & j & l^{\prime}\\
m & 0 & -m
\end{array}
\right)  \left(
\begin{array}
[c]{ccc}
l & j & l^{\prime}\\
\sigma & 0 & -\sigma
\end{array}
\right)  P_{j}\left(  \cos\theta\right)  . \label{gad7_c}
\end{align}
To simplify further Eq.  \eqref{gad7_b} we use (see Ref.  [AW75], p. 441, Eq. 
II.A.61)
\begin{align}
&  \sum_{M_{g}}\left(
\begin{array}
[c]{ccc}%
I_{g} & l & I_{f}\\
M_{g} & m & -M_{f}%
\end{array}
\right)  \left(
\begin{array}
[c]{ccc}%
I_{g} & l^{\prime} & I_{f}^{\prime}\\
M_{g} & m^{\prime} & -M_{f}^{\prime}%
\end{array}
\right) \nonumber\\
&  =\left(  -1\right)  ^{2l^{\prime}-I_{g}}\sum_{k,\kappa}\left(  -1\right)
^{k+m-M_{f}^{\prime}}\left(  2k+1\right)  \left(
\begin{array}
[c]{ccc}%
l & l^{\prime} & k\\
m & -m^{\prime} & \kappa
\end{array}
\right)  \nonumber \\
&\times \left(
\begin{array}
[c]{ccc}%
I_{f} & I_{f}^{\prime} & k\\
M_{f} & -M_{f}^{\prime} & \kappa
\end{array}
\right)  \left\{
\begin{array}
[c]{ccc}%
l & l^{\prime} & k\\
I_{f}^{\prime} & I_{f} & I_{g}%
\end{array}
\right\}  \label{gad7_d}%
\end{align}
and
\begin{align}
&\sum_{\sigma=\left(  -1,1\right)  }\left(
\begin{array}
[c]{ccc}%
l & j & l^{\prime}\\
\sigma & 0 & -\sigma
\end{array}
\right)  \Delta_{\pi l}\Delta^{*}_{\pi l^{\prime}}\nonumber\\
&=\left\{
\begin{array}
[c]{c}%
2\left(
\begin{array}
[c]{ccc}%
l & j & l^{\prime}\\
1 & 0 & -1
\end{array}
\right)  \Delta_{\pi l}\Delta^{*}_{\pi l^{\prime}}\;,\;\;\;\;\;\mathrm{for}%
\;\;\;j=\mathrm{even}\\
0\;,\;\;\;\;\;\;\;\;\;\mathrm{for}\;\;\;\;\;j=\mathrm{odd},
\end{array}
\right.
\end{align}
where use has been made of the parity selection rule
\[
\Pi_{1}\Pi_{2}=\left\{
\begin{array}
[c]{c}%
\left(  -1\right)  ^{l}\;,\;\;\;\mathrm{for}%
\;\;\;\mathrm{electric\;transitions}\\
\left(  -1\right)  ^{l+1}\;,\;\;\;\mathrm{for}%
\;\;\;\mathrm{magnetic\;transitions \ .}%
\end{array}
\right.
\]

Eq. \eqref{gad7_b} becomes
\begin{align}
W\left(  \theta\right)   &  =\sum_{ \overset{M_{i},k,\kappa, M_{f}%
,M_{f}^{\prime},}{l,l^{\prime},m,m^{\prime}} } \left(  -1\right)
^{2m^{\prime}+k+M_{f}}a_{i\longrightarrow f}a_{i\longrightarrow f^{\prime}%
}^{\ast}\nonumber \\
&\times \sqrt{\left(  2l+1\right)  \left(  2l^{\prime}+1\right)  \left(
2I_{f}+1\right)  }
\left(  2j+1\right)  \left(  2k+1\right)  \nonumber \\
&\times \left(
\begin{array}
[c]{ccc}%
I_{f} & I_{f} & k\\
M_{f} & -M_{f}^{\prime} & \kappa
\end{array}
\right)  \left(
\begin{array}
[c]{ccc}%
I_{f} & l^{\prime} & k\\
M_{f} & -m^{\prime} & \kappa
\end{array}
\right) \nonumber\\
&  \times\left(
\begin{array}
[c]{ccc}%
l & j & l^{\prime}\\
1 & 0 & -1
\end{array}
\right)  \left(
\begin{array}
[c]{ccc}%
l & j & l^{\prime}\\
m & 0 & -m^{\prime}%
\end{array}
\right)  \left\{
\begin{array}
[c]{ccc}%
l & l^{\prime} & k\\
I_{f}^{\prime} & I_{f} & I_{g}%
\end{array}
\right\}  \nonumber \\
&\times \Delta_{l}\Delta^{*}_{l^{\prime}}P_{j}\left(  \cos\theta\right)  \;.
\end{align}

Using
\begin{align}
&\sum_{m,m^{\prime}}\left(  -1\right)  ^{2m^{\prime}}\left(
\begin{array}
[c]{ccc}%
l & j & l^{\prime}\\
m & 0 & -m^{\prime}%
\end{array}
\right)  \left(
\begin{array}
[c]{ccc}%
l & j & l^{\prime}\\
1 & 0 & -1
\end{array}
\right)  
\nonumber \\
&=\left(  -1\right)  ^{l+l^{\prime}+k}\frac{1}{2k+1}\delta_{kj}%
\delta_{\kappa0},
\end{align}
one gets
\begin{align*}
W\left(  \theta\right)   &  =\sum_{ \overset{k=\mathrm{even},M_{i},}{M_{f},
l,l^{\prime}} } \left(  -1\right)  ^{l+l^{\prime}+k}\sqrt{\left(  2l+1\right)
\left(  2l^{\prime}+1\right)  \left(  2I_{f}+1\right)  }\nonumber \\
&\times \left(  2k+1\right)
\left|  a_{i\longrightarrow f}\right|  ^{2} \left(
\begin{array}
[c]{ccc}%
I_{f} & I_{f} & k\\
M_{f} & -M_{f} & 0
\end{array}
\right)\\
&  \times  \left(
\begin{array}
[c]{ccc}%
l & j & l^{\prime}\\
1 & 0 & -1
\end{array}
\right)  \left\{
\begin{array}
[c]{ccc}%
l & l^{\prime} & k\\
I_{f}^{\prime} & I_{f} & I_{g}%
\end{array}
\right\}  \Delta_{l}\Delta^{*}_{l^{\prime}}P_{j}\left(  \cos\theta\right)  ,
\end{align*}
or in a more compact form
\begin{eqnarray}
W\left(  \theta\right)  &=&\sum_{ \overset{ k=\mathrm{even},M_{i},M_{f},}{
l,l^{\prime}} } \left(  -1\right)  ^{M_{f}}\left|  a_{i\longrightarrow
f}\right|  ^{2} F_{k}\left(  l,l^{\prime},I_{g},I_{f}\right)  \nonumber \\
&\times& \left(
\begin{array}
[c]{ccc}%
I_{f} & I_{f} & k\\
M_{f} & -M_{f} & 0
\end{array}
\right)  \sqrt{2k+1}P_{k}\left(  \cos\theta\right)  \Delta_{l}\Delta^{*}
_{l^{\prime}}\;,\label{watheta}\nonumber \\
\end{eqnarray}
where
\begin{align}
F_{k}\left(  l,l^{\prime},I_{g},I_{f}\right)   &  =\left(  -1\right)
^{I_{f}-I_{g}-1}\nonumber \\
&\times \sqrt{\left(  2l+1\right)  \left(  2l^{\prime}+1\right)
\left(  2I_{f}+1\right)  \left(  2k+1\right)  }\nonumber\\
&  \times\left(
\begin{array}
[c]{ccc}%
l & l^{\prime} & k\\
1 & -1 & 0
\end{array}
\right)  \left\{
\begin{array}
[c]{ccc}%
l & l^{\prime} & k\\
I_{f} & I_{f} & I_{g}%
\end{array}
\right\}  \ .
\end{align}

The angular distribution of $\gamma$-rays described above is in the reference
frame of the excited nucleus. To obtain the distribution in the laboratory one
has to perform the transformation
\begin{equation}
\theta_{L}=\arctan\left\{  {\frac{\sin\theta}{\gamma\left[  \cos\theta+
\beta\right]  }}\right\}  \ ,
\end{equation}
and
\begin{equation}
W (\theta_{L}) = \gamma^{2} \left(  1+ \beta\cos\theta\right)  ^{2} W
(\theta)\ ,
\end{equation}
where $\gamma=(1-\beta^2)^{-1/2}$  and $\beta=v/c$, where $v$ is the nucleus-nucleus relative velocity. The photon energy in the laboratory is $E^{ph}_{L}=\gamma E^{ph}_{cm}
\left(  1 +\beta\cos\theta\right)  $.

\subsection{Second-order perturbation theory}

To second-order, the amplitude  for a two-step excitation to a state $|2>$
via intermediate states $|1>$ is given by
\begin{eqnarray}
a_{20}^{2nd}&=&\sum_{1}\
{1\over (i\hbar)^2} \ \int_{-\infty}^\infty dt \
e^{i\omega_{21} t} \ V_{21}(t)\ \nonumber \\
&\times&\int_{-\infty}^t \ dt' \ e^{i \omega_{10} t'}
V_{10}(t') \, ,
\label{h}
\end{eqnarray}
where $V_{21}(t)$ is a short
notation for the interaction potential inside brackets of the integral of
Eq. (\eqref{h}) for the transition $\vert 1>
\rightarrow \vert 2>$.

Using the integral representation of the step function
\begin{equation}
\Theta (t-t') = - \lim_{\delta \rightarrow 0^+}
\ {1\over 2 \pi i} \int_{-\infty}^\infty {e^{-iq(t-t')}\over q+i\delta} dq = \
\left\{ \begin{array}{ll}
           1,               & \mbox{if $t>t'$ \ , }\\
           0,               & \mbox{if $t<t'$},
\end{array}
\right.
\label{i}
\end{equation}
one finds 
\begin{eqnarray}
a^{2nd}_{20}&=& {1\over 2} \ \sum_{1} a_{21}^{1st} (\omega_{21})
\ a^{1st}_{10}(\omega_{10}) \nonumber \\
&+& {i\over 2 \pi} \ \sum_{1} {\cal P} \int_{-\infty}^\infty
\ {dq \over q}\ a^{1st}_{21}(\omega_{21}-q)\
a_{10}^{1st}(\omega_{10}+q) \nonumber \\
\label{j}
\end{eqnarray}
where ${\cal P}$ stands for the principal value of the integral.
For numerical evaluation it is more appropriate to rewrite the
principal value integral in Eq. (\ref{j}) as
\begin{eqnarray}
&&{\cal P} \int_{-\infty}^\infty
\ {dq \over q}\ a^{1st}_{21} (\omega_{21}-q)\
a_{10}^{1st}(\omega_{10}+q) =
\nonumber\\
& &\int_0^\infty
\ {dq \over q}\ \biggl[ a^{1st}_{21}
(\omega_{21}-q)\
a_{10}^{1st}(\omega_{10}+q)\nonumber \\
&-&a^{1st}_{21} (\omega_{21}+q)\
a_{10}^{1st}(\omega_{10}-q)
\biggr] \ . \nonumber \\
& & \  \
\label{k}
\end{eqnarray}
To calculate $a^{1st}(\omega)$ for negative values of $\omega$, we note that
the interaction potential can be written as a sum of an even and an odd
part. This implies that $a^{1st}(-\omega)=
-\Bigl[a^{1st}(\omega)\Bigr]^*$.

\subsection{Coulomb excitation of a harmonic oscillator}

Let us consider the Coulomb excitation of a linear harmonic
oscillator. For dipole excitations the interaction Hamiltonian has the form
\begin{equation}
\mathcal{H}_{int}^{E1}(t)=xf(t)+zg(t)\;,
\end{equation}
where $f(t)$ and $g(t)${  \ are time-dependent
functions and }$x\;${  and }$z${  \ are the intrinsic nuclear
coordinates perpendicular and parallel to the beam, respectively. This field
induces independent oscillations  in the perpendicular and parallel directions.

{Introducing the occupation numbers }$n_{x}
=0,\;1,\;2,\;\cdots ${  \ and }$n_{z}=0,\;1,\;2,\;\cdots ${  \ the total number of occupied states is} 
\begin{equation}
N=n_{x}+n_{z}.
\end{equation}

{Considering only the oscillations along one of the directions, e.g.,
along the x-direction, \ the full Hamiltonian is} 
\begin{equation}
\mathcal{H=H}_{0}-xf(t)\;,
\end{equation}
{where the minus sign is arbitrary and is introduced to simplify
the equations.}

{The solutions for the intrinsic Hamiltonian, }$H_{0}=m\omega
^{2}x^{2}/2,${  \ are obtained in terms of the Hermite polynomials
(here, }$n=n_{x}${  )} 
\begin{equation}
{  \psi (x)=}\left( \frac{m\omega }{\pi \hbar }\right) ^{1/4}
{1\over (2^{n}n!)^{1/2}}{ H}_{n}\left( \sqrt{\frac{m\omega }{\hbar }}%
x\right) \exp \left( -\frac{m\omega x^{2}}{2\hbar }\right).
\end{equation}

{Since this forms a complete basis, the wavefunction at time }$t$ 
{can be written as} 
\begin{equation}
\psi (x,t)=\sum_{n=0}^{\infty }\exp \left[ -i\omega \left( n+1/2\right) t%
\right] \;a_{n}(t)\;\psi _{n}(x)
\end{equation}
{with the condition }$a_{n}(0)=\delta _{n,0}${  .
Inserting this expansion into the time-dependent Schr\"{o}dinger equation} 
\begin{equation}
i\hbar \frac{\partial }{\partial t}\psi (x,t)=\left[ -\frac{\hbar ^{2}}{2m}%
\frac{\partial ^{2}}{\partial x^{2}}+\frac{1}{2}m\omega ^{2}x^{2}+xf(t)%
\right] \psi (x,t)
\end{equation}
{and using the identity (which can be derived from }$%
H_{n+1}(x)=2xH_{n}(x)-2nH_{n-1}(x)${  )} 
\begin{equation}
x\psi _{n}(x)=\left( \frac{\hbar }{2m\omega }\right) ^{1/2}\left[ \sqrt{n}%
\psi _{n-1}(x)+\sqrt{n+1}\psi _{n+1}(x)\right]
\end{equation}
{we get} 
\begin{eqnarray}
&&i\hbar \sum_{n=0}^{\infty }\exp \left[ -i\omega \left( n+1/2\right) t%
\right] \;\frac{da_{n}}{dt}(t)\;\psi _{n}(x)=  \notag \\
&&-\left( \frac{\hbar }{2m\omega }\right) ^{1/2}f(t)\sum_{n=0}^{\infty }\exp %
\left[ -i\omega \left( n+1/2\right) t\right] \;a_{n}(t)\;  \notag \\
&&\times \left[ \sqrt{n}\psi _{n-1}(x)+\sqrt{n+1}\psi _{n+1}(x)\right] .
\end{eqnarray}

{Using the orthogonality conditions of }$\psi _{n}${  \ one
obtains} 
\begin{equation}
\frac{da_{n}}{dt}(t)=if_{0}(t)\left[ \mathrm{e}^{-i\omega t}\sqrt{n+1}%
a_{n+1}(t)+\mathrm{e}^{-i\omega t}\sqrt{n}a_{n-1}(t)\right] \;,
\end{equation}
{where } 
\begin{equation}
f_{0}(t)=\left( 2m\hbar \omega \right) ^{-1/2}f(t).
\end{equation}

{The above equation allows to obtain }$a_{n}(t)${  \ by
iteration, starting with }$n=0$. The solution is} 
\begin{equation}
a_{n}(t)=i^{n}\frac{\chi ^{n}(t)}{\sqrt{n!}}\exp \left[ -\int_{0}^{t}dt^{%
\prime }f_{0}(t^{\prime })\mathrm{e}^{-i\omega t^{\prime }}\chi (t^{\prime })%
\right]
\end{equation}
{where} 
\begin{equation}
{  \chi (t)=}\int_{0}^{t}{  dt}^{\prime }{  f}_{0}{  (t}%
^{\prime }{  )e}^{-i\omega t^{\prime }}{  \chi (t}^{\prime }%
{  )\;.}
\end{equation}

{But,} 
\begin{equation}
\int_{0}^{t}dt^{\prime }f_{0}(t^{\prime })\mathrm{e}^{-i\omega t^{\prime
}}\chi (t^{\prime })=\int_{0}^{t}dt^{\prime }\frac{d\chi ^{\ast }}{%
dt^{\prime }}(t^{\prime })\;\chi (t^{\prime })=\frac{1}{2}\left| \chi
(t)\right| ^{2}
\end{equation}
{and the probability to excite the }$n${th state is} 
\begin{equation}
{  P}_{n}{  (t)=}\frac{\left| \chi (t)\right| ^{2n}}{n!}\exp \left[
-\left| \chi (t)\right| ^{2}\right] .
\end{equation}

{For a nucleus, }$n${  can also be considered as the number of
phonons for the vibration along the x-direction. The probability to have }$%
n_{z}${   vibrations along the z-direction and }$n_{x}${  
vibrations along the x-direction is} 
\begin{equation}
{  P}_{n_{z},n_{x}}{  (t)=}\frac{\left| \chi _{z}(t)\right|
^{2n_{z}}\left| \chi _{x}(t)\right| ^{2n_{x}}}{n_{z}!n_{x}!}\exp \left[
-\left| \chi _{z}(t)\right| ^{2}-\left| \chi _{x}(t)\right| ^{2}\right] .
\end{equation}

\begin{figure}
[ptb]
\begin{center}
\includegraphics[
width=3.5in
]%
{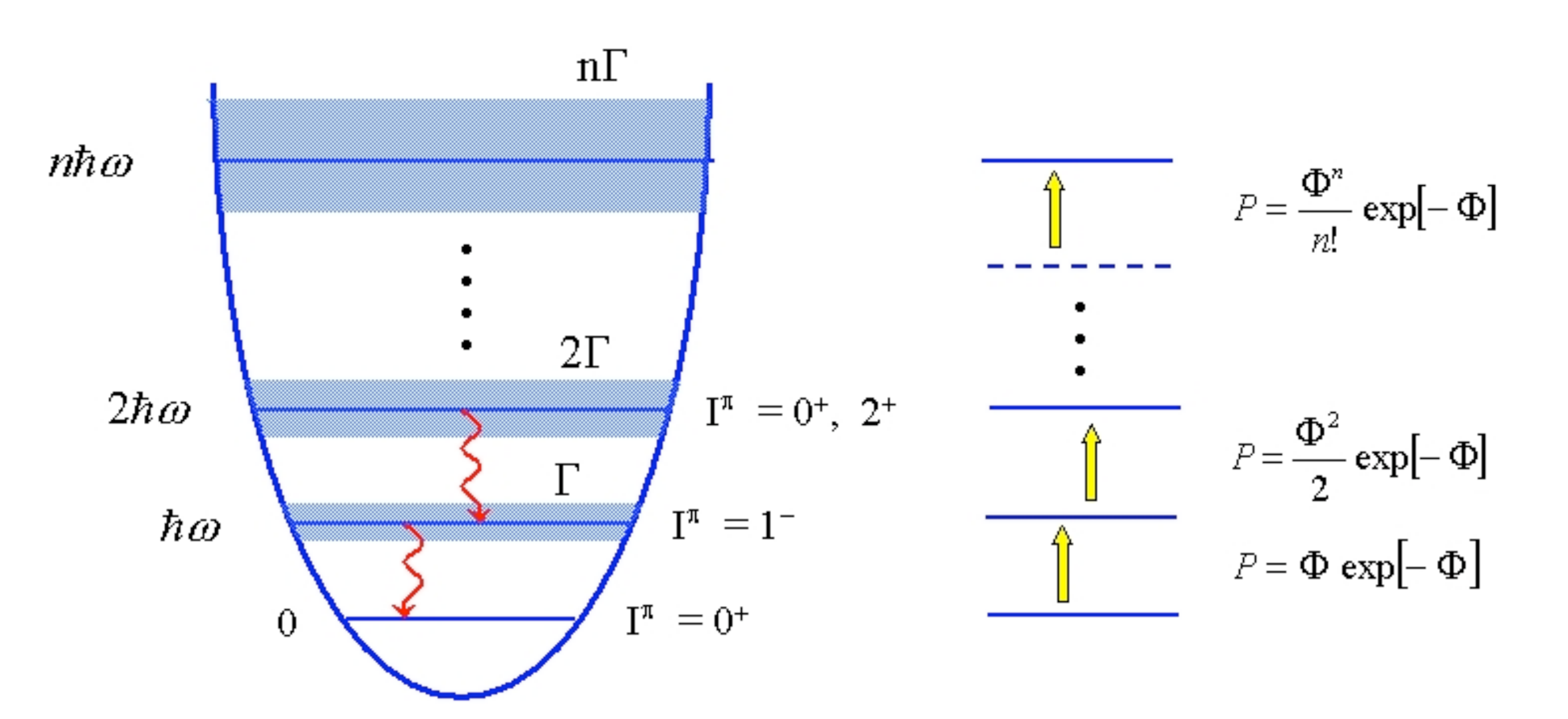}%
\caption{\sl  The harmonic oscillator model for
the excitation of collective states in nuclei.}%
\label{harmosc}%
\end{center}
\end{figure}

{The probability that N-phonons will be excited is} 
\begin{eqnarray}
P_{N}
&=&\sum_{n_{z},\;n_{x}(N=n_{z}+n_{x})}P_{n_{z},n_{x}}=%
\sum_{n_{z}=0}^{N}P_{n_{z},N-n_{z}}  \notag \\
&=&\sum_{n_{z}=0}^{N}\frac{\left| \chi _{z}(t)\right| ^{2n_{z}}\left| \chi
_{1}(t)\right| ^{2\left( N-n_{z}\right) }}{n_{z}!\left( N-n_{z}\right) !}\nonumber
\\ &\times& \exp \left[ -\left| \chi _{z}(t)\right| ^{2}-\left| \chi _{x}(t)\right| ^{2}%
\right]  \notag \\
&=&\frac{1}{N!}\exp \left[ -\left| \chi _{z}(t)\right| ^{2}-\left| \chi
_{x}(t)\right| ^{2}\right] \nonumber \\
&\times&\sum_{n_{z}=0}^{N}\frac{N!\left| \chi
_{z}(t)\right| ^{2n_{z}}\left| \chi _{x}(t)\right| ^{2\left( N-n_{z}\right) }%
}{n_{z}!\left( N-n_{z}\right) !}\;.
\end{eqnarray}

{The last sum yields }$\left| \phi (t)\right| ^{2N}${  , where } 
\begin{equation}
\left| \phi (t)\right| ^{2}\equiv \left| \chi _{z}(t)\right| ^{2}+\left|
\chi _{x}(t)\right| ^{2}.
\end{equation}
{Thus, one gets the Poisson formula} 
\begin{equation}
P_{N}=\frac{1}{N!}\;\left| \phi (t)\right| ^{2N}\;\exp \left[ -\left| \phi
(t)\right| ^{2}\right] .
\end{equation}
{for dipole excitations of a harmonic oscillator. In Refs. 
[AW66,AW75] it was shown that the same result can be obtained for
quadrupole excitations. Thus, the above result is very general (see Figure 
\ref{harmosc}).}

{The Poisson equation is very useful. It shows that all one needs
to obtain the probability to excite the nth state is to calculate the
probability to excite the }$N=1${  \ state with first-order
perturbation, }$P_{0\rightarrow 1}^{first}=\left| \phi (t)\right| ^{2}$%
{  \ , and use it in the Poisson formula. Although this only holds for
a harmonic oscillator system, it has been largely used in many theoretical
considerations of more complicated systems. The factor }$\exp \left[ -\left|
\phi (t)\right| ^{2}\right] ${  \ accounts for the loss of occupation
probability due to the excitation to all states from a given one. It also is
responsible for conservation of unitarity, so that }$\sum_{N=0}^{\infty
}P_{N}=1${  . }

\subsection{Schr\"{o}dinger equation in a space-time lattice}

{The operations }$\left[ {}\right] u_{\alpha }\;${ in the
r.h.s. of Eq. \eqref{(3.12)} is easy. The operation }$\left[ {}\right]
^{-1}u_{\alpha }^{\prime }\;${ is more complicated. The problem is to
find the vector }$u\;${ in the equation} 
\begin{equation}
\mathbf{v}=\mathcal{A}^{-1}\;\mathbf{v,\;\;\;\;\;\;\;}\mathrm{where}\mathbf{%
\;\;\;\;\;v}=\mathcal{A}\;\mathbf{u}  \label{(D.1)}
\end{equation}
$\mathbf{u}\;${ is a vector composed with the }$u^{(j)}=u(r_{j},t)\;$%
{ components of the wave-function }$u(r,t)${ . In Eq. \ref
{(3.12)} }$\mathcal{A}\;${ is a tri-diagonal operator (matrix). In
matrix notation} 
\begin{equation}
\left( 
\begin{array}{c}
v_{1} \\ 
v_{2} \\ 
. \\ 
. \\ 
. \\ 
, \\ 
. \\ 
v_{N}
\end{array}
\right) { =}\left( 
\begin{array}{ccccccc}
A_{1} & 0 & 0 & . & . & 0 & . \\ 
A_{2}^{-} & A_{2} & A_{2}^{+} & 0 & . & . & . \\ 
0 & A_{3}^{-} & A_{3} & A_{3}^{+} & 0 & . & . \\ 
. &  &  &  &  &  &  \\ 
. &  &  &  &  &  &  \\ 
0 & . & . & . & . & . & A_{\substack{ N  \\ }}
\end{array}
\right) \left( 
\begin{array}{c}
u_{1} \\ 
u_{2} \\ 
u_{3} \\ 
. \\ 
. \\ 
. \\ 
u_{N}
\end{array}
\right) .  \label{(D.3)}
\end{equation}
{ This involves the following relations} 
\begin{equation}
{ A}_{i}^{-}{ u}_{i-1}{ +A}_{i}{ u}_{i}{ +A}%
_{i}^{+}{ u}_{i+1}{ =v}_{i}.  \label{(D.4)}
\end{equation}

{ Assuming a solution of the form} 
\begin{equation}
u_{i+1}=\alpha _{i}u_{i}+\beta _{i}  \label{(D.5)}
\end{equation}
{ and inserting in \eqref{(D.4)} we find the recursion relations for }$%
\alpha _{i}\;${ and }$\beta _{i}:$%
\begin{equation}
{ \alpha }_{i-1}{ =\gamma }_{i}{ \;A}_{i}^{-}\;,\;\;\;\;%
\beta _{i-1}=\gamma _{i}\left( A_{i}^{+}\beta _{i}-v_{i}\right)
\label{(D.6)}
\end{equation}
{ where } 
\begin{equation}
\gamma _{i}=-\frac{1}{A_{i}+\alpha _{i}A_{i}^{+}}.
\end{equation}
{ At the end of the lattice we assume }$u_{N}=v_{N}\;${ . This
implies that }$\alpha _{N-1}=0\;${ and }$\beta _{N-1}=v_{N}\;${ .%
}

{ For the problem defined by Eq. \eqref{(3.12)} we have} 
\begin{equation}
A_{k}=\frac{1}{i\tau }+2+\frac{\Delta t}{2\hbar \tau }v_{\alpha }^{\left(
k\right) },\;\;\;\mathrm{and}\;\;\;\;\;\;\;A_{k}^{-}=A_{k}^{+}=-1.
\label{(D.7)}
\end{equation}

{ We can now determine }$\alpha _{i}${ \ and }$\beta _{i}$%
{ \ by running Eqs. \eqref{(D.6)} backwards from }$i=N-2${ \ down
to }$i=1${ . Then we use Eq. \eqref{(D.5)} running forward from }$i=2$%
{ \ to }$N${ , assuming that }$u_{1}\;${ at the other
extreme of the lattice is given by }$u_{1}=v_{1}/A_{1}\;${ .}

{ Another way to solve the problem \eqref{(D.3)} for }$u\;${ is by
a LU-decomposition, followed by a forward and backward substitution. This
method does not need to involve the Dirichlet condition.}

{ Let us assume that the }$A${ \ matrix in \eqref{(D.3)} can be
written as a product of }$B\;${ and }$U\;${ matrix, where} 
\begin{equation}
{ A=LU=}\left( 
\begin{array}{cccccc}
b_{1} & c_{1} & 0 & 0 & . & . \\ 
a_{2} & b_{2} & c_{2} & 0 & . & . \\ 
0 & a_{3} & b_{3} & c_{3} & . & . \\ 
. & .. & . & .. & . & .
\end{array}
\right)  \label{(D.8)}
\end{equation}

\begin{equation}
L=\left( 
\begin{array}{cccccc}
1 & 0 & 0 & 0 & . & . \\ 
\alpha _{2} & 1 & 0 & 0 & . & . \\ 
0 & \alpha _{3} & 1 & 0 & . & . \\ 
0 & 0 & \alpha _{4} & 1 & . & . \\ 
. & . & . & . & . & . \\ 
. & . & . & . & . & .
\end{array}
\right) \;\;\;U=\left( 
\begin{array}{cccccc}
\beta _{1} & \alpha _{1} & 0 & 0 & . & . \\ 
0 & \beta _{2} & \gamma _{2} & 0 & . & . \\ 
0 & 0 & \beta _{3} & \gamma _{3} & . & . \\ 
. & . & . & . &  &  \\ 
. & . & . & . &  & 
\end{array}
\right) .  \label{(D.9)}
\end{equation}
{ Then, }$AU=V\;\rightarrow L(Uu)=v${ , or }$Ly=V${ . \
The elements of \ }$y${ \ \ are} 
\begin{equation}
\left( 
\begin{array}{c}
y_{1} \\ 
y_{2} \\ 
. \\ 
. \\ 
. \\ 
y_{N}
\end{array}
\right) =\left( 
\begin{array}{ccccccc}
\beta _{1} & \gamma _{1} & 0 & 0 & . & . & . \\ 
0 & \beta _{2} & \gamma _{2} & 0 & . & . & . \\ 
0 & 0 & \beta _{2} & \gamma _{3} & . & . & . \\ 
0 & 0 & . & . & . & . & .
\end{array}
\right) \;\left( 
\begin{array}{c}
u_{1} \\ 
u_{2} \\ 
u_{3} \\ 
. \\ 
. \\ 
. \\ 
u_{N}
\end{array}
\right) .  \label{(D.10)}
\end{equation}
{ This can be solved by backwards substitution} 
\begin{equation}
u_{N}=\frac{y_{N}}{\beta _{N}},\;\;\;\;\;\;\;\;\;\;\beta _{i}u_{i}+\gamma
_{i}u_{i+1}=y_{i}\;,  \label{(D.11)}
\end{equation}
{ or} 
\begin{equation}
{ u}_{i}=\left( y_{i}-\gamma _{1}u_{i}+1\right) { \;/\;\beta }%
_{i}.  \label{(D.12)}
\end{equation}

{ The other matrix equation} 
\begin{equation}
\left( 
\begin{array}{c}
\begin{array}{cccccc}
1 & 0 & 0 & 0 & . & . \\ 
\alpha _{1} & 1 & 0 & 0 & . & . \\ 
0 & \alpha _{3}^{{}} & 1 & 0 & . & . \\ 
0 & 0 & \alpha _{4} & 1 & 0 & . \\ 
. & . & . & . &  & .
\end{array}
\end{array}
\right) \;\;\left( 
\begin{array}{c}
y_{1} \\ 
y_{2} \\ 
. \\ 
. \\ 
. \\ 
y_{N}
\end{array}
\right) = 
\begin{array}{c}
\left( 
\begin{array}{c}
v_{1} \\ 
v_{2} \\ 
. \\ 
. \\ 
. \\ 
v_{N}
\end{array}
\right)
\end{array}
\label{(D.13)}
\end{equation}
{ can be solved by forward substitution} 
\begin{equation}
{ y}_{1}{ =v}_{1},\;\;\;\;\;\;\;\;\;\;\;\;\alpha
_{i}y_{i-1}+y_{i}=v_{i}\;,  \label{(D.14)}
\end{equation}
{ or} 
\begin{equation}
{ y}_{i}{ =v}_{i}{ -\alpha }_{i}{ y}_{i-1.} 
\tag{(D.15)}
\end{equation}

{ Now, we need to find }$\alpha _{i}\;${ and }$\beta _{i}\;$%
{ as a function of the original elements of }$\mathcal{A}$%
\begin{equation*}
\left( 
\begin{array}{ccccccc}
1 & 0 & 0 & 0 & . & . & . \\ 
\alpha _{2} & 1 & 0 & 0 & . & . &  \\ 
0 & \alpha _{3} & 1 & 0 & . & . &  \\ 
. & . &  & . & . &  &  \\ 
. & . &  & . &  & . &  \\ 
. & . &  & . &  &  & .
\end{array}
\right) \left( 
\begin{array}{ccccccc}
\beta _{1} & \gamma _{1} & 0 & 0 & 0 & . & . \\ 
0 & \beta _{2} & \gamma _{2} & 0 & 0 & . & . \\ 
0 & 0 & \beta _{3} & \gamma _{3} &  & . & . \\ 
. & . &  &  & . &  &  \\ 
. & . &  &  &  & . &  \\ 
. & . &  &  &  &  & \beta _{N}
\end{array}
\right)
\end{equation*}

\begin{equation*}
=\left( 
\begin{array}{ccccccc}
b_{1} & c_{1} & 0 & 0 & 0 & . & . \\ 
b_{2} & b_{2} & c_{2} & 0 & 0 & . & . \\ 
0 & a_{3}^{{}} & b_{3} & c_{3} & 0 & . & . \\ 
& . & . &  & . & . &  \\ 
& . & . &  & . &  & .
\end{array}
\right)
\end{equation*}
{ which implies} 
\begin{equation*}
b_{1}=\beta _{1},\;\;\;\;b_{2}=\alpha _{2}\gamma _{1}+\beta _{2},\;\;\cdots
,\;\;b_{i}=\gamma _{i-1}\alpha _{i}+\beta _{i}.
\end{equation*}
{ or} 
\begin{equation}
{ \beta }_{i}{ =b}_{i}{ -\gamma }_{i-1}{ \alpha }%
_{i}\;,\;\;\;\;\;\;\;\;\;\;c_{i}=\gamma _{i}.  \label{(D.15)}
\end{equation}

{ Also,} 
\begin{equation*}
a_{2}=\beta _{1}\alpha _{2},\;\;\;\;\;a_{3}=\beta _{2}\alpha _{3},\;\;\cdots
,\;\;a_{i}=\beta _{i-1}\alpha _{i},
\end{equation*}
{ or} 
\begin{equation}
\alpha _{i}=\frac{a_{i}}{\beta _{i-1}}.  \label{(D.17)}
\end{equation}
{ Thus, knowing }$a_{i,}\;b_{i}${ \ and }$c_{i}${ , one
can go upwards with this set of equations to solve the problem.}

{ In our particular case,} 
\begin{equation}
a_{i}=c_{i}=-1.  \label{(D.17b)}
\end{equation}
{ Thus, the above equations simplify to} 
\begin{eqnarray}
{ \beta }_{1} &=&{ b}_{1},\;\;\;\;\;\;\;\;\;\;\;\;{ \beta }%
_{i}{ =b}_{i}{ -}\frac{1}{\beta _{i-1}}{ %
\;,\;\;\;\;\;\;\;\;i=2,...,N}  \notag \\
{ y}_{1} &=&{ v}_{1},\;\;\;\;\;\;\;\;\;\;\;{ y}_{i}{ %
=v}_{i}{ +\;}\frac{y_{i-1}}{\beta _{i-1}}{ \;\;,\;\;\;\;\;\;\;%
\;i=2,...,N}  \notag \\
{ u}_{N} &=&\frac{y_{N}}{\beta _{N}},\;\;\;\;\;\;\;\;\;\;\;\;\;{ %
u}_{i}{ =\;}\frac{\left( y_{i}+u_{i+1}\right) }{\beta _{i}}{ %
\;\;,\;\;i=N-1,...,1.}  \nonumber \\
\label{(D.18)}
\end{eqnarray}

\section{Acknowledgments}
This work was partially supported by the U.S. DOE
grants DE-FG02-08ER41533 and DE-FC02-07ER41457
(UNEDF, SciDAC-2), the 
the Research Corporation under Award No. 2009-7123, and the JUSTIPEN/DOE grant
DEFG02- 06ER41407.

\section{References}

\begin{description}

\item[AAB90] C.E. Aguiar, A.N.F. Aleixo and C.A. Bertulani, Phys. Rev. 
C 42 (1990) 2180.

\item[AB89]  A.N.F. Aleixo and C.A. Bertulani, Nucl. Phys. A 505
(1989) 448.

\item[ABE98] T. Aumann, P.F. Bortignon, H. Emling, Annu. Rev. Nucl.
Part. Sci. 48, 351 (1998).

\item[ABS93] T. Aumann, CA. Bertulani and K. S\"ummerer., Phys. Rev. C 47 (1993) 1728.

\item[AC79]L.G. Arnold and B.C. Clark, Phys. Lett. B 84, 46 (1979).

\item[AGB82]Y. Alhassid, M. Gai and G.F. Bertsch, Phys. Rev. Lett.
49 (1982) 1482.

\item[Arn73]R.A. Arndt, D.L. Long, L.D. Roper, Nucl. Phys. A 209
(1973) 429.

\item[Aum99] T. Aumann, et al.,
  Phys. Rev. C 59 (1999) 1252. 
   
\item[Aum06]
T.~Aumann, Europ.~Phys.~Journ.~A 26 (2006) 441.

\item[Ax62] P. Axel, {Phys. Rev.} {126} (1962) 671.

\item[AW79] A. Winther and K. Alder, Nucl. Phys. A 319 (1979) 518.

\item[Ald56] K. Alder, A. Bohr, T.Huus, B.Mottelson and A.Winther, Rev. 
Mod. Phys. 28 (1956) 432.

\item[AS64] M. Abramowitz  and I.A.  Stegun, ``Handbook of
Mathematical Functions", (Washington, DC: National Bureau of Standards ), 1964

\item[AW66] K. Alder  and A. Winther, ``Coulomb
Excitation", (New York: Academic Press), 1966.

\item[AW75] K. Alder  and A. Winther, ``Electromagnetic
Excitation" (Amsterdam: North-Holland), 1975.

\item[Ba62]W.C. Barber, Ann. Rev. Nucl. Sci. \textbf{12}, 1 (1962).

\item[Bar88] J. Barrette et al., Phys. Lett.
B 209 (1988) 182.

\item[BB85] C.A. Bertulani and G. Baur, Nucl. Phys. A 442 (1985) 739. 

\item[BB86a] G. F. Bertsch and R. A. Broglia, {Physics Today}, August
1986, p. 44.

\item[BB86] G. Baur and C.A. Bertulani, Phys. Lett. B 174 (1986) 23.

\item[BB88] C.A. Bertulani and G. Baur, Phys. Rep.
163  (1988) 299.

\item[BB90]F.E. Bertrand   and J.R. Beene, Nucl.
Phys. A 520  627c (1990).

\item[BB93]  G.F. Bertsch and C.A. Bertulani, Nucl. Phys. A 556
(1993) 136.

\item[BB94]  C.A. Bertulani and G.F. Bertsch, Phys. Rev. C 49
(1994) 2839.

\item[BBR86] G. Baur, C.A. Bertulani and H. Rebel, Nucl. Phys. A458 (1986) 188. 

\item[BC92]C.A. Bertulani and L.F. Canto, Nucl. Phys. A 539
(1992) 163.

\item[BP99] C.A. Bertulani and V. Ponomarev, Phys. Rep. 321, 139 (1999).

\item[BD04] C.A. Bertulani and P. Danielewicz, ``Introduction to Nuclear Reactions", IOP, London, 2004.

\item[Bee90] J. Beene et al., Phys. Rev. C 41,
920 (1990).

\item[Ber04]COULINT: a numerical code for Coulomb excitation at all bombarding energies, C.A. Bertulani, unpublished. 

\item[Ber05]C. A. Bertulani, Phys. Rev. Lett. 94, 072701 (2005).

\item[Ber07] C.A. Bertulani, ``Nuclear Physics in a Nutshell", Princeton Press, Princeton, 2007.

\item[BHK09]C.A. Bertulani, Jun-Ting Huang and Plamen Krastev, Mod. Phys. Lett. A 24, 1109 (2009).

\item[BM75] A. Bohr and B. Mottelson, ``Nuclear Structure", vols. I and II, Benjamin, New York, 1975.

\item[BM93]C.A. Bertulani and K.W. McVoy, Phys.
Rev. C 48 (1993) 2534.

\item[BP99] C.A. Bertulani  and V.Yu. Ponomarev, Phys.
Rep. 321  139 (1999).

\item[BR53] L.C. Biederharn and
M.E. Rose, Rev. Mod. Phys. 25, 729 (1953).

\item[BS92]C.A. Bertulani and A. Sustich, Phys. Rev. C46 (1992) 2340.

\item[BS94]
D. Brink and G.R. Satchler, ``Angular Momentum", Oxford University
Press, Oxford, 1953.

\item[Ca96] C.A. Bertulani, L.F. Canto, M.S. Hussein and A.F.R. de Toledo Piza, Phys. Rev. C 53 (1996) 334.

\item[Cob97] A. Cobis, D. Fedorov, A. Jensen, Phys. Rev. Lett. 79, (1997)
  2411 . 

\item[Dan98] B.V. Danilin, I.J. Thompson, J.S. Vaagen, M.V. Zhukov,
  Nucl. Phys. A 632 (1998) 383. 
  
\item[DB88] S. S. Dietrich and B. L. Berman, At. Data and Nucl. Data Tables,
{38} (1988) 199.

\item[EB92]H. Esbensen and G. F. Bertsch, Nucl. Phys. A 542, 310 (1992).

\item[EBB95]  H. Esbensen, G.F. Bertsch and C.A. Bertulani, Nucl. Phys. 
A 581 (1995) 107.

\item[EB05] H. Esbensen and C.A. Bertulani, Phys. Rev. C { 65}, 024605 (2002).

\item[Ed60]  A.R. Edmonds, ``Angular Momentum in Quantum Mechanics'',
Princeton University Press, Princeton, New Jersey, 1960.

\item[Efr96] V. Efros, W. Balogh, H. Herndl, R. Honger,
  H. Oberhummer, Z. Phys. A 355 (1996) 101.

\item[EG88] J.M. Eisenberg and W. Greiner, ``Excitation
Mechanisms of the Nucleus", (North-Holland, Amsterdam, 1988).

\item[Fe24] E. Fermi, Z. Physik, 29,  315  (1924).

\item[Fe25] E. Fermi, Nuovo Cimento 2 (1925) 143.

\item[Few84]M.P. Fewell, Nucl. Phys. A 425 (1984) 373.

\item[FS65] H. Frauenfelder and
R.M. Steffen, ``Alpha-, beta- and gamma-ray spectroscopy", Vol.
2, ed. K. Siegbahn.

\item[Gla01] T. Glasmacher, Nucl. Phys. A 693,
90 (2001).

\item[Go02] H. Goldstein,  C.P.  Poole,  C.P. Poole Jr.  and  J.L.  Safko,
``Classical Mechanics" (Prentice Hall, 3rd edition), 2002.

\item[GR80]I.S. Gradshteyn and I.M. Ryzhik, ``Table of
Integrals, Series, and Products" (New York: Academic Press ), 1980.

\item[GT48]M. Goldhaber and E. Teller, Phys. Rev. 74, 1046 (1948).

\item[GZK66]K . Greisen, Phys. Rev. Lett. 16 (1966) 748;
G .T. Zatsepin and V. A. Kuzmin, Zh. Eksp. Teor. Fiz. 4 (1966) 114, JETP Lett. 4 (1966) 78.

\item[Haik05]H. Simon, \textit{Technical Proposal for the Design,
Construction, Commissioning, and Operation of the ELISe setup}, GSI Internal
Report, Dec. 2005.

\item[Iek93]  K. Ieki et al., Phys. Rev. Lett. 70 (1993) 730.

\item[Ja75] J.D.  Jackson, ``Classical Electrodynamics"
 (New York: Wiley), 1975.

\item[Kam87]
M. Kamimura {et al}., Prog. Theor. Phys. Suppl. { 89}, 1 (1986);
N. Austern { et al}., Phys. Rep. { 154}, 125 (1987).

\item[Le74] A. Lepr\^etre et al.,  Nucl. Phys. A 219
(1974) 39.

\item[LL89] L.D. Landau and E.M. Lifshitz, The classical theory of fields, 4th ed. (Pergamon, Oxford, 1979).

 \item[Ma75] N. Marty, M. Morlet, A. Willis, V. Comparat and R. Frascaria,
 {Nucl. Phys.} {A 238} (1975) 93.

\item[Mer74]S.P. Merkuriev, Sov. J. Nucl. Phys. 19 (1974) 222.

\item[Mo88] S. Mordechai {et al.}, {Phys. Rev. Lett.} {61} (1988)
531.

\item[Mot94]
T.~Motobayashi et~al., Phys. Rev. Lett. 73 (1994) 2680.

\item[Mye77]W.D. Myers, W.G. Swiatecki, T. Kodama, L.J. El-Jaick and E.R.
Hilf, Phys. Rev. C 15, 2032 (1977).

\item[Nak06]T. Nakamura et al., Phys. Rev. Lett. 96, 252502 (2006).

\item[Oga03]
K. Ogata, M. Yahiro, Y. Iseri, T. Matsumoto and M. Kamimura,
Phys. Rev. C { 68}, 064609 (2003).

\item[Oga06]
K. Ogata, S. Hashimoto, Y. Iseri, M. Kamimura and M. Yahiro,
Phys. Rev. C { 73}, 024605 (2006).

\item[OB09]K. Ogata and C.A. Bertulani, Prog. Theor. Phys. (Letter) 121 (2009) 1399.

\item[OB09b]K. Ogata and C.A. Bertulani, in preparation.

\item[Pry48]M. H. L. Pryce, Proc. Roy. Soc. A195, 62 (1948).

\item[Pus96]A Pushkin, B Jonson and M V Zhukov, J. Phys. G 22
(1996) L95.

\item[PW71] R. Pitthan and Th. Walcher, {Phys. Lett.} {36 B} (1971) 563.

\item[Rit93] J. Ritman et al., Phys. Rev. Lett. 70 (1993) 533.

\item[Sa51]R.G. Sachs and N. Austern, Phys. Rev. 81, 705 (1951).

\item[Sch54]L.I. Schiff, Phys. Rev.  96, 765 (1954).

\item[Sch93] R. Schmidt et al., Phys. Rev. Lett. 70 (1993) 1767.

\item[She54] R. Sherr,  C.W. Li   and R.F. Christy, Phys.
Rev. 96  (1954) 1258.

\item[Shi95]S. Shimoura et al., Phys. Lett. B 348 (1995) 29.

\item[Sie37]A.J.F. Siegert, Phys. Rev. 52, 787 (1937).

\item[SIS90]Y. Suzuki, K. Ikeda, and H. Sato, Prog. Theor. Phys.
83, 180 (1990).

\item[SJ50]H. Steinwedel and H. Jensen, Z. Naturforschung 5A, 413 (1950).

\item[Sud01]T. Suda, K. Maruayama, ``Proposal for the RIKEN beam
factory", RIKEN, 2001; M. Wakasugia, T. Suda, Y. Yano, Nucl. Inst. Meth. Phys.
A 532, 216 (2004).

\item[Tem55] G.M. Temmer and N.P. Heydenburg, Phys.
Rev. 98,  1198 (1955).

\item[Thoe99]M. Thoennessen et al., Phys. Rev. C 59, 111 (1999).

\item[Ti04]D. R. Tilley et al., Nucl. Phys. A 745, 155 (2004).

\item[Ve70] A. Veyssi\`{e}re, H.  Beil, R. Berg\`{e}re, P. Carlos  
and A. Lepr\^{e}tre, Nucl. Phys. A 159 561 (1970).

\item[VGB90] A.C. Vasconcellos-Gomes and C.A. Bertulani, Nucl. Phys. A 517, 639 (1990).

\item[WB65]  A. Winther and J. De Boer, ``A Computer Program for Multiple
Coulomb Excitation'', Caltech, Technical report, November 18, 1965. Reprinted
in [AW75].

\item[WW34] C.F. Weizsacker, Z. Physik, 612 (1934);
E.J. Williams, Phys. Rev. 45,  729 (1934).

\item[Zhu93]M.V. Zhukov, B.V. Danilin, D.V. Fedorov, J.M. Bang, I.J.
Thompson and J.S. Vaagen, Phys. Rep. {231}, 151 (1993).

\end{description}

\end{document}